\documentclass[aps,preprint,prd,nofootinbib,superscriptaddress]{revtex4}

\usepackage{graphicx}
\usepackage{amsmath,latexsym}
\usepackage{epsfig}
\usepackage{axodraw}

\newcommand{\nc}{\newcommand}
\nc{\renc}{\renewcommand}

\nc{\half}{{\textstyle{1\over2}}}
\nc{\etal}{\mbox{\it et al. }}
\nc{\ie}{{\it i.e.}}
\nc{\eg}{{\it e.g.}}

\renc{\thefootnote}{\arabic{footnote}}
\nc{\capt}[1]{{\bf Figure.} {\small\sl #1}}

\nc{\eqs}[2]{\mbox{Eqs.~(\ref{#1},\,\ref{#2})}}
\nc{\eq}[1]{\mbox{Eq.~(\ref{#1})}}
 \nc{\figs}[2]{\mbox{Figs.~(\ref{#1},\,\ref{#2})}}
\nc{\fig}[1]{\mbox{Fig~.(\ref{#1})}}

\nc{\mtag}[1]{\label{#1} \mbox{\marginpar{{\footnotesize #1}}}}
\renc{\baselinestretch}{1.5}
\newlength{\overeqskip}
\newlength{\undereqskip}
\nc{\be}[1]{\begin{equation} \mbox{$\label{#1}$}}
\nc{\bea}[1]{\begin{eqnarray} \mbox{$\label{#1}$}}
\nc{\Section}[2]{\section{#2}\label{#1}}
\nc{\Bibitem}[1]{\bibitem{#1}}
\nc{\Label}[1]{\label{#1}}

\nc{\eea}{\vspace{\undereqskip}\end{eqnarray}}
\nc{\ee}{\vspace{\undereqskip}\end{equation}}
\nc{\bdm}{\begin{displaymath}}
\nc{\edm}{\end{displaymath}}
\nc{\dpsty}{\displaystyle}
\nc{\bc}{\begin{center}}
\nc{\ec}{\end{center}}
\nc{\ba}{\begin{array}}
\nc{\ea}{\end{array}}
\nc{\bab}{\begin{abstract}}
\nc{\eab}{\end{abstract}}
\nc{\btab}{\begin{tabular}}
\nc{\etab}{\end{tabular}}
\nc{\bit}{\begin{itemize}}
\nc{\eit}{\end{itemize}}
\nc{\ben}{\begin{enumerate}}
\nc{\een}{\end{enumerate}}
\nc{\bfig}{\begin{figure}}
\nc{\efig}{\end{figure}}
\nc{\ls}{\lesssim}
\def\gs{\mathrel{\lower4pt\vbox{\lineskip=0pt\baselineskip=0pt
           \hbox{$>$}\hbox{$\sim$}}}}

\nc{\arreq}{&\!=\!&}
\nc{\arrmi}{&\!-\!&}
\nc{\arrpl}{&\!+\!&}
\nc{\arrap}{&\!\!\!\approx\!\!\!&}
\nc{\non}{\nonumber\\*}

\nc{\DOT}{\hspace{-0.08in}{\bf .}\hspace{0.1in}}
\nc{\Laada}{\hbox {$\sqcap$ \kern -1em $\sqcup$}}
\nc\loota{{\scriptstyle\sqcap\kern-0.55em\hbox{$\scriptstyle\sqcup$}}}
\nc\Loota{{\sqcap\kern-0.65em\hbox{$\sqcup$}}}
\nc\laada{\Loota}
\nc{\qed}{\hskip 3em \hbox{\BOX} \vskip 2ex}
\nc{\real}{{\rm I \! R}}
\nc{\Z}{{\sf Z \!\!\! Z}}
\nc{\complex}{{\rm C\!\!\! {\sf I}\,\,}}
\nc{\slask}{\!\!\!/}
\nc{\bis}{{\prime\prime}}
\nc{\pa}{\partial}
\nc{\na}{\nabla}
\nc{\ra}{\rangle}
\nc{\la}{\langle}
\nc{\goto}{\rightarrow}
\nc{\swap}{\leftrightarrow}

\nc{\EE}[1]{ \mbox{$\cdot10^{#1}$} }
\nc{\abs}[1]{\left|#1\right|}
\nc{\at}[2]{\left.#1\right|_{#2}}
\nc{\norm}[1]{\|#1\|}
\nc{\abscut}[2]{\Abs{#1}_{\scriptscriptstyle#2}}
\nc{\vek}[1]{{\rm\bf #1}}
\nc{\integral}[2]{\int\limits_{#1}^{#2}}
\nc{\inv}[1]{\frac{1}{#1}}
\nc{\dd}[2]{{{\partial #1}\over{\partial #2}}}
\nc{\ddd}[2]{{{{\partial}^2 #1}\over{\partial {#2}^2}}}
\nc{\dddd}[3]{{{{\partial}^2 #1}\over
        {\partial #2 \partial #3}}}
\nc{\dder}[2]{{{d #1}\over{d #2}}}
\nc{\ddder}[2]{{{d^2 #1}\over{d {#2}^2}}}
\nc{\dddder}[3]{{d^2 #1}\over
        {d #2 d #3}}
\nc{\dx}[1]{d\,^{#1}x}
\nc{\dy}[1]{d\,^{#1}y}
\nc{\dz}[1]{d\,^{#1}z}
\nc{\dl}[1]{\frac{d\,^{#1}l}{(2\pi)^{#1}}}
\nc{\dk}[1]{\frac{d\,^{#1}k}{(2\pi)^{#1}}}
\nc{\dq}[1]{\frac{d\,^{#1}q}{(2\pi)^{#1}}}

\nc{\cc}{\mbox{$c.c.$ }}
\nc{\hc}{\mbox{$h.c.$ }}
\nc{\cf}{cf.\ }
\nc{\erfc}{{\rm erfc}}
\nc{\Tr}{{\rm Tr\,}}
\nc{\tr}{{\rm tr\,}}
\nc{\pol}{{\rm pol}}
\nc{\sign}{{\rm sign}}
\nc{\bfT}{{\bf T }}
\nc{\eV}{\rm\ eV}
\nc{\GeV}{\rm\ GeV}
\nc{\MeV}{\rm\ MeV}
\nc{\keV}{\rm\ keV}
\nc{\TeV}{\rm\ TeV}

\nc{\cA}{{\cal A}}
\nc{\cB}{{\cal B}}
\nc{\cD}{{\cal D}}
\nc{\cE}{{\cal E}}
\nc{\cG}{{\cal G}}
\nc{\cH}{{\cal H}}
\nc{\cL}{{\cal L}}
\nc{\cO}{{\cal O}}
\nc{\cT}{{\cal T}}
\nc{\cN}{{\cal N}}
\nc{\rvac}[1]{|{\cal O}#1\rangle}
\nc{\lvac}[1]{\langle{\cal O}#1|}
\nc{\rvacb}[1]{|{\cal O}_\beta #1\rangle}
\nc{\lvacb}[1]{\langle{\cal O}_\beta #1 |}
\nc{\bb}{\bar{\beta}}
\nc{\bt}{\tilde{\beta}}
\nc{\ctH}{\tilde{\cal H}}
\nc{\chH}{\hat{\cal H}}
\nc{\1}{\aa}
\nc{\2}{\"{a}}
\nc{\3}{\"{o}}
\nc{\4}{\AA}
\nc{\5}{\"{A}}
\nc{\6}{\"{O}}
\nc{\al}{\alpha}
\nc{\g}{\gamma}
\nc{\Del}{\Delta}
\nc{\e}{\epsilon}
\nc{\eps}{\epsilon}
\nc{\lam}{\lambda}
\nc{\om}{\omega}
\nc{\Om}{\Omega}
\nc{\ve}{\varepsilon}
\nc{\mn}{{\mu\nu}}
\nc{\vp}{\varphi}

\nc{\rf}[1]{(\ref{#1})}
\nc{\nn}{\nonumber \\*}
\nc{\bfB}{\bf{B}}
\nc{\bfv}{\bf{v}}
\nc{\bfx}{\bf{x}}
\nc{\bfy}{\bf{y}}
\nc{\vx}{\vec{x}}
\nc{\vy}{\vec{y}}
\nc{\oB}{\overline{B}}
\nc{\oI}{\overline{I}}
\nc{\oR}{\overline{R}}
\nc{\rar}{\rightarrow}
\nc{\ti}{\times}
\nc{\slsh}{\hskip-5pt/}
\nc{\sm}{Standard~Model~}
\nc{\MP}{M_{\rm Pl}}
\nc{\tp}{t_{\rm Pl}}
\nc{\ave}{\bar{E}}

\nc{\eff}{{\rm eff}}
\nc{\kk}{\vek{k}}
\nc{\pp}{{\rm p}}
\nc{\ga}{g_{a\gamma}}
\nc{\vv}{\\}
\nc{\eee}{{\bf E}}
\nc{\bbb}{{\bf B}}
\nc{\qcd}{T_{\rm QCD}}
\def\vec#1{{\bf #1}}

\newcommand{\ggut}{G_\mathrm{GUT}}

\newcommand{\gps}{G_{\mathrm PS}} 
\newcommand{\gsm}{G_{\mathrm SM}} 
\begin{document}


\title{Particle physics models of inflation and curvaton scenarios}

\author{Anupam Mazumdar}
\affiliation{Lancaster University, Physics Department, Lancaster LA1 4YB, UK,}
\affiliation{Niels Bohr Institute,  Blegdamsvej-17,  DK-2100, Denmark.}

\author{Jonathan Rocher}
\affiliation{Service de Physique Th\'eorique, Universit\'e Libre de Bruxelles, 
CP225, Boulevard du Triomphe, 1050 Brussels, Belgium}


\begin{abstract}
We review the particle theory origin of inflation and curvaton mechanisms for generating large scale structures and the
observed temperature anisotropy in the cosmic microwave background (CMB) radiation. Since  inflaton or curvaton 
energy density creates all matter, it is important to understand the process of reheating and preheating into the relevant
degrees of freedom required for the success of Big Bang Nucleosynthesis. We discuss two distinct classes 
of models, one where inflaton and curvaton belong to the hidden sector, which are coupled to the Standard Model gauge sector 
very weakly. There is another class of models of inflaton and curvaton, which are embedded within Minimal Supersymmetric Standard Model 
(MSSM) gauge group and beyond, and whose origins lie within {\it gauge invariant} combinations of supersymmetric quarks and leptons. Their masses and couplings are all well motivated from low energy physics, therefore such models provide us with a unique opportunity that they can be
verified/falsified by the CMB data and also by the future collider and non-collider based experiments. We then briefly discuss stringy origin of inflation, alternative cosmological scenarios, and bouncing universes.
\end{abstract}

\maketitle 



\tableofcontents

\newpage

\section{Introduction}\label{sec:introduction}
%
%

The paradigm of primordial inflation has met with glorious successes over the past three decades since its conception~\cite{Guth:1980zm,Starobinsky:1980te,Linde:1981mu,Linde:1983gd,Albrecht:1982wi,Linde:1986fd} (for some excellent reviews on inflation, see \cite{Starobinsky:1985ww,Linde:2005dd,Olive:1989nu,Lyth:1998xn,Boyanovsky:2009xh}).  In the most general scenario, inflation occurs because a slowly rolling scalar field, the inflaton, dynamically gives rise to an epoch of accelerated expansion dominated by a false vacuum (for a review on inflaton models, see~\cite{Lyth:1998xn}). During inflation, quantum fluctuations imprinted on space-time are stretched outside the Hubble patch. These primordial fluctuations eventually re-enter the Hubble patch, whence their form can be extracted by observing the perturbations in the  Cosmic Microwave Background (CMB) radiation. Slow-roll inflationary scenarios
generically predict almost Gaussian {\it adiabatic} perturbations with a {\it nearly} flat spectrum (for a review on generating quantum fluctuations during inflation, see~\cite{Mukhanov:1990me}), which have met with an unprecedented success with the latest observations, see the recent data from WMAP~\cite{Komatsu:2008hk}~\footnote{See \texttt{http://map.gsfc.nasa.gov/}.}. Future CMB experiments such as PLANCK~\footnote{See \texttt{http://www.sciops.esa.int/index.php?project=PLANCK\&page=index}} will improve the current data, and also provide useful constraints on the scale of inflation in terms of primordial gravity waves~\cite{Grishchuk:1974ny,Grishchuk:1989ss,Allen:1987bk,Sahni:1990tx}, departure from random Gaussian fluctuations~\cite{Allen:1987vq,Falk:1992sf,Gangui:1993tt,Maldacena:2002vr,Acquaviva:2002ud,Bartolo:2004if},  isocurvature perturbations~\cite{Kodama:1985bj,Liddle:1993fq}, etc.

The end of inflation can be considered as a paradigm for the origin of matter, since all matter arises from the vacuum energy stored in the inflaton field. However the present models do not give clear predictions as to what sort of matter there is to be found in the early universe. Theoretical and observational successes of the Big Bang Nucleosynthesis (BBN) have constrained the degrees of freedom around the temperature of $T\geq 1-5$~MeV,
which contains the Standard Model (SM) quarks and leptons and three generations of neutrinos~\cite{Sarkar:1995dd,Burles:2000zk,Olive:1998vj,Fields:2006ga}. The present observational uncertainties allow {\it only} one extra species of relativistic particle at the time of BBN~\cite{Copi:1995cb,Olive:1998vj}. From the current observations we also know that SM baryons constitute about 4.6\%\ of the total energy density, almost 23\%\ of the total energy density is in non-luminous, non-baryonic dark matter, and the rest of the energy density is in the form of dark energy~\cite{Komatsu:2008hk}.

Besides the cosmological issues, one of the theoretical challenges is to understand why the mass scale of the SM, of ${\cal O}(100)$~GeV, is much lower than the scale of gravity, $M_{\rm P}=(8\pi G_{N})^{-1/2}=2.436\times 10^{18}$~GeV.  Unfortunately, the SM masses are not protected from the quantum corrections, which is  known as the {\it hierarchy problem}. The most popular remedy is the supersymmetry (SUSY)~(for a review, see~\cite{Nilles:1983ge,Haber:1984rc, Sohnius:1985qm,Martin:1997ns,Chung:2003fi}), which is believed to be broken at a 
scale $\sim {\cal O}(100-1000)$~GeV.  The SUSY is presumably broken first at high scales in some hidden sector then transmitted to the minimal SUSY extension of the SM, known as the MSSM, by gravitational~\cite{Nilles:1983ge,Haber:1984rc} or gauge interactions~\cite{Giudice:1998bp,Shadmi:1999jy}.
The Large Hadronic Collider (LHC)~\footnote{\texttt{http://lhc.web.cern.ch/lhc/}} at CERN has a potential to discover the MSSM particles. Theoretical estimations of radiative corrections of gauge couplings also suggest that MSSM  enables grand unification (GUT) of the gauge interactions at scales $M_{GUT}\sim 10^{16}$~GeV (for a GUT review, see~\cite{Raby:2008gh,Nath:2006ut}). When SUSY is broken locally, like any other gauge symmetry, an intimate connection with gravity emerges, known as the supergravity (SUGRA), which is valid below the Planck scale~\cite{Nilles:1983ge}. Furthermore, the unification of gravity with the other gauge interactions seems to require viewing fundamental ÔparticlesÕ as, instead, excitations of extended objects in the framework of string theory~\cite{Green:1987sp,Green:1987mn,Polchinski:1998rq}. Therefore, it is important to ask whether 
beyond the SM physics can provide all the right ingredients for inflation to occur or not.

Since the origin of baryons and dark matter bring inflation closer to the particle physics, the models of inflation 
must rely on an effective field theory treatment, where the inflaton belongs to the {\it hidden sector}, or the 
{\it observable sector}. In the former case, the coupling between the inflaton and the (MS)SM sector is either through gravitational strength, or via small unknown Yukawa coupling. Unfortunately, in the case of a hidden sector inflation the particle origin, mass, and couplings are largely unconstrained and at times unmotivated from theoretical point of view. Therefore, the inflationary predictions from such hidden sector models are also highly model dependent. 

String theory lends strong support to hidden sector models of inflation. There are plenty of absolute gauge singlet moduli, which mainly arise in the gravitational sector upon compactiÞcations~\cite{Giddings:2001yu,Grana:2005jc,Douglas:2006es}. There are many attempts to embed inflation within string theory, for a review, see~\cite{McAllister:2007bg,Burgess:2007pz,Linde:2005dd,Kallosh:2007ig,HenryTye:2006uv,Cline:2006hu,Kachru:2003sx,Quevedo:2002xw,Baumann:2009ds,Baumann:2009ni}.  The most exciting phenomenological revelation from string theory is that it can stabilize all the moduli~\cite{Giddings:2001yu}, and also the volume modulus~\cite{Kachru:2003aw}, with large and small positive cosmological constants, which has lead us to believe in a stringy landscape~\cite{Bousso:2000xa,Douglas:2004zg,Susskind:2003kw,Douglas:2006es}. At low energies, the number of vacua could be humongous, $10^{500}-10^{1000}$~\cite{Douglas:2004zg,Susskind:2003kw,Douglas:2006es}, and one in $10^{10}$ could be SM like~\cite{Blumenhagen:2004xx,Gmeiner:2005vz,Douglas:2006xy}. False vacuum inflation in such a landscape is generic with all possible scales of inflation down to the current cosmological constant~\cite{Kachru:2003aw}. However, our patch of the universe must have had a graceful exit from inflation at least before BBN. Exiting from such eternally inflating regime and exciting the SM degrees of freedom pose a new challenge for string theory.

In order to seek an observable sector origin of inflation, it is important to ask whether inflation can happen within the GUT theory~\cite{Dvali:1994ms,Lazarides:1995vr,Jeannerot:1997is,Davis:1995bx,Dvali:1997mh,Jeannerot:2000sv,Lazarides:2000ck,Gomez:2002tj,Jeannerot:2002wt,Senoguz:2004vu,BasteroGil:2006cm,Jeannerot:2006jj,Rocher:2006nh,Kyae:2003vn,Kyae:2005vg,Kyae:2002ss,Kyae:2005nv,Antusch:2004hd,Antusch:2005kf,Battye:2006pk,urRehman:2006hu,Lazarides:2007fh,Lazarides:2008nx,Antusch:2008gw}. Invariably all of  the models of inflation require an {\it absolute gauge singlet} inflaton couplings to the GUT/MSSM fields to drive the first phase of inflation, or to prepare the initial conditions for inflation. There is also an interesting proposal to realize inflation within the SM, with a non-trivial Higgs coupling to the  Ricci scalar~\cite{Bezrukov:2007ep}. The advantage is that inflation occurs within an observable sector physics, therefore the origin of matter creation is ascertained. But 
this idea does not rely on SUSY at all, and assumes the SM to be valid all the way up to the Planck scale.

In a recent development, it has been shown that within MSSM parameters allow a unique possibility to realize inflation with the help of {\it gauge invariant flat directions}~\cite{Allahverdi:2006iq,Allahverdi:2006cx,Allahverdi:2006we,Allahverdi:2008bt}. In MSSM there are many scalars, which span into a moduli space of {\it gauge invariant} $F$-and $D$-flat directions made up of squarks and sleptons (SUSY partners of quarks and leptons) (for a review, see~\cite{Enqvist:2003gh,Dine:2003ax}), which carry the SM charges, i.e.  baryon and/or lepton number. These inflatons have an {\it enhanced symmetry point} near the origin (at a VEV defined by zero). Away from the origin the inflatons break wholly or partly the SM gauge symmetry depending on the flat direction. But such a spontaneous breaking of charge and color in the early universe is not considered to be dangerous, provided they all settle down to their minimum before the electroweak phase transition. Note that in all these cases inflation occurs within an observable sector, where their mass and couplings are all well motivated from low energy physics.

In any inflationary scenario, it is important to understand the mechanisms of how to excite the SM quarks and leptons, known as the
reheating~\cite{Albrecht:1982mp,Turner:1983he,Dolgov:1989us,Kolb:1988aj} and preheating~\cite{Traschen:1990sw,Shtanov:1994ce,Kofman:1994rk,Kofman:1997yn,Micha:2002ey,Micha:2004bv, Podolsky:2005bw,Felder:2006cc,Khlebnikov:1996mc,Khlebnikov:1996wr,Khlebnikov:1996zt,Tranberg:2003gi,Arrizabalaga:2004iw,Aarts:1998td,Aarts:1999zn,Baacke:1996se,Baacke:1996kj,Baacke:1997rs,Greene:1997ge,Cormier:2001iw,GarciaBellido:2003wd,Felder:1998vq,Felder:1999pv},  and how to thermalize the universe with the MS(SM) degrees of freedom~\cite{Davidson:2000er,Allahverdi:2002pu,Allahverdi:2005mz,Allahverdi:2007zz,GarciaBellido:2008ab}, for a review see~\cite{Allahverdi:2010xz}. In this regard the observable sector models of inflation have an advantage, since the inflaton couplings to the matter fields are all known.

There is yet another paradigm for generating the amplitude for the CMB perturbations, known as the curvaton scenario~\cite{Lyth:2001nq,Enqvist:2001zp,Mollerach:1989hu,Moroi:2001ct,Moroi:2002rd,Lyth:2002my}. The curvaton is a light scalar field, which obtains its quantum fluctuations induced by the vacuum energy of the inflaton potential. However, the curvaton 
being light does not decay as rapidly as the inflaton, albeit its slow dynamics leads to its late decay. The relative 
perturbations between the fields give rise to entropy perturbations, which feed the curvature perturbations. Once the curvaton decays, it converts all its entropy perturbations into the adiabatic and nearly scale invariant perturations. One advantage of the curvaton scenario is that it is possible to generate significant non-Gaussianity~\cite{Linde:1996gt,Lyth:2001nq,Lyth:2002my}, however the present non-Gaussianity bounds are also tied to the residual isocurvature perturbations~\cite{Beltran:2008ei}. The challenges for the curvaton paradigm are the same as that of the inflaton, if the curvaton decays, then it must excite the (MS)SM degrees of freedom~\cite{Enqvist:2002rf,Enqvist:2003mr,Allahverdi:2006dr}. There are also alternative mechanisms to understand the temperature anisotropy in the CMB data {\it without invoking inflation}, such as in the case of a bouncing cosmology~\cite{Gasperini:1992em,Gasperini:2002bn,Khoury:2001wf,Steinhardt:2001st,Nayeri:2005ck,Biswas:2006bs},
we will briefly discuss some of these scenarios.

The main goal of this review is to address the origin of infation and curvaton, where they explain the large scale structures, and also the microphysical origin of the inflaton and curvaton. This can be achieved provided they
belong to a well-motivated sector of particle physics. Our aim is to review such models in details and some of their cosmological consequences.

The review is organized as follows. In section (Sec.) II, we recapitulate some basic inflationary cosmology, in particular
quantum fluctuations during inflation, multi-field perturbations, curvaton scenario, non-Gaussianity, challenges 
and requirements for a successful inflation. In Sec. III, we present 
some background material for particle physics tools required for inflation and curvaton, we particularly focus on MSSM, 
renormalization group equations, moduli space of flat directions within MSSM, properties of flat directions 
and their cosmological consequences, SUGRA and their role in building inflationary models, and a brief 
discussion on cosmic strings and grand unified theories (GUT). In Sec. IV, we discuss models of inflation, particularly
highlighting the connections with particle theory, we discuss non-SUSY models of inflation, SM Higgs inflation and implications for collider and non-collider experiments, SUSY $F$-and $D$-term inflation models, and embedding inflation within SUSY GUTs. In Sec. V, we discuss various attempts to gauge the inflaton with the SM charges, we discuss MSSM inflation models where we recognize the  inflaton candidates.  We also consider  thermal production of  dark matter in conjunction with inflationary parameter space.  
In Sec. VI, we discuss inflaton decay, reheating and thermalization. We focus on gauge singlet inflaton couplings to the SM and the MSSM. We discuss thermalization in perturbative decay of inflaton, basics of preheating and its applications to SUSY inflationary models.  In Sec. VII, we discuss models of curvaton where the curvaton is a MSSM flat direction. We then single out curvaton candidates, and discuss predictions for non-Gaussianity. In Sec. VIII, we discuss inflationary models within a string theory setup. We briefly describe alternative mechanisms for generating primordial perturbations in the context of a non-singular bouncing cosmology, and discuss various challenges they face.

\section{Inflation}\label{sec:inflationgeneral}
%
%

\subsection{Slow-roll inflation}\label{sec:generalprop:slowroll}

A completely flat potential would render inflation future eternal (but not 
past~\cite{Borde:1993xh,Borde:2001nh,Linde:1986fd,Linde:1983gd,Linde:1996hg,Linde:1993xx}),
provided the energy density stored in the flat direction dominates.
The inflaton direction is however not completely flat but has
a potential $V(\phi)$ with some slope. An inflationary phase is
obtained when the expansion rate evolution satisfies $\ddot a >0$. 
Slow-roll inflation assumes that the potential dominates over the 
kinetic energy of the inflaton $\dot\phi^2 \ll V(\phi)$, and $\ddot 
\phi \ll V^{\prime}(\phi)$, therefore the Friedmann and the
Klein-Gordon equations can be approximated as:
\begin{eqnarray}
\label{slowr1}
H^2 &\approx & \frac{1}{3M_{\rm P}^2}V(\phi)\,, \\
\label{slowr2}
3H\dot\phi &\approx &-V^{\prime}(\phi)\,,
\end{eqnarray}
where prime denotes derivative with respect to $\phi$. 
The slow-roll conditions are give by:
\begin{eqnarray}
\label{ep1}
\epsilon(\phi)&\equiv&\frac{M_{\rm P}^2}{2}\left(\frac{V^{\prime}}{V}\right)^2
\ll 1\,,\\
\label{eta1}
|\eta(\phi)|&\equiv&{M_{\rm P}^2}\left |\frac{V^{\prime\prime}}{V}\right| \ll 1\,.
\end{eqnarray}
Note that $\epsilon$ is positive by definition. These conditions are necessary 
but not sufficient for inflation. They only constrain the shape of the potential but not the velocity of
the field $\dot\phi$. Therefore, a tacit assumption behind the success
of the slow-roll conditions is that the inflaton field should not have
a large initial velocity.

Slow-roll inflation comes to an end when the slow-roll conditions are violated,
$\epsilon \sim 1$, and $\eta \sim 1$. However, there are certain models
where this need not be true, for instance in hybrid inflation models
\cite{Linde:1993cn}, where inflation comes to an end via a phase transition,
or in oscillatory models of inflation where slow-roll conditions are
satisfied only on average \cite{Damour:1997cb,Liddle:1998pz}, or inflation happens on average over
every oscillations of a bouncing universe~\cite{Biswas:2009fv}, or in fast roll inflation where 
the slow-roll conditions are never met~\cite{Linde:2001ae}~\footnote{A phase of fast roll inflation
prior to a slow roll phase of inflation has been invoked in order to suppress the power spectrum on large scales~\cite{Contaldi:2003zv,Boyanovsky:2006pm,Destri:2008fj}.}. The K-inflation where only the kinetic term 
dominates where there is no potential at all~\cite{ArmendarizPicon:1999rj}.

One of the salient features of the slow-roll inflation is that there exists a
late time attractor behavior~\footnote{A more rigorous set of parameters are presented describing
slow-roll inflation in Eq.~(\ref{eq:epsilonn}), which is independent of slow-roll conditions or the number 
of fields.}. This means that during inflation the
evolution of a scalar field at a given field value has to be independent
of the initial conditions.  Therefore slow-roll inflation should provide
an attractor behavior which at late times leads to an identical field
evolution in the phase space irrespective of the initial conditions
\cite{Salopek:1990jq}. In fact the slow-roll solution does not give
an exact attractor solution to the full equation of motion but is nevertheless
a fairly good approximation \cite{Salopek:1990jq}. A similar statement
has been proven for multi-field exponential potentials without slow-roll
conditions (i.e. assisted inflation) \cite{Liddle:1998jc}.

The standard definition of the number of e-foldings between time, $t$, and the end of inflation, $t_{end}$, is given by
\begin{equation}
N\equiv \ln\frac{a(t_{end})}{a(t)}=\int_{t}^{t_{end}}Hdt\approx
\frac{1}{M_{\rm P}^2}\int_{\phi_{end}}^{\phi}\frac{V}{V^{\prime}}d\phi\,,
\end{equation}
where $\phi_{end}$ is defined by $\epsilon(\phi_{end})\sim 1$, provided
inflation comes to an end via a violation of the slow-roll conditions.
The number of e-foldings can be related to the Hubble crossing
mode $k=a_{k}H_{k}$ by comparing with the present Hubble length $a_{0}H_{0}$.
The final result is \cite{Liddle:2000cg,Liddle:2003as}
\begin{equation}
\label{efoldsk}
N(k)=62~-~\ln\frac{k}{a_0H_0}~-~\ln\frac{10^{16}~{\rm GeV}}{V_{k}^{1/4}}~+~
\ln\frac{V_{k}^{1/4}}{V_{end}^{1/4}}~-~\frac{1}{3}\ln\frac{V_{end}^{1/4}}
{\rho_{rh}^{1/4}}\,,
\end{equation}
where the subscripts $end$ ($rh$) refer to the end of inflation (end of
reheating)~\footnote{For the particular scale of today's Hubble length, we define:
$N_Q\equiv N(k=a_0H_0)$. The corresponding slow-roll parameters at that scale will be denoted by 
$\epsilon_Q,~\eta_Q$.}.

A simple generalization of the above formula has been derived in~\cite{Burgess:2005sb},
where there are mutiple stages of inflation, with potentials $V_{I}, V_i, V_{L}$ one after the 
other separated by the matter epochs, whose
expansions are parameterized by $t^{n_1},~t^{n_2},...,~t^{n_L}$,
with $n_1,~n_2,...,n_L \leq 1$.  
We can assume instant reheating after the last phase of inflation,
which yields, $\rho_{L,end}=\rho_{rh,L}$.  Concentrating on the
present horizon scale $N_Q\equiv N_I(a_0H_0)$, we obtain~\cite{Burgess:2005sb}
\begin{equation}\label{lnefolds2app}
        \sum_{i=1}^L N_i=62+\frac
        14\ln\Big({V^2_{I}\over V_{L}M_{\rm P}^4}\Big)+
        \sum_{i=1}^{L-1}{n_i\over 2}\ln\Big({V_{i+1}\over V_i}\Big)\,.
\end{equation}
One can impose new constraints arising from the fact that there should be no
reprocessing of modes in between the two phases of inflation on the scales 
probed by CMB experiments. This requirement constraints~\cite{Burgess:2005sb}
\begin{equation}\label{noreprosapp}
         \ln\Big({1\over a_{k_c}H_{i}}\Big)>
         \ln\Big({1\over a_{rh,i}H_{i+1}}\Big)\,,
         \ \ i=2,...,L-1\,
\end{equation}
Hence, in general we obtain the following set of constraints~\cite{Burgess:2005sb}:
\begin{eqnarray}\label{genconstsapp}
         & & N_Q>6.9+{1-n_1\over 2}\ln\Big({V_I\over V_{II}}\Big)\,,
         \nonumber\\ & & \sum_{j=2}^{i}N_j  >
         \frac 12 \ln\Big({V_{II}\over V_{i+1}}\Big)
         -\sum_{j=2}^{i}{n_j\over 2}\ln\Big({V_{i}\over V_{i+1}}\Big)\,,
         \ \ i=2,...,L-1\,.
\end{eqnarray}
In the special case $n_i=1$, it is easy to see that these constraints are
trivially satisfied ($\sum_jN_j>0$). The details of  the thermal history of the universe determine the
precise number of e-foldings required to solve the horizon problem, but for most practical purposes (for high scale inflation with 
large reheating temperature) it is sufficient to assume that $N_Q\approx 50-60$, keeping all the uncertainties
such as the scale of inflation and the end of inflation within a margin of
$10$~e-foldings. A significant modification can take place only if there
is an epoch of late inflation such as thermal inflation
\cite{Lyth:1995hj,Lyth:1995ka}, or in theories with a low quantum gravity scale
\cite{Mazumdar:1999tk,Mazumdar:2001ya,ArkaniHamed:1999gq}, or if there are phase transitions 
during inflation~\cite{Adams:1997de,Hunt:2007dn,Hunt:2004vt}.

\subsection{Primordial density perturbations}\label{sec:generalprop:primorperturb}

Initially, the theory of cosmological perturbations has been developed
in the context of FRW cosmology \cite{Lifshitz:1945du}, and for models of 
inflation in  \cite{Mukhanov:1981xt,Guth:1982ec,Hawking:1982cz,Starobinsky:1982ee,Bardeen:1983qw,Lyth:1984gv,Linde:1984ir,Vilenkin:1982de,Vilenkin:1983xp,Kodama:1985bj,Sasaki:1986hm,Stewart:1993bc,Lyth:1991bc,Kolb:1995iv,Lyth:2004gb,Mukhanov:1989rq}. For a complete review on this topic, see \cite{Mukhanov:1990me}.
For a real single scalar field there arise only adiabatic density perturbations. In case of several fluctuating fields there will in
general also be isocurvature perturbations. Irrespective of the nature of perturbations, any light scalar field obtains quantum
fluctuations in a de Sitter space, which has many applications.


\subsubsection{Fluctuations in de Sitter}\label{FIDS}

By solving the Klein-Gordon equation for a light scalar field in a conformal metric:
$ds^2=g_{\mu\nu}dx^{\mu}dx^{\nu}=a^2(\tau,x)(d\tau^2-dx^2)$,
one can find the plane wave solution, $\phi(\vec x, \tau)= \frac{1}{(2\pi)^{3/2}}\int d^3k \left(\phi_{k}(\tau)e^{ik\cdot\vec x}+
{\rm h.c.}\right)$, for the mode function:
\cite{Bunch:1978yq,Linde:1982uu,Vilenkin:1982de,Vilenkin:1983xq,Vilenkin:1982wt,Mottola:1984ar,Allen:1985ux,Allen:1983dg}:
\begin{eqnarray}
\phi_{k}(\tau)&=&\left(\frac{\pi}{4}\right)^{1/2}H|\tau|^{3/2}\left(
c_{1}H^{(1)}_{\nu}(k\tau)+c_{2}H^{(2)}_{\nu}(k\tau)\right)\,, \nonumber \\
\tau &=&-H^{-1}e^{-Ht}\,,~~{\rm and}~~\nu^2=\frac{9}{4}-\frac{m^2}{H^2}\,,
\end{eqnarray}
where $m$ is the mass of the scalar field, $H^{(1)}_{\nu}$ and $H^{(2)}_{\nu}$
are the Hankel functions and $c_1,c_2$ are constants. By using a point splitting
regularization scheme, it is possible to obtain a Bunch-Davies vacuum
for a de Sitter background which actually corresponds to taking $c_1=0$,
and $c_2=1$. 

Generically, in a de Sitter phase, the main contribution to the two point
correlation function comes from the long wavelength modes; $k|\tau| \ll 1$
or $k\ll H\exp(Ht)$, determined by the Hubble expansion rate~\cite{Vilenkin:1982de,Vilenkin:1982wt}.
\begin{equation}
\label{variance}
\langle \phi^2\rangle\approx \frac{1}{(2\pi)^{3}}\int_{H}^{He^{Ht}}
d^3k|\phi_{k}|^2\,.
\end{equation}
The integration yields an
indefinite increase in the variance with time
\begin{equation}
\label{variance1}
\langle \phi^2\rangle \approx \frac{H^3}{4\pi^2}t\,.
\end{equation}
This result can also be obtained by considering the Brownian motion
of the scalar field \cite{Linde:1993xx,Linde:1993nz,GarciaBellido:1993wn,GarciaBellido:1994ci,Linde:1994gy}.
For a massive field with $m\ll H$, and $\nu\neq 3/2$, one does not obtain an
indefinite growth of the variance of the long wavelength fluctuations, but
\cite{Linde:1982uu,Vilenkin:1982wt,Vilenkin:1983xq,Vilenkin:1983xp,Enqvist:1987au}:
\begin{equation}
\langle \phi^2\rangle=\frac{3H^4}{8\pi^2m^2}\left(1-e^{-(2m^2/3H^2)t}
\right)\,.
\end{equation}
In the limiting case when $m\rightarrow H$, the variance goes as
$\langle \phi^2\rangle \approx H^2$. In the limit $m\gg H$,
the variance goes as $\langle \phi^2\rangle \approx (H^3/12\pi^2m)$. Only in a massless case $\langle \phi^2\rangle$ can
be treated as a homogeneous background field with a long wavelength mode.


\subsubsection{Adiabatic perturbations and the Sachs-Wolfe effect}

Let us consider small inhomogeneities,
$\phi(\vec x,t) =\phi(t)+\delta\phi(\vec x,t)$, such that $\delta\phi\ll\phi$.
Perturbations in matter densities automatically induce perturbations in
the background metric, but the separation between the background metric
and a perturbed one is not unique. One needs to choose a gauge. A simple choice would be to
fix the observer to the unperturbed matter particles, where the observer
will detect a velocity of matter field falling under gravity; this is
known as the Newtonian or the longitudinal gauge because the observer
in the Newtonian gravity limit measures the gravitational potential
well where matter is falling in and clumping. The induced metric
can be written as
\begin{equation}
\label{gauge}
ds^2=a^2(\tau)\left[(1+2\Phi)d\tau^2-(1-2\Psi)\delta_{ik}dx^{i}dx^{k}\right]\,,
\end{equation}
where $\Phi$ has a complete analogue of Newtonian gravitational potential.
In the case when the spatial part of the energy momentum tensor is diagonal,
i.e. $\delta T^{i}_{j}=\delta^{i}_{j}$, it follows that $\Phi=\Psi$,
\cite{Mukhanov:1990me}. Right at the time of horizon crossing one finds a
solution for $\delta\phi$ as
\begin{equation}
\label{pertphi}
\langle |\delta\phi_{k}|^2\rangle=\frac{H(t_{\ast})^2}{2k^3}\,,
\end{equation}
where $t_\ast$ denotes the instance of horizon crossing. Correspondingly,
we can also define a power spectrum
\begin{equation}
\label{spect}
{\cal P}_{\phi}(k)=\frac{k^3}{2\pi^2}\langle |\delta\phi_{k}|^2\rangle=
\left[\frac{H(t_{\ast})}{2\pi}\right]^2 \equiv
\left. \left[\frac{H}{2\pi}\right]^2\right|_{k=aH}\,.
\end{equation}
Note that the phase of $\delta\phi_{k}$ can be arbitrary, and therefore,
inflation has generated a Gaussian perturbation.

In the limit $k\rightarrow 0$, one can find an exact solution for the long
wavelength inhomogeneities $k\ll aH$ \cite{Starobinsky:1985ww,Mukhanov:1990me}, which
reads:
\begin{eqnarray}
\label{pertPhi}
\Phi_{k}&\approx&c_{1}\left(\frac{1}{a}\int_{0}^{t}a~dt^{\prime}\right)^{\cdot}
+ c_{2}\frac{H}{a}\,, \\
\label{pertphi1}
\frac{\delta\phi_{k}}{\dot\phi}&=&\frac{1}{a}\left(c_{1}\int_{0}^{t}
a~dt^{\prime}-c_2\right)\,,
\end{eqnarray}
where  the dot denotes derivative with respect to physical time. The growing
solutions are proportional to $c_1$, the decaying  proportional
to $c_2$. Concentrating upon the growing solution, it is possible to obtain
a leading order term in an expansion with the help of the slow-roll conditions:
\begin{eqnarray}
\label{pertPhi1}
\Phi_{k}&\approx &-c_{1}\frac{\dot H}{H^2}\,, \\
\label{pertphi2}
\frac{\delta\phi_{k}}{\dot\phi}&\approx &\frac{c_{1}}{H}\,.
\end{eqnarray}
Note that at the end of inflation, which is indicated by $\ddot a =0$, or
equivalently by $\dot H =-H^2$, one obtains a constant Newtonian
potential $\Phi_{k}\approx c_{1}$. This is perhaps the most
significant result for a single field perturbation.

In a long wavelength limit one obtains a constant of motion $\zeta$
\cite{Bardeen:1983qw,Brandenberger:1984cz,Mukhanov:1990me} defined as~\footnote{If the equation of state for 
matter remains constant, there exists a simple relationship which connects the metric perturbations
at two different times: $\Phi_{k}(t_f)= \frac{1+\frac{2}{3}\left(1+w(t_f)\right)^{-1}}{1+\frac{2}{3}\left(1+w(t_i)\right)^{-1}}\Phi_{k}(t_i)$
\cite{Bardeen:1983qw,Brandenberger:1984cz,Mukhanov:1990me}.}~:
\begin{equation}
\label{zeta1}
\zeta=\frac{2}{3}\frac{H^{-1}\dot\Phi_{k}+\Phi_{k}}{1+w}+\Phi_{k}\,,~~
w=\frac{p}{\rho}\,.
\end{equation}

This is also known as a comoving curvature perturbation \cite{Lukash:1980iv} reads in the
longitudinal gauge \cite{Mukhanov:1990me} for the slow-roll inflation as
\begin{equation}
\label{curvpert}
{\zeta}_{k}=\Phi_{k}-\frac{H^2}{\dot H}\left(H^{-1}\dot\Phi_{k}+\Phi_{k}
\right)\,.
\end{equation}
For CMB and structure formation we need to know the metric perturbation
during the matter dominated era when the metric perturbation is
$\Phi(t_f)\approx (3/5)c_{1}$. Substituting the value of $c_{1}$
from Eq.~(\ref{pertphi2}), we obtain
\begin{equation}
\label{curv1}
\Phi_{k}(t_f) \approx \left.\frac{3}{5}H\frac{\delta \phi_{k}}{\dot\phi}
\right|_{k=aH}\,.
\end{equation}
In a similar way it is also possible to show that the comoving curvature
perturbations is given by
\begin{equation}
{\zeta}_{k}\approx \left.\frac{H}{\dot \phi}\delta \phi \right|_{k=aH}\,,
\end{equation}
where $\delta_{\phi}$ denotes the field perturbation on a spatially flat
hypersurfaces, because on a comoving hypersurface $\delta\phi=0$,
by definition. Therefore, on flat hypersurfaces
\begin{equation}
\delta\phi_{k}=\dot\phi\delta t\,,
\end{equation}
where $\delta t$ is the time displacement going from flat to comoving
hypersurfaces \cite{Liddle:1993fq,Liddle:2000cg}. As a result
\begin{equation}
\label{curvpert1}
{\zeta}_{k} \equiv H \delta t\,.
\end{equation}
Note that during matter dominated era the curvature perturbation and the
metric perturbations are related to each other
\begin{equation}
\label{curmet}
\Phi_{k}=-\frac{3}{5}{\zeta}_{k}\,.
\end{equation}
In the matter dominated era the photon sees this potential well
created by the primordial fluctuation and the redshift in the
emitted photon is given by
\begin{equation}
\frac{\Delta T_{k}}{T}=-\Phi_{k}\,.
\end{equation}
At the same time, the proper time scale inside the fluctuation becomes slower
by an amount $\delta t/t =\Phi_{k}$. Therefore, for the scale factor
$a\propto t^{2/3}$, decoupling  occurs earlier with
\begin{equation}
\frac{\delta a}{a}=\frac{2}{3}\frac{\delta t}{t}=\frac{2}{3}\Phi_{k}\,.
\end{equation}
By virtue of $T\propto a^{-1}$ this results in a temperature which is hotter by
\begin{equation}
\label{sachs}
\frac{\Delta T_{k}}{T} =-\Phi_{k} +\frac{2}{3}\Phi_{k} =-\frac{\Phi_{k}}{3}\,.
\end{equation}
This is known as the Sachs-Wolfe effect \cite{Sachs:1967er}.


\subsubsection{Spectrum of adiabatic perturbations}

Now, one can immediately calculate the spectrum of the metric perturbations.
For a critical density universe
\begin{equation}
\delta_{k}\equiv \left.\frac{\delta\rho}{\rho}\right|_{k}=-\frac{2}{3}
\left(\frac{k}{aH}\right)^2\Phi_{k}\,,
\end{equation}
where $\nabla^2\rightarrow -k^2$, in the Fourier domain. Therefore,
with the help of Eqs.~(\ref{spect},\ref{curv1}), one obtains
\begin{equation}
\label{pspect}
\delta_{k}^2\equiv \frac{4}{9}{\cal P}_{\Phi}(k)\, =\frac{4}{9}\frac{9}{25}
\left(\frac{H}{\dot\phi}\right)^2\left(\frac{H}{2\pi}\right)^2\,,
\end{equation}
where the right hand side can be evaluated at the time of horizon exit
$k=aH$. In fact the above expression can also be expressed in terms of
curvature perturbations \cite{Liddle:1993fq,Liddle:2000cg}
\begin{equation}
\delta_{k}=\frac{2}{5}\left(\frac{k}{aH}\right)^2{\zeta}_{k}\,,
\end{equation}
and following Eq.~(\ref{curvpert}), we obtain
$\delta_{k}^2=({4}/{25}){\cal P}_{\zeta}(k)=({4}/{25})(H/\dot\phi)^2(H/2\pi)^2$,
exactly the same expression as in Eq.~(\ref{pspect}).
With the help of the slow-roll equation $3H\dot\phi=-V^{\prime}$,
and the critical density formula $3H^2M^2_{\rm P}=V$, one obtains
\begin{equation}\label{eq:powerspectrum1}
\delta_{k}^2\approx \frac{1}{75\pi^2 M_{\rm P}^6}\frac{V^3}{V^{\prime 2}}\,
=\frac{1}{150\pi^2 M_{\rm P}^4}\frac{V}{\epsilon}\,,~~~{\rm and}~~~{\cal P}_{\zeta}(k)=\frac{1}{24\pi^2M_{\rm P}^4}\frac{V}{\epsilon}\,,
\end{equation}
where we have used the slow-roll parameter
$\epsilon\equiv (M_{\rm P}^2/2)(V^{\prime}/V)^2$.
The COBE satellite measured the CMB anisotropy and fixes the normalization
of ${\cal P}_{\zeta}(k)$ on a very large scales. For a critical density universe,
if we assume that the primordial spectrum can be approximated by a power law
(ignoring  the gravitational waves and the $k-$dependence of the power 
$n_s$)~\cite{Komatsu:2008hk}
\begin{equation}\label{eq:primordialpowerspectrum}
{\cal P}_{\zeta}(k)\simeq (2.445\pm 0.096)\times 10^{-9}\left(\frac{k}{k_0}\right)^{n_s-1}~,
\end{equation}
where $n_s$ is called the spectral index (or spectral tilt), the reference 
scale is: $k_0=7.5a_{0}H_{0}\sim 0.002~{\rm Mpc}^{-1}$, and the error bar on the 
normalization is given at $1\sigma$. The spectral index $n(k)$ is defined as
\begin{equation}\label{spectind}
n(k)-1\equiv \frac{d\ln{\cal P}_{\zeta}}{d\ln k}\,.
\end{equation}
This definition is equivalent to the power law behavior if $n(k)$ is close to
a constant quantity over a range of $k$ of interest. One particular value of 
interest is $n_s\equiv n(k_0)$.  If $n_s=1$, the spectrum is flat and known as Harrison-Zeldovich 
spectrum~\cite{Harrison:1969fb,Zeldovich:1969sb}. For $n_s\neq 1$, the spectrum 
is tilted, and $n_s>1$ ($n_s<1$) is known as a blue (red) spectrum.
In the slow-roll approximation, this tilt can be expressed in terms of the 
slow-roll parameters and at first order~\footnote{At second order, $$n_s-1=
2\left[-3\epsilon +\eta -\left(\frac{5}{3}+12C\right)\epsilon^2+(8C-1)\epsilon
\eta +\frac{1}{3}\eta^2-(C-\frac{1}{3})\xi^2\right]~,$$ where $C$ is a numerical 
constant $C=\ln2+\gamma_E-2\simeq-0.7296$~\cite{Stewart:1993bc}.}
\begin{equation}
\label{spectind4}
n_s-1=-6\epsilon +2\eta +\mathcal{O}(\epsilon^2,\eta^2,\epsilon\eta,\xi^2)~.
\end{equation}
The running of these parameters are given by~\cite{Salopek:1990jq}
\begin{equation}
\label{spectind3}
\frac{\mathrm{d}\epsilon}{\mathrm{d}\ln k}=2\epsilon\eta-4\epsilon^2~,\quad \frac{\mathrm{d}\eta}{\mathrm{d}\ln k}
=-2\epsilon\eta +\xi^2~,\quad \frac{\mathrm{d}\xi^2}{\mathrm{d}\ln k}=-2\epsilon\xi^2+\eta\xi^2
+\sigma^3~,
\end{equation}
where
\begin{equation}
\xi^2\equiv M_{\rm P}^4\frac{V^{\prime}(\mathrm{d}^3V/\mathrm{d}\phi^3)}{V^2}\,,
\quad\sigma^3\equiv M_{\rm P}^6\frac{V^{\prime 2}(\mathrm{d}^4V/\mathrm{d}\phi^4)}{V^3}\,.
\end{equation}
Slow-roll inflation requires that
$\epsilon \ll 1, |\eta|\ll 1$, and therefore naturally predicts small
variation in the spectral index within $\Delta \ln k\approx 1$~\cite{Kosowsky:1995aa}
\begin{equation}
\frac{\mathrm{d} n(k)}{\mathrm{d}\ln k}=-16\epsilon\eta+24\epsilon^2+2\xi^2\,.
\end{equation}
Independently of slow-roll considerations or of the number of 
fields involved in the dynamics of inflation, a new set of 
parameters, known as the Hubble-flow parameters, were 
discussed in~\cite{Schwarz:2001vv,Schwarz:2004tz}~\footnote{It is 
possible to extend the calculation of metric perturbation beyond the
slow-roll approximations based on a formalism similar to that developed 
in Refs.~\cite{Mukhanov:1985rz,Sasaki:1986hm,Mukhanov:1989rq,Kolb:1995iv}.
}:
\begin{equation}\label{eq:epsilonn}
\epsilon_0\equiv H~,
\qquad \epsilon_{n+1}\equiv \frac{\ln |\epsilon_{n}|}{ N}~.
\end{equation}
It gives for the slow-roll parameter 
$\epsilon_1= -\dot{H}/H^2$, and inflation takes place 
only when $\ddot{a}>0$ which is equivalent to $\epsilon_1<1$. 
Slow-roll inflation takes place when $\forall\,n,\,\, \epsilon_n 
\ll 1$. In the slow-roll limit, these parameters can be 
related to the slow-roll parameters,
\begin{equation}
\epsilon_1 \simeq \epsilon +\mathcal{O}(\epsilon^2,\eta^2,\xi^2)~,\quad 
\epsilon_2 \simeq 4\epsilon - 2\eta +\mathcal{O}(\epsilon^2,\eta^2,\xi^2)~.
\end{equation}
Transient violation of slow-roll conditions were studied in hybrid inflation \cite{Clesse:2008pf},
for computation of the power spectrum, see \cite{Martin:2006rs,Ringeval:2007am}.
Models of inflation with large $\eta$ were also considered in~\cite{Linde:2001ae}.

\subsubsection{Gravitational waves}

Gravitational waves are linearized tensor
perturbations of the metric and do not couple to the energy
momentum tensor. Therefore, they do not give rise a gravitational
instability, but carry the underlying geometric structure of the space-time.
The first calculation of the gravitational wave production was made in
\cite{Grishchuk:1974ny}, and the topic has been considered by many authors
\cite{Grishchuk:1989ss,Allen:1987bk,Sahni:1990tx}. For reviews on gravitational waves, see
\cite{Mukhanov:1990me,Maggiore:1999vm}.

The gravitational wave perturbations are described by a line element
$ds^2+\delta ds^2$, where
\begin{equation}
{d}s^2=a^2(\tau)(\mathrm{d}\tau^2-\mathrm{d}x^{i}\mathrm{d}x_{i})\,,\quad \delta 
{d}s^2=-a^2(\tau)h_{ij} \mathrm{d}x^{i} \mathrm{d}x^{j}\,.
\end{equation}
The gauge invariant and conformally invariant $3$-tensor $h_{ij}$ is symmetric,
traceless $\delta^{ij}h_{ij}=0$, and divergenceless $\nabla_{i}h_{ij}=0$
($\nabla_{i}$ is a covariant derivative). Massless spin
$2$ gravitons have two degrees of freedom and as a result are also
transverse. This means that in a Fourier domain the gravitational wave
has a form
\begin{equation}
h_{ij}=h_{+}e^{+}_{ij}+h_{\times}e^{\times}_{ij}\,.
\end{equation}
For the Einstein gravity, the gravitational wave equation of motion follows
that of a massless Klein Gordon equation \cite{Mukhanov:1990me}. Especially,
for a flat universe
\begin{equation}
\ddot h^{i}_{j}+3H\dot h^{i}_{j}+\left(\frac{k^2}{a^2}\right)h^{i}_{j}=0\,,
\end{equation}
As  any massless field, the gravitational waves also feel the quantum
fluctuations in an expanding background. The spectrum mimics that of
Eq.~(\ref{spect})
\begin{equation}
{\cal P}_{\rm grav}(k)=\left.\frac{2}{M_{\rm P}^2}\left(\frac{H}{2\pi}\right)^2
\right|_{k=aH}\,.
\end{equation}
Note that the spectrum has a Planck mass suppression, which suggests that the
amplitude of the gravitational waves is smaller compared to that of the
adiabatic perturbations. Therefore it is usually assumed that their
contribution to the CMB anisotropy is small. The corresponding
spectral index can be expanded in terms of the slow-roll parameters at first 
order as~\footnote{At second order, $r\simeq 16\epsilon\left[1+\frac{2}{3}
(3C-1)(2\epsilon-\eta)\right]$, where $C$ is a numerical constant 
$C=\ln2+\gamma_E-2\simeq-0.7296$~\cite{Stewart:1993bc}.}
\begin{equation}
r\equiv \frac{{\cal P}_{\rm grav}}
{ {\cal P}_{\zeta}}=16\epsilon~,\quad {\rm and}\quad 
n_t=\frac{\mathrm{d}\ln{\cal P}_{\rm grav}(k)}{\mathrm{d}\ln k}\simeq -2\epsilon,\,.
\end{equation}
Note that the tensor spectral index is negative. It is expected that PLANCK could 
detect gravity waves if $r\gtrsim 0.1$, however the spectral index will be 
hard to measure in forthcoming experiments. The primordial gravity waves 
can be generated for large field value inflationary models.  Using the 
definition of the number of e-foldings it is possible to derive the range 
of $\Delta \phi$ (see for 
instance~\cite{Lyth:1996im,Lyth-new-book,Hotchkiss:2008sa})
\begin{equation}
16\epsilon = r<0.003\left(\frac{50}{N}\right)^2\left(\frac{\Delta \phi}
{M_{\rm P}}\right)~.
\end{equation}
%
\subsection{Multi-field perturbations}\label{sec:generalprop:multifieldperturb}
In multi-field inflation models contributions to the density
perturbations come from all the fields. However unlike in a single scalar
case, in the multi-field case there might not be a unique late time
trajectory corresponding to all the fields. In a very few cases
it is possible to obtain a late time attractor behavior of all the
fields; an example is assisted inflation~\cite{Liddle:1998jc}. 
If there is no late time attractor then different trajectories inherit 
difference in perturbations, known as entropy perturbations, which
opens new set of constraints which we will discuss 
below~\cite{Sasaki:1995aw,Lyth:1998xn,Langlois:1999dw,Gordon:2000hv,Lyth:2004gb}.

\subsubsection{Adiabatic and isocurvature conditions}

There are only two kinds of perturbations that can be generated.
The first one is the adiabatic perturbation discussed previously;
it is a perturbation along the late time classical trajectories
of the scalar fields during inflation.
When the primordial perturbations enter our horizon they
perturb the matter density with a generic {\it adiabatic condition}, which is
satisfied when the density contrast of the individual species is related
to the total density contrast $\delta_{k}$~\cite{Kodama:1985bj,Liddle:1993fq}
\begin{equation}\label{adia}
\frac{1}{3}\delta_{k b}=\frac{1}{3}\delta_{k c}=\frac{1}{4}\delta_{k\nu}=
\frac{1}{4}\delta_{k\gamma}=\frac{1}{4}\delta_{k}\,,
\end{equation}
where $b$ stands for baryons, $c$  for cold dark matter, $\gamma$
for photons and $\nu$  for neutrinos.

The other type is the isocurvature perturbation. During
inflation this can be viewed as a perturbation orthogonal to the unique
late time classical trajectory. Therefore, if there were $N$ fluctuating
scalar fields during inflation, there would be $N-1$ degrees of freedom
which would contribute to the isocurvature perturbation~\cite{Polarski:1994rz,Peter:1994dx,Starobinsky:2001xq,Starobinsky:1994mh}.

The {\it isocurvature condition} is known as $\delta\rho=0$:
the sum total of all the energy contrasts must be zero. The
most general density perturbations is then given by a linear
combination of an adiabatic and an isocurvature density perturbations.


\subsubsection{Adiabatic perturbations due to multi-field}

In a comoving gauge, see Eq.~(\ref{curvpert}),
${\zeta}=-H\delta\phi/\dot\phi$ holds good even for multi-field
inflation models, provided we identify each field component of $\phi$ along
the slow-roll direction. There also exists a relationship between
the comoving curvature perturbations and the number of e-foldings, $N$, given by
\cite{Salopek:1995vw,Sasaki:1995aw}
\begin{equation}
\label{multi1}
{\zeta} =\delta N=\frac{\partial N}{\partial \phi_{a}}\delta\phi_{a}\,,
\end{equation}
where $N$ is measured by a comoving observer while passing from flat
hypersurface (which defines $\delta\phi$) to the comoving hypersurface
(which determines $\zeta$, where it remains conserved on large scales
even for multi-field case~\cite{Lyth:2004gb}.). The repeated indices
are summed over and the subscript $a$ denotes a component of the inflaton.
A more intuitive discussion has been given in \cite{Liddle:2000cg,Lyth:1998xn,Lyth-new-book}.

If again one assumes that the perturbations in $\delta\phi_{a}$ have random
phases with an amplitude $(H/2\pi)^2$, one obtains:
\begin{equation}
\label{multi5}
\delta_{k}^2=\frac{V}{75\pi^2~M_{\rm P}^2}\frac{\partial N}{\partial\phi_{a}}
\frac{\partial N}{\partial\phi_{a}}\,.
\end{equation}
For a single component
$\partial N/\partial \phi\equiv (M_{\rm P}^{-2}V/ V^{\prime})$, and then
Eq.~(\ref{multi5}) reduces to Eq.~(\ref{eq:powerspectrum1}). By using 
slow-roll equations we can again define the spectral index
\begin{equation}
\label{multi6}
n-1=-\frac{M_{\rm P}^2V_{,a}V_{,a}}{V^2}-\frac{2}{M_{\rm P}^2 N_{,a}
N_{,a}}+2\frac{M_{\rm P}^2N_{,a}N_{,b}V_{,ab}}{V N_{,c}N_{,c}}\,,
\end{equation}
where $V_{,a}\equiv \partial V/\partial \phi_{a}$, and similarly
$N_{,a}\equiv\partial N/\partial \phi_{a}$. For a 
single component we recover Eq.~(\ref{spectind4}) from Eq.~(\ref{multi6}).


\subsubsection{Isocurvature perturbations and CMB}

One may of course simply assume a purely
isocurvature initial condition. For any species the entropy
perturbation is defined by
\begin{equation}
S_{i}=\frac{\delta n_{i}}{n_{i}}-\frac{\delta n_{\gamma}}{n_{\gamma}}\,,
\end{equation}
Thus, if initially there is a radiation bath with
a common radiation density contrast $\delta_{r}$, a baryon-density contrast
$\delta_{b}=3\delta_{r}/4$, and a CDM density contrast $\delta_{c}$, then
\begin{equation}
\label{cdm}
S_c=\delta_{c}-\frac{3}{4}\delta_{r}=\frac{\rho_{r}\delta\rho_{c}-(3/4)
\rho_{c}\delta\rho_{r}}{\rho_{r}\rho_{c}}=\frac{\rho_{r}+(3/4)\rho_{c}}
{\rho_{r}\rho_{c}}\delta\rho_{c}\approx \delta_{c}\,,
\end{equation}
where we have used the isocurvature condition
$\delta\rho_{r}+\delta\rho_{c}=0$, and the last equality holds in a
radiation dominated universe. Similarly the baryon isocurvature is given 
by: $S_{B}=\delta_B-(3/4)\delta_r$ and the neutrino (or any other 
relativistic species) isocurvature component is given by:
$S_{\nu}=(3/4)\delta_{\nu}-(3/4)\delta_r$.

However a pure isocurvature perturbation
gives five times larger contribution to the Sachs-Wolfe effect compared
to the adiabatic case \cite{Bond:1984fp,Kodama:1986ud,Liddle:1993fq}.
This result can be derived very easily in a matter dominated era with
an isocurvature condition $\delta\rho_{c}=-\delta\rho_{r}$, which gives
a contribution ${\zeta}_{k}=(1/3)S_{k}$. Therefore from
Eqs.~(\ref{curmet},\ref{sachs}), we obtain $\Delta T_{k}/T =-S_{k}/15$.
There is an additional contribution from radiation because we are in a
matter dominated era, see Eq.~(\ref{cdm}),
$S\approx \delta_{c} \equiv -(3/4)\delta_{r}$. The sum total isocurvature
perturbation $\Delta T_{k}/T = -S/15 -S/3= -6S/15$, where $S$ is measured
on the last scattering surface. The Sachs-Wolfe effect for isocurvature
perturbations fixes the {\em slope} of the perturbations, rather than
the amplitude \cite{Linde:1996gt,Liddle:1999pr}. Present CMB data rules 
out pure isocurvature perturbation spectrum~\cite{Stompor:1995py,Enqvist:2000hp,Enqvist:2001fu,Pierpaoli:1999zj,Beltran:2004uv,Beltran:2005xd,Beltran:2005gr,Beltran:2006sq,Komatsu:2008hk},
although a mixture of adiabatic and isocurvature perturbations remains
a possibility.

In the latter case it has been argued that the adiabatic and
isocurvature perturbations might naturally turn out to be 
correlated~\cite{Langlois:1999dw,Bucher:2000hy,KurkiSuonio:2004mn,Trotta:2006ww}. 
It is sometimes useful to consider $\alpha$ defined by: 
\begin{equation}\label{eq:ratioisocurvature}
\frac{\alpha}{1-\alpha} = \frac{{\cal P}_{\cal S}(k_0)}{{\cal P}_{\zeta}(k_0)}~,
\end{equation}
where ${\cal P}_{\cal S}(k_0)$ is the power spectrum of the entropy perturbation $S_c$ 
at the pivot scale. This parameter $\alpha$ is constrained by observations, see 
Sec.~\ref{sec:generalprop:confrontationdata} below.

 
\subsubsection{Non-Gaussianity}\label{sec:nongaussianities}
 
The inflaton inevitably has to have interactions with other scalars, fermions and 
gauge fields for a successful reheating. Furthermore, there could be more than one 
light scalar dynamics involved during and after inflation. The collective dynamics 
of more than one light field can source 
non-Gaussianity~\cite{Allen:1987vq,Falk:1992sf,Gangui:1993tt,Maldacena:2002vr,Acquaviva:2002ud,Seery:2005wm,Seery:2005gb,Enqvist:2004bk} (for a review 
see~\cite{Bartolo:2004if,Lyth-new-book}).  The non-Gaussianity can also be generated by invoking initial
conditions which depart from Bunch-Davies vacuum~\cite{Collins:2009pf}, non-canonical kinetic term~\cite{Alishahiha:2004eh,Huang:2007hh}, breaking slow-roll conditions abruptly for a brief period~\cite{Kofman:1991qx,Chen:2008wn}, multi-field inflationary models~\cite{Seery:2005gb}, curvaton 
scenarios~\cite{Lyth:2001nq,Lyth:2002my} and large non-Gaussianity
during prehating~\cite{Jokinen:2005by,Enqvist:2004ey,Enqvist:2005qu,Enqvist:2005nc}. In~\cite{Kohri:2009ac}, non-Gaussianity
during preheating has been found less significant when studied in the context of  $\delta N$ formalism. The simplest form for the 
local non-Gaussianity can be written as~\cite{Salopek:1990jq,Sasaki:1995aw,Sasaki:1998ug,Lyth:2004gb}: 
\begin{equation}\label{ng}
{\zeta}(x)\equiv g(x)+\frac{3}{5}f_{NL}~g^2(x)+\frac{9}{25}g_{NL}~g^3(x)+\dots \,,
\end{equation}
where $g(x)$ is the Gaussian random fluctuations. In general, the non-Gaussianity can be calculated by studying the
bispectrum (three point correlator $\langle g_{k_1},g_{k_2},g_{k_3}\rangle \neq 0$) and the trispectrum 
(four point correlator $\langle g_{k_1},g_{k_2},g_{k_3},g_{k_4}\rangle \neq 0$)~\footnote{Assuming that $\zeta$ is constant and 
dominated by the Gaussian perturbations, $g$, the power spectrum ${\cal P}_{\zeta}$ is determined by, $\langle \zeta_{\vec k_1} \zeta_{\vec k_2} \rangle =(2\pi)^3 {\cal P}_\zeta (k_1)\delta ({\vec k_1}+{\vec k_2})$, the bispectrum ${\cal B}_{\zeta}$ is determined by~$\langle \zeta_{\vec k_1} \zeta_{\vec k_2} \zeta_{\vec k_3} \rangle={(2\pi)}^3 {\cal B}_\zeta (k_1,k_2,k_3) \delta ({\vec k_1}+{\vec k_2}+{\vec k_3})$, and the trispectrum ${\cal T}_{\zeta}$ is given by $\langle \zeta_{\vec k_1} \zeta_{\vec k_2} \zeta_{\vec k_3} \rangle={(2\pi)}^3 {\cal B}_\zeta (k_1,k_2,k_3) \delta ({\vec k_1}+{\vec k_2}+{\vec k_3})$, where ${\cal B}_\zeta$ and ${\cal T}_\zeta$ can be written as: ${\cal B}_\zeta (k_1,k_2,k_3)=({6}/{5}) f_{\rm NL}\left({\cal P}_\zeta (k_1) {\cal P}\zeta (k_2)+ {\cal P}_\zeta (k_2) {\cal P}_\zeta (k_3)+ {\cal P}_\zeta (k_3) {\cal P}_\zeta (k_1)\right)$, and ${\cal T}_\zeta (k_1,k_2,k_3,k_4)=\tau_{\rm NL} \left({\cal P}_\zeta(k_{13}) {\cal P}_\zeta (k_3) {\cal P}_\zeta (k_4)+11~{\rm permutations.}\right) + ({54}/{25}) g_{\rm NL}\left( {\cal P}_\zeta (k_2) {\cal P}_\zeta (k_3) {\cal P}_\zeta (k_4)+3~{\rm permutations.} \right).$  Here $f_{\rm NL},~\tau_{\rm NL}$ and $g_{\rm NL}$ are non-linearity
parameters where $\tau_{\rm NL} =({36}/{25})f_{\rm NL}^2$, see for instance~\cite{Lyth:2005du,Lyth:2005fi}.}.


The $\delta N$ formalism developed in Refs.~\cite{Sasaki:1995aw,Lyth:2005du,Lyth:2005fi} provides a powerful tool to study 
the non-Gaussianity. It assumes that light fields contribute to the local evolution of the number of 
e-foldings~\cite{Wands:2000dp,Sasaki:1995aw,Salopek:1995vw,Lyth:2004gb}:
\begin{equation}
{\zeta}(x,t)=\delta N(\phi_1(x),\phi_2(x),...,t)\equiv
N(\phi_1(x),\phi_2(x),...,t)-N(\phi_1,\phi_2,...,t)\,,
\end{equation}
where $N(x,t)$ is the number of e-foldings of expansion starting from an initial flat slice ending
to a slice of uniform density. For instance, upto the first order in field perturbations, 
${\zeta}(x,t)=\sum N_{i}(t)\delta\phi_{i}(x)$ leads to ${\cal P}_{\zeta}=(H_{k}/2\pi)^2\sum N_{i}^2(k)$.
For a single field~\cite{Lyth:2005du,Lyth:2005fi}, 
\begin{equation}
{\zeta}=N^{\prime}\delta\phi+\frac{1}{2}N^{\prime\prime}(\delta\phi)^2=N^{\prime}\delta\phi+
\frac{1}{2}\frac{N^{\prime\prime}}{N^{\prime 2}}(N^{\prime}\delta \phi)^2\,.
\end{equation}
where $\prime\equiv \delta/\delta\phi$ and $(3/5)f_{NL}=(1/2)(N^{\prime\prime}/N^{\prime2})$. Given the fact that
$N^{\prime}=H(t)/\dot\phi$, we can evaluate $N^{\prime},~N^{\prime\prime}$ in terms of the slow-roll parameters,
which yields~\cite{Maldacena:2002vr}
\begin{equation}
\label{fNLMalda}
\frac{3}{5}f_{NL}=\frac{\eta-2\epsilon}{2}\,.
\end{equation}
The value of $f_{NL}$ in the case of slow-roll inflation is always bounded by the slow-roll parameters. During 
inflation these parameters $\epsilon,~\eta \ll 1$, therefore non-Gaussianity is negligible. Similar conclusion holds for
more that one fields during inflation~\cite{Seery:2005gb,Kim:2006ys}. 


\subsection{Curvaton and fluctuating inflaton coupling/mass scenarios}

The curvaton paradigm involves at least two fields, the inflaton and a light field curvaton, which are 
not coupled to each other, we will discuss a slightly variant scenario when the fields have coupling.
It is essential that (1) the curvature perturbations created by the inflaton are negligible compared to
the total curvature perturbations, (2) the curvaton field is very light during inflation therefore, 
it obtains random fluctuations of order $H_{inf}/2\pi$, and (3) the curvaton oscillations dominates the universe and 
its decay generates the total curvature perturbations~\cite{Lyth:2001nq,Lyth:2002my,Enqvist:2001zp,Moroi:2001ct,Moroi:2002rd}, see also \cite{Linde:1996gt}. 

Let us assume a curvaton field, $\sigma$, whose equation of motion and the perturbations read as:
\begin{equation}
\ddot\sigma+3H\dot\sigma+V_{\sigma}=0\,,~~~~~~~~~~~~~
\ddot\delta\sigma_{k}+3H\dot\delta\sigma_k+((k/a)^2+V_{\sigma\sigma})\delta\sigma_k=0\,.
\end{equation}
where $V_{\sigma\sigma}\ll H^2$ during inflation, therefore the VEV is $\sigma\approx \sigma_{\ast}$
nearly a constant. The perturbations in$\sigma$ field is given by: 
$\delta\sigma/\sigma \sim (H_{inf}/2\pi\sigma_*)$  for $H_{inf}\ll \sigma_{\ast}$, therefore 
${\cal P}^{1/2}_{\delta\sigma/\sigma}\sim (H_{\inf}/ 2\pi\sigma_*)$.  It is assumed that the curvaton field
rolls slowly as the universe becomes radiation dominated after the inflaton decay.

On large scales the curvature perturbations are given by: $\zeta=-H\delta t=-H\delta\rho/\dot\rho$, 
$\zeta_r=(1/4)\delta\rho_r/\rho_r$ and  $\zeta_\sigma=(1/3)\delta\rho_{\sigma}/\rho_{\sigma}\equiv (1/3)\delta_{\sigma}$, they all
evolve independently~\cite{Wands:2000dp}. The value of $\zeta_\sigma$ has been calculated assuming that the curvaton is 
oscillating with a pressureless equation of state. During these oscillations the curvaton converts its 
fluctuations into the curvature perturbations. The total curvature perturbations is given by~\cite{Lyth:2001nq,Lyth:2002my}:
\begin{equation}
\zeta=\frac{4\rho_r\zeta_r+3\rho_\sigma\zeta_{\sigma}}{4\rho_r+3\rho_\sigma}\,.
\end{equation}
Since prior to the curvaton oscillations, the curvature perturbations in the universe is dominated by that of 
the inflaton decay products, i.e. radiation, therefore, $\zeta=\zeta_r$, which simplifies the above expression:
\begin{equation}
\zeta=\frac{\rho_\sigma}{4\rho_r+3\rho_\sigma}\delta_\sigma\,.
\end{equation}
If the curvaton energy density dominates over radiation, then $\zeta=(1/3)\delta_\sigma$, otherwise, the fraction,
$r\sim \rho_{\sigma}/\rho_r<1$, would signify the curvaton energy density at the time of decay. In which case 
$\zeta=(1/4)r\delta_{\sigma}$, and
\begin{equation}
{\cal P}_{\zeta}\approx r^2\left(\frac{H_{\inf}}{2\pi\sigma_*}\right)^2\,,
\end{equation}
and the spectral tilt is given by
\begin{equation}
n_s\equiv 1-6\epsilon+2\eta = 1+\frac{\dot H_{\inf}}{H^2_{inf}}+\frac{2}{3}\frac{V_{\sigma\sigma}}{H_{inf}^2}\,.
\end{equation}
Since $\dot H_{inf}/H^2_{inf},~V_{\sigma\sigma}/H^2_{inf}\ll 1$, the spectral tilt is fairly close to one. The coherent
oscillations of the curvaton generates non-Gaussian perturbations. A small perturbations around the minimum leads to
\begin{equation}
\frac{\delta\rho_{\sigma}}{\rho_{\sigma}}=2\frac{\delta\sigma}{\sigma}+\frac{(\delta \sigma)^2}{\sigma^2}\,,
\end{equation}
averaged over many-many oscillations over Hubble period. In some realistic curvaton 
scenarios, the curvaton may 
decay {\it almost instantly} via SM gauge couplings in less than one Hubble time 
scale. In which case the curvaton oscillations may not last long enough to generate 
any significant non-Gaussianity.

By using $\zeta=(1/3)\delta_\sigma$ and Eq.~(\ref{ng}), the non-Gaussianity parameter 
can be determined by (for a quadratic potential) for $f_{NL}\gg 1$~\cite{Lyth:2002my}~\footnote{For 
a departure from quadratic potential, the form of $f_{NL}$ modifies by Eq.~(\ref{modeqfNL}) (see Sec.~\ref{See: Sec.CurvatonandNG}). For values of $f_{NL}\sim {\cal O}(1)$, one should employ the $\delta N$ formalism
\cite{Lyth:2005du,Lyth:2005fi} or equivalently the second order perturbation theory, for a review see~\cite{Bartolo:2004if}.}:
\begin{equation}
\label{fNLcurvaton}
f_{NL}=\frac{5}{4r}\,.
\end{equation}

Another interesting proposal is that the perturbations could be
generated from the fluctuations of the inflaton coupling to the
SM degrees of 
freedom~\cite{Kofman:2003nx,Dvali:2003em,Dvali:2003ar,Enqvist:2003uk,Mazumdar:2003iy}. 
It has been argued that the coupling strength of the inflaton to ordinary matter
or the inflaton mass, instead of being a constant, could depend on the
VEV of various fields in the
theory. If these fields are light during inflation their quantum
fluctuations will lead to spatial fluctuations in the inflaton decay
rate. As a consequence, when the inflaton decays, adiabatic density
perturbations will be created because fluctuations in the decay rate
translate into fluctuations in the reheating temperature.

This can be understood intuitively from the fact that fluctuations in
the inflaton decay rate leads to fluctuations in the reheat
temperature of the universe, given by 
$T_{\rm rh}\sim \lambda\sqrt{m_{\phi}M_{\rm P}}$, where $m_{\phi}$ is 
the mass of the inflaton. The
fluctuations in the decay rate, $\Gamma\sim m_\phi\lambda^2$ can be 
translated into fluctuations in the energy density of a thermal bath with 
$\delta \rho_{\gamma}/\rho_{\gamma}=-(2/3)\delta\Gamma/\Gamma$~\cite{Dvali:2003em,Dvali:2003ar}. 
The factor $2/3$ appears due to red-shift of the modes during the decay of the
inflaton whose coherent oscillations still dominates the energy density of the universe. 
The inflaton decay rate fluctuates if either $\lambda$ or $m_\phi$
is a function of a fluctuating light field.

The fluctuation in the decay rate for the various cases is given by:
\begin{equation}
\label{decay}
\frac{\delta \Gamma}{\Gamma} = \left \{ 
\begin{array}{llll} 
& 2\frac{\delta S}{M} 
& =\frac{H_{inf}}{\pi M}, & \qquad {\rm direct~decay}.
\\
& 2 \frac{\delta S}{S} 
& =\frac{H_{inf}}{\pi S}, &\qquad {\rm indirect~decay}.
\\
& \frac{\delta S}{S}
&=\frac{H_{inf}}{2\pi S}, & \qquad {\rm fluctuating~mass}.
\end{array}
\right.  \end{equation}
Various examples do predict non-Gaussianity within a range of $f_{NL}\sim 5$~\cite{Dvali:2003em,Dvali:2003ar}.


\subsection{Confrontation to the CMB and other observational data}\label{sec:generalprop:confrontationdata}

The CMB data is currently providing (including WMAP, 
CBI~\footnote{See \texttt{http://www.sciops.esa.int/index.php?project=PLANCK\&page=index}}, 
VSA~\footnote{See \texttt{http://www.mrao.cam.ac.uk/telescopes/vsa/}}, 
ACBAR~\footnote{See \texttt{http://cosmology.berkeley.edu/group/swlh/acbar/}}, 
Boomerang~\footnote{See \texttt{http://www.astro.caltech.edu/$~$lgg/boomerang\_front.htm}}) and will 
provide (including PLANCK) stringent observational data to constrain the power spectrum of density fluctuations. There are 
other data set  which can be used in conjunction; type Ia supernovae (SN), Baryon 
Acoustic Oscillations (BAO), large scale structures, Lyman-$\alpha$ forest, etc. 
In this section, we briefly review the current bounds on the amplitude of the power spectrum, 
spectral index, and tensor to scalar ratio, running of the spectral index, non-Gaussianities, cosmic 
strings, and isocurvature perturbations. 


\subsubsection{Primordial power spectrum for scalar and tensor}

Most of the observational tests of inflation models  arise from
the 2 point correlation function, related to the power spectrum of the 
primordial perturbations, both for scalar and tensor perturbations.
The most recent update on the WMAP results, combined with SN and BAO data 
confirmed that so far the minimal $6$-parameter $\Lambda$CDM model provides 
a very good fit of the combined observations. It contains
the baryon, CDM, and dark energy fractions; $\Omega_b h^2$, 
$\Omega_c h^2$, $\Omega_\Lambda$, the spectral index $n_s$, the optical depth 
of reionization, $\tau_{\rm reion}$, and the normalization of the power-spectrum
$\mathcal{P}_\zeta(k_0)$, with central values and $1\sigma$ error bars for the inflation related 
parameters given by~\cite{Komatsu:2008hk}:
\begin{equation}
\Delta_\mathcal{R}^2(k_0)=(2.445\pm 0.096)\times 10^{-9}\;\mathrm{at}\;k_0= 
0.002 \mathrm{Mpc}^{-1}~,\quad n_s(k_0)=0.960\pm 0.13~.
\end{equation}
In this minimal model, the primordial power spectrum is approximated by the 
expression of Eq.~(\ref{eq:primordialpowerspectrum}) and the tensor 
contribution or the $k-$dependence of the spectral index are neglected~\footnote{The confrontation 
to the data presented here relies on the slow-roll conditions. Alternatively, as proposed by several teams, one can  
reconstruct the primordial power spectrum~\cite{Spergel:2006hy,Verde:2008zza} as 
well as the inflationary potential \cite{Lesgourgues:2007gp,Powell:2007gu}
(see also~\cite{Komatsu:2008hk} and Refs. therein). These approaches are limited 
as only a small range in $\phi\in [\phi_Q,~\phi_e]$ is observable and therefore 
accessible to this reconstruction.}.
It is important to stress that these central values and error bars vary 
significantly when other parameters are introduced to fit the data, in part 
because of degeneracies between parameters (in particular $n_s$ with 
$\Omega_b h^2$, the optical depth $\tau$, its running, the tensor-to-scalar ratio, $r$, and the 
fraction of cosmic strings). 

If the tensor-to-scalar ratio $r$ and/or a running $\alpha_s$ are 
introduced, the best fit and error bars~\footnote{Note that the
results vary significantly when WMAP data only or combined observations are used, 
see Ref.~\cite{Komatsu:2008hk} for details.} (at $1\sigma$)~\cite{Komatsu:2008hk}
\begin{equation}
\begin{split}
n_s&=1.017^{+0.042}_{-0.043}~,\quad \alpha_s=-0.028\pm 0.020~,\\
n_s&=0.970\pm 0.015~,\quad r< 0.22~~(\mathrm{at}~2\sigma) ~,\\
n_s&=1.089^{+0.070}_{-0.068}~,\quad r< 0.55~~(\mathrm{at}~2\sigma)~, \quad 
\alpha_s=-0.053 \pm 0.028~.
\end{split}
\end{equation}
These data therefore suggest that a red spectrum is favored ($n_s=1$ 
excluded at $2.5\sigma$ from WMAP and at $3.1\sigma$ when other data sets are included) 
if there is no running. Although, the reader and the model builder should keep in mind 
that a lot more data is necessary before these results can be used as bench-mark points 
(see for e.g.~\cite{Trotta:2008qt} for a pedagogical presentation about Bayesian model 
selection).

\begin{figure}[h!] 
\scalebox{0.8}{\includegraphics{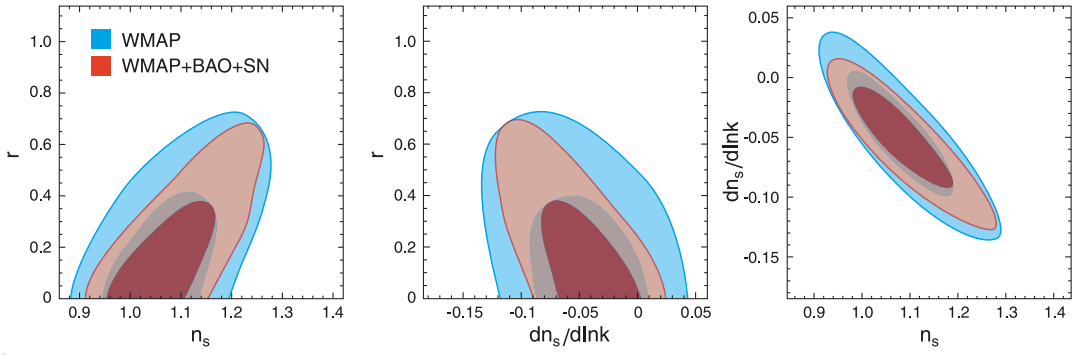}}
\caption{\label{fig:MCMCWMAP}
\small Two-dimensional likelihood for the spectral index, the running and the ratio 
tensor/scalar, from the WMAP data only (blue) and the WMAP data combined with the BAO 
and supernovae data sets (red) at $1\sigma$ and $2\sigma$. 
Figures are taken from~\cite{Komatsu:2008hk}.}
\end{figure}

These various best fit values  are consistent with                      
a model that predicts a non-negligible level of tensor or a running of the 
spectral index. These results are summarized in Fig.~\ref{fig:MCMCWMAP}.


The confrontation of inflationary models to data can also be done 
by directly constraining the parameters of the potential for each 
model~\cite{Martin:2006rs} (see also~\cite{Ringeval:2007am} for a pedagogical 
review). These methods have the advantage of not relying on a generic 
parameterization of the power spectrum/potential, or the slow-roll conditions, 
as some models violate those assumptions temporarily~\cite{Clesse:2008pf} 
or constantly~\cite{Linde:2001ae}. It is one of the best methods to carry a bayesian 
analysis for a model selection based on the data. Their disadvantage is that they require to treat 
each model individually and cannot provide ways to use current constraints to build new models.


\subsubsection{Cosmic strings and CMB fluctuations}

\begin{figure}[h!] 
\scalebox{0.35}{\includegraphics{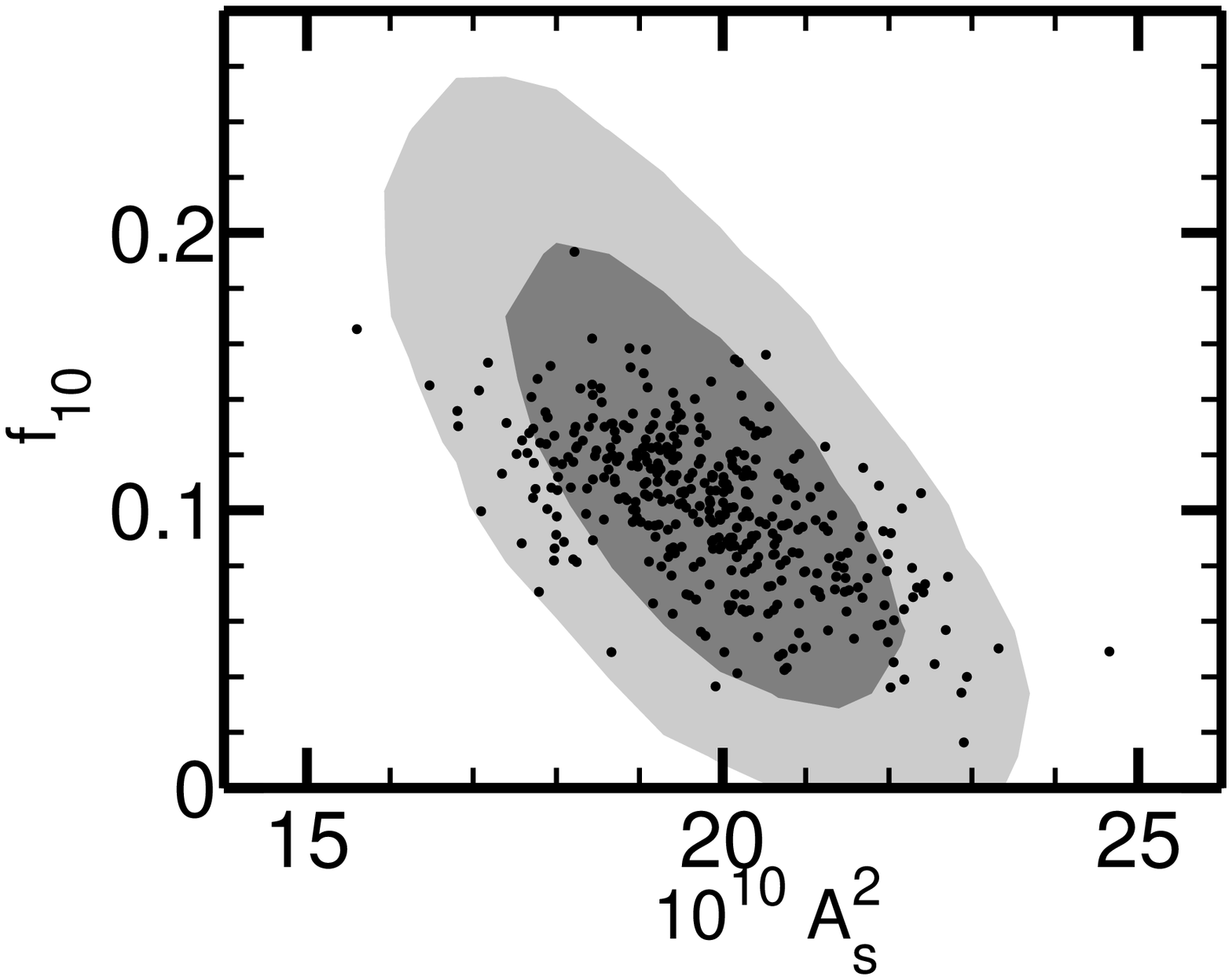}}
\scalebox{0.35}{\includegraphics{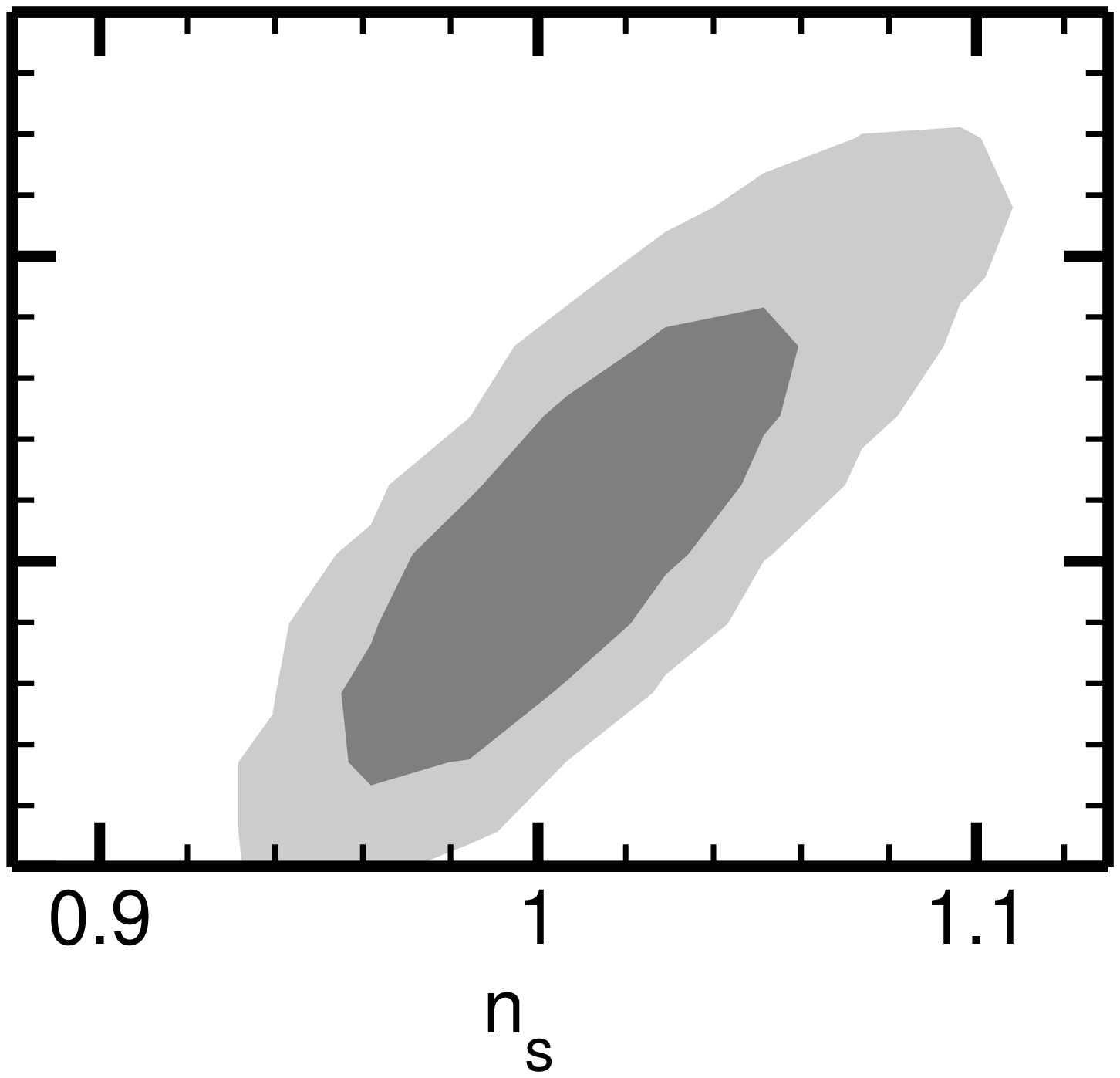}}
\caption{\label{fig:MCMCwithStrings}
\small Two-dimensional likelihood between the fraction of cosmic strings $f_{10}$ 
and the normalization and tilt of the power spectrum. Figures are taken from~\cite{Bevis:2007gh}.}
\end{figure}

As we shall also see below (at Sec.~\ref{sec:modelsinflation}) that several models of 
inflation can produce cosmic strings (see Sec.~\ref{sec:partphys:symmbreak}). This 
has important consequences when confronting the model to the data as some 
degeneracy has been observed between the spectral index and the fraction of cosmic 
strings responsible for the temperature anisotropies at the 10th multipole $f_{10}$~\cite{Bevis:2007gh}. 
The 2D likelihood function when $f_{10}$ is included is represented in Fig.~\ref{fig:MCMCwithStrings}. 


When a certain fraction of cosmic strings is present, the spectral index best fit and 
error bars - from WMAP 3years data - are shifted and enlarged, and (at $1\sigma$) 
we read approximately:
\begin{equation}
n_s\simeq 1.01\pm 0.05~,\quad f_{10}\simeq 0.11\pm 0.9~.
\end{equation}
Once other data sets are taken into account (BBN, large scale structures), the current 
data can only put an upper constraint on the fraction of cosmic 
strings $f_{10}<0.11$ at $2\sigma$~\cite{Bevis:2007gh}. This constraint should be 
improved by the future PLANCK data, both at large multipole and from confrontation to 
the polarized data, notably 
the B-modes. Luckily, if cosmic string are present then they should contribute to 
them~\cite{Seljak:2006hi,Pogosian:2007gi,Bevis:2007qz}, 
and their fraction is not degenerate with the primordial tensor signal from 
inflation~\cite{Urrestilla:2008jv}.


The current CMB fluctuations generated from strings involve the simplest 
Nambu-Goto strings, the presence of currents inside cosmic strings 
can possibly 
affect their precise signature in CMB~\cite{Davis:1997bs}. More generally, the properties of cosmic strings 
arising from non-SUSY theories, from SUSY $F$-and $D$-term or $N=2$ SUSY P-term hybrid inflation, 
or from brane inflation are different. By adding only 1 parameter to fit the data 
($f_{10}$) and confronting all models to the posterior probabilities might not be the 
best strategy in such a case (see Sec.~\ref{sec:partphys:symmbreak}).


\subsubsection{Isocurvature perturbations}

The isocurvature perturbations measure the deviation from the adiabaticity of the primordial 
fluctuations, denoted by the quantity $S$ (in the context of cold dark matter in Eq.~(\ref{cdm}), it was denoted by $S_c$). 
The isocurvature perturbations arise if there are light scalar fields fluctuating during inflation, which do not thermalize with the inflaton 
decay products, i.e. the SM degrees of freedom, after the end of inflation. Usually this deviation is measured by the parameter $\alpha$ 
related to the ratio between the entropy power spectrum, $\mathcal{P}_S$, over the 
curvature perturbation, $\mathcal{P}_\zeta$, via the Eq.~(\ref{eq:ratioisocurvature}). There could be some 
correlations between $S$ and $\zeta$, the parameter:
\begin{equation}
\beta(k_0)\equiv -\frac{\mathcal{P}_{S,\zeta}}{\sqrt{\mathcal{P}_{S}(k_0)\mathcal{P}_{\zeta}(k_0)}}\,,
\end{equation}
where $\mathcal{P}_{S,\zeta}$ is the cross-correlated power spectrum between $S$ 
and $\zeta$, distinguish between the totally correlated case ($\beta=-1$) and the 
totally anti-correlated case ($\beta=0$). Two parameters $\alpha_{-1}$ (for $\beta=-1$) and $\alpha_0$ (for $\beta=0$)
are commonly used to describe each case, which are typically encountered in the curvaton
scenario and in the axion dark matter  scenario, respectively.

The most recent observations from WMAP 5-years data lead to the values for $\alpha_{-1}$ 
and $\alpha_0$ compatible with zero, and respectively, slightly and strongly 
degenerate with the spectral index~\cite{Komatsu:2008hk}. Marginalizing over other 
parameters, it was found that, at $2\sigma$:
\begin{equation}
\alpha_{-1} < 0.0041~,  \quad \quad \quad \quad \alpha_0 < 0.072~, 
\end{equation}
when the WMAP data were combined with BAO and SN data. These constraints suggest 
that the deviation from $S=0$ is smaller than $2.1\%$ and $8.9\%$ respectively at $95\%$ confidence level.


\subsubsection{Higher order correlation functions}

Higher order correlations, such as bispectrum ${\cal B}_{\zeta}$ and trispectrum ${\cal T}_{\zeta}$ 
(defined in Sec.~\ref{sec:nongaussianities}) can also
constrain the inflationary dynamics and the interactions, (see \cite{Bartolo:2004if} for a review). 
The amount of non-gaussianities has been recently constrained by the WMAP 
data~\cite{Komatsu:2008hk}. As pointed out in Eq.~(\ref{fNLMalda}), even a single field inflation model 
in a slow-roll regime generates small non-Gaussianities at the level,  
$f_{NL} \sim \epsilon,\eta \sim 10^{-2}$~\cite{Salopek:1990jq,Maldacena:2002vr}, though the 
current limits are around four orders of magnitude above this level. 

The first constraint on {\it full} bispectrum was computed for the COBE data~\cite{Komatsu:2001wu} 
\begin{equation}\label{eq:constraintsfnlCOBE}
-3500 \leq f_{NL} \leq 2000~,~~~~~~ \quad \mathrm{at}~~ \quad 2\sigma~,
\end{equation}
but due to the computational cost the WMAP employ the ``KSW estimator''~\cite{Komatsu:2003iq} 
that combine {\it squeezed} triangular configurations in the harmonic space to construct an optimal estimator 
for $f_{NL}$. For example inflation, curvaton, and preheating induced non-Gaussianity belong 
to this category, where $\zeta$ and $\zeta^2$ are evaluated at the same 
location in space~\cite{Komatsu:2008hk}. In addition, the {\it equilateral non-linear coupling} parameter 
$f_{NL~eq}$ provides a complementary description of the bispectrum, combining triangular 
configurations in the harmonic space that are equilateral~\footnote{Indeed, models can generate a 
large amount of non-Gaussianities only in one configuration. Note that 
several other configurations and statistics to estimate $f_{NL}$ have been proposed and 
searched for, such as the point-source bispectrum ${\cal B}_{src}$ measured in the WMAP 
5-years data~\cite{Komatsu:2008hk}, the flattened triangle configurations from models with departure from 
the Bunch-Davies initial condition~\cite{Chen:2006nt}. }.

The constraints from the most recent observations are currently given by WMAP 5-years 
data~\cite{Komatsu:2008hk}~\footnote{Note that there is no known constraint on the
trispectrum parameter, $\tau_{NL}$. It is expected to be of order $\tau_{\rm NL}\sim (f_{NL})^2$.}:
\begin{equation}\label{eq:constraintsfnl} 
\begin{split}
-9 < f_{NL} < 115~,\quad~\mathrm{at} \quad 2\sigma~,\quad\quad\quad\quad
-151 < f_{NL~eq} < 253~, \quad~ \mathrm{at} \quad 2\sigma~,
\end{split}
\end{equation}
meaning that there is a hint in favor of a non-vanishing positive $f_{NL}$. The lower 
bound on $f_{NL}$ is even raised above zero when the bispectrum maps are less restrictive. 
This is the origin of the discrepancy between these results and the prior claimed 
detection of non-Gaussianities from~\cite{Yadav:2007yy} using the WMAP 3-years data. 
Note that using another statistics (Minkowski functional) has lead so far a negatively preferred 
value for $f_{NL} \sim -60 \pm 60$. This discrepancy has not yet been fully resolved~\cite{Komatsu:2008hk}.
The future observations from PLANCK and the galaxy distribution should be able to 
constrain a deviation up to $|f_{NL}|\gtrsim 5$~\cite{Komatsu:2001rj,Komatsu:2009kd,Seljak:2008xr,Slosar:2008hx}.




\subsection{Dynamical challenges for inflation}\label{sec:generalprop:dynamicchalleng}

Inflation has several dynamical challenges which have been discussed in the literature.

\subsubsection{Initial conditions for inflation}

The question of initial condition is a worrisome factor. universe could have started either cold or hot. 
Whether universe began hot or cold, once vacuum energy density takes over it would always  
yield a cold universe. However there are nontrivial initial conditions to be satisfied.

\begin{itemize}

\item{Homogeneity problem:\\
In an Einstein gravity inflation does not solve the homogeneity problem, instead inflation requires an initial patch of the 
universe, $r$, to be sufficiently homogeneous on scales larger than the Hubble patch, $r\gg H^{-1}$,
before inflation could begin~see Refs.~\cite{Goldwirth:1989pr,Goldwirth:1991rj,Albrecht:1986pi,Brandenberger:1990wu,Calzetta:1992bp,Vachaspati:1998dy}. Initial conditions if set at the Planckian scale do not suffer through this problem as shown by
Refs.~\cite{Linde:1983gd,Linde:1985ub,Linde:1996hg}. Low scale models of inflation require earlier phases of
inflation in order to set the initial conditions.}

\item{Chaotic initial conditions:\\
For sufficiently flat potential the only constraint is given by: $(1/2)\dot\phi^2+(1/2)(\partial_{i}\phi)^2+V(\phi)\leq M_{\rm P}^4$, see
Refs.~\cite{Linde:1983gd,Linde:1985ub,Linde:1996hg,Vilenkin:1983xq}.  The initial conditions are set by: 
$(1/2)\dot\phi^2 \sim (1/2)(\partial_{i}\phi)^2\sim V(\phi)\sim {\cal O}(M_{\rm P})$. If by any chance 
$(1/2)\dot\phi^2+(1/2)(\partial_{i}\phi)^2\leq V(\phi)$ in a particular domain, the inflation begins and within a 
Planck time the potential energy density, $V(\phi)$, starts dominating over kinetic term. In domains where
$(1/2)\dot\phi^2+(1/2)(\partial_{i}\phi)^2> V(\phi)$, inflation does not take place and do not exist classically. 
The above mentioned conditions are naturally satisfied when $\phi\geq M_{\rm P}$ for a simple chaotic
type potential, $V\sim m^2\phi^2$, where there exists a window, 
${\cal O}(100-10) M_{\rm P}>\phi > {\cal O}(M_{\rm P})$, where universe enters a  process of eternal 
self-reproduction~\cite{Linde:1985ub,Linde:1996hg}. In the self-production regime new regions 
of $H_{inf}^{-1}$ prop up on a timescale of one e-folding with field values $\sim \phi\pm \Delta \phi/2$, where 
$\Delta \phi\sim H_{inf}/2\pi$. In such regions quantum fluctuations dominate over the classical slow-roll of the field. After
few e-foldings, ${N}$, these regions are locally homogeneous and grow almost independently. The correlation 
between the two regions $\langle \phi+\Delta \phi/2,\phi-\Delta\phi/2\rangle\sim e^{-{N}}$, die exponentially.  Such 
self-reproduction regions in the inflationary potential can solve the initial homogeneity problem without any trouble.
This is also known as eternal inflation~\footnote{Note that inflation is not past eternal. The first argument is
based on singularity theorems due to Hawking-Penrose~\cite{Borde:1993xh}. The large stochastic fluctuations
cannot take the field forever up the potential. When the field fluctuations become non-linear in one Hubble region, the 
entire region would collapse to form a blackhole. The second reason is  due to~\cite{Borde:2001nh}, where it 
has been argued that null and time like geodesics are past-incomplete during inflation as long as the
averaged expansion rate is such that $H_{av}>0$ holds along the past directed geodesics. The observers along these geodesics would take finite amount of time to hit the singularity. }.

Also note that chaotic initial conditions can be obtained for low scale models of inflation provided the potential
is extremely flat. For instance, near the saddle point $\phi_0$ of a potential, $V^{\prime}(\phi_0)=0,~V^{\prime\prime}(\phi_0)=0$,
the quantum fluctuations will dominate over classical motion in a range $\Delta \phi\ll M_{\rm P}$. 
Such regions will support self-reproduction of space-time with locally homogeneous regions. However as we shall
argue that in order to reach a plateau of such a low scale inflationary potential one requires stochastic 
jumps of the inflaton field during a prior phase of inflation~\cite{Allahverdi:2007wh,Allahverdi:2008bt}.

Furthermore, if inflation is driven by a collection of scalar fields as in the case of {\it assisted inflation}, then the initial 
condition problem for a single field chaotic inflation model can be ameliorated, as one would not require super-Planckian VEVs
for the inflatons~\cite{Kanti:1999ie,Jokinen:2004bp,Dimopoulos:2005ac}.}

\item{Problems with a large VEV:\\
A natural question arises for many models: can we trust the effective field theory treatment of an inflaton potential 
when the VEV of the field is super-Planckian.  The answer is no, in particular when the inflaton 
has couplings to the SM or MSSM degrees of freedom. An effective field theory treatment is trustable {\it only}
when the momentum is bounded by the cut-off as well as the total energy density is below the Planckian
energy density.  For any renormalizable coupling between the inflaton and any matter field would render  
super-Planckian VEV dependent mass to the matter field ( for a reasonable gauge or Yukawa coupling $\leq {\cal O}(1)$).
The effective field theory prescription is bound to break down when super-heavy quanta is running in the loops 
of an inflaton field. For a gauge invariant inflaton, i.e. MSSM inflaton, it is impossible to consider VEVs above the 
cut-off, as the inflaton has SM gauge interactions. Similarly, embedding inflation in SUGRA or in string theory 
will always provide inflaton VEV below the Planck scale, this remains true for any realistic  potential arising from 
the low energy effective theory~\cite{Burgess:2003jk,Weinberg:2008hq,Burgess:2009ea}. In a limit when the 
inflaton coupling to matter vanishes, or in a free field theory case, it is possible to obtain VEVs above the 
cut-off maintaing the effective field theory arguments given in Refs.~\cite{Linde:1983gd,Linde:1985ub,Linde:1996hg}.

}

\item{Quantum initial conditions:\\
If inflation lasts long enough, i.e. $N\sim 60-70$ e-foldings, then it is inevitable that the present {\it observable}  
mode would originate from sub-Planckian length scales. Potentially quantum gravity corrections at those length scales 
can leave some imprint in the CMB perturbations, it is  known as the trans-Planckian problem for an inflationary cosmology
\cite{Brandenberger:2000wr,Martin:2000xs,Danielsson:2002kx,Kaloper:2002cs,Martin:2003kp,Collins:2005cm}.
There are two pertinent questions, the first one is related to the choice 
of Bunch-Davies vacuum as an initial state~\cite{Bunch:1977sq,Bunch:1978yq} in order to evolve the quantum perturbations. 
Second one has to do with an adiabatic evolution of the state throughout the dynamics of inflation. Both the questions have
been raised in the literature. It was observed that if either the vacuum or the evolution of a state would violate Lorentz-invariance 
the corrections to the amplitude of the CMB perturbations would be as large as order one. Typically the corrections will be 
$\propto (H_{inf}/M_{\ast})$, where $M_{\ast}$ is the cut-off above which either the initial state is modified or the evolution.
In Ref.~\cite{Danielsson:2002kx}, the initial state was chosen to be alpha-vacuum (a variant of Bunch-Davies vacuum with large excitations) and the modification to the amplitude of the CMB perturbations were found to be $(H_{inf}/M_{\ast})^2$. It was argued in~\cite{Burgess:2002ub}
that as long as the state evolves adiabatically and Lorentz-invariance is maintained, the trans-Planckian corrections would be small of order $\sim(H_{inf}/M_{\ast})^2$, which is a good news for any low scale inflation. The quantum corrections to the inflaton potential
arising from trace-anomaly and light scalars yield similar corrections to the CMB perturbations~\cite{Boyanovsky:2005px,Boyanovsky:2006qi}. }

\end{itemize}



\subsubsection{Choice of a vacuum where inflation ends}

In order to realize our observable universe, inflation must come to an end in the right vacuum. The exit must happen such that 
the relevant degrees of freedom required for the BBN, i.e. the relativistic SM degrees of freedom, and right abundance for 
the cold dark matter can be excited. There are only few models where inflation can 
end right in the SM vacuum; the SM Higgs inflation~\cite{Bezrukov:2007ep}, and the MSSM inflation~\cite{Allahverdi:2006iq}. 
In many particle physics models of inflation, the existence of a hidden sector coupling to the MSSM or the SM sector is 
common, see~\cite{Lyth:1998xn}. All these models require extra set of assumptions in order to make sure that 
the inflaton energy density gets transferred into the MSSM or the SM degrees of freedom.  Note that a hidden sector 
inflaton can excite hidden degrees of freedom as the couplings between the two hidden sectors are not barred by any symmetry.
Furthermore, gravity will always couple one such sector to the another. Therefore, it is desirable to end inflation where one can directly
excite the SM quarks and leptons~\footnote{Suppose inflation ends in a GUT vacuum, it does not guarantee automatically that the GUT would be broken down to the SM vacuum. There are many ways to break it and one requires special care in realizing
such a scenario. See Sec. \ref{sec:partphys:SUSYGUTs}}.

In the case of stringy models, there exists no construction where inflation ends right in the MSSM or the SM sector. 
The problem becomes more challenging  with an introduction of a string landscape, since there are nearly
$10^{500}$ to even $10^{1000}$ vacua \cite{Douglas:2004zg,Susskind:2003kw,Douglas:2006es},
with the vast majority of those having large cosmological constants. In such cases exiting inflation from the 
string landscape and exiting the inflation in our own vacuum  becomes even more challenging task~\cite{Allahverdi:2007wh}.


\subsubsection{Quantum to classical transition}

The initial sub-Hubble perturbations are quantum in nature. 
The perturbations are then stretched outside the Hubble patch during inflation, therefore the 
correlation function, $\langle \delta\phi(x)\delta\phi(x^{\prime})\rangle$,  evolves during inflation. It has been shown that
the evolution is similar to that of a squeezed state~\cite{Brandenberger:1990bx,Polarski:1995jg,Lesgourgues:1996jc,Kiefer:1998qe,Martineau:2006ki,Burgess:2006jn}, the squeezing happening due to the exponential expansion. 
For very long wavelength (super Hubble) modes the quantum correlation between the two inflating regimes 
dies away exponentially by the number of e-foldings of inflation~$\sim e^{-N}$. This lends some support to this idea that two distinct Hubble patches behave for all good purposes classical~\cite{Guth:1982ec,Linde:1985ub,Linde:1996hg}. But it is still unclear whether the short wavelength modes have any role to play in decohering the long wavelength modes~\cite{Burgess:2006jn}?

The density matrix of the fluctuations within the causal horizon, 
$\rho[\delta\phi(x),\delta\phi^{\prime}(y)]=\langle \delta\phi(x)|\rho|\delta\phi^{\prime}(y)\rangle=\Psi[\delta\phi(x)]\Psi^{\ast}[\delta\phi(y)]$, evolves from pure state to a mixed state, $\rho[\delta\phi(x),\delta\phi^{\prime}(y)]=P[\delta\phi(x)]\delta[\delta\phi(x)-\delta\phi^{\prime}(y)]$, under the influence of a time dependent interaction Hamiltonian arising from (a) time dependent evolution of the inflaton, and (b) the
inflaton interactions to matter. It was pointed out in  \cite{Burgess:2006jn} that short wavelength modes can play a role in 
decohering the long wavelength modes, once reheating takes place. The thermal bath produced from the inflaton decay can act as an environment for the long wavelength modes when these modes re-enter the Hubble patch after the end of inflation.
The decoherence effects during inflation are still an open issue~\cite{Burgess:2006jn,Kiefer:2008ku,Koksma:2009wa}.


\subsubsection{Inflaton decay, reheating and thermalization}

Reheating takes place due to the perturbative decay of the inflaton
\cite{Albrecht:1982mp,Turner:1983he,Dolgov:1989us,Kolb:1988aj}.
After the end of inflation,  when $H \leq m_{\phi}$, the inflaton field
oscillates about the minimum of the potential. Averaging
over one oscillation results in pressureless equation of state where
$\langle p\rangle =\langle \dot\phi^2/2 -V(\phi)\rangle$
vanishes~\cite{Albrecht:1982mp,Turner:1983he},
so that the energy density starts evolving like matter domination (in a quadratic potential) with
$\rho_{\phi}=\rho_{i}(a_{i}/a)^{3}$ (subscript $i$ denotes the
quantities right after the end of inflation)~\footnote{For $\lambda\phi^4$ potential the coherent oscillations yield an 
effective equation of state similar to that of a radiation epoch.}. If $\Gamma_{\phi}$ represents
the {\it total} decay width of the inflaton to pairs of fermions. This releases the energy 
into the thermal bath of relativistic particles
when $H(a)=\sqrt{(1/3M_{\rm P}^2)\rho_{i}}(a_{i}/a)^{3/2}\approx \Gamma_{\phi}$.
The energy density of the thermal bath is determined by the reheat
temperature $T_{R}$, given by:
\begin{equation}
T_{R}=\left(\frac{90}{\pi^2 g_{\ast}}\right)^{1/4}\sqrt{\Gamma_{\phi}
M_{\rm P}}=0.3\left(\frac{200}{g_{\ast}}\right)^{1/4}\sqrt{\Gamma_{\phi}
M_{\rm P}}\,,
\end{equation}
where $g_{\ast}$ denotes the effective relativistic degrees of freedom in the
plasma.
However the inflaton might not decay instantaneously. In such a case
there might already exist a thermal plasma of {\it some}relativistic species
at a temperature higher than the reheat temperature already before the
end of reheating~\cite{Kolb:1988aj}. If the inflaton decays with a
rate $\Gamma_\phi$, then the instantaneous plasma temperature is found
to be~\cite{Kolb:1988aj}:
\begin{equation}
\label{instT}
T_{inst}(a)\sim \left(g_{\ast}^{-1/2}H\Gamma_{\phi}M_{\rm P}^2\right)^{1/4}\,.
\end{equation}
The temperature of the universe reaches its maximum $T_{max}$ soon after the inflaton
field starts oscillating around the minimum. Once the maximum temperature is
reached, then $\rho_{\psi} \sim a^{-3/2}$, and $T\sim a^{-3/8}$ until
reheating and thermalization is completely over. Thermalization is achieved when both {\it kinetic} and
{\it chemical} equilibrium are reached, for a review see~\cite{Allahverdi:2010xz}. 

Note that the above analysis is solely based on energetic argument. It claims that 
$T_{\rm R}\leq \rho_{inf}^{1/4}$, but ignored the microphysical aspects such as what degrees of freedom are
excited after the end of inflation. For a successful cosmology one needs to ask how the inflaton energy gets 
converted into the SM degrees of freedom. This will be discussed in chapters V and VI.

For large reheat temperatures, $T_{\rm R}\sim 10^{9}$~GeV, the universe could abundantly 
create thermal relics of unstable gravitinos with a mass of order $100-1000$~GeV, which could spoil the success of 
BBN~\cite{Ellis:1984eq,Moroi:1993mb,Moroi:1995fs,Bolz:2000fu,Kawasaki:2004yh,Kawasaki:2004qu}.
For extremely low reheat temperatures, i.e. $T_{\rm R}\sim {\cal O}(1-10)$~MeV, it becomes a great challenge to 
obtain matter-anti-matter asymmetry and the right abundance for the dark matter. Only a few particle physics
scenarios  can successfully create baryons and dark matter at such a low temperature, see for instance~\cite{Kohri:2009ka,Mazumdar:2003vg,Allahverdi:2001dm,Benakli:1998ur}.


\subsection{Requirements for a successful inflation}\label{sec:generalprop:requirements}

The success of inflation is closely tied to the success of BBN~\cite{Sarkar:1995dd,Burles:2000zk,Fields:2006ga} 
as the epoch constraints the number of relativistic degrees of freedom beyond the SM, and the baryonic asymmetry. Furthermore 
the CMB data and galaxy formation also constraints the abundance of cold dark matter and baryons.


\subsubsection{Baryons and nucleosynthesis}

For a successful BBN, which takes place within the first few hundred seconds, the abundances of light elements 
${^2}H,~^{3}He, ~^{4}He$ and $^{7}Li$ crucially depends on the baryon-to-photon ratio:
\begin{equation}
\eta \equiv \frac{n_{B}}{n_{\gamma}}\,.
\end{equation}
All the relevant physical processes take place essentially
in the range from a few MeV $\sim 0.1$~sec down to $60-70$~KeV
$\sim 10^{3}$~sec. During this period only photons, $e^{\pm}$ pairs,
and the  three neutrino flavors contribute significantly
to the energy density. Any additional energy density may be parameterized
in terms of the effective number of light neutrino species $N_{\nu}$,
so that
\begin{equation}
g_{\ast}=10.75+\frac{7}{4}(N_{\nu}-3)\,.
\end{equation}
BBN constraints the number of light neutrino species by $N_{\nu} \leq 4$~\cite{Copi:1995cb,Olive:1998vj}.
The four LEP experiments combined give the best fit as $N_{\nu}=2.994 \pm 0.12$~\cite{Amsler:2008zzb}.
The likelihood analysis which includes all the three elements $({\rm D}, ^4He, {\rm and}~^7Li)$ yields
the baryon to photon ratio~\cite{Cyburt:2002uv}
\begin{equation}
4.7\times 10^{-10} <\eta < 6.2\times 10^{-10}\,, \quad \quad \quad
0.017 < \Omega_{b}h^2<0.023\,.
\end{equation}
Despite the uncertainties there appears to be a general concordance
between theoretical BBN predictions and observations, which is now
being bolstered by the CMB data $\Omega_bh^2=0.02229\pm 0.00073$~\cite{Spergel:2006hy}.


\subsubsection{Baryogenesis}

The baryon asymmetry of the universe (BAU) parameterized as $\eta_{\rm
B}\equiv(n_{\rm B}-n_{\bar{\rm B}})/s\approx \eta$ is determined to be $0.9 \times
10^{-10}$ by the recent analysis of WMAP data~\cite{Komatsu:2008hk}. 
As pointed out by Sakharov \cite{Sakharov:1967dj}, baryogenesis requires three ingredients:
$(1)$ baryon number non-conservation, $(2)$ $C$ and $CP$ violation,
and $(3)$ out-of-equilibrium condition. 

All these three conditions are believed to be met in the very early universe. Baryogenesis  during the
electroweak phase transition~\cite{Kuzmin:1985mm} has been studied widely, see~\cite{Rubakov:1996vz}.
Another mechanism known as Affleck-Dine baryogenesis, which happens due to the non-trivial dynamics of 
a light scalar condensate is a natural outcome of  
inflation~\cite{Affleck:1984fy,Dine:1995kz,Dine:1995uk}. It is also possible to convert leptonic asymmetry into 
baryonic asymmetry, $B=a(B-L)$~\cite{Fukugita:1986hr,Luty:1992un,Flanz:1994yx,Covi:1996wh}, for a review 
see~\cite{Buchmuller:2005eh}, where
$a=28/79$ in the case of SM and $a=8/23$ for the MSSM~\cite{Khlebnikov:1988sr}.

A lepton asymmetry can be generated from
the out-of-equilibrium decay of the right handed (RH) neutrinos into Higgs bosons and
light leptons, provided $CP-$violating phases exist in the neutrino
Yukawa couplings.  The created lepton
asymmetry will be converted into a baryonic asymmetry via sphaleron
processes. This scenario works most comfortably if
$T_{\rm R} \geq M_{1} \geq 10^{9}$~GeV, where $M_1$ is the lightest RH neutrino~\cite{Giudice:2003jh}~\footnote{Thermal leptogenesis can work 
below the reheat temperature $T_{\rm R} <10^{9}$~GeV in the case of  a
resonant leptogenesis~\cite{Pilaftsis:1997jf,Pilaftsis:2003gt}.}.
There exist various scenarios of non-thermal
leptogenesis~\cite{Lazarides:1991wu,Asaka:1999jb,Giudice:1999fb,Allahverdi:2002gz,Allahverdi:2003tu} which can
work for $T_{\rm R} \leq M_N$. 


\subsubsection{Cold dark matter}

The WMAP data, galaxy clusters and large scale structure data pin down the dark matter
abundance to be : $\Omega_{DM}=0.22$~\cite{Komatsu:2009kd}.  It is important to note that at the end of inflation 
right abundance of dark matter must be created. There are many well motivated particle physics candidates for 
cold dark matter~\cite{Jungman:1995df,Bertone:2004pz}.
All the plausible candidates arise from beyond the SM physics. Within MSSM the lightest SUSY particle
(LSP) is an excellent candidate by virtue of R-parity~\cite{Nilles:1983ge}, the non-topological solitons such as 
Q-balls~\cite{Enqvist:2003gh,Dine:2003ax}, and Kaluza-Klein dark matter particles in theories with extra 
dimensions~\cite{Bertone:2004pz} are the most popular ones.

The dark matter particles can be created via thermal scatterings in the case of thermal cold relics, and 
non-thermally in the process of decay of heavier particles. The thermal relic abundance is easy to calculate,
the number density of dark matter $X$, $n_X$, gets exponential suppression in comparison with the 
number density of relativistic degrees of freedom determined by the freeze-out temperature, $T_f< m_X$.
The abundance is calculated by solving the Boltzmann equation~\cite{Kolb:1988aj}:
\begin{equation}
\frac{dn_X}{dt}+3Hn_X=-\langle \sigma v\rangle (n_X^2-(n_X^{eq})^2)\,,
\end{equation}
where $\sigma $ is the total annihilation cross section, $v$ is the velocity and bracket denotes 
thermally averaged quantities. In the case of heavy $X$, the cross section can be expanded 
with respect to the velocity in powers of $v^2$, 
$\langle \sigma v\rangle =a+b\langle v^2\rangle +{\cal O}(\langle v^4\rangle)+...\approx a+6b/x$, where
$x=m_X/T$ and $a, b$ are expressed in ${\rm GeV}^{-2}$. In the regime where the dark matter abundance 
is frozen-out for $x\gg x_f\equiv m_X/T_f$, the relic density
can be expressed in terms of the critical density:
\begin{equation}
\Omega_Xh^2\approx \frac{1.07\times 10^{9}}{M_{\rm P}}\frac{x_f}{\sqrt{g_{\ast}}(a+3b/x_f)}~{\rm GeV}^{-1}\,,
\end{equation}
where $g_{\ast}$ is the relativistic degrees of freedom and $x_f\sim 25-30$ (in the standard LSP case)
are evaluated at the time of freeze-out. An approximate order of magnitude estimation of the abundance can be written as:
\begin{equation}
\Omega_Xh^2\approx\frac{3\times 10^{-27}~\mathrm{cm}^3~\mathrm{s}^{-1}}{\langle \sigma v \rangle}\,.
\end{equation}
%

\section{Particle physics tools for inflation}\label{sec:particlephysics}
%
%

\subsection{Standard Model of particle physics}\label{sec:partphys:SM}

The Glashow-Weinberg-Salam model of electroweak interactions~\cite{Glashow:1961tr,Weinberg:1967tq,Salam:1969as},
for details see~\cite{Burgess:2007zi},
is based on $SU(2)_L \times U(1)_Y$ gauge theory containing
three $SU(2)_L$ gauge bosons, $W_\mu^i$, $i=1,2,3$, and one $U(1)_Y$
gauge boson, $B_\mu$, with  kinetic energy terms,
${\cal L}_{\rm KE} =-{1\over 4}W_{\mu\nu}^i W^{\mu\nu i}-{1\over 4} B_{\mu\nu} B^{\mu\nu}$,
where $W_{\mu\nu}^i= \partial_\nu W_\mu^i-\partial _\mu W_\nu^i
+g \epsilon^{ijk}W_\mu^j W_\nu^k $ and $
B_{\mu\nu}=\partial_\nu B_\mu-\partial_\mu B_\nu$.
Coupled to the gauge fields is a complex scalar $SU(2)$
doublet, $H$,
\begin{equation}
H
= \left(\begin{array}{c}
 H^{+}   \\
 H^0   \end{array}\right)
\end{equation}
with a  scalar Higgs potential given by
\begin{equation}
 V=\mu^2 \mid H^\dagger H\mid +\lambda
\biggl(\mid H^\dagger H\mid\biggr)^2\quad ,
\label{wspot}
\end{equation}
where $\lambda>0$. For $\mu^2<0$, the minimum energy configuration is given by
the Higgs VEV, which mediates the symmetry breaking: $SU(2)_L\times U(1)_Y\rightarrow U(1)_{em}$
such that the electromagnetism is unbroken.
\begin{equation}
\langle H\rangle
= {1\over\sqrt{2}} \left(\begin{array}{c}
 0   \\
 v   \end{array}\right)\quad .
\label{VEVdef}
\end{equation}
The Higgs VEV, also known as the Higgs mechanism, generates masses for the $W$ and $Z$ 
gauge bosons 
\begin{equation}
{\cal L}_s=(D^\mu H)^\dagger (D_\mu H)-V(H)\,,~~~
{\rm where}~~~
D_\mu=\partial_\mu +i {g\over 2}\tau\cdot W_\mu+i{g^\prime\over 2}
B_\mu Y.
\end{equation}
Since, after symmetry breaking, in the unitary gauge there are no Goldstone
bosons left, only the physical Higgs scalar remains in the spectrum.
The mass to the gauge boson arises from the scalar kinetic energy term,
\begin{equation}
{1\over 2} (0 ,v )
\biggl({1\over 2}g \tau\cdot W_\mu
+{1\over 2} g^\prime B_\mu
\biggr)^2 \left(\begin{array}{c}  0 \\  v \end{array}
\right).
\end{equation}
The physical gauge fields are two
charged fields, $W^\pm$, and two  neutral gauge
bosons, $Z$ and $\gamma$.  
The  masses of the gauge bosons are given by:
\begin{eqnarray}
M_W^2 = {1\over 4} g^2 v^2\,,~~~~
M_Z^2 = {1\over 4} (g^2 + g^{\prime~2})v^2\,,~~~~
M_A =  0.
\end{eqnarray}
Since the massless photon must couple with electromagnetic
strength, $e$, the coupling constants 
define the weak mixing angle $\theta_W$, 
$e= g \sin\theta_W $ and $e= g^\prime \cos\theta_W $.


One can also include 
fermions, let us consider the electron and its neutrino as an example.  
The fermions in terms of their left-and right-handed projections, $\psi_{L,R}={1\over 2}(1\mp\gamma_5)\psi$.
We need to couple the fermions with the $SU(2)_L$ doublets, therefore:
\begin{equation}
L_L=
\left(\begin{array}{c}
\nu_L\\ e_L \end{array}\right).
\end{equation}
The known matter content of the SM has the following charges:
$Q_L=(3,2,1/6), (B=1/3, L=0)$,~ $u_{R}=(\bar 3, 1,-2/3), (B=-1/3, L=0)$,~
$d_R=(\bar 3,1, 1/3), (B=-1/3, L=0)$,~$L_L=(1,2,-1/2), (B=0, L=1)$,~
$e_R=(1,1,1), (B=0, L=-1)$,~$H=(1,2,1/2), (B=0, L=0)$, using the notation
$(n_3, n_2, Y)$,  with color $SU(3)$, weak $SU(2)$ and hypercharge $U(Y)$, respectively.
In MSSM, anomaly cancelation will require additional Higgs with an opposite hypercharge, 
$(1,2,-1/2)$.

Using the hypercharge assignments of the fields, the leptons can be coupled
in a gauge invariant manner to the $SU(2)_L\times U(1)_Y$ gauge fields,
\begin{equation}
{\cal L}_{lepton}
=i {\overline e}_R \gamma^\mu
\biggl(\partial_\mu+i{g^\prime\over 2}Y_e B_\mu\biggr)e_R+
i{\overline L}_L\gamma^\mu
\biggl(\partial_\mu+i {g\over 2}\tau\cdot W_\mu
+i{g^\prime\over 2}Y_L
B_\mu\biggr)L_L\,\,.
\end{equation}
All of the known fermions can be accommodated in the Standard
Model in an identical manner as was done for the leptons. 

A fermion mass term,
${\cal L}_{mass}=-m{\overline{\psi}}\psi=-m
({\overline{\psi}}_L\psi_R+
{\overline{\psi}}_R\psi_L) $, is not gauge invariant under
$SU(2)_L$ and $U(1)_Y$.  The
gauge invariant  Yukawa coupling of the
Higgs boson to the up and down quarks  is given by, ${\cal L}_d\sim -\lambda_d\overline Q_LHd_R+h.c.$:
\begin{equation}
-\lambda_d {1\over\sqrt{2}}
({\overline u}_L,~ {\overline d}_L)\left(
\begin{array}{c}  0 \\
v \end{array} \right) d_R + h.c.
\end{equation}
which provides mass term for the down quark if
$\lambda_d = {m_d \sqrt{2}/ v}$, while the up quark mass is determined by:
${\cal L}_u=-\lambda_u {\overline Q}_L \Phi^c u_R + h.c.$, since 
fact that $H^c \equiv - i \tau_2  H^*$ ($\tau$ is a Paui matrix)
Similar couplings are used to generate mass terms for the charged leptons,
the neutrino has no right handed partner, it remains massless within SM.

In order to obtain neutrino masses, one would have to introduce right handed neutrinos with
a Yukawa coupling, i.e. ${\cal L}\sim h(LH)\nu_{R}$.  From the VEV of the Higgs the neutrinos
will obtain masses $\propto hv$. In order to get neutrino masses in the interesting
range $m_\nu \sim 10^{-1}$~eV, for solar and atmospheric neutrino mixing, the Yukawa coupling 
has to be very tiny, i.e. $h\sim 10^{-12}$.
The origin of neutrino masses can also arise from higher dimensional lepton number violating
operator~\cite{Minkowski:1977sc,Gell-Mann,Yanagida,Mohapatra:1980yp}:
\begin{equation}
{\cal L} \sim \frac{1}{M} (LH)(LH).
\end{equation}
When the Higgs gets a VEV, these gives rise to Majorana
masses for the neutrinos of order
\begin{equation}
m_\nu \sim \frac{v^2}{M}.
\end{equation}
In order to get neutrino masses in the interesting range
we require $M \sim 10^{14}\GeV$, remarkably close to the GUT scale.


\subsection{Radiative corrections in an effective field theory}
\label{SPEFT}

An effective field theory is a powerful tool to study the loop corrections of the scalars coupled to gravity~\cite{Donoghue:1995cz,Donoghue:1996mt,Burgess:2003jk,Burgess:2003zw,Burgess:2007pt,Cheung:2007st,Boyanovsky:2005sh,Burgess:2009ea,Weinberg:2009bg}.
The most general action for $N$ dimensionless scalar fields, $\theta^{i}$, and the metric $g_{\mu\nu}$ can be written in 
terms of derivative expansion, with terms involving up to two derivatives, see~\cite{Burgess:2003jk,Burgess:2003zw,Burgess:2007pt,Burgess:2009ea}:
\begin{eqnarray}
 \label{Leffdef}
 - \frac{{\cal L}_\eff}{\sqrt{-g}} &=& v^4 V(\theta) + \frac{M_{\rm P}^2}{2}
 \, g^{\mu\nu} \left[  W(\theta) \, R_{\mu\nu}
 + G_{ij}(\theta) \, \partial_\mu \theta^i
 \partial_\nu \theta^j \right] \\
 && \quad + A(\theta) (\partial \theta)^4 + B(\theta)
 \, R^2 + C(\theta) \, R \, (\partial \theta)^2
 + \frac{E(\theta)}{M^2} \, (\partial \theta)^6
 + \frac{F(\theta)}{M^2} \, R^3 + \cdots \,, \nonumber
\end{eqnarray}
where $M\ll M_{\rm P}$ is regarded as the
lightest mass scale of particles which would be integrated out,
the coefficient functions, $V(\theta)$,
$G_{ij}(\theta), A(\theta), B(\theta), C(\theta), E(\theta), F(\theta)$ are dimensionless, and 
$R^3$ collectively represents
all possible independent invariants constructed from three Riemann
tensors, or two Riemann tensors and two of its covariant
derivatives; $R (\partial \theta)^2$ denotes all possible
invariants involving one power of the Riemann tensor and two
derivatives acting on $\theta^i$.

An effective action for a scalar field expanded about  a classical solution 
$ \theta^i(x) = \vartheta^i(x) + {\phi^i(x)}/{M_{\rm P}},~~
 g_{\mu\nu} (x) = \hat g_{\mu\nu} (x) +{h_{\mu\nu}(x)}/{M_{\rm P}}$ can be given by~\cite{Burgess:2009ea}:
\begin{equation}\label{lagIPI}
 {\cal L}_{\eff} = \hat {\cal L}_\eff + M^2 M_{\rm p}^2 \sum_{n}
 \frac{c_{n}}{M^{d_{n}}} \; {\cal O}_{n} \left(
 \frac{\phi}{M_{\rm P}} , \frac{ h_{\mu\nu}}{M_{\rm P}} \right)
\end{equation}
where $\hat{\cal L}_\eff $ is the classical Lagrangian density evaluated at the background configuration,
$\phi$ and $h_{\mu\nu}$ are the small perturbations around the background scalar field and the metric.
The sum over $n$ runs over the labels for a complete set of
interactions, ${\cal O}_{n}$, each of which involves $N_n = N^{(\phi)}_n
+ N^{(h)}_n \ge 2$ powers of the fields $\phi^i$ (for N fields) and $h_{\mu\nu}$
(with $N_n \neq 1$). The parameter
$d_{n}$ counts the number of derivatives appearing in ${\cal O}_n$, and
$c_{n}/M^{d_n}$ is a dimensionless quantity, where $M \ll M_{\rm P}$ denotes the 
scale at which heavy degrees of freedom have been integrated out. The overall prefactor, $M^2
M_p^2$, is to keep the kinetic terms canonical for individual fields. 

In terms of dimensionless couplings, $\lambda_{n}$, the scalar part of the potential can be 
expanded as~\footnote{Similar potential can also be derived in the context 
of supergravity (SUGRA), see Refs.~\cite{Adams:1996yd,Adams:1997de}.}:
\begin{equation}
\label{IPIexample}
V(\phi)=v^4\left[ \lambda_0 + \lambda_2 
\left( \frac{\phi}{M_{\rm P}} \right)^2 + \lambda_4 
\left( \frac{\phi}{M_{\rm P}} \right)^4 + \cdots \right] \,,
\end{equation}
Note that the natural scale for the scalar masses under the
above assumptions is $m \simeq v^2/M_{\rm P}$. The quartic coupling
constant, $\lambda_4 (v/M_{\rm P})^4$, is similarly Planck suppressed.
For the purpose of inflation, the scale of the scalar potential will be governed by $V\sim v^4\ll M^4$.
Such small masses and couplings are needed to keep the successes of inflation.

In general, in order to study the evolution of couplings, $c_n$, and how the two scales $M$ and $M_{\rm P}$
appear in a given problem, one has to study the $1-$ particle irreducible (1PI) graphs perturbatively in
the interaction part of the Lagrangian density. To this end, one can obtain the leading order expression for the 
amplitude of such graphs, by assuming that $E$ denotes the largest of physical scales
that appear explicitly in the propagators or vertices of the
calculation. One can neglect any other smaller scales compared with $E$ when estimating the size
of a particular Feynman graph~\cite{Burgess:2003jk,Burgess:2007pt,Burgess:2009ea}.
\begin{equation}\label{AmpIPI}
 {\cal A}_{\epsilon} (E) \simeq E^2 M_{\rm P}^2 \left( \frac{1}{M_{\rm P}} \right)^{\epsilon}
 \left( \frac{E}{4 \pi \, M_{\rm P}}
 \right)^{2L} \prod_n \left[ c_n \left( \frac{E}{M}
 \right)^{d_n-2} \right]^{V_n} \,,
\end{equation}
where $\epsilon$ denotes the number of external lines, $L$ loops and $V_n$ vertices with $d_n$ derivatives.
The factor $E/4\pi M_{\rm P}\ll 1$ ensures that successive insertions of interactions to be smaller than preceding
ones. 

One particular application is integrating out a particle of mass $m\gg M$. In the inflationary context, 
$M\sim H_{inf}\sim v^2/M_{\rm P}$ for the above potential Eq.~(\ref{IPIexample}), and $\lambda_{n}\sim {\cal O}(1)$.
Expanding the above amplitude Eq.~(\ref{AmpIPI}) for arbitrary derivatives and comparing the coefficients with
Eq.~(\ref{lagIPI}), one finds $v^4/(E^2M_{\rm P}^2) \simeq v^4/(m^2 M_{\rm P}^2) \simeq H_{inf}^2/m^2 \ll 1$. 
In addition, the generic loop factor $(m / 4 \pi M_{\rm P})^2$, the $d_n \ge 4$ interactions  are 
suppressed by at least two powers of $m^2/M_{\rm P}^2$, and $\lambda_n$  are additionally
suppressed by powers of $H_{inf}^2/m^2$. Only the $d_n = 2$ interactions
remain unsuppressed beyond the basic loop factor if $\lambda_n \lesssim {\cal O}(1)$. On the 
other hand, if there are interactions in the scalar potential that are unsuppressed by
powers of $M_{\rm P}$, such as if $\lambda_n \simeq (M_{\rm P}/v)^{N_n} \lambda_n$, i.e. higher
powers of $\phi$ and $h_{\mu\nu}$, then loops involving these
interactions can modify the inflaton potential~\cite{Burgess:2003zw,Burgess:2009ea}~\footnote{There are two types of operators; (i) {\it relevant operators}, which are proportional to positive powers of $M$, and (ii) {\it marginal operators}, which grow logarithmically with $M $, have been considered in studying the trans-Planckian physics, see for instance~\cite{Kaloper:2002uj,Kaloper:2002cs}.}.

One such known example is in the case of Higgs inflation~\cite{Bezrukov:2007ep}, where 
the presence of $\xi H^{\dagger}HR$ (interaction vertex involves $d_n=2$ derivatives) 
term helps flattening the Higgs potential~\footnote{In an Einstein frame the Higgs potential with canonical kinetic terms
is given by  Eq.~(\ref{eq:sminflationpoten}), see section \ref{sec:SMinflaton}.}. 
A successful inflation requires 
$\xi \sim 5\times 10^{4}\sqrt{\lambda}$, where $\lambda$
is the Higgs self-coupling. The unitarity bound on interactions such as Higgs-Higgs scattering, 
graviton-Higgs scattering (through graviton exchange) leads to $E<E_{max}\approx M_{\rm P}/\xi$, which 
constraints the validity range for an effective field theory treatment for the Higgs inflation to be within a narrow range
$\sqrt{\lambda}\ll H_{inf}/M \ll 1$, where $H_{inf}\sim \sqrt{\lambda M_{\rm P}}/\xi$. Any Higgs coupling to the matter 
field will induce corrections to the Higgs potential and alter the predictions for inflation, see~\cite{Burgess:2009ea}.

In other extreme limit, when a light particle is running in the loop, i.e. $m\ll H_{inf}$ and $m\ll \dot\phi/\phi $, the above analysis
when applied to $2$-point correlations, yields a well known result, $\langle \phi^2\rangle \sim H_{inf}^2/4\pi^2$~\cite{Burgess:2009ea}.

The one-loop effective potential, which can be generated when the heavy fields are explicitly integrated out, 
serves as an useful tool to lift a generic flat direction, which is helpful to obtain inflationary potential. A simple 
calculation of the virtual effects of the heavy scalars and fermions on the light scalar potential, which can
be the inflaton, can be obtained by matching the one-loop corrected effective potential for the full theory.
This gives the following result following Coleman-Weinberg~\cite{Coleman:1973jx}~\footnote{A supersymmetric (SUSY) generalization of the Coleman-Weinberg formula will be presented in Sec.~\ref{See:SUSYgeneralization of one-loop effective potential}.}: 
\begin{eqnarray}\label{induced-one-loop}
V_{eff}(\phi) &=&V_{inf}(\phi)+\Delta V\,\nonumber \\
\Delta V &=& \frac{1}{64\pi^2}\sum_{i}(-)^{F_i}M_{i}(\phi)^4\ln\frac{M_{i}(\phi)^2}{\Lambda(\phi)^2}\,,
\end{eqnarray}
where $V_{inf}$ is now the renormalized potential, $\Lambda(\phi)$ is the
renormalization mass scale. The sum extends over all helicity states $i$, $F_{i}$ is the fermion number, and $M(\phi)$
is the mass of the i-th state.

As an example, let us consider a chiral Lagrangian for $N$ fermions invariant under $Z_N$ symmetry~\cite{Frieman:1995pm}. In general, $Z_N$ symmetry can be defined for an Abelian and non-Abelian gauge group, 
which keeps the action, $S$, invariant under the symmetry transformation: $S\longrightarrow e^{2\pi ij/N}\times S$, where
$j=1,2,\cdots , N$. The relevant Lagrangian is: 
\begin{equation}
{\cal L}=(1/2)\partial_{\mu}\phi\partial^{\mu}\phi+\sum_{j=0}^{N-1}
\bar\psi_j i\gamma^{\mu}\partial_{\mu}\psi_{j}+[m_0+\epsilon e^{i(\phi/f+2\pi j/N)}]\bar\psi_{j~L}\psi_{j~R}+h.c.\,, 
\end{equation}
where 
$\psi_{(R,~L)}=(1\pm \gamma^{5})\psi/2$, ~$m_0$  is an explicit breaking term, and the scalar VEV responsible for generating
$\epsilon$ via some Yukawa interaction between scalar and $N$ fermions is given by: $\langle \phi\rangle=e^{i\phi/f}f/\sqrt{2}$. 
Under the $Z_N$ discrete symmetry: $\psi_j\rightarrow \psi_{j+1},~\psi_{N-1}\rightarrow \psi_0,~\phi\rightarrow\phi+2\pi j f/N$. 
The induced one-loop potential can be calculated from Eq.~(\ref{induced-one-loop}), with a cut-off: $\lambda < f$. The potential is:
\begin{equation}
\Delta V=-\sum_{j=0}^{N-1}\frac{M_j^4}{16\pi}\ln\left(\frac{M_j^2}{\Lambda^2}\right)\,,
\end{equation}
where $M_{j}^2=m_0^2\epsilon^2+2m_0\epsilon \cos\left(\phi/f+2\pi j/N\right)$. The scalar field, $\phi$, is 
a simple example of pseudo-Nambu Goldstone Boson (pNGB), which can protect its potential naturally due to 
symmetries~\footnote{ Technically natural small mass scales are protected by symmetries, such that when the small mass vanishes it can not be generated in any order of perturbation theory.}. 
For $N=2$ the pNGB mass is $m_{\phi}\sim m_0\epsilon/f$.


\subsection{Supersymmetry (SUSY)}\label{sec:partphys:SUSY}

The SM physics has number of pressing issues, the most compelling one
is the quadratically divergent contributions to the Higgs mass, which arise in 
one-loop computation from the fermion contribution and quartic self interaction of the Higgs boson. 
The quadratic divergence is independent of the mass of the Higgs boson and cancel, exactly 
if $\lambda_s=\lambda_f^2$, where $\lambda_f$ is the fermion Yukawa and $\lambda_s$  is the 
quartic scalar coupling. However this procedure fails at 2-loops and one requires fine tuning of
the couplings order by order in perturbations to a precision of roughly one part in $10^{17}$ (for the 
scale of gravity at $M_{\rm P}\sim 10^{18}$~GeV),  often known as the {\it hierarchy problem} or 
the {\it naturalness problem}. 

In SUSY, there is a scalar of same mass associated with every fermion and the couplings are 
such that $\lambda_s=\lambda_f^2$. The electroweak symmetry is still broken by the Higgs 
mechanism, but the quadratic divergences in the scalar sector are absent. The minimal extension 
of the SM in SUSY is known as MSSM (minimal SUSY SM).

The matter fields of $N=1$ SUSY are chiral superfields 
$\Phi=\phi+\sqrt{2}\theta\psi+\theta\theta F$, which
describe a scalar $\phi$, a fermion $\psi$ and a scalar auxiliary field $F$.  The 
SUSY scalar potential $V$ is the sum of the $F$- and $D$-terms are:
\begin{equation}
\label{fplusd}
V= \sum_i |F_i|^2+\frac 12 \sum_a g_a^2D^aD^a
\end{equation}
where
\begin{equation}
F_i\equiv {\partial W\over \partial \phi_i},~~D^a=\phi^\dagger T^a \phi~\,,
\label{fddefs}
\end{equation}
where $W$ is the superpotential, and we have assumed that $\phi_i$ transforms 
under a gauge group $G$ with the generators of the Lie algebra given by $T^{a}$. Note that all the
kinetic energy terms are included in the $D$-term.


\subsubsection{Minimal Supersymmetric Standard Model (MSSM) }\label{sec:partphys:MSSMnExtensions}

In addition to the  usual quark and lepton superfields,
MSSM has two Higgs fields, $H_u$ and $H_d$. Two Higgses are needed
because $H^\dagger$ is forbidden in the superpotential. 
The superpotential for the MSSM is given by, see~\cite{Nilles:1983ge,Haber:1984rc, Martin:1997ns,Chung:2003fi}
\begin{equation}
\label{mssm}
W_{MSSM}=\lambda_uQH_u u+\lambda_dQH_d d+\lambda_eLH_d e~
+\mu H_uH_d\,,
\end{equation}
where $H_{u}, H_{d}, Q, L, u, d, e$ in
Eq.~(\ref{mssm}) are chiral superfields, and the dimensionless Yukawa couplings
$\lambda_{u}, \lambda_{d}, \lambda_{e}$ are $3\times 3$ matrices in the family
space. We have suppressed the gauge and family indices. The $H_{u}, H_{d}, Q, L$ fields are
$SU(2)$ doublets, while $u, d, e$ are  $SU(2)$ singlets. The last term is the $\mu$
term, which is a SUSY version of the SM Higgs boson mass.
Terms proportional to $H_{u}^{\ast}H_{u}$ or $H^{\ast}_{d}H_{d}$ are
forbidden in the superpotential, since $W_{MSSM}$ must be analytic in the
chiral fields. $H_{u}$ and $H_{d}$ are required not only because they  give
masses to all the quarks and leptons, but also for the cancellation of
gauge anomalies. The Yukawa matrices determine the masses and CKM mixing
angles of the ordinary quarks and leptons through the neutral components of
$H_{u}=(H^{+}_{u},H^{0}_{u})$ and $H_{d}=(H^{0}_{d}H^{-}_{d})$.
Since the top quark, bottom quark and tau lepton are the heaviest fermions
in the SM, we assume that only the third family, $(3,3)$ element of the matrices
$\lambda_{u}, \lambda_{d}, \lambda_{e}$ are important.

The $\mu$ term provides masses to the Higgsinos
\begin{equation}
{\cal L} \supset -\mu(\tilde H_{u}^{+}\tilde H_{d}^{-}-\tilde H^{0}_{u}
\tilde H^{0}_{d})+{c.c} \,,
\end{equation}
and contributes to the Higgs $(mass)^2$ terms in the scalar potential  through
\begin{equation}
\label{higgsmass}
-{\cal L} \supset V \supset |\mu|^2(|H^{0}_{u}|^2+|H^{+}_{u}|^2+|H^{0}_{d}|^2+
|H^{-}_{d}|^2)\,.
\end{equation}
Note that Eq.~(\ref{higgsmass}) is positive definite. Therefore, it
cannot lead to electroweak symmetry breaking without including
SUSY breaking $(mass)^2$ soft terms for the Higgs fields,
which can be negative. Hence, $|\mu|^2$ should almost cancel the
negative soft $(mass)^2$ term in order to allow for a Higgs VEV
of order $\sim 174$~GeV. That the two different sources of masses
should be precisely of same order is a puzzle for which
many solutions has been suggested~\cite{Kim:1983dt,Giudice:1988yz,Casas:1992mk,Dvali:1996cu}.

The most general gauge invariant and renormalizable superpotential
would also include baryon number $B$ or lepton number $L$ violating terms, with 
each violating by one unit: $W_{\Delta L=1}=\frac{1}{2}\lambda^{ijk}L_{i}L_{j}e_{k}+
\lambda^{\prime ijk}L_{i}Q_{j}d_{k}+\mu^{\prime i}L_{i}H_{\mu}$ and
$W_{\Delta B=1}=\frac{1}{2}\lambda^{\prime \prime ijk}u_{i}d_{j}d_{k}$,
where $i=1,2,3$ represents the family indices. The chiral supermultiplets
carry baryon number assignments $B=+1/3$ for $Q_{i}$, $B=-1/3$ for
$u_{i},  d_{i}$, and $B=0$ for all others. The total lepton number
assignments are $L=+1$ for $L_{i}$, $L=-1$ for $e_{i}$, and $L=0$ for
all the others. Unless $\lambda^{\prime}$ and $\lambda^{\prime\prime}$ terms are very
much suppressed, one would obtain rapid proton decay which violates
both $B$ and $L$ by one unit.

There exists a discrete $Z_2$ symmetry, which can forbid baryon and lepton number violating terms,
known as $R$-parity \cite{Fayet:1979yb}. For each particle:
\begin{equation}
P_{R}=(-1)^{3(B-L)+2s}\,
\end{equation}
with $P_{R}=+1$ for the SM particles and the Higgs bosons, while $P_{R}=-1$
for all the sleptons, squarks, gauginos, and Higgsinos.
Here $s$ is spin of the particle. Besides forbidding $B$ and $L$ violation from the 
renormalizable interactions, $R$-parity has  interesting phenomenological and cosmological consequences.
The lightest sparticle with $P_{R}=-1$, the LSP, must be absolutely stable.
If electrically neutral, the LSP is a natural
candidate for cold dark matter \cite{Ellis:1983ew,Dimopoulos:1988jw}~\footnote{Symmetries with the property that fields within the same supermultiplet have different transformations are called $R$ symmetries;  
they do not commute with SUSY.  Continuous $U(1)$ $R$ symmetries are often employed in  inflationary model-building literature. Under this symmetry a general chiral  superfield transforms as $\Phi\rightarrow e^{iR\alpha}\Phi$, and in order to
keep the theory R-invariant, the superpotential must have $R=2$. }.


\subsubsection{Soft SUSY breaking Lagrangian}

In the MSSM there are several proposals for SUSY breaking,
which we shall discuss below. However most of the time it is not
important to know the exact mechanism of low energy SUSY
breaking. This ignorance of the origin of SUSY breaking
can always be hidden by simply writing down explicitly the soft
breaking terms.

The most general soft SUSY breaking terms in the MSSM
Lagrangian can be written as (see e.g. \cite{Haber:1984rc,Nilles:1983ge})
\begin{equation}
{\cal L}_{soft}=-\frac{1}{2}\left(M_{\lambda}\lambda^{a}\lambda^{a}+
{\rm c.c.}\right)-(m^2)^{i}_{j}\phi^{j\ast}\phi_{i}-\left(\frac{1}{2}
b_{ij}\phi_{i}\phi_{j}+\frac{1}{6}a^{ijk}\phi_{i}\phi_{j}\phi_{k}+{\rm c.c.}
\right)\,,
\end{equation}
where $M_{\lambda}$ is the common gaugino mass $(m^2)^{j}_{i}\sim m^2_{0}\sim ({\cal O}(100){\rm GeV})^2$
are $3\times 3$ matrices determining the masses for
squarks and sleptons, denoted as
$m^2_{Q},m^2_{u},m^2_{d},m^2_{L},m^2_{ e}, m^2_{H_u},m^2_{H_d}, b\sim m_{0}^2\sim ({\cal O}(100))^2~{\rm GeV^2}$;
$b_{ij}$ is the mass term for the combination $H_{u}H_{d}$; and finally,
$a^{ijk}$ are complex $3\times 3$ matrices in the family space which yield the
$A$-terms $a_{u},~a_{d},~a_{e}\sim m_{0}\sim {\cal O}(100)$~GeV.  There are a total of $105$ new entries in the
MSSM Lagrangian which have no counterpart in the SM. However the arbitrariness
in the parameters can be partly removed by the experimental constraints on
flavor changing neutral currents (FCNC) and $CP$ violation
\cite{Dimopoulos:1995ju}~\footnote{Within global SUSY there exists a mass formula
$Str{\bf M^2}\equiv \sum_{j=0}^{1}(-1)^{j}tr{\bf M_j^2}$ , which prevents all the squarks and sleptons to 
have masses larger than those of quarks and leptons. This constraints on SUSY breaking scenarios in the global
case, but by introducing SUGRA the above relationship will be modified. }

There are a number of possibilities for the origin of SUSY 
breaking~\cite{Haber:1984rc,Nilles:1983ge,Chung:2003fi}.
Fayet-Iliopoulos mechanism \cite{Fayet:1974jb} provides SUSY breaking by
virtue of a non-zero $D$-term but requires a $U(1)$ symmetry. However, this
mechanism does not work in the MSSM because some of the squarks and sleptons
will get non-zero VEVs which may break color, electromagnetism, and/or
lepton number without breaking SUSY.  Therefore the contribution
from the Fayet-Iliopoulos (FI) term should be negligible at low scales.

There are models of SUSY breaking by $F$-terms, known as
O'Raifeartaigh models \cite{O'Raifeartaigh:1975pr}, where the idea is to pick
a set of chiral supermultiplets $\Phi_{i} \supset (\phi_{i},\psi_{i}F_{i})$
and a superpotential $W$ in such a way that
$F_{i}=-\delta W/\delta\phi^{\ast}_{i}=0$ have no simultaneous solution.
The model requires a linear gauge singlet superfield in the
superpotential. Such singlet chiral supermultiplet is not present in the
MSSM. The scale of SUSY breaking has to be set by hand.

The only mechanism of SUSY breaking where the breaking
scale is not introduced either at the level of superpotential or in
the gauge sector is through dynamical SUSY breaking
\cite{Witten:1982df,Affleck:1983mk,Dine:1993yw,Dine:1995ag}, see 
also~\cite{Intriligator:2006dd}. In these models a small
SUSY breaking scale arises by dimensional transmutation.
It is customary to treat the SUSY breaking sector as a hidden
sector which has no direct couplings to the visible sector represented
by the chiral supermultiplets of the MSSM. The only allowed interactions
are those which mediate the SUSY breaking in the hidden sector to the
visible sector.

The main contenders are  gravity mediated SUSY breaking, which is
associated with new physics which includes gravity at the string scale
or at the Planck scale \cite{Chamseddine:1982jx,Nilles:1982dy,Nilles:1983ge}, and gauge mediated
SUSY breaking, which is transmitted to the visible sector
by the ordinary electroweak and QCD gauge interactions
\cite{Dine:1981gu,Dine:1993yw,Dine:1996xk,Dimopoulos:1996vz}.  There
are other variants of SUSY breaking based upon ideas on
gravity and gauge mediation with some extensions, such as dynamical
SUSY breaking (see \cite{Shadmi:1999jy}, and references therein),
and anomaly mediation (see \cite{Randall:1998uk,Giudice:1998bp}).


\subsubsection{Next to MSSM (NMSSM)}

The simplest  extension of the MSSM can be 
obtained by adding a new gauge-singlet chiral supermultiplet with even 
matter parity. The superpotential reads as
\cite{Nilles:1982mp,Frere:1983ag,Derendinger:1983bz,King:1995vk}, see also~\cite{Munoz:2007fa}:
\begin{equation}
W_{\rm NMSSM} \>=\> W_{\rm MSSM}
+ \lambda S H_u H_d + \frac{1}{3} \kappa S^3 + \frac{1}{2} \mu_S S^2  ,
\label{NMSSMwww}
\end{equation}
where $S$ is the new chiral supermultiplet. It  is often 
called the next-to-minimal SUSY standard model (NMSSM). 
The NMSSM introduces extra coefficients, by choosing them correctly
it is possible to realize a successful electroweak symmetry breaking.

One of the virtues of the NMSSM is that it can provide a solution to the
$\mu$ problem. An effective $\mu$-term for $H_u H_d$ will arise from
eq.~(\ref{NMSSMwww}), with $ \mu_{\rm eff} \>=\> \lambda s.$ It is
determined by the dimensionless couplings and the soft terms of order
$m_{\rm soft}$, instead of being a free parameter conceptually independent
of SUSY breaking. In general, NMSSM also provides extra sources  for large CP 
violation and conditions for electroweak baryogenesis~\cite{Menon:2004wv,Balazs:2004ae}.

The singlet $S$ contains a real $P_R=+1$,  and a $P_R=-1$ Weyl fermion 
singlino, $\tilde S$. These fields have no gauge couplings of their own, so they 
can only couple to the SM particles via mixing with the neutral 
MSSM fields with the same spin and charge.
The odd $R$-parity singlino $\tilde{S}$ mixes with the four MSSM 
neutralinos.. The singlino could be the LSP and the
dark matter candidate~\cite{Flores:1991rx,Abel:1992ts,Panagiotakopoulos:1999ah}, in some parameter
space neutralino type dark matter is also possible~\cite{Cerdeno:2004xw,Cerdeno:2007sn}.  For collider 
signatures, see~\cite{Balazs:2007pf}.


\subsubsection{Gravity mediated SUSY breaking}

The gravity mediated SUSY breaking, which is
associated with new physics which includes gravity at the string scale
or at the Planck scale \cite{Chamseddine:1982jx,Nilles:1983ge}. It is  assumed that SUSY is broken by the VEV
$\langle F\rangle \neq 0$ and is communicated to the MSSM by
gravity. On dimensional grounds, the soft terms in the visible
sector should then be of the order $m_{0}\sim \langle F\rangle/M_{\rm P}$, see \cite{Nilles:1983ge}.

The SUGRA Lagrangian must contain the non-renormalizable terms
which communicate between the hidden and the observable sectors. For
the cases where the kinetic terms for the chiral and gauge fields are
minimal, one obtains the following soft terms
\cite{Chamseddine:1982jx,Nilles:1983ge}
\begin{equation}
m_{1/2}\sim \frac{\langle F\rangle}{M_{\rm P}}\,, \quad m_0^2\sim
\frac{|\langle F\rangle|^2}{M_{\rm P}^2}\,, \quad A_{0}\sim \frac{\langle F
\rangle}{M_{\rm P}}\,, \quad B_0 \sim \frac{\langle F\rangle}{M_{\rm P}}\,.
\end{equation}
The gauginos get a common mass $M_{1}=M_{2}=M_3=m_{1/2}$,
the squark and slepton masses are $m^2_{Q}=m^2_{ u}=m^2_{ d}=m^2_{L}=
m^2_{ e} =m^2_{0}$, and for the Higgses $m^2_{H_{u}}=m^2_{H_{d}}=m^2_{0}$.
The $A$-terms are proportional to the Yukawa couplings
while $b=B_{0}\mu$.

Note that $m_{soft} \rightarrow 0$ as
$M_{\rm P} \rightarrow \infty$.
In order to obtain a phenomenologically acceptable soft SUSY
mass $m_{soft}\sim {\cal O}(100)$~GeV, one therefore requires
the scale of SUSY breaking in the hidden sector to be
$\sqrt{\langle F\rangle} \sim 10^{10}-10^{11}$~GeV.

Another possibility is that the SUSY is broken via gaugino condensate
$\langle 0|\lambda^{a}\lambda^{b}|0\rangle=\delta^{ab}\Lambda^3\neq 0$,
where $\Lambda$ is the condensation scale \cite{Nilles:1982mp,Nilles:1982dy,Nilles:1983ge}. If
the composite field $\lambda^{a}\lambda^{b}$ belongs to the
$\langle F\rangle \sim \Lambda^3/M_{\rm P}$-term, then again on dimensional
grounds one would expect the soft SUSY mass contribution to be
\cite{Chamseddine:1982jx,Nilles:1983ge}
\begin{equation}
m_{soft} \sim \frac{\Lambda^3}{M_{\rm P}^2}\,.
\end{equation}
In this case the nature of SUSY breaking is dynamical and the scale
is given by $\Lambda \sim 10^{13}$~GeV.

Commonly gravity mediated SUSY breaking scenario is also known as minimal SUGRA
(mSUGRA). In mSUGRA the number of independent parameters reduce a lot,
there are $m_{0}, m_{1/2}, A_0$, the GUT scale value for $\mu_0, b_0$, and the gravitino mass.
Further more $b_0=A_0-m_0$ and $m_{3/2}=m_0$ further reduces the parameter space.
Nowadays, the popular choice of parameters is known as CMSSM (constrained MSSM), they 
are: $\tan\beta=\langle H_u\rangle/\langle H_d\rangle$, $m_0,~A_0,~m_{1/2}$ and $\rm sgn(\mu_0)$.
Within CMSSM the LSP is the lightest neutralino (the neutral higgsinos
($\tilde{H}_u^0$ and $\tilde{H}_d^{0}$) and the neutral gauginos $(\tilde{B}, \tilde W^{0})$ combine to form 
four mass eigenstates called neutralinos), $m_{\tilde N_1}<m_{\tilde N_2}<m_{\tilde N_3}<m_{\tilde N_4}$
is known to be the LSP, unless gravitino is the lightest, or R-parity is not conserved.

There are variants of gravity mediation, known as anomaly mediated SUSY 
breaking (AMSB) scenarios, where at tree level gaugino masses are
not present. The masses for gauginos arise from one-loop whose origin can be 
traced to the super-conformal (super-Weyl) anomaly which is 
common to all SUGRA models~\cite{Randall:1998uk,Giudice:1998xp,Bagger:1999rd}. At low 
energies gaugino mass parameters are given by:
\begin{equation}
M_{i}\approx \frac{b_i g_i^2}{16\pi^2}m_{3/2}\,.
\end{equation}
where $m_{3/2}\sim 100-1000$~GeV is the gravitino mass, and $b_i$ are the MSSM 
gauge beta-functions, i.e. for $SU(3),~SU(2),~U(1)$ gauge groups: $(b_3=-3, b_2=1, b_1=33/5)$. 
AMSB can naturally suppress the flavor changing processes, however at a cost of negative squared 
masses for the sleptons. Therefore AMSB cannot alone be the main source for SUSY breaking in
the slepton sector.


\subsubsection{Gauge mediated SUSY breaking}

In gauge mediated SUSY breaking one employs a heavy
messenger sector which couples directly to the SUSY
breaking sector but indirectly to the observable sector via
standard model gauge interactions 
only~\cite{Dine:1981gu,Dine:1993yw,Dine:1996xk,Dimopoulos:1996vz,Nappi:1982hm}.
As a result the soft terms in the MSSM arise through ordinary
gauge interactions. There will still be gravitational communication,
but it is a weak effect.

The simplest example is a messenger sector with a pair of $SU(2)$ doublet
chiral fields $l,~\bar l$ and a pair of $SU(3)$ triplet fields
$q,~ \bar q$, which couple to a singlet field $z$ with Yukawa
couplings $\lambda_2,~\lambda_3$, respectively. The superpotential
is given by
\begin{equation}
W_{mess}=\lambda_{2} zl\bar l+\lambda_{3}zq\bar  q\,.
\end{equation}
The singlet acquires a non-zero VEV and a non-zero F-term
$\langle F_z\rangle$. This can be accomplished either substituting $z$ into
an O'Raifeartaigh type model \cite{Dine:1981gu,Nappi:1982hm}, or by a dynamical
mechanism \cite{Dine:1993yw,Dine:1996xk}. One may parameterize
SUSY breaking in a superpotential $W_{break}$ by
$\langle \partial W_{break}/\partial z\rangle =-\langle F_{z}^{\ast}\rangle$.
As a consequence, the messenger fermions acquire masses and a scalar
potential with $\langle \partial W_{mess} /\partial z\rangle =0$.

SUSY breaking is then mediated to the observable fields by
one-loop corrections, which generate masses for the MSSM gauginos
\cite{Dine:1993yw}. The $q, \bar{q}$ messenger loop diagrams provide masses
to the gluino and the bino, while $l,\bar{l}$ messenger loop diagrams
provide masses to the wino and the bino, i.e.,
$M_{a=1,2,3}=(\alpha_{a}/4\pi)\Lambda$, where
$\Lambda =\langle F_{z}\rangle/\langle z\rangle$.

For squarks and sleptons the leading term comes from two-loop diagrams,
e.g. $m^2_{\phi}\propto \alpha^2$. The $A$-terms get negligible
contribution at two-loop order compared to the gaugino masses, they
come with an extra suppression of $\alpha/4\pi$ compared with the gaugino
mass, therefore $a_{u}=a_{d}=a_{e}=0$ is a good approximation. The Yukawa
couplings at the electroweak scale are generated by evolving the
RG equations.

One can estimate~\cite{Dine:1993yw,Dine:1996xk} the soft SUSY breaking masses
to be of order
\begin{equation}
m_{soft}\sim \frac{\alpha_{a}}{4\pi}\frac{\langle F\rangle}{M_{s}}\,.
\end{equation}
If $M_s\sim \langle z\rangle$ and $\sqrt{\langle F\rangle}$ are 
comparable mass scales, then the SUSY breaking can take 
place at about $\sqrt{\langle F\rangle} \sim 10^{4}-10^{6}$~GeV. 

In gauge mediated SUSY breaking the gravitino could be the LSP, and the next-to-LSP (NLSP)  could
be either stau, bino-like neutralino. The NLSP decay into gravitino
could be very long ranging from seconds to years, the decay process: 
$\tilde\tau\rightarrow \tau \tilde G$ is governed by the gravitational coupling. The
decay of long lived staus are constrained by the BBN.


\subsubsection{Split SUSY}

An advantage of a low scale (TeV) SUSY is the gauge coupling unification,
and the LSP, such as neutralino or gravitino as a dark matter candidate. Both of these
virtues can be kept without any need of a TeV scale supersymmetry as shown 
in Refs.~\cite{Wells:2003tf,Wells:2004di,ArkaniHamed:2004fb,Giudice:2004tc}.
Split SUSY has light SM like Higgs bosons, the $A$-term and the $\mu$-term, but 
super heavy squarks and sleptons. If the SUSY breaking is mediated via gravity, then
the gravitino mass comes out to be:
\begin{equation}
m_{3/2}\geq \frac{m_{0}^2}{M_{\rm P}}\,,
\end{equation}
where $m_{0}$ correspond to the scalar masses,  
which suppresses flavor changing neutral currents and  
CP violations mediated via heavy squarks.  While the $A$-term, $\mu$-term, and
 gauginos are light~\cite{ArkaniHamed:2004fb, Giudice:2004tc}, 
\begin{equation}
A,~\mu,~ m_{1/2}\sim \frac{m_{3/2}^3}{M_{\rm P}^2}\,.
\end{equation}
Therefore the neutralino like LSP can be realizable with a very long lived gluino, which decays via
off-shell squark to quark, anti-quark and LSP.

The reasoning for such a split spectrum is due to an accidental R-symmetry, which can protect the masses for 
gauginos, the $A$-term and the $\mu$ term, which 
gets broken via non-renormalizable interactions. The model also allows the Higgs mass to be fine tuned to be light 
at the weak scale. However there are other advantages, there will be no gravitino problem for BBN, as the 
gravitino mass can be made higher than TeV, and the proton decay upper limits can also be satisfied as the squarks are heavy.


\subsubsection{Renormalization group equations in the MSSM}

In many cosmological applications of flat directions, it is important to
consider the running of $(\rm mass)^2$
below $M_{\rm GUT}$. For simplicity we can also assume that it is the scale where
SUSY breaking is transmitted to the visible sector.
The running of low-energy soft breaking masses has
been studied in great detail in the context of MSSM phenomenology
\cite{Nilles:1983ge,Martin:1997ns,Haber:1984rc}, see also~\cite{Drees:1995hj}, in particular in connection with radiative electroweak
symmetry breaking \cite{Ibanez:1982fr}. A general form of  RG equations, which to one loop can be written as:
\begin{equation}
\label{rge}
{\partial m_i^2\over \partial t}=\sum_{g}a_{ig} m_g^2+
\sum_a h^2_a(\sum_j b_{ij}m_j^2+A^2)~,
\end{equation}
where $a_{ig}$ and $b_{ij}$ are constants, $m_g$ is the gaugino mass,
$h_a$ the Yukawa coupling, $A$ is the $A$-term, and $t=\ln M_X/q$. The
full RG equations have been listed in \cite{Nilles:1983ge,Martin:1997ns,Haber:1984rc}.

However, let us consider some of the salient features of the MSSM one-loop
RG equations. The one-loop RG equations for the three gaugino mass parameters are
determined by:
\begin{equation}
\frac{d}{dt}m_i=\frac{1}{8\pi^2}b_i g_i^2m_{i}\,,~~~~(b_{i}=33/5,~1,~-3)
\end{equation}
where $i=1,2,3$ correspond to $U(1),~SU(2),~SU(3)$. An interesting property
is that the three ratios $m_i/g_i^2$ are RG scale independent. Therefore at the
GUT scale, where the gauge couplings unify at $M_{GUT}\sim 2\times 10^{16}$~GeV,
it is assumed that gauginos masses also unify with a value $m_{1/2}$. Then at any scale:
\begin{equation}
\frac{m_{i}}{g_{i}^2}=\frac{m_{1/2}}{g^2_{GUT}}\,,
\end{equation}
where $g_{GUT}$ is the unified gauge coupling at the GUT scale. The RG evolution
due to Yukawa interactions are small except for top. The ones relevant to flat directions, involving the Higgs
doublet $H_u$ which couples to the top quark, the right-handed stop
$\widetilde{u}_3$, the left-handed doublet of third generation squarks
$\widetilde{Q}_3$ and the $A-$parameter $A_t$ associated with the top
Yukawa interaction. The RG equations read \cite{Nilles:1983ge,Martin:1997ns,Haber:1984rc}
\begin{eqnarray}
\label{betafct}
{d \over dq} m^2_{H_u} &=& {3h^2_t \over 8 \pi^2}
\left(m^2_{H_u} + m^2_{{Q}_3} + m^2_{{u}_3} +
|A_t|^2 \right)
- {1 \over 2 \pi^2} \left({1 \over 4} g^2_1 |m_1|^2 + {3 \over 4}
g^2_2 |m_2|^2 \right) \,, \nonumber \\
{d \over dq}m^2_{{u}_3} &=& {2h^2_t \over 8 \pi^2}
\left(m^2_{H_u} + m^2_{{Q}_3} + m^2_{{u}_3} + |A_t|^2
\right)
- {1 \over 2 \pi^2}\left({4 \over 9} g^2_1 |m_1|^2 + {4 \over 3}
g^2_3 |m_3|^2 \right) \,, \nonumber \\
{d \over dq} m^2_{{Q}_3} &=& {h^2_t \over 8 \pi^2}
\left(m^2_{H_u} + m^2_{{Q}_3} + m^2_{{u}_3} + |A_t|^2
\right)\,
- {1 \over 2 \pi^2}\left({1 \over 36} g^2_1 |m_1|^2 + {3 \over 4}
g^2_2 |m_2|^2 + {4 \over 3} g^2_3 |m_3|^2 \right) \, ,
\nonumber \\
{d \over dq} A_t &=& {3 h^2_t \over 8 \pi^2} A_t - {1 \over 2 \pi^2}
\left( {13 \over 36} g^2_1 m_1 + {3 \over 4} g^2_2 m_2 + {4 \over 3}
g^2_3 m_3 \right) \, .
\end{eqnarray}
Here $q$ denotes the logarithmic scale; this could be an external
energy or momentum scale, but in the case at hand the relevant scale
is set by the VEV(s) of the fields themselves. $h_t$ is the
top Yukawa coupling, while $g_i$ and $m_i$ are respectively the gauge
couplings and soft breaking gaugino masses of
$U(1)_Y \times SU(2)\times SU(3)$ . If $h_t$ is the only
large Yukawa coupling (i.e. as long as $\tan\beta$ is not very large),
the beta functions for $({\rm mass})^2$ of squarks of the first and
second generations and sleptons only receive significant
contributions from gauge/gaugino loops. A review of these effects can
be found in \cite{Drees:1995hj,Martin:1997ns}.


\subsection{$F$-and $D$-flat directions of MSSM}

Field configurations satisfying simultaneously:
\begin{equation}
\label{fflatdflat}
D^{a}\equiv X^{\dagger}T^{a}X=0\,, \quad \quad
F_{X_{i}}\equiv \frac{\partial W}{\partial X_{i}}=0\,.
\end{equation}
for $N$ chiral superfields $X_{i}$, are called respectively
$D$-flat and $F$-flat. $D$-flat directions are parameterized by gauge invariant monomials of
the chiral superfields. A powerful tool for finding the flat directions
has been developed in
\cite{Buccella:1982nx,Affleck:1983mk,Affleck:1984fy,Dine:1995uk,Dine:1995kz,Luty:1995sd,Gherghetta:1995dv},
for a review see~\cite{Enqvist:2003gh,Dine:2003ax}
where the correspondence between gauge invariance and flat directions has been
employed. 

A single flat direction necessarily carries a global $U(1)$ quantum
number, which corresponds to an invariance of the effective Lagrangian for the
order parameter $\phi$ under phase rotation $\phi\to e^{i\theta}\phi$.
In the MSSM the global $U(1)$ symmetry is $B-L$. For example, the
$LH_u$-direction (see below) has $B-L=-1$.
A flat direction can be represented by a composite gauge invariant
operator, $X_m$,  formed from the product of $k$ chiral superfields
$\Phi_i$ making up the flat direction: $X_m=\Phi_1\Phi_2\cdots \Phi_m$.
The scalar component of the superfield $X_m$ is related to the
order parameter $\phi$  through $X_m=c\phi^m$.

An example of a $D$-and $F$-flat direction is provided by
\begin{equation}
\label{example}
H_u=\frac1{\sqrt{2}}\left(\begin{array}{l}0\\ \phi\end{array}\right),~
L=\frac1{\sqrt{2}}\left(\begin{array}{l}\phi\\ 0\end{array}\right)~,
\end{equation}
where $\phi$ is a complex field parameterizing the flat direction,
or the order parameter, or the AD field. All the other fields are
set to zero. In terms of the composite gauge invariant operators,
we would write $X_m=LH_{u}~(m=2)$.

From Eq.~\ref{example} one clearly obtains $
F_{H_u}^*=\lambda_uQ u +\mu H_d=F_{L}^*=\lambda_dH_d e\equiv 0$
for all $\phi$. However there exists a non-zero F-component given
by $F^*_{H_d}=\mu H_u$. Since $\mu$ can not be much larger than the
electroweak scale $M_W\sim {\cal O}(1)$~TeV, this contribution is of
the same order as the soft SUSY breaking masses, which are
going to lift the degeneracy. Therefore, following \cite{Dine:1995kz}, one may
nevertheless consider $LH_u$ to correspond to a F-flat direction.

The relevant $D$-terms read
\begin{equation}
\label{Dterm0}
D^a_{SU(2)}=H_u^\dagger\tau_3H_u+L^\dagger\tau_3L=\frac12\vert\phi\vert^2
-\frac12\vert\phi\vert^2\equiv 0\,.
\end{equation}
Therefore the $LH_u$ direction is also $D$-flat.

The only other direction involving the Higgs fields and thus soft
terms of the order of $\mu$ is $H_uH_d$. The rest are purely leptonic,
such as $LLe$, or baryonic, such as $udd$, or
mixtures of leptons and baryons, such as $QLd$. These combinations
give rise to several independent flat directions that can be obtained
by permuting the flavor indices. For instance, $LLe$ contains
the directions $L_1L_2e_3$, $L_2L_3e_1$, and $L_1L_3e_2$, let
us consider a particular configuration~\cite{Dine:2003ax}:
\begin{eqnarray} 
L_{1} = \frac{1}{\sqrt{3}} \left( \begin{array}{c} \phi  \\
0 \end{array}  \right), ~~~
L_{2} = \frac{1}{\sqrt{3}} \left(\begin{array}{c}
0 \\
\phi \end{array} \right), ~~~
{e_{3}} =\frac{1}{\sqrt{3}} \phi,  
\end{eqnarray}
The SU(2)$\times$U(1) $D$-terms are
\begin{eqnarray}
\label{true} 
V_{D}   &=& \frac{g^{2}}{8}\left( |L_{1}|^{2} - |L_{2}|^{2}\right)^{2} +
\frac{g'^{2}}{72}\! \! \left ( |L_{1}|^{2} - 3|L_{2}|^{2} + 2
|{e_{3}}|^{2}\right )^{2} , \end{eqnarray}
where $g = e/ \sin \theta_{w}$ is the SU(2) coupling and $g' = e/ \cos \theta_{\rm w}$ is the
U(1)$_{\rm Y}$ coupling. When $L_{1}^{1} = L_{2}^{2} = {e_{3}} = \phi$ the $D$-terms in the 
potential vanish as they must.

Along a flat direction gauge symmetries get broken, with the gauge
supermultiplets gaining mass by super-Higgs mechanism with
$m_g=g\langle\phi\rangle$. Several chiral supermultiplets typically
become massive by virtue of Yukawa couplings in the superpotential;
for example, in the $LH_u$ direction one finds the mass terms
$W_{\rm mass}=\lambda_u\langle\phi\rangle Q u+\lambda_e\langle\phi\rangle H_d e$.
In this respect when the flat direction VEV vanishes, i.e. $\phi=0$, the gauge symmetry
gets enhanced.

Vacuum degeneracy along a flat direction can be broken in two ways:
by SUSY breaking, or by higher order non-renormalizable
operators appearing in the effective low energy theory. Let us first
consider the latter option. 


\subsubsection{Non-renormalizable superpotential corrections}

Non-renormalizable superpotential terms in the MSSM can be viewed
as effective terms that arise after one integrates out fields with very
large mass scales appearing in a more fundamental (say, string) theory.
Here we do not concern ourselves with the possible restrictions
on the effective terms due to discrete symmetries present
in the  fundamental theory, but assume that all operators
consistent with symmetries may arise. Thus in terms of the
invariant operators $X_m$, one can have terms of the type \cite{Dine:1995uk,Dine:1995kz}
\begin{equation}
\label{Xton}
W=\frac{h}{d M^{d-3}} X^{k}_{m}=\frac{h}{d M^{d-3}}\phi^d\,,
\end{equation}
where the dimensionality of the effective scalar operator $d=mk$,
and $h$ is a coupling constant which could be complex with
$|h|\sim {\cal O}(1)$. Here $M$ is some large mass, typically of
the order of the Planck mass or the string scale (in the heterotic
case $M \sim M_{GUT}$). The lowest value of $k$ is $1$ or $2$, depending
on whether the flat direction is even or odd under $R$-parity.

A second type of term lifting the flat direction would be of the form
\cite{Dine:1995uk,Dine:1995kz}
\begin{equation}
\label{2ndtype}
W={h^{\prime}\over M^{d-3}}\psi\phi^{d-1}~\,,
\end{equation}
where $\psi$ is not contained in $X_m$. The superpotential term
\eq{2ndtype} spoils F-flatness through $F_\psi \neq 0$. An example
is provided by the direction $ u_1 u_2 u_3 e_1 e_2$,
which is lifted by the non-renormalizable term
$W=(h'/M) u_1 u_2 d_2 e_1$. This superpotential term
gives a non-zero contribution
$F_{ d_2}^*=(h'/M) u_1 u_2 e_1\sim (h^{\prime}/M)\phi^3$
along the flat direction.

Assuming minimal kinetic terms, both types discussed above
in Eqs.~(\ref{Xton},\ref{2ndtype}) yield a generic non-renormalizable
potential contribution that can be written as
\be{nrpot}
V(\phi)={\vert\lambda\vert^2\over M^{2d-6}}(\phi^*\phi)^{d-1}~,
\ee
where we have defined the coupling $|\lambda|^2\equiv |h|^2+|h'|^2$.
By virtue of an accidental $R$-symmetry under which $\phi$ has a charge
$R=2/d$, the potential \eq{nrpot} conserves the $U(1)$ symmetry carried
by the flat direction, in spite of the fact that at the superpotential
level it is violated, see Eqs.~(\ref{Xton},\ref{2ndtype}).

All the non-renormalizable
operators can be generated from SM gauge monomials with $R$-parity
constraint which allows only even number of odd matter parity fields
($Q,L, u, d, e$) to be present in each superpotential term.
At each dimension $d$, the various $F=0$ constraints are separately imposed in
order to construct the basis for monomials.

As an example, consider flat directions involving the Higgs fields
such as $H_{u}H_{d}$ and $LH_{u}$ directions. Even though they
are already lifted by the $\mu$ term, since $\mu$
is of the order of SUSY breaking scale, for cosmological purposes
they can be considered flat.
At the $d=4$ level the superpotential reads~\cite{Dine:1995uk,Dine:1995kz,Gherghetta:1995dv}
\begin{equation}
\label{exnon}
W_{4} \supset \frac{\lambda}{M}(H_{u}H_{d})^2+\frac{\lambda_{ij}}{M}
(L_{i}H_{u})(L_{j}H_{u})\,.
\end{equation}
Let us assume $\lambda, \lambda_{ij}\neq 0$.
Note that $F_{H_{d}}=0$ constraint implies
$\lambda H^{\alpha}_{u}(H_{u}H_{d})=0$, which acts as a basis for the
monomials. An additional constraint can be obtained by contracting
$F_{H_{d}}=0$ by $\epsilon_{\alpha \beta}H^{\beta}_{d}$, which forms the
polynomial $H_{u}H_{d}=0$ in the same monomial basis. Similarly the
constraint $F_{H_{u}}=0$, along with the contraction yields
$\lambda^{ij}(L_{i}H_{u})(L_{j}H_{u})=0$. This implies that $L_{i}H_{u}=0$
for all $i$. Therefore the two monomials $LH_{u}$ and $H_{u}H_{d}$ can
be lifted by $d=4$ terms in the superpotential Eq.~(\ref{exnon}).

The other renormalizable flat directions are
$LLe, u u d, Q dL, QQQL, Q uQ d, u u d e$
and $Q uL e,  d d dLL,  u u u e e, Q uQ u e, QQQQ u,  u u dQdQ d$,
and $(QQQ)_{4}LLL e$. The unique flat 
directions involving $(Q,u,e)$ is lifted by $d=9$, $(L,d)$ by $d=7$, and
$(L,d,e)$ by $d=5$. The flat directions involving $(L,e),(u,d)$ and $(L,d,e)$
are all lifted by $d=6$ terms in the superpotential, while the rest of the
flat directions are lifted already by $d=4$ superpotential 
terms~\cite{Gherghetta:1995dv}.

Vacuum degeneracy will also be lifted by SUSY
breaking soft terms. The full flat direction potential in the simplest case reads~\cite{Dine:1995uk,Dine:1995kz}
\begin{equation}
\label{susybreak}
V(\phi)=m_0^2\vert\phi\vert^2+\left[{\lambda A\phi^d\over dM^{d-3}}
+{\rm h.c.}\right] +\lambda^2\frac{|\phi|^{2d-2}}{M^{2d-6}}\,,
\end{equation}
where the SUSY breaking mass $m_0,~A \sim 100-1000$~GeV.  

While considering the dynamics of a flat direction, in a cosmological setting, 
the superpotential of Eq.~(\ref{2ndtype}) generates a vanishing $A$-term. This 
is due to the fact that $\phi$ being light during inflation, i.e. $H_{inf}\gg m_0$, 
it obtains large VEV during inflation due to random walk. As a result $\psi$ field
gets a large mass induced by the VEV of $\phi$, which drives $\psi$ to roll down to 
its minimum in less than one Hubble time, i.e. $\langle \psi\rangle =0$. The $A$-term 
being proportional to $\psi$ vanishes in this limit, and does not play any dynamical role
during and after inflation. In other words, the $\psi$ field being super massive decouple 
from the dynamics.

The $A$-term in \eq{susybreak} violates the $U(1)$ carried by the flat direction and thus provides the
necessary source for $B-L$ violation in AD baryogenesis. In
general, the coupling $\lambda$ is complex and has an associated  phase
$\theta_\lambda$. Writing
$\phi=\vert\phi\vert \exp(i\theta)$, one obtains a potential
proportional to $\cos(\theta_\lambda+n\theta)$ in the angular
direction. This has $n$ discrete minima for the phase of $\phi$,
at each of which $U(1)$ is broken.

A very generic one-loop quantum corrections result in a logarithmic running of the soft
SUSY breaking parameters . The {\it effective} potential for the flat direction
is then given by~\cite{Enqvist:1997si,Enqvist:1998en,Enqvist:2003gh}:
\begin{eqnarray}
V_{eff}(\phi) &=& \frac{1}{2} m_0^2 \phi^2 \left[
  1+ K_1 \log \left(\frac{\phi^2}{\mu_0^2}\right) \right] -
  \frac{\lambda_{d,0} A_0}{dM^{d-3}} \phi^d \left[ 1+ K_2
  \log\left(\frac{\phi^2}{\mu_0^2}\right) \right] \nonumber \\ & & +
  \frac{\lambda_{d,0}^2}{M^{2(d-3)}} \phi^{2(d-1)} \left[ 1 + K_3 \log
  \left(\frac{\phi^2}{\mu_0^2}\right) \right]\,.
\end{eqnarray}
where $m_0$, $A_0$, and $\lambda_{d,0}$ are the values of $m_{\phi}$,
$A$ and $\lambda_n$ given at a scale $\mu_0$. Here $A_0$ is chosen to
be real and positive (this can always be done by re-parameterizing the
phase of the complex scalar field $\phi$), and $|K_i|<1$ are
coefficients determined by the one-loop renormalization group
equations. We will provide an explicit example in section in
\ref{Rad-correct-MSSM-pot}.


\subsubsection{Spontaneous symmetry breaking and the physical degrees of freedom}

A flat direction VEV spontaneously breaks symmetry and gives 
masses to the gauge bosons/gauginos similar to the Higgs 
mechanism~\cite{Dine:1995kz,Allahverdi:2005mz,Allahverdi:2007zz,Allahverdi:2008pf,Allahverdi:2006xh}.
A crucial point is to identify the physical degrees of freedom and
their mass spectrum in presence of a non-zero flat direction VEV.
Let us consider the simplest flat direction, which includes only two
fields: ${H_u H_d}$. This is also familiar from the electroweak
symmetry breaking in MSSM. A clear and detailed discussion is given
in~\cite{Haber:1984rc}. 

One can always rotate the field configuration to a basis where, up to
an overall phase, $H^1_{u} = H^2_{d} = 0$ and $H^2_{u} = H^1_{d} =
\phi_0/\sqrt{2}$. Here superscripts denote the weak isospin components
of the Higgs doublets. In this basis the complex scalar field is
defined by:
\begin{equation} \label{flat}
\varphi = {(H^2_{u} + H^1_{d}) \over \sqrt{2}}\,,
\end{equation}
represents a flat direction. Its VEV breaks the $SU(2)_W \times
U(1)_Y$ down to $U(1)_{\rm em}$~(in exactly the same fashion as in the
electroweak vacuum). The $W^{\pm}$ and $Z$ gauge bosons then obtain
masses $m_W,~m_Z \sim g \phi_0$ from their couplings to the Higgs
fields via covariant derivatives ($g$ denotes a general gauge
coupling). There are also:
\begin{equation} \label{gold}
\chi_1 = {(H^2_{u} - H^1_{d}) \over \sqrt{2}}\,,~~
{\rm and}~~
\chi_2 = {(H^1_{u} + H^2_{d}) \over \sqrt{2}}\,.
\end{equation}
Then $\chi_2$ and $\chi_{1,~R}$~($R$ and $I$ denote the real and
imaginary parts of a complex scalar field respectively) acquire masses
equal to $m_W$ and $m_Z$, respectively, through the $D-$term part of
the scalar potential. Note that
\begin{equation}
\chi_3 = {(H^1_{u} - H^2_{d}) \over \sqrt{2}}\,,
\end{equation}
and $\chi_{1,~I}$ are the three Goldstone bosons, which are eaten by
the gauge fields via the Higgs mechanism. Therefore, out of $8$ real
degrees of freedom in the two Higgs doublets, there are only two light
{\it physical} fields: $\varphi_{R},~\varphi_{I}$. They are exactly
massless when SUSY is not broken (and there is no $\mu$ term either).

An important point is that the masses induced by the flat direction
VEV are SUSY conserving. One therefore finds the same mass spectrum in
the fermionic sector. More specifically, the Higgsino fields
${\widetilde H}^1_{u}$ and ${\widetilde H}^2_{d}$ are paired with the
Winos, while $({\widetilde H}^2_{u} - {\widetilde H}^1_{d})/\sqrt{2}$
is paired with the Zino to acquire masses equal to $m_W$ and $m_Z$,
respectively, through the gaugino-gauge-Higgsino interaction terms.
The fermionic partner of the flat direction $({\widetilde H}^2_{u} +
{\widetilde H}^1_{d})/\sqrt{2}$ remains massless (note that the photon
and photino are also massless, but not relevant for our discussion).

SUSY being broken, $\varphi$ obtains a mass
$m_{\varphi} \sim {\cal O}({\rm TeV})$ from soft SUSY breaking term
(the same is true for the gauginos). However, for $g \varphi_0 \gg
{\cal O}({\rm TeV})$, which is the situation relevant to the early
Universe, the mass spectrum is hierarchical: $\chi_{1,~R}$,
$\chi_{2}$, and gauge fields (plus their fermionic partners) are
superheavy.

In a general case the total number of light scalars, $N_{light}$, is
given by~\cite{Allahverdi:2008pf,Allahverdi:2006xh}:
\begin{equation} \label{lightnumb}
N_{light} = N_{total} - (2 \times N_{broken}),
\end{equation}
where $N_{total}$ is the total number of scalar degrees of freedom,
and $N_{broken}$ is the number of spontaneously broken symmetries.
Note that the factor $2$ counts for the number of eaten Goldstone
bosons plus the number of degrees of freedom which have obtained large
masses equal to those of the gauge bosons. In the case of $H_u H_d$
direction, Eq.~(\ref{lightnumb}) reads:~$N_{light}=2 = 2\times 2\times 2 - (2 \times 3)$. Similarly,
for $LH_u$ flat direction, $N_{light}= 2\times 2\times 2-(2\times 3)=2$.


\subsection{$N=1$ Supergravity (SUGRA)}

At tree level, $N=1$ SUGRA potential in four dimensions is
given by the sum of $F$ and $D$-terms \cite{Nilles:1983ge}
\begin{equation}
\label{eq:partphys:sugrapotential}
V=e^{K(\phi_i,\phi^{\ast i}) /M_{\rm P}^2}\left[\left(K^{-1}\right)^{j}_{i}F_{i}F^{j}-3\frac{
|W|^2}{M_{\rm P}^2}\right]+
\frac{g^2}{2}{\rm Re}f^{-1}_{ab}{\hat D}^{a}{\hat D}^{b}\,,
\end{equation}
where
\begin{equation}
F^{i}=W^{i}+K^{i}\frac{W}{\MP^2}\,, \quad \quad
{\hat D}^{a}=-K^{i}(T^{a})^{j}_{i}\phi_{j}+\xi^{a}\,.
\end{equation}
where we have added the Fayet-Iliopoulos contribution $\xi^{a}$ to the
$D$-term, and $\hat D^{a}=D^{a}/g^{a}$, where $g^a$ is gauge coupling.
Here $K(\phi_i,\phi^{\ast i})$ is the K\"ahler potential, which is a function of the
fields $\phi_i$, and $K^i\equiv \partial K/\partial \phi_i$. In the simplest case, at tree-level
$K=\phi^{\ast i}\phi_{i}$ (and $K^{j}_{i}=(K^{-1})^{j}_{i}=\delta^{j}_{i}$)~\footnote{In general the superpotential can have non-renormalizable contributions. Similarly, the 
K\"ahler potential can be expanded as: 
$K=\phi_{i}\phi^{\ast i}+ (k^{ij}_k \phi_i\phi_j\phi^{\ast k}+c.c.)/M_{\rm P}+
(k^{ij}_{kl}\phi_i \phi_j\phi^{\ast k}\phi^{\ast l}\phi^{\ast k}\phi^{\ast l}+k^{ijk}_{l}\phi_i\phi_j\phi_k\phi^{\ast l}
+c.c.)/M_{\rm P}^2+\cdots )$. We will discuss no-Scale SUGRA where the choice of K\"ahler potential
plays an important role in maintaining flat potential during inflation.}. The kinetic terms for the scalars 
take the form: 
\begin{equation}
\frac{\partial^2K}{\partial \phi_{i}\partial \phi_{j}^{\ast}}D_{\mu}\phi_{i} D^{\mu}\phi^{\ast}_{j}\,.
\end{equation}
The real part of the gauge kinetic function matrix is given by ${\rm Re}f_{ab}$. 
In the simplest case, it is just a constant, 
$f_{ab}=\delta_{ab}/g_a^2$, and the kinetic terms for the gauge potentials, $A^{a}_{\mu}$, are given by~\footnote{In general, 
$f_{ab}=\delta_{ab}(1/g_a^2+f^{i}_{a}\phi_{i}/M_{\rm P}+\cdots)$. The gauginos masses are typically 
given by $m_{\lambda^{a}}={\rm Re}[f^{i}_{a}]\langle F_{i}\rangle/2M_{\rm P}$. For a universal gaugino
masses, $f^{i}_{a}$ are the same for all the three gauge groups of MSSM.}: 
\begin{equation}
\frac{1}{4}({\rm Re} f_{ab})F^{a}_{\mu\nu}F_{a}^{\mu\nu}\,.
\end{equation}
SUGRA will be broken if one or more of the $F_{i}$ obtain a VEV. The gravitino, spin $\pm 3/2$ 
component of the graviton, then absorb the goldstino component to become massive. Requiring
classically $\langle V\rangle=0$, as a constraint to obtain the zero cosmological constant,
one obtains 
\begin{equation}
m_{3/2}^2=\frac{\langle K^{i}_{j}F_{i}F^{\ast j}\rangle }{3M_{\rm P}^2}=e^{\langle K\rangle/M_{\rm P}^2}
\frac{|\langle W\rangle|^2}{M_{\rm P}^4}\,.
\end{equation}
In case of SUGRA at tree-level with minimal kinetic term, the super-trace formula
modifies to (for $D$-flat directions):
\begin{equation}
Str{\bf M^2 } \equiv \sum_{spin~J} (-1)^{2J} (2J+1)tr{\bf M}_{J}^2
\approx 2(n_\phi-1)m_{3/2}^2\,,
\end{equation}
where $n_{\phi}$ is the number of the chiral multiplets in the spontaneously broken SUGRA.
 
 
\subsubsection{SUSY generalization of one-loop effective potential} 
\label{See:SUSYgeneralization of one-loop effective potential}
 
A SUSY generalization of one-loop effective potential, Eq.~(\ref{induced-one-loop}), 
is given by~\cite{Ferrara:1994kg,Ferrara:1979wa}:
\begin{equation}
\Delta V = \frac{1}{64 \pi^{2}}Str\left({\bf M}^{0}\right)\Lambda_{c}^{4}\ln\left(\frac{\Lambda_{c}^{2}}{\mu^{2}}\right) +
\frac{1}{32\pi^{2}}Str\left({\bf M}^{2}\right)\Lambda_{c}^{2}+\frac{1}{64 \pi^{2}}Str\left({\bf M}^{4}\ln\left(\frac{{\bf M}^{2}}{\Lambda_{c}^{2}}\right)\right)+... \; ,
\label{eqf-oneloop}
\end{equation}
with $\Lambda_{c}$ being a momentum cut-off and $\mu$ the scale parameter. The renormalized potential will not depend on
$\Lambda_c$, and the dots stand for $\Lambda_{c}$-independent contributions.

The first term in Eq.~(\ref{eqf-oneloop}), being field-independent, this term can affect the cosmological constant 
problem in SUGRA, but does not affect the discussion of the gauge hierarchy problem. However, 
this term is always absent in SUSY theories, which possess equal numbers of bosonic and
fermionic degrees of freedom. In Ref.~\cite{Zumino:1974bg,Ferrara:1979wa}, it was shown that for
unbroken $N=1$ global SUSY, $Str {\bf M}^n$ is identically
vanishing for any $n$, due to the fermion-boson degeneracy within
SUSY multiplets. It was argued that the term $Str {\bf M}^2$ vanishes~\cite{Zumino:1974bg}, as
a field identity, if global SUSY is spontaneously broken in
the absence of anomalous $U(1)$ factors~\cite{Ferrara:1979wa}. Anyway, the third term in
Eq.~(\ref{eqf-oneloop}), plays the most important role in inflationary models in lifting the flat potential for the inflaton.
 
It was argued in \cite{Clauwens:2007wc}, that in the derivation of \cite{Ferrara:1979wa}, an explicit use has
been made of the fact that all first order partial derivatives of the tree-level effective potential to the fields vanish. 
This limits the region of applicability of the simple Coleman-Weinberg formula in the context of inflationary models.
Only at extremum (or saddle) points it is justified to throw away the second term of
Eq.~(\ref{eqf-oneloop})~\footnote{For a particular case of SUSY model of inflation, known as the hybrid inflation 
(see the discussion in section IVE.1, the Coleman-Weinberg formula given by the third term in Eq.~(\ref{eqf-oneloop}) is 
applicable  only in the global minimum and along the inflationary valley, because here all derivatives in the tree-level 
potential vanish. If one wishes to study the field dynamics throughout the phase space, one requires special 
care~\cite{Clauwens:2007wc}.}.


\subsubsection{Inflaton-induced SUGRA corrections}

Since non-zero inflationary potential gives rise to
SUSY breaking, the scale of which is given by the time dependent Hubble
parameter. It is important to know how this will affect any other light scalar field during and after inflation,
in particular the flat directions of MSSM. At early times this breaking is dominant over breaking from
the hidden sector~\footnote{Note that this does not replace the soft SUSY breaking terms required to solve the problems 
for the low energy physics, such as addressing the hierarchy problem or the electroweak symmetry breaking within MSSM, etc.}. After the end of inflation, in most models the inflaton oscillates and its finite energy density still dominates and breaks SUSY in the visible sector. A particular class of non-renormalizable interaction terms induced by the
inflaton arise if the K\"ahler potential has a form \cite{Gaillard:1995az,Bagger:1995ay,Dine:1995uk,Dine:1995kz}
\begin{equation}
\label{Iphicoupl}
K =\int d^4\theta \frac{1}{M_{\rm P}^2}(I^{\dagger}I)(\phi^{\dagger}\phi)\,,
\end{equation}
where $I$ is the inflaton whose energy density
$\rho \approx \langle \int d^4\theta I^{\dagger}I\rangle$ dominates
during inflation, and $\phi$ is the flat direction. The interaction,
Eq. (\ref{Iphicoupl}), will generate an effective mass term in the Lagrangian in the global
SUSY limit, given by
\begin{equation}
{\cal L}= \frac{\rho_{I}}{M_{\rm P}^2}\phi^{\dagger}\phi =3H_I^2
\phi^{\dagger}\phi \,,
\end{equation}
where $H_I$ is the Hubble parameter during inflation.

For a minimal choice of flat direction K\"ahler potential
$K(\phi^{\dagger},\phi)=\phi^{\dagger}\phi$, during inflation
the effective mass for the flat direction is found  to be \cite{Dine:1995uk,Dine:1995kz}~\footnote{If the K\"ahler 
potential has a shift symmetry, or its of type no-scale model, the Hubble induced mass correction to the flat direction
does not arise at the tree-level potential.}:
\begin{equation}
m^2_{\phi}=\left(2+\frac{F^{\ast}_{I}F_{I}}{V(I)}\right)H^2\,.
\end{equation}
Here it has been assumed that the main contribution to the inflaton
potential comes from the $F$-term. If there were $D$-term contributions
 $V_{D}(I)$ to the inflationary potential, then a correction of order
$V_{F}(I)/(V_{F}(I)+V_{D}(I))$ must be taken into account.
In purely $D$-term inflation there is no Hubble
induced mass correction to the flat direction during inflation
because $F_{I}=0$. However, when $D$-term inflation ends, the energy density
stored in the $D$-term is converted to an $F$-term and to kinetic energy of
the inflaton. Thus again a mass term $m_{\phi}^2 =\pm {\cal O}(1)H^2$
appears naturally, however the overall sign is undetermined \cite{Kolda:1998kc}.


\subsubsection{No-scale SUGRA}

There exists a choice of  K\"ahler potential for which there is no
Hubble induced correction to the mass of the lights scalars. An example of
this is provided by no-scale models, for which
$K\sim \ln (z+z^*+\phi_i^\dagger\phi)$, where $z$ belongs to SUSY
breaking sector, and $\phi_{i}$ belongs to the matter sector, and both are
measured in terms of reduced Planck mass~\cite{Cremmer:1983bf,Witten:1982hu,Ellis:1983sf}
(for a review, see~\cite{Lahanas:1986uc}). In no-scale models there
exists an enhanced symmetry known as the Heisenberg symmetry
\cite{Binetruy:1987xj}, which is defined on the chiral fields
as $\delta z=\epsilon^{\ast}\phi^{i}$, $\delta\phi^{i}=\epsilon^{i}$,
and $\delta y^{i}=0$, where $y^{i}$ are the hidden sector fields, such
that the combinations $\eta =z+z^{\ast}-\phi_{i}^{\ast}\phi^{i}$,
and $y_{i}=0$ are invariant. For a especial choice
\begin{equation}
K=f(\eta)+\ln[W(\phi)/M_{\rm P}^3]^2+g(y)\,,
\end{equation}
The $N=1$ SUGRA potential reads~\cite{Gaillard:1995az,Campbell:1998yi}
\begin{equation}
V_F=e^{f(\eta)+g(y)}\left[\left(\frac{f^{\prime 2}}{f^{\prime \prime}}-3\right)
\frac{|W|^2}{M_{\rm P}^2}-\frac{1}{f^{\prime 2}}\frac{|W_{i}|^2}{M_{\rm P}^2}
+g_{a}(g^{-1})^{a}_{b}g^{b}\frac{|W|^2}{M_{\rm P}^2}\right]\,.
\end{equation}
Note that there is no cross term in the potential such as
$|\phi^{\ast}_{i}W|^2$. As a consequence any tree level flat direction
remains flat even during inflation \cite{Gaillard:1995az} (in fact
it is the Heisenberg symmetry which protects the flat directions from obtaining
Hubble induced masses~\cite{Binetruy:1987xj}).  

A particular choice of K\"ahler potential, i.e. $$K=-3\ln(\varphi+\bar\varphi),$$ arises 
quite naturally from string compactifications~\cite{Witten:1985xb,Burgess:1985zz}. For a constant 
superpotential, $W_0$, the F-term of the potential yields;
\begin{equation}
V_F=e^{K/M_{\rm P}^2}\left[(K^{-1})^{j}_{i}K^{i}K_{j}-3\right]|W_0|^2\,.
\end{equation}
and for the above choice of the K\"ahler potential, $V_F=0$ for all $\varphi$, because the
K\"ahler potential satisfies $(K^{-1})^{j}_{i}K^{i}K_{j}=3$, this is a property of no-scale model.  The symmetry
is broken by gauge interactions or by coupling in the renormalizable
part of the K\"ahler potential. Then the mass of the flat direction
condensate arises from the running of the gauge couplings.

\subsection{(SUSY) Grand Unified Theories}\label{sec:partphys:SUSYGUTs}

The observation that within SUSY the value of three gauge couplings nearly meet at $\sim 2\times 10^{16}$~GeV
has led to the idea~\footnote{Initial attempts were made by Pati and Salam~\cite{Pati:1974yy} in
$SU(2)_L\times SU(2)_R\times SU(4)_C$ model, where quarks and
leptons are unified-a lepton becomes the fourth color of a quark.
Although the model did not have gauge unification, but later
models unified the couplings in a left-right symmetric
model~\cite{Mohapatra:1974hk,Mohapatra:1979ia,Mohapatra:1980yp}.}
that the three gauge groups emerge from a single (Grand Unified)
group $G_{GUT}$ with a single gauge coupling $g_{GUT}$~\footnote{The requirement of a simple gauge group can be
relaxed. It is also possible to have a single high energy gauge
coupling constant with a non-simple gauge group made up of products
of identical groups, i.e. $G_{GUT} = H\times H\times \dots $. The single
gauge coupling constant is ensured by imposing an additional
symmetry that render the theory invariant under exchange of the
factors of $H$.}. Another motivation for such a unification beyond
the SM is to explain the quantification of the electric charge,
that is to explain why $3Q(\rm{quark})=Q(\rm{electron})$.
Various Lie group candidates are $SU(n+1)$,
$SO(4n+2)$ families and $E_6$, when imposing that the groups must
contain anomaly free complex representations to accommodate 
the known fermions of the SM \cite{Slansky:1981yr}. The smallest and most studied of
these candidates are $SU(5)$, $SO(10)$, and $E_6$. 
For reviews, see~\cite{Langacker:1980js,Kounnas:1985cj,Mohapatra:1986uf,Mohapatra:1999vv,Raby:2008gh}
and more specifically~\cite{Nath:2006ut} for proton decay,
and~\cite{Slansky:1981yr} for group theory and useful tables on
symmetry breaking.

Although the idea of gauge unification is very appealing, but this concept has additional
constraints. Clearly the GUT group must contain $G_{GUT} \supset
G_{SM}$ and must break spontaneously into its subgroups and finally
$G_{SM}$ at some high scales. During this breaking the extra gauge bosons
of $G_{GUT}$, which carry quantum numbers of several groups and $G_{SM}$,
acquire a superheavy mass, $M_{GUT}$, which prevent from a too rapid
decay of the nucleons. The choice of $G_{GUT}$ and the presence of SUSY
affect the predictions for proton lifetime of a given model as the scale of SUSY breaking and the particles
involved in Feynman graphs modify the results. 
Finally, the electroweak precision measurements require the
particles beyond the SM to be heavy enough, for instance
the famous doublet-triplet splitting of the SM higgs from the additional Higgs fields
(see the $SU(5)$ example below).

\subsubsection{$SU(5)$ and $SO(10)$ GUT}

The most studied candidate is based on $SU(5)$ whose
generators are represented by five-by-five traceless, Hermitian
matrices. $SU(3)\times SU(2)\times U(1)$ is one of its maximal
subgroup (of same rank) together with $SU(4)\times U(1)$. Let us
highlight the main features of the fermionic, bosonic and the Higgs
sectors. If the gauge bosons necessarily belong to the adjoint
representation of dimension 24, the fermions and the Higgs can be
accommodated in any representations that lead to the right
phenomenology. The smallest representations for $SU(5)$ are the
fundamental $\bf{5}$, $\bf{10}$, $\bf{15}$,
$\bf{24},~\dots$.  First,  the 15 fermions can be put in a
fundamental $\bf{\bar{5}}$ and $\bf{10}$, a anti-symmetric
traceless $5\times 5$ matrix~\cite{Georgi:1974sy}. The
decomposition of these representations under $SU(3)_c\times
SU(2)_L$ lead to the quantum numbers of the fermions under the SM:
\begin{equation}
{\bf {\bar 5}=({\bar 3},1)+(1,2)} \quad { \rm and} \quad  {\bf
10}=(\bf {3,2})+({\bf \bar 3,1})+({\bf 1,1})~.
\end{equation}
This accommodates a quark triplet, $d_R$, and a leptonic
doublet, $(e^-_L,\nu_L)$, respectively in the $\bf{5}$, and the
quark triplets, $(u_L,d_L)$ $u_R$, and $e^+$, respectively in the
${\bf 10}$. The construction of Q as a combination of two diagonal
\emph{traceless} generators of $SU(5)$ requires that the sum of all
the charges of the fermions in the fundamental representation vanishes, 
and thus explains the quantification of the charges.

There are $24$ gauge bosons of $SU(5)$ and their quantum number
under the SM are given by the decomposition under
$SU(5)\supset SU(3)\times SU(2)$~\cite{Slansky:1981yr}:
\begin{equation}\label{eq:partphys:decomp24}
{\bf 24= (1,1)+(1,3)+(3,2)+(\bar 3,2)+(8,1)}~,
\end{equation}
which are the photon, the triplet $W^{\pm},~Z$ bosons,
the so-called $X$ and $\bar X$, and the gluon octet. The 12 gauge bosons
$X$ and $\bar X$ carry both $SU(3)_c$ and $SU(2)_L$ quantum
numbers, therefore they violate the baryon quantum number, leading
to a proton decay.

The Higgs sector of $SU(5)$ must contain a Higgs field breaking
$SU(5)$ into $G_{SM}$, and at least another Higgs for the electroweak
breaking, as the two scales are different. The smallest
representation that contains a singlet under the SM is the adjoint
$\Sigma={\bf 24}$, see Eq.~(\ref{eq:partphys:decomp24}). Its
potential is a generalization of the usual mexican hat potential:
\begin{equation}
V(\Sigma)=-\frac{1}{2}m^2_{\Sigma}{\rm Tr} \Sigma^2
+\frac{1}{4}a({\rm Tr} \Sigma^2)^2 +\frac{1}{2} b{\rm
Tr}\Sigma^4~,
\end{equation}
where $m^2_{\Sigma}>0$ and $a, b$ are constants. The most general
potential for $\Sigma$ would also contain a term of the form 
${\rm Tr} \Sigma^3$, if the invariance $\Sigma \rightarrow -\Sigma$ is
not assumed. To break $SU(5)$ into the SM, the VEV has to be taken
in the direction $\Sigma_{24}={\rm Diag}(-1,-1,-1,3/2,3/2)$~\footnote{We
remind the reader that any field in the adjoint can be decomposed
into a linear combination of the the generators of $SU(5)$, the
factors are gauge singlet fields.}, though depending on the sign of
$b$, the minimum of the above potential can be obtained for VEVs
$\propto {\rm Diag}(1,1,1,1,-4)$, instead. The breaking in this case is the other maximal subgroup,
$SU(5)\rightarrow SU(4)\times U(1)$. The SM Higgs can be embedded
into a fundamental, $H={\bf 5}$, as it contains a component (the
$H_2=\bf{(1,2)}$), a singlet under $SU(3)_c$ but doublet under
$SU(2)_L$ has a potential:
\begin{equation}
V(H)=-\frac{m_5^2}{2}H^\dagger H+\frac{\lambda}{4}(H^\dagger
H)^2~.
\end{equation}
It is important to note that summing the two potentials
$V=V(\Sigma)+V(H)$ is not sufficient to construct a realistic
theory. As a general rule, the potential of all the Higgs fields
contain cross-terms that are allowed by symmetries. The correct
approach is to consider a field content and minimize the
multi-dimensional potential to be sure that the minima give
the right phenomenology. Embedding the SM Higgs into a fundamental leads
to the well-known doublet-triplet splitting problem, as a mass term
for $H_5$ would give the same mass to the SM doublet and the
additional triplet $H_3=\bf{(3,1)}$ contained in $H$. This is
catastrophic as $H_3$ possesses the right quantum number to
contribute to the proton decay, thus requiring that its mass is higher
than $\sim 10^{15}$ GeV. Interestingly one possible solution of
the second problem is to allow cross-terms between $H$ and
$\Sigma$.

Non-SUSY $SU(5)$ unification has many
problems. The three gauge couplings do not exactly meet at 
the unification scale at a single value. In addition, the unification scale $M_{GUT}\sim 10^{14} GeV$ is now
incompatible with the most recent observations of the proton decay~\cite{Shiozawa:1998si}.

SUSY $SU(5)$ introduces new low mass particles in the spectrum
and the existence of a SUSY breaking scale affect the running
of the gauge couplings, which allow for a better unification at a
larger scale $M_{GUT} \simeq 2\times 10^{16}$ GeV. Now the proton decay is 
even more suppressed in the dimensional $6$ channel. However the proton decay can now 
happen through dimensional 4 operators, that would be more dramatic. 
In order to suppress them one requires the R-parity (see the discussion in Sect.~\ref{sec:partphys:MSSMnExtensions}).

One of the major problems of SUSY $SU(5)$ is the doublet-triplet
splitting. Part of the superpotential which generates the masses
to the fermions is given by:
\begin{equation}
W\sim a{\rm Tr}(\Sigma)+b{\rm Tr}(\Sigma^2)+c{\rm
Tr}(\Sigma^3)+\lambda(H\Sigma\bar H+\mu H\bar H)\,,
\end{equation}
where $\Sigma$ belongs to ${\bf 45}$, $H$ belongs to ${\bf 5}$ and
$\bar H$ to ${\bf \bar 5}$. Out of three degenerate vacua, the
right one with the SM gauge group is given by; ${\rm Diag}(\langle
\Sigma\rangle)=(2,2,2,-3,-3)M_{GUT}$. Substituting the VEV in the
above superpotential generates an effective superpotential;
\begin{equation}
W\sim \lambda(2M_{GUT}+\mu)\bar H(\bar
3)H(3)+\lambda(-3M_{GUT}+\mu)H(2)\bar H(\bar 2)\,.
\end{equation}
Since, the Higgs mass ought to be around $\sim 100$~GeV, the
cancellation in $\mu\sim 3M_{GUT}$ has to be within one part in
$10^{12}$ to match the observed expectation. This fine tuning is
also related to the $\mu$-problem in MSSM. This fine tuning can be
replaced by a see saw mechanism in the case of SUSY $SO(10)$, where
the triplet can be made naturally heavy as compared to the
doublet~\cite{Dimopoulos:1981xm}. On the cosmological front,
$SU(5)$ conserves $B-L$, and baryogenesis via leptogenesis cannot
take place. This is another reason why it is desirable to go
beyond $SU(5)$.


The next to the simplest gauge group candidate for the unification is
$SO(10)$~\cite{Fritzsch:1974nn} (see some recent 
reviews~\cite{Babu:1992ia,Barr:1997hq,Bajc:2004xe,Aulakh:2006hs}), for which the entire generation
of fermions can be fit into a single representation, ${\bf 16}$, of
$SO(10)$ which decompose under $SU(5)\times U(1)$ into ${\bf \bar
5 + 10+ 1}$. It can therefore accommodate all the fermions of the MSSM
along with an additional singlet fermion, with the quantum numbers
of a right-handed neutrino which can fit into the fundamental spinorial
representation $\bf{16}$~\cite{Minkowski:1977sc,Gell-Mann,Yanagida,Mohapatra:1980yp}.

The SM leptons are now identified as a fourth color of quarks
after the inclusion of right handed neutrino. The direct product
of spinor representation gives, ${\bf 16\bigoplus 16=10\bigoplus
120\bigoplus 126}$, therefore the Higgs sector must contain either
of these three, ${\bf 10, 120}$, or ${\bf 126}$, in order to
generate mass terms for the fermions, for detailed discussion
see Refs.~\cite{Langacker:1980js,Nath:2006ut,Raby:2008gh}. In particular
when ${\bf 126}$ develops a VEV, it gives rise to the Majorana
right handed neutrino masses as large as the GUT scale, in order
to generate the neutrino masses one would then have to invoke the
see saw mechanism~\cite{Mohapatra:1980yp,Goh:2003hf}~\footnote{The $SO(10)$ 
can be broken down to the MSSM via many routes, but
the most popular one is through left-right symmetric group,
$SO(10)\rightarrow SU(4)_c\times SU(2)_L\times SU(2)_R\rightarrow
SU(3)_c\times SU(2)_L\times SU(2)_R\times U(1)_{B-L} \rightarrow
SU(3)_c\times SU(2)_L\times U(1)_Y$ (see for e.g.
\cite{Jeannerot:2003qv} and references therein). This can happen
by a scalar belonging to ${\bf 210}$ or a combination of the
fields belonging to  ${\bf 45+54}$. The next stage of breaking,
i.e. left-right symmetry, can happen via the Higgs of $126$
dimensional representation or ${\bf 16}$ of $SO(10)$,
see for a recent review~\cite{Raby:2008gh}. }. This also
opens up naturally the possibility to realize baryogenesis via
leptogenesis \cite{Fukugita:1986hr}. On the proton decay front,
$SO(10)$ is also attractive as it contains a (gauged) $B-L$, the
kernel of which is a $Z_2$ that automatically plays the role of
R-parity. This requires however to break $SO(10)$ only with {\it safe}
representations, such as ${\bf 10,~45,~54,~120,~126,~210,\dots}$ but not
${\bf 16,~144,~560,\dots}$~\cite{Martin:1992mq}. Finally in the
doublet triplet front, the situation is improved compared to the
$SU(5)$ case by employing the Dimopoulos-Wilczek scenario
\cite{Dimopoulos:1981zu}.


\subsubsection{Symmetry breaking in SUSY GUT}

The most simple superpotential to break a  GUT symmetry is simply the generalization of
the Higgs, one particular type of superpotential which can
break $G_{GUT}$ can be written as:
\begin{equation}
W=\lambda X({\rm Tr}\Sigma^2-M_{\Sigma}^2)~,
\end{equation}
where $X$ is a GUT singlet. Such superpotentials play important role 
in inflationary cosmology, in the context
of hybrid inflation at the GUT scale. In particular the most
general Higgs sector (containing many scalar fields that could
play constructive roles in inflation) can be constructed once the
representations of all the fields present are known. 
For instance, $SO(10)$ can contain Higgs field in many representations,
such as a fundamental $\bf{10}$ $H_i$, a four index
$\bf{210}$ field $\Phi_{ijkl}$, or a 3 index $\bf{120}$
field $\Omega_{ijk}$. It is possible to construct a mass term, for example $\Phi^2$ symbolizing,
$\Phi_{ijkl}\Phi_{ijkl}$, but it is not possible to construct a
scalar term (with all indices contracted) using three factors
of $H$, or $\Omega$~\footnote{The same conclusion can be made by looking at whether
the product of three $\bf{10}$ contains a singlet of $SO(10)$ or not.} . 
For example,  in the minimal $SO(10)$~\cite{Bajc:2004xe}, the field content which realize the
breaking of $SO(10)$ are
$\Phi_{ijkl}$ in $\bf{210}$, $H_i$ in $\mathbf{210}$,
$\Sigma_{ijklm}$ and $\bar{\Sigma}_{ijklm}$ in $\bf{126}$ and
$\bf{\overline{126}}$. The most general Higgs 
superpotential is then given by~\footnote{The fermionic sector is also written in the form of a
superpotential. The only Higgs coupling to the fermions generate Yukawa terms are in 
a representation $\bf{R}$ such that $\bf{16}\times \bf{16}\times R$ contains a singlet.
In the case of a minimal $SO(10)$, this is the case only for $H$ and
$\overline{\Sigma}$ and the superpotential generating the fermion mass
matrix: $W_F = \Psi_a (Y_H H+Y_\Sigma \overline{\Sigma}) \Psi_b$,
where $a,b$ are family indices and the $Y$ are the symmetric
Yukawa matrices. The most general fermionic sector contains only
one additional term involving a ${\bf 120}$ as ${\bf 16\times 16
\times 120}$.
One can also enlarge the minimal $SO(10)$ model to have a more realistic fermionic
mass matrices~\cite{Aulakh:2005mw,Aulakh:2006hs,Senjanovic:2006nc}.}:
\begin{equation}\label{eq:superpotminso10}
W_H =  m_\Phi\Phi^2+\lambda_\Phi \Phi^3 + m_H H^2 +m_\Sigma \Sigma
\overline{\Sigma} +\eta \Phi \Sigma \overline{\Sigma}+\Phi H (\alpha
\Sigma+\overline{\alpha}\overline{\Sigma})
\end{equation}
The possible symmetry breaking depends on the minima
of the full superpotential, but only MSSM singlets inside each
Higgs fields can take a non-vanishing VEV. For example, in the
minimal $SO(10)$, there are three MSSM singlets
inside $\Phi$. The decomposition is given by: ${\bf 210=(15,1,1)+(1,1,1)+(15,1,3)+(15,3,1)+(6,2,2)+(10,2,2)+(\overline{10},2,2)}$ under the Pati-Salam subgroup $SU(4)_c\times SU(2)_L\times SU(2)_R$~\cite{Slansky:1981yr}.  The MSSM singlets are found in $(1,1,1)$,
$(15,1,1)$ and $(15,1,3)$.


The minimal SUSY $SO(10)$ model contains $2$ additional MSSM singlets in $\Sigma$ and 
$\overline{\Sigma}$. Candidates for the inflaton should be searched for within these
components. All the minima can be found by solving for the values
of the singlet fields minimizing all the $F$-terms. The symmetries
of these minima are given by the number of invariant generators of $SO(10)$.



\subsection{Symmetry breaking and topological defects}\label{sec:partphys:symmbreak}

The formation of topological defects during symmetry breaking has been 
found by Kibble~\cite{Kibble:1976sj}, and has rapidly gained popularity
in the 80's and the 90's as it was realized that the cosmic strings - the line-like topological 
defects - formed at the GUT scale could generate temperature anisotropies at the level of $10^{-5}$ as 
observed by the COBE. They form during symmetry breaking, if generated via 
the Higgs mechanism, then their nature depend on the actual 
symmetry breaking and the fields involved. 
As presented earlier in this chapter, these symmetry 
breaking are assumed to have occurred in the SM of 
particle physics as well as many of its extensions: MSSM, (SUSY) GUTs, etc.
Even string theories often give rise to a $4D$ effective theories that 
possesses symmetries that are larger than those of the SM, requiring 
then some symmetry breaking~\cite{Green:1987sp,Green:1987mn}.
In this section we will give a rapid presentation on 
some general properties of topological defects useful for inflation 
model builders, and we refer the reader to the more specialized 
literature for more 
details~\cite{Kibble:1976sj,Vilenkin:1984ib,Vachaspati:1993pj,Vilenkin:1994,Achucarro:1999it,Durrer:2001cg,Sakellariadou:2002mq,Sakellariadou:2007bv}, for a review see~\cite{Hindmarsh:1994re}.


\subsubsection{Formation of cosmic defects during or after inflation in $4D$}

As we will describe in the chapter~\ref{sec:modelsinflation}, many 
semi-realistic models of inflation within regular $4D$ field theory 
consider the interaction between the inflaton $\phi$ and a second scalar field $\psi$ that acquires 
during or at the end of inflation a non-vanishing VEV. As a consequence, depending on the precise model and 
inflation dynamics, it is frequent to break symmetries during or at 
the end of inflation, if the Higgs-type field $\psi$ is charged under 
some symmetries. The hybrid inflation models is the most common class 
of such models (see Sec.~\ref{sec:modelsinflation}). 

Let us for a moment forget about the phase of inflation and illustrate the 
formation of topological defects. If $\psi$ is non-trivially charged under 
some gauge symmetry $G$, a non-vanishing VEV of $\psi$ will realize the 
symmetry breaking $G\rightarrow H$. The manifold of all the vacua 
accessible to $\psi$ is given by the quotient group $\mathcal{M}=G/H$. For 
example, for the simplest abelian Higgs model, the symmetry breaking is 
$U(1)\rightarrow I$ and the manifold of vacua is $\mathcal{M}=U(1)$, 
corresponding to the circle of constant radius in the complex plane 
$|\phi|={\rm constant}$. What govern the formation and the type of topological 
defects are the topological properties of 
$\mathcal{M}$~\cite{Kibble:1976sj,Vilenkin:1994}. The homotopy groups 
$\pi_n$ of order $n$ are the most efficient way to study these properties. 
Each group $\pi_n(\mathcal{M})$ is composed of all classes of 
hypersurfaces of dimension $n$ that can shrink to a point while staying 
inside $\mathcal{M}$~\cite{Nakahara:2003nw}. If any hypersurface can 
shrink to a point, the homotopy group contains only one element and is 
said to be trivial. In particular, $\pi_0(\mathcal{M})$ is trivial if and only 
if $\mathcal{M}$ is connected, $\pi_1(\mathcal{M})$ is trivial if and only 
if $\mathcal{M}$ is simply connected. One can easily visualize that 
if $\mathcal{M}$ is not connected (for example during the breaking of a 
discrete group $Z_n \rightarrow I$), uncorrelated regions of the universe 
will fall in different vacua and will necessarily be separated by domain 
walls~\cite{Kibble:1976sj,Vilenkin:1984ib}. With the same reasoning, if the 
universe undergo a phase transition satisfying $\pi_1[G/H]\neq I$, 
\emph{cosmic strings}, that is line-like defects will necessarily form, 
with a density given by the correlation length or the mass scale of the 
Higgs field responsible for the symmetry breaking. 
It is for example the case for $U(1)\rightarrow I$ or all 
symmetries of the type $G\rightarrow H\times Z_n$. In general, the 
formation of topological defects of space-time dimension $d$ is governed 
by the non-triviality of the homotopy group:
\begin{equation}
\pi_{3-d}\neq I~. 
\end{equation}
One can show that 
any symmetry breaking of the form $G\rightarrow H\times U(1)$ gives rise to 
the formation of monopole (point-like defects). This is the origin of the 
well-known monopole problem, since the Standard Model group contains a 
$U(1)$ factor. This formation of unwanted defects was one of the original 
motivation to introduce a phase of inflation \cite{Linde:1981mu}. 

Note that the above topological conditions of formation of defects only 
govern the formation of topologically stable defects. It was however found 
that defects solutions can form even when the topology is 
trivial~\cite{Vachaspati:1993pj,Achucarro:1999it}. 
The most well-known example are the electro-weak strings, formed during the 
eletroweak symmetry breaking which are perturbatively stable
for a range of parameters which are not realized in nature,
and belong to the broader class of embedded defects. 

These defects are a priori unstable though mechanisms (such as plasma 
effects) have been found to stabilize them. They are of interest for 
inflation model builders since this mechanism can allow lift the 
constraints from the formation of cosmic strings (see 
Sec.~\ref{sec:models:Dterminflation} on $D$-term inflation)


\subsubsection{Formation of cosmic (super)strings after brane inflation}
\label{See: Formation of cosmic (super)strings after brane inflation}

Recently, a new class of models of inflation was proposed, mimicking 
hybrid inflation within extra-dimensional theories (see~\ref{BI}). 
This model like $D$-term hybrid inflation produces cosmic string-like 
objects called $F$- and 
$D$-strings~\cite{Sarangi:2002yt,Copeland:2003bj,Jones:2003da}. The 
nature of these objects are distinct from regular cosmic strings, as 
$F$-strings are Fundamental strings of cosmic size and $D$-strings are 
$D$-brane of spatial dimension 1. Fundamental strings are expected to
have a Planckian size and therefore a Planckian mass-per-unit length 
$\mu \sim \MP^2$, leading to $G\mu\sim 1$ that is incompatible with observation. 
However in the context of the recently proposed large extra dimension, 
the fundamental Planck mass can be reduced by large warp factors and 
these object can be formed with a cosmic size. In fact the range of 
mass-per-unit length in Planckian unit for these objects is:
$10^{-13}<G\mu < 10^{-6}$.

The $D$-strings can be formed at the end of brane inflation when a 
brane collide another brane of different dimension or an anti-brane, 
giving rise to the production of $Dp$-branes, with $p$ dimensions, 
of which 1 is in the 
non-compact dimensions. This mechanism is considered as a generalization 
of the production of regular cosmic strings at the end of $D$-term 
inflation in $N=1$ SUGRA~\cite{Binetruy:2004hh}. 
The energy per unit length (or the tension) of a $D1$-brane is  given by 
$\mu = M_s^2/(2\pi g_s)$, where $M_s$ is the string scale and $g_s$ the 
string coupling. But for $g_s\gtrsim 1$ this can give rise to too large 
of a tension, considering the CMB bounds (see below). Therefore some 
$D(p-2)$-branes are assumed with $(p-3)$ dimensions compactified to the 
volume $V_c$. Then the string tension reads the generic value~\cite{Sarangi:2002yt}
\begin{equation}
\mu = \frac{M_s^{p-1}V_c}{(2\pi)^{p-2}g_s}=\frac{M_s^2}{4\alpha \pi}\simeq 2M_s^2 ~,
\end{equation}
if the gauge coupling $\alpha\simeq 1/25$.
The CMB constraint, $\mathcal{P}_{\zeta}\sim 10^{-10}$, requires from brane 
inflation that $M_s\sim 10^{15}$ GeV leading to $G\mu\simeq 10^{-7}$.


\subsubsection{Cosmological consequences of (topological) defects}

Let's turn to the consequences of the formation of topological defects for 
observations, keeping in mind that a phase of inflation has to take place 
at some energy to explain the most recent CMB observations and solve the 
horizon problem.  For reviews on defect evolution, 
see~\cite{Hindmarsh:1994re,Vilenkin:1994,Durrer:2001cg,Sakellariadou:2002mq,Sakellariadou:2007bv}.
Domain walls, that is topological defects of space-time dimension 3, 
are cosmologically disastrous as they evolve following $\rho_{\rm DW} 
\propto t^{-1}$ and would 
dominate the energy density of the universe unless they form at an 
energy lower than $\sim 100$ MeV. They cannot even have formed at an 
energy higher than $\sim 1$ MeV without producing temperature 
fluctuations larger than $\delta T/T\gtrsim 10^{-5}$ in the 
CMB~\cite{Zeldovich:1974uw}, and without being in contradiction with 
CMB observations~\footnote{The defects in higher dimensions can give rise to inflation, see Sec.~\ref{BI}.}.

Point like topological defect, called (magnetic) 
monopoles~\cite{'tHooft:1974qc,Polyakov:1974ek}, with a mass of order of 
the GUT scale are expected to form during the early phase transition 
from $G_{GUT}$ down to $SU(3)\times SU(2)\times U(1)$ if $G_{GUT}$ is 
assumed simple. It was argued that unless their abundance is $n_{M}/s > 
10^{-10}$ at the time of phase transition, their abundance can not be 
diluted in an adiabatic expansion of the universe~\cite{Preskill:1979zi}, 
these ideas ultimately propelled the birth of inflation in order to 
dilute them. 

On the other hand cosmic strings do not suffer from these problems,
as a network of cosmic strings can inter commute~\footnote{The probability 
of inter-commutation, $p$, that is an exchange of partners when two strings 
intersect of one another is very close to $p=1$ for 
cosmic strings from $4D$ field theory. This probability become much smaller 
when considering cosmic superstrings appearing from brane inflation. This 
is the main origin of the differences in the observational signature between 
cosmic strings and cosmic (super)strings.} and lead to the formation 
of closed loops from long strings. The loops 
oscillate due to their tension and decay into gravitational (and 
potentially particle) radiation, leading to a global scaling regime: their 
relative energy density  evolve as, $\rho_{\rm CS} \propto t^{-1}$, and do 
not dominate the universe (see~\cite{Kibble:1984hp} for the analytical 
proof and for example~\cite{Albrecht:1989mk,Ringeval:2005kr,Martins:2005es} 
for numerical confirmations, see also for more recent 
discussion~\cite{Polchinski:2006ee,Dubath:2007mf}. Their observational 
signatures are numerous (see \cite{Pshirkov:2009vb} for a recent 
discussion); contribution to the CMB temperature fluctuations 
via the Kaiser-Stebbins effect~\cite{Kaiser:1984iv}, gravitational 
lensing~\cite{Vilenkin:1984ea,Sazhin:2006kf}, 
generation of gravity wave 
background~\cite{Damour:2001bk,Damour:2004kw,Polchinski:2007rg,DePies:2007bm}. 
The amplitude of these effects are all mainly related to one property -
their tension $T$. If the strings do not carry currents, this tension 
is equal to their energy per unit length in Planck units:
\begin{equation}
G\mu = 2\pi \epsilon(\beta) v^2/M_{\rm P}^2~. 
\end{equation}
where $v$ is the VEV of the Higgs far away from the string. 
In the case of non-SUSY strings, or if the strings saturate the Bogomolnyi 
bound (they are then called ``BPS strings'' for Bogomolnyi-Prasad-Sommerfeld), $\epsilon(\beta)=1$. 
In the context of SUSY/SUGRA, this function becomes 
dependent on the ratio of the Higgs to the gauge boson mass, 
$\beta\equiv m_\phi /m_A$. For non-BPS strings, the function 
$\epsilon(\beta)$ is given in~\cite{Hill:1987ye}. This is the case 
for $F$- and $D$-strings originating from brane inflation.

Currently the constraints on their effects on CMB is among the most 
stringent observation and is directly relevant to inflationary physics when 
they are formed close to or at the end of inflation. In this case, 
their formation affects the 
normalization of the fluctuation power spectrum by imposing an 
additional contribution. This can be described by an additional 
contribution to the temperature quadrupole 
anisotropy~\cite{Kallosh:2003ux,Rocher:2004my,Rocher:2004et,Jeannerot:2006jj}
\begin{equation}
\left.\frac{\delta T}{T}\right|_Q^2=\left.\frac{\delta T}{T}
\right|_{\rm infl}^2+\left.\frac{\delta T}{T}\right|_{\rm CS}^2~,
\end{equation}
where $(\delta T/T)_{\rm CS}= y 2\pi \epsilon(\beta) v^2/M_{\rm p}^2$. 
The constant $y$ parametrizes the density of the 
string network at the last scattering surface and has to be 
extracted from numerical simulations. The most recent simulations 
computing this parameter predicts $y=8.9 \pm 2.7$ \cite{Landriau:2003xf}, 
though older simulations or semi-analytical calculations give $y\in 
[3 - 6]$ (see for e.g.~\cite{Jeannerot:2005mc} and references therein). 
The current constraints from this normalization (for example in 
the case of their formation at the end of $F$-term inflation) lead
to~\cite{Rocher:2004et,Jeannerot:2005mc}:
\begin{equation}
v\lesssim 2\times 10^{15}\;\mathrm{GeV}~,\quad
G\mu \lesssim 8\times 10^{-7}~,
\end{equation}
In more recent studies, the effect of a cosmic string network on the 
CMB anisotropies have been computed at $\ell =10$ instead of the 
quadrupole $\ell=2$~\cite{Bevis:2006mj} as the latter is polluted by 
a large cosmic variance error. These simulations are improved 
compared to previous analysis since cosmic strings are described using 
field theory instead of modeled.
Analyzing of the presence and the impact of cosmic strings in the CMB 
data assuming a model of inflation can therefore be made fully 
consistently using Monte-Carlo methods~\cite{Bevis:2007gh}. At $\ell 
=10$, a fraction $f_{10} < 0.11$ of the temperature anisotropies 
from cosmic strings are found compatible with the 3 years WMAP data 
(at $2\sigma$), assuming a 7-parameter 
$\Lambda$CDM model and it is shown that the fraction of cosmic strings 
$f_{10}$ is strongly degenerate with the spectral index (see 
Sec.~\ref{sec:generalprop:confrontationdata}). This constraint can be 
translated into $G\mu < 7\times 10^{-7}$.

Other CMB searches, such as  direct searches in spatial map for line 
discontinuity gives a bound on $G\mu \lesssim 3\times 10^{-7}$~\cite{Jeong:2004ut}. 
To search for cosmic string signals in future CMB data, one should also 
incorporate recent developments on the small-scale signal in temperature 
anisotropies~\cite{Fraisse:2007nu,Pogosian:2008am}, 
on the CMB B-mode polarization signal~\cite{Seljak:2006hi,Pogosian:2007gi,Bevis:2007qz} 
or from the generation of non-Gaussianities~\cite{Hindmarsh:2009qk,Hindmarsh:2009es}. 
Recent work have also investigated the cosmological evolution and 
CMB signatures of meta-stable semi-local cosmic 
strings~\cite{Urrestilla:2007sf} as their formation 
is very motivated from a particle physics point of 
view, especially at the end of $F$-term hybrid inflation~\cite{Jeannerot:2003qv} 
or $D$-term hybrid inflation (see Sec.~\ref{sec:dtermwithoutstrings}).

Other very stringent constraints on the presence of cosmic strings in 
the universe comes from the amplitude of gravity wave background arises from 
the timing of the millisecond pulsar, which gives~\cite{DePies:2007bm}
$$G\mu < 10^{-8}-10^{-9}~.$$ Note that these 
constraints are more model-dependent than the CMB or lensing constraints.

\section{Models of inflation}\label{sec:modelsinflation}
%
%

A detailed account on inflation model building can be found in many
reviews~\cite{Lyth:1998xn,Linde:2005ht,Olive:1989nu,Liddle:2000cg,Lyth:2007qh}.
In this chapter and the next, we will review some inflationary
models~\footnote{The very first attempt to build an inflation model was
made in~\cite{Starobinsky:1980te}, where one-loop quantum correction to
the energy momentum tensor due to the space-time curvature were taken
into account, resulting in terms of higher order in curvature
invariants. Such corrections to the Einstein equation admit a de Sitter
solution~\cite{Dowker:1975tf}, which was presented
in~\cite{Starobinsky:1980te,Vilenkin:1985md}.
Inflation in Einstein gravity with an additional $R^2$ term
was considered in~\cite{Starobinsky:1982ee}. A similar situation arises
in theories with a variable Planck mass, i.e. in scalar tensor
theories~\cite{Will:1993ns}.} that are motivated or that originate from
particle physics, and present their successes and their challenges. By
no means this section is an exhaustive description of all the models,
their number being huge and still growing gradually.

\subsection{What is the inflaton ?}

There are two classes of models of inflation, which have been discussed
extensively in the literature. In the first one, the inflaton field
belongs to some hidden sector (not charged under the SM), such models
will have at least one SM {\it gauge singlet} component, whose couplings
to other fields and mass are chosen to match the CMB observations.  This
section will review some of the important ones. Models involving
{\it gauge invariant} inflatons, charged under the SM gauge group or its extensions will be
discussed in Sec.~\ref{sec:mssminflation}.

All the inflationary models are tested by the required amplitude of 
density perturbations for the observed large scale structures~\cite{Mukhanov:1990me}. 
Therefore the predictions for the CMB fluctuations are the most important ones to judge the merits of the
models, especially the spectral index, tensor to scalar ratio, and running of the spectral tilt with the help 
of slow-roll parameters $\epsilon$, $\eta$, and $\xi^2$, as defined in Sec.~\ref{sec:generalprop:slowroll}~\footnote{Review of some of these models using the Hubble-flow parameters of Eq.~(\ref{eq:epsilonn}) has been performed in~\cite{Martin:2003bt,Martin:2006rs}.}. We will also discuss generation of non-Gaussianities and cosmic strings (see Sec.~\ref{sec:generalprop:confrontationdata} for more details about the current status of the cosmological data), in order to visualize how well a given model matches the observations.
Although we might not be able to pin down {\it the} model(s) of
inflation, but certainly we would be able to rule them out from observations. From particle physics point of view, 
there exists an important criteria for a successful inflation; which is to end inflation in the right vacuum where the 
SM baryons can be excited naturally after the end of inflation in order to 
have a successful baryogenesis and BBN.


\subsection{Non-SUSY one-field models}

The most general form for the potential of a gauge singlet scalar field $\phi$
contains an infinite number of terms,
\begin{equation}
\label{simplesinglepot}
V=V_0+\sum_{\alpha=2}^\infty \frac{\lambda_\alpha}{M_{\rm P}^{\alpha-4}}\phi^{\alpha}~.
\end{equation}
In $4d$, restricting to renormalizable terms allows to prevent all terms
with $\alpha\geq 4$. Furthermore, assuming extra symmetries can ensure that certain
neglected terms in this series are not generated at the loop level.
Note that this is the case for the constant term $V_0$ also. The usual example of
such a symmetry is the parity $Z_2$, under which $\phi\rightarrow -\phi$, which
allows to prevent all terms with $\alpha$ odd. Most phenomenological models of
inflation proposed initially assume that one or two terms in
Eq.~(\ref{simplesinglepot}) dominate over the others, though some do contain an
infinite number of terms. In Refs.~\cite{Destri:2007pv,Cirigliano:2004yh}, terms proportional to $\phi^2,~\phi^3$ and $\phi^4$
with adjustable coefficients were included in studying the inflationary dynamics and the perturbations, and in Ref.~\cite{Destri:2009wn}
non-renormalizable terms up to the 6th order were also taken into account in order to put bound on tensor to scalar
ratio.



\subsubsection{Large field models}\label{sec:models:largefieldmodels}

\paragraph{Power-law chaotic inflation:}
The simplest inflation model by the number of free parameters is perhaps the chaotic
inflation~\cite{Linde:1983gd} with the potential dominated by only one
of the terms in the above series
\begin{equation}\label{eq:potentialchaotic}
V=\frac{\lambda_\alpha}{M_{\rm P}^{\alpha-4}}\phi^{\alpha}~,
\end{equation}
with $\alpha$ a positive integer. The first two slow-roll parameters are
given by
\begin{equation}
\epsilon= \frac{\alpha^2}{2}\frac{M_{\rm P}^2}{\phi^2}~, \quad \quad
\eta =\alpha(\alpha-1)\frac{M_{\rm P}^2}{\phi^2}\,.
\end{equation}
Inflation ends when $\epsilon = 1$, reached for $\phi_e=\alpha M_{\rm P}/
\sqrt{2}$. The largest cosmological scale becomes super-Hubble when
$\phi_Q=\sqrt{2N_Q\alpha}M_{\rm P}$, which is super Planckian; this is
the first challenge for this class of models (see discussion
Sec.~\ref{sec:generalprop:dynamicchalleng}). The spectral index for
the scalar and tensor to scalar ratio read:
\begin{equation}
n_s=1-\frac{2+\alpha}{2N_Q+\alpha/2}~, \qquad
r=\frac{4\alpha}{N_Q+\alpha/4}\,.
\end{equation}
The amplitude of the density perturbations, if normalized at the COBE
scale, yields to extremely small coupling constants;
$\lambda_\alpha \ll 1$ (for e.g. $\lambda_4 \simeq 3.7\times 10^{-14}$).
The smallness of the coupling, $\lambda_\alpha / M_{\rm P}^{\alpha-4}$, is often
considered as an unnatural fine-tuning. Even when dimension full, for
example if $\alpha=2$, the generation (and the stability) of a mass scale,
$\sqrt{\lambda_2} M_{\rm P} \simeq 10^{13}$~GeV, is a challenge in theories
beyond the SM, as they require unnatural cancellations.  
These class of models have an interesting behavior for initial conditions
with a large phase space distribution where there exists a late attractor
trajectory leading to an end of inflation when the slow-roll conditions
are violated close to the Planck
scale~\cite{Linde:1983gd,Linde:1985ub,Brandenberger:1990wu,Kofman:2002cj}.

Recently, these models also suffer from an observational challenge
which comes from the prediction of a high tensor to scalar ratio that is
under tension by the most recent CMB+BAO+SN data~\cite{Komatsu:2008hk}.
The cases $\alpha\geq 4$ lie outside the $2\sigma$ region, though
the case $\alpha=2$ is still in the $1\sigma$ region for $N_Q\gtrsim 60$ (see
Sec.~\ref{sec:generalprop:confrontationdata}).

Note that the above mentioned monomial potential can be a good approximation
to describe in a certain field range  for various models of inflation
proposed and motivated from particle physics; natural inflation
when the inflaton is a pseudo-Goldstone boson \cite{Freese:1990rb},
or the Landau-Ginzburg potential when the inflaton is a Higgs-type
field~\cite{Bezrukov:2007ep}. Some of these potentials will be discussed below. 
Chaotic inflation was also found to emerge from SUGRA
theories~\cite{Murayama:1993xu,Kadota:2007nc,Kadota:2008pm} (see
Sec.~\ref{sec:models:chaoticSUGRA})
as well as in certain classes of brane-world models~\cite{Maartens:1999hf},
where it has been claimed to give rise to a larger expansion rate due to
modification in the Hubble equation rate. The necessity of super Planckian
VEVs represents though a challenge to such embedding in particle physics.

Variants of these simple models have been constructed, based on the
radiative corrections appearing at the loop level.
First, the radiative corrections due to the inflaton self-interaction can
induce a running of the mass of the inflaton with its VEV (see
Sec.~\ref{SPEFT}). For example for $\alpha=2$, a ``running mass'' potential of
the form  $V(\phi)=m^2(1+\alpha\ln(\phi/\mu))\phi^2$ has been considered
in~\cite{Enqvist:2002si}. Though the dynamics is not significantly affected
by this running, it was found to be able to affect the decay of the inflaton
during reheating via fragmentation.


In \cite{NeferSenoguz:2008nn}, the chaotic inflation
potential was extended to include a Yukawa coupling $(h/2)\phi \bar{N}_RN_R$
to the right-handed neutrino $N_R$. This introduces a one-loop correction to the
inflation potential of the form $V_{\rm 1-loop}\simeq \kappa \phi^4\ln(h\phi/\mu)$,
where $\kappa=h^4/16\pi^2$ and $\mu$ is a renormalization scale.
It was found that the radiative corrections can affect significantly the
predictions of the chaotic models, namely by reducing the level of
tensor-to-scalar ratio. For example, for the quartic potential, $h=0$ gives rise
to $n_s\simeq 0.95$ and $r\simeq 0.25$, whereas for $h\simeq 1.7\times
10^{-3}$, $n_s\simeq 0.95$ and $r\simeq 0.084$ which is within the $1\sigma$
contour of the WMAP data. Note that these results are very sensitive to
the value of $h$.\\


\paragraph{Exponential potential:}
An exponential potential also belongs to the large field models:
\begin{equation}
V(\phi)=V_{0}\exp\left(-\sqrt{\frac{2}{p}}\frac{\phi}{M_{\rm P}}\right)\,.
\end{equation}
It would give rise to a power law expansion $a(t)\propto t^p$,
so that inflation occurs when $p>1$. The case $p=2$ corresponds to the
exactly de Sitter evolution and a never ending accelerated expansion. Even for
$p\neq 2$, violation of slow-roll never takes place, since
$\epsilon(\phi)=1/p$ and inflation has to be ended by a phase transition
or gravitational production of particles~\cite{Lyth:1998xn,Copeland:2000hn}.

The confrontation to the CMB data yields: $n_s=1-2/p$ and $r=16/p$; the
model predicts a hight tensor to scalar ratio and it is within the one
sigma contour-plot of WMAP (with non-negligible $r$) for $p\in[73-133]$.
Multiple exponentials with differing slopes give rise to what has been
dubbed as {\it assisted inflation} \cite{Liddle:1998jc}. Such potentials might
arise in string theories and theories with extra dimensions.\\

\paragraph{A combination of exponential and power law potential:}

Another form of potential has been found emerging from SUGRA~\cite{Copeland:1994vg,Linde:1997sj,Panagiotakopoulos:1997if,Panagiotakopoulos:1997ej,Clesse:2009ur}, which is given by:
\begin{equation}
V(\phi)=V_0\exp\left(\frac{\phi^2}{2M_{\rm P}^2}\right)\left[1-\frac{\phi^2}
{2M_{\rm P}^2}+\frac{\phi^4}{2M_{\rm P}^4}\right]~,
\end{equation}
This potential can emerge from hybrid inflation when SUGRA corrections
dominate over radiative corrections, that is in the small coupling
limit. In this class of models, the spectral tilt tends to exceed unity
$n_s>1$ depending on the number of e-foldings of
inflation~\cite{Panagiotakopoulos:1997if,Panagiotakopoulos:1997ej}. It is
therefore disfavored at $2\sigma$ at least, unless cosmic strings are
formed at the end of inflation (see Sec.~\ref{sec:originalhybridinflation}
on hybrid inflation and Sec.~\ref{sec:partphys:symmbreak} on topological
defects).


\subsubsection{Small field models}

Contrary to the models of the previous section, the small field
models~\footnote{Sometimes they are also known by the name of {\it modular
inflation}~\cite{Lyth:1998xn,Kadota:2003fs,Lyth:2007qh}} take place for
VEVs much smaller than the Planck scale, which is their main
motivation. Their potential is of the form
\begin{equation}\label{eq:generalformsmallfield}
V(\phi)=V_0\left[1+f\left(\frac{\phi}{\mu}\right)\right]~,
\end{equation}
with the potential dominated by the constant $V_0$ ($\phi/\mu \ll 1$).
Below we discuss some of the variants.

The most studied potential of this form is the effective hybrid model,
based on
\begin{equation}\label{eq:potenhyb1d}
V(\phi)=M^4\left[1+\left(\frac{\phi}{\mu}\right)^p\right]~,
\end{equation}
since, $p=2$ corresponds to the effective potential for the two
field hybrid model (see Sec.~\ref{sec:models:multi-field}). In
that case, a second field triggers the end of inflation by
fast-rolling due to a waterfall at a critical value $\phi=\phi_c$. In
the large field limit $\phi\gg \mu$, one recovers the large field
potential of Eq.~(\ref{eq:potentialchaotic}). The slow-roll
parameters read
\begin{equation}
\epsilon(\phi)=\frac{M_{\rm P}^2}{2\mu^2}\frac{p^2(\phi^2/\mu^2)^{p-1}}
{\left[1+(\phi/\mu)^{p}\right]^2}~, \quad
\eta(\phi)=\frac{M_{\rm P}^2}{\mu^2}\frac{p(p-1)(\phi/\mu)^{p-2}}{1+(\phi/\mu)^{p}}~.
\end{equation}
The parameter $\epsilon(\phi)$ is small both for $\phi\gg \mu$
and $\phi\rightarrow 0$, which identifies two phases of slow-roll
inflation at large field and at small field. The spectral index
and the ratio tensor to scalar in the slow-roll approximations
read
\begin{equation}\label{eq:nsandrhybrid1field}
\begin{split}
n_s(\phi)-1&=-\frac{p\,M_{\rm P}^2}{\mu^2}\,\left(\frac{\phi}{\mu}\right)^{p-2}\,
\frac{2-2p+(2+p)\left(\frac{\phi}{\mu}\right)^p}
{\left[1+\left(\frac{\phi}{\mu}\right)^p\right]^2}~,\\
r & = \frac{8p^2M_{\rm P}^2}{\mu^2}\,\left(\frac{\phi}{\mu}\right)^{2p-2}\frac{1}
{\left[1+\left(\frac{\phi}{\mu}\right)^p\right]^2}\,.
\end{split}
\end{equation}
Let us first discuss the small field limit. For $p>1$ and $p\neq 2$,
the model is considered slightly disfavored by the recent data,
since $n_s \gtrsim 1$ with negligible tensor to scalar ratio, and running of the spectral index is
predicted ($r\propto(\phi/\mu)^{2p} $,
$\alpha_s\propto(\phi/\mu)^{2p}$). It was pointed out
however that to confront this model with CMB data, one could
include the concordance model parameters and a weight for cosmic
strings, since this model in its two-field version can produce
cosmic strings. In that case, $n_s\sim 1$ is in agreement with
observations at $1\sigma$~\cite{Bevis:2007gh}. In the case $p=2$,
$\eta$ is constant and $n_s$ is predicted to be:
\begin{equation}
n_s(\phi)-1\simeq 2\eta \simeq \frac{4M_{\rm P}^2}{\mu^2}~.
\end{equation}
which is well above unity unless $\mu \gg M_{\rm P}$. This is
equivalent to having an extreme fine-tuning on a coupling
constant, which is considered unnatural. The model is therefore disfavored
at more than $2\sigma$ if the data are analyzed with the minimal
6-parameter model. However, if $\mu \gtrsim 7M_{\rm P}$ then it predicts
a spectral index around $1 < n_s \lesssim 1.1$, which is still
in agreement with current observations if cosmic strings have a
non-negligible contribution to the CMB anisotropies
(see~\cite{Bevis:2007gh} and Sec.~\ref{sec:partphys:symmbreak}).

The case $p=2$ has  been studied away from the
small field regime and/or without the slow-roll
approximations~\cite{Copeland:1994vg,Cardoso:2006wf,Clesse:2008pf}.
It was shown that the model is in agreement with the CMB
data if inflation is realized in the large field regime. This
can be achieved by some mechanism (such as a waterfall triggered
by an external field) independently of the parameter
$\mu$~\cite{Copeland:1994vg,Cardoso:2006wf,Clesse:2008pf} or
when $\mu\lesssim 0.32M_{\rm P}$~\cite{Clesse:2008pf}. In the last
case, the Hubble-flow parameter $\epsilon_1$
differs from $\epsilon$ at small VEVs and violation of
slow-roll is responsible for forcing inflation to take place in
the large field model without any condition on a waterfall
parameter~\cite{Clesse:2008pf}. In both cases, inflation is
necessarily realized for inflaton VEVs large compared to $\mu$,
which reduces the appeal of the model.

An inverted hybrid (or hiltop) inflation model has also been
proposed where the curvature of the slope is negative by
construction, in order to predict a red spectrum, \cite{Binetruy:1986ss}
\begin{equation}
V(\phi)=M^4\left[ 1-\left(\frac{\phi}{\mu}\right)^p\right]~.
\end{equation}
This model emerges naturally when studying natural
inflation~\cite{Freese:1990rb}, modular inflation~\cite{Kadota:2003fs}
in the small field regime or from
generic SUGRA theories, see~\cite{Lyth:1998xn,Boubekeur:2005zm,Kohri:2007gq}.
Again the model differs from the chaotic limit
if $\phi\ll \mu$, which is therefore the limit of interest.
Another motivation to assume that the form of the potential is
only valid at small VEVs is the fact the potential is not
bounded from below; at large VEV, it must therefore be compensated
by other terms for field theory to be well defined.
In this limit, $\eta$ is negative and so is $n_s-1$, since $\epsilon
\ll |\eta|$ if $p>1$. However, $n_s$ is very close to unity for $p\neq
2$ or well below for $p=2$ and therefore $p=2$ is incompatible with
data~\cite{Lyth:1998xn}. Note that in the case $p=2$, the limit of small
fields could be abandoned and super-Planckian VEVs would then be necessary
to obtain CMB predictions in agreement with the WMAP5 data. But in addition
to the problem of super-Planckian VEV, inflation is then realized
in a sector where the potential is not trustable because not
bounded from below. As an example of the completion of an inverted
hybrid model, in Ref.~\cite{King:1997ig} authors have analyzed
the potential:
\begin{equation}
V=V_0-\frac{1}{2}m^2\phi^2+\lambda\phi^4~,
\end{equation}
where the amplitude of the CMB predictions can be matched, with almost
a flat spectral tilt. In this model $\epsilon$ remains negligible and
$n_s-1\approx 2|\eta|$.

The running of one (or more) parameters
of the general scalar potential can also be the origin of the function
$f(\phi)$ in Eq.~(\ref{eq:generalformsmallfield}). The most common one,
the ``hybrid running mass'' model is driven by a potential of the
form~\cite{Stewart:1996ey,Stewart:1997wg,Covi:1998jp,Covi:1998mb}
\begin{equation}
V(\phi)=M^4\left[ 1+\frac{\eta_0\phi^2}{2M_{\rm P}^2}\ln\left(
\frac{\phi}{\phi_*}-\frac{1}{2}\right)\right]~.
\end{equation}
A reasonable fit to the CMB data has been found in Ref.~\cite{Covi:2002th},
with a significant running of the spectral tilt for $\eta_0<0$, and $\phi_*$
the scale of the RG flow.
The current WMAP data actually disfavors running of the spectral tilt
and therefore these models are now very well constrained. The validity of the
running of a quartic coupling constant has also been
proposed in Ref.~\cite{Lyth:1998xn}.

We will see below that another form of inflationary potential
can also emerge of type~\cite{Stewart:1994pt}:
\begin{equation}\label{eq:models:potensmooth1field}
V(\phi) = M^4\left[1-\left(\frac{\phi_*}{\phi}\right)^n\right]~,
\end{equation}
only valid in the large field limit $\phi\gg \phi_*$, since in the
small field limit, the potential is not bounded from below and
should be completed.

\subsection{Non-SUSY models involving several fields}\label{sec:models:multi-field}

\subsubsection{Original hybrid inflation}\label{sec:originalhybridinflation}

The most studied multi-field inflation model is the hybrid inflation first discussed in
Ref.~\cite{Linde:1993cn} (and studied extensively in~\cite{Copeland:1994vg}) as a model that differs from chaotic
inflation on two main properties; it ends inflation with a waterfall
triggered by a Higgs (not necessarily the SM Higgs) field coupled
to the inflaton and it does not necessarily require an extremely small coupling to
account for the normalization of the power-spectrum. The model is based on
the potential given by~\cite{Linde:1993cn}
\begin{equation} \label{eq:potenhyb2d}
V(\phi,\psi) = \frac{1}{2} m^2 \phi^2 + \frac{\lambda}{4} \left(\psi^2
- M^2 \right)^2 +\frac{\lambda'}{2} \phi^2 \psi^2~,
\end{equation}
where $\phi$ is the inflaton and $\psi$ is the Higgs-type field.
$\lambda$ and $\lambda'$ are two positive coupling constants, $m$
and $M$ are two mass parameters.  It is the most general form
(omitting a quartic term $\lambda'' \phi^4$) of renormalizable
potential satisfying the symmetries: $\psi \leftrightarrow -\psi$
and $\phi \leftrightarrow -\phi $. Inflation
is assumed to be realized in the false-vacuum along the
$\psi=0$ valley and ends with a tachyonic instability for the
Higgs-type field. The critical point of instability below which
the potential develops non-vanishing minimum is at
\begin{equation}
\phi_{\rm c} = M \sqrt{\frac {\lambda}{\lambda'}}\,.
\end{equation}
The system then evolves toward its true minimum at $V=0$,
$\langle\phi\rangle=0$, and $\langle\psi\rangle=\pm M$~\footnote{The hybrid
inflation models were also considered in Refs.~\cite{Kaloper:1998sw,Mazumdar:1999tk,Mohapatra:2000cm,Mazumdar:2001ya,Mazumdar:2003vg,Green:2002wk,Allahverdi:2001dm,Lyth:1998bn,Lyth:1999ty,Mazumdar:2003jv} in the
context of large extra dimensions at TeV scale~\cite{ArkaniHamed:1998rs}.}.

The inflationary valley, for $\langle\psi\rangle=0$, is usually
assumed to be where the last 60 e-foldings take place. This is supported
by numerical and analytical simulations~\cite{Tetradis:1997kp,Lazarides:1997vv,Panagiotakopoulos:1997if,Mendes:2000sq,Clesse:2008pf}, where the fine-tuning of the initial
conditions were discussed. In Ref.~\cite{Clesse:2008pf} it was found that when
the initial VEV of the inflaton, $\phi$, is sub-Planckian, a subdominant
but non-negligible part of the initial conditions for the phase space leads to a successful
inflation, i.e. around less than $15\%$ depending on the model parameters.
Initial conditions with super-Planckian VEVs also lead to automatically
successful inflation similarly to chaotic inflation.
In the inflationary valley, $\langle\psi\rangle=0$, the effective potential is given by:
\begin{equation}
V_{\rm eff}(\phi) \simeq \frac{\lambda M^4 }{4}+\frac{1}{2} m^2 \phi^2~,
\end{equation}
whose phenomenology has been studied around Eq.~(\ref{eq:potenhyb1d}),
using $p=2$, $V_0=\lambda M^4/4$, and $V_0/M_{\rm P}^2=m^2/2$.
Therefore, the predictions are disfavored by the data
because of a blue tilt in the spectrum, i.e. $n_s>1$, in the small field regime, 
$\phi_Q < M_{\rm P}$. Note however that if its end is accompanied by the
formation of cosmic strings, a slightly blue spectrum is found in
agreement with the data, if these cosmic strings contribute to the CMB
anisotropies around $10\%$ \cite{Bevis:2007gh}. It was also suggeste
in~\cite{NeferSenoguz:2008nn} that loop corrections to the hybrid tree level
potential due to a Yukawa coupling to the right handed neutrino can render
the spectral index of the model below 1, like for the chaotic model. This
was confirmed recently in~\cite{Rehman:2009wv}, where a red spectral tilt was
found, even in the small field regime. These Yukawa couplings were also found
suitable for a successful reheating phase and the generation of lepton/baryon
asymmetry after inflation.

Note that a more realistic version of the model would include a quartic term
for $\phi$ allowed by symmetries and generated by the Feynman diagrams
involving loops of $\psi$ fields. If this term dominates over the
quadratic term in the inflationary valley, it was found that the
spectral index would then be very close to unity~\cite{Clesse:2008pf}.
Note that the coupling to a Higgs-type waterfall field has potentially important
cosmological consequences~\cite{GarciaBellido:1996qt,GarciaBellido:1997wm},
as topological defects generically form during this symmetry breaking
\emph{after} inflation. This will be discussed  in Sec.~\ref{sec:ftermhybridinflation}.


\subsubsection{Mutated and smooth hybrid inflation}
\label{MSHI}

Two variations of the hybrid inflation idea were proposed soon after the
original model, both assuming that the term $\phi^2$ is negligible. 

The
two-field scalar potentials are of the form:
\begin{equation}\label{eq:models:shiftedvalleypotential}
V_{pq}(\phi,\psi)=M^4\left[1-\left(\frac{\psi}{m}\right)^p\right]^2+\lambda \phi^2\psi^q~.
\end{equation}
They share the common feature of having an inflationary trajectory during
which $\langle\psi\rangle$ is varying and not vanishing. They also both reduce,
in the one-field approximation to the form of
Eq.~(\ref{eq:models:potensmooth1field}).

The {\bf mutated hybrid inflation} is one of them with a potential of
the form of~Eq.~(\ref{eq:models:shiftedvalleypotential}) with
$(p,q)=(1,2)$~\cite{Stewart:1994pt}
\begin{equation}
V^{\rm mut}(\phi,\psi) = \frac{1}{2}m^2(\psi-M)^2 +\frac{\lambda}{4}\phi^2\psi^2 ~.
\end{equation}
Inflation in this case always happens for sub-Planckian VEVs. If one assumes chaotic initial
values $\phi\gg \psi$, like for hybrid inflation, the potential is minimized at
$\psi=0$. While $\phi$ slow-rolls to smaller values, $\psi$ settles in the
local minimum satisfying, $\psi = M\alpha(\phi)/[1+\alpha(\phi)]$, with
$\alpha(\phi)=2m^2/(\lambda \phi^2)$. At large $\phi$, $\alpha\ll 1$,
and its effective potential is of the form of
Eq.~(\ref{eq:models:potensmooth1field}), with $n=2$:
\begin{equation}
V_{eff}^{\rm mut} \simeq \frac{1}{2}m^2 M^2\left(1-\frac{2m^2}{\lambda \phi^2}\right)
+\mathcal{O}[\alpha^2(\phi)]~,
\end{equation}
while the kinetic terms, though modified, are close to minimal.
In this approximation, the model predicts a red spectral index
and negligible tensor to scalar ratio,
\begin{equation}
n_s -1 \simeq -\frac{3}{8N_Q}\simeq 0.97~, \quad
r\simeq \frac{3m}{2\lambda N_Q^{3/2}}\ll \frac{3}{8N_Q^2}\sim 10^{-4}~,
\end{equation}
if we assume $N_Q\simeq 60$.
It is worth noting also that the model can emerge from a SUSY theory, e.g.,  
from a superpotential of the form~\cite{Stewart:1994pt}
\begin{equation}
W=\Lambda^2f(\Psi)\Sigma_1+\lambda\Phi\Psi\Sigma_2~.
\end{equation}
$f(\Psi)$ should be generated by non-perturbative process, such
as gaugino condensation~\cite{Nilles:1983ge} in some hidden sector, in order to
generate a $\Lambda$ much smaller than the Planck scale. It also needs to
satisfy $f(0)=1$ and $f'(0)<0$.\\


The {\bf smooth hybrid inflation} model~\cite{Lazarides:1995vr} also belongs
to the similar class of model, with a potential of the form
of~Eq.~(\ref{eq:models:shiftedvalleypotential}) with $(p,q)=(4,6)$. It
therefore involves non-renormalizable terms of order $M_{\rm P}^{-2}$.
This model is also characterized by a $\phi$-dependent minimum for
$\psi$ and, therefore a realization of inflation along a multi-field
trajectory. The motivation presented in \cite{Lazarides:1995vr} is to
avoid the formation of topological defects since the symmetry breaking
occurs \emph{during} inflation when all topological defects are
inflated away. This is necessary when the symmetry breaking gives rise
to monopoles or domain walls.
Along the inflationary trajectory, the effective one-field potential is
of the form of Eq.~(\ref{eq:models:potensmooth1field}), with $n=4$.
The end of slow-roll inflation in this model is necessarily triggered
by a violation of the conditions; $\epsilon,\eta \ll 1$, since no waterfall transition
takes place. This allows the predictions for the spectral
index to be~\cite{Lazarides:1995vr}
\begin{equation}
n_s -1 \simeq -\frac{5}{3N_Q}\simeq 0.97~,
\end{equation}
and the ratio for tensor to scalar is found to be negligible.
This model can also emerge from a SUSY framework, which helps
protect the form of the potential, as discussed in
Sec.~\ref{sec:models:SUSYmodelswithNR}. The embedding of the model in
particle physics and  its predictions for reheating and leptogenesis
will be discussed in that section too~\footnote{Reheating in presence of a SM gauge 
singlet within SUSY is quite different, for the discussion see Sec.~\ref{GSICM}. In many {\it gauge singlet} 
SUSY models of inflation, the role of MSSM squarks and sleptons are not taken into account appropriately. This also
affects leptogenesis and in general thermal history of the universe~\cite{Allahverdi:2007zz}.}.


\subsubsection{Shifted and other variants of hybrid inflation}
{\label{SAOVHI}

The {\bf shifted hybrid inflation} model~\cite{Jeannerot:2000sv} shares also the
features of shifting the VEV away from $\langle\psi\rangle=0$ in the inflationary
valley. Like the smooth hybrid inflation, the introduction of
non-renormalizable terms in $V$ is employed to shift the inflationary
valley:
\begin{equation}\label{eq:models:shiftedinflationpotential}
V^{\rm shift}(\phi,\psi)=M^4\left[\left(1-\tilde{\psi}^2
+\xi\tilde{\psi}^4\right)^2 + \tilde{\phi}^2\tilde{\psi}^2
\left(1-2\xi\tilde{\psi}^2\right)^2\right]~.
\end{equation}
The parameter $\xi$ controls the existence,  locations, and number of
valleys where inflation can occur.
The model was proposed in the context of SUSY GUTs and will be studied in
details in Sec.~\ref{sec:models:SUSYmodelswithNR}. Let us for now only
mention that the predicted spectral index lies within: $n_s\in [0.89,0.99]$, and the
tensor to scalar ratio $r\lesssim 10^{-5}$ can be in agreement with the
observations for certain values of $\xi$.

The {\bf inverted hybrid inflation} has also been proposed in a 2-field
version~\cite{Lyth:1996kt}
\begin{equation}
V(\phi,\psi) = V_0-\frac{1}{2}m_{\phi}^2\phi^2+\frac{1}{2}m_{\psi}^2\psi^2
-\frac{\lambda}{2}\psi^2\phi^2+\dots ~.
\end{equation}
Clearly this potential is not bounded from below at large VEVs,
and the model is therefore only complete once non-renormalizable terms
(contained in the dots) are assumed. Another modified version of hybrid
inflation, the {\bf complex hybrid inflation} has been proposed recently
by~\cite{Delepine:2006rn}. In this model, the potential is not invariant
under the change of phase of the waterfall field, assumed complex. This
modification was proposed as a new way to generate a baryonic asymmetry.

Finally, the {\bf thermal hybrid inflation} model has also been
considered~\cite{Lyth:1995ka}. In this model, thermal corrections are
assumed to generate part of the hybrid potential~\footnote{In order to estimate 
the coefficients of thermal corrections one has to understand the exact particle contents. However 
note that $\phi$ belongs to a hidden sector, therefore  exact particle contents are model dependent and 
sometimes chosen just to meet the desired results.}
\begin{equation}
V(\phi,\psi) = V_0+T^4+T^2\psi^2-\frac{1}{2}m_\psi^2\psi^2+T^2\phi^2 ~.
\end{equation}
However, as it is well known, the rapid expansion due to inflation
dilute the content of the universe, exponentially rapidly driving the
temperature to 0. This model therefore requires that a period of hot
big-bang evolution takes place immediately before the phase of
inflation. This condition can be considered rather fine tuned. In
addition, inflation starts with $T\sim V_0^{1/4}$ and the temperature
triggers the waterfall transition taking place at $T\sim m_\psi$. With $T\sim 1/a$
during inflation, inflation will last only 10 e-foldings and to be
acceptable, this mechanism should be invoked many times.

Another way temperature effects could influence inflation has been proposed
in a number of recent articles (see~\cite{Berera:1995ie,Berera:2008ar}), known as {\bf thermal inflation}. This
mechanism is based on the assumption that the couplings between the
inflaton field and other particles (independently required for a successful
reheating) can generate a constant decay of the inflaton \emph{during} inflation
(assuming $\Gamma\sim H_{\rm infl}$). This particle production would induce a
thermal bath with a finite temperature that back-reacts on the inflaton
dynamics and induces finite temperature effects on the potential. In
particular this mechanism introduces a new viscosity term $C_w\dot{\phi}$ in the
field dynamics Eq.~(\ref{slowr2}), which slows down the rolling of the inflaton~\footnote{To our knowledge the viscosity term 
has never been introduced consistently in the equations of motion in an expanding background.
Note that the decay term usually introduced phenomenologically during the inflaton oscillations
is valid {\it only} when the frequency of the inflaton oscillations is larger than the Hubble expansion rate. Therefore
one can safely use the decay of the inflaton as in the case of a flat space-time~\cite{Kofman:1997yn}.}.
This idea was used in a number of articles to realize inflation with potentials that
would be too steep for inflation without temperature effects, for example in string
theory (see for example~\cite{Ramos:2001zw}).
This possibility is however ad-hoc and still
debated and some authors \cite{Yokoyama:1998ju,Kofman:1997yn} argue that these effects are
unlikely to take place, as the viscosity term is expected to be negligible
during slow-roll inflation. They argue that this mechanism could only be considered
as a phenomenological idea that still lacks some theoretical support and an
explicit regime where this can be realized.


\subsubsection{Assisted inflation}
\label{AINF}

There could be many more light fields during inflation, they could
collectively assist inflation by increasing the effective friction term
for all the individual fields~\cite{Liddle:1998jc,Copeland:1999cs,Kanti:1999ie,Kanti:1999vt,Kaloper:1999gm,Green:1999vv,Jokinen:2004bp}. This idea can be illustrated with the help of $'m'$ identical scalar fields with an exponential potentials~\cite{Liddle:1998jc}:
\begin{equation}
\label{assist-pot}
V(\phi_i) = V_0 \exp \left( - \sqrt{\frac{2}{p}} \,
    \frac{\phi_i}{M_{\rm P}} \right) \,.
\end{equation}
For a particular solution; where all the scalar fields are equal:
$\phi_1 = \phi_2 = \cdots = \phi_m$.
\begin{eqnarray}
H^2 & = & \frac{1}{3 M_{\rm P}^2} \, m \, \left[ V(\phi_1) +
    \frac{1}{2} \dot{\phi}_1^2 \right] \,; \\
\ddot{\phi}_1 & = & - 3 H \dot{\phi}_1 - \frac{dV(\phi_1)}{d\phi_1} \,.
\end{eqnarray}
These can be mapped to the equations of a model with a single scalar field
$\tilde{\phi}$ by the redefinitions
\begin{equation}
\label{scal}
\tilde{\phi}_1^2 = m \, \phi_1^2 \quad ; \quad  \tilde{V} = m\, V
    \quad ; \quad \tilde{p} = mp \,,
\end{equation}
so the expansion rate is $a \propto t^{\tilde{p}}$, provided that $\tilde{p} > 1/3$.
The expansion becomes quicker the more scalar fields there are. In particular,
potentials with $p < 1$, which for a single field are unable to support
inflation, can do so as long as there are enough scalar fields to make
$mp>1$.

In order to calculate the density perturbation produced in
multi-scalar field models, we recall the results from \cite{Sasaki:1995aw}:
\begin{equation}
{\cal P}_{\cal R} = \left( \frac{H}{2\pi} \right)^2 \,
    \frac{\partial N}{\partial \phi_i} \,
    \frac{\partial N}{\partial \phi_j}  \, \delta_{ij} \,,
\end{equation}
where ${\cal P}_{\cal R}$ is the spectrum of the curvature perturbation
${\cal R}$, $N$ is the number of
$e$-foldings of inflationary expansion remaining, and there is a summation
over $i$ and $j$. Since $N = - \int H \, dt$, we have
\begin{equation}
\sum_i \frac{\partial N}{\partial \phi_i} \dot{\phi}_i = -H \,,
\end{equation}
where in our case each term in the sum is the same, yielding
\begin{equation}
{\cal P}_{\cal R} = \left( \frac{H}{2\pi} \right)^2 \, \frac{1}{m} \,
    \frac{H^2}{\dot{\phi}_1^2} \,.
\end{equation}
Note that this last expression only contains one of the scalar fields, chosen
arbitrarily to be $\phi_1$. This estimation for the spectral tilt is given by \cite{Sasaki:1995aw}:
\begin{equation}
\label{Spectral}
n -1 = 2\frac{\dot{H}}{H^{2}} - 2\frac{\frac{\partial N}{\partial \phi_i}
    \left(\frac{\dot{\phi_i}
    \dot{\phi_j}}{M_{\rm P}^2H^{2}} \, -
    \frac{M_{\rm P}^2V_{,i,j}}{V} \right) \, \frac{\partial N}
    {\partial \phi_j} }{ \delta_{ij} \frac{\partial N}{\partial
    \phi_i} \, \frac{\partial N}{\partial \phi_j}} \,,
\end{equation}
where there is a summation over repeated indices and the commas indicate
derivatives with respect to the corresponding field component.
Under our assumptions, the complicated second term on the right-hand side of
the above equation cancels out, and
Eq.~(\ref{Spectral}) reduces to the simple form
\begin{equation}
\label{Spec}
1-n = - 2\frac{\dot{H}}{H^{2}} = \frac{m^{2}_{\rm Pl}}{8\pi} \,
    \left[\frac{\frac{\partial V(\phi_1)}{\partial \phi_1}}{V(\phi_1)}
    \right]^2 = \frac{2}{m p} \,.
\end{equation}

This result shows that the spectral index also matches that produced by a
single scalar field with $\tilde{p} = mp$. The more scalar fields there are,
the closer to scale-invariance is the spectrum that they produce. Note
however that if the fields have such steep potentials as to be individually
non-inflationary, $p < 1$, then many fields are needed before the spectrum is
flat enough. The above calculation can be repeated for arbitrary slopes, $p_{i}$
in Eq.~(\ref{assist-pot}). In which case the spectral tilt would have been given by
$n=1-2/\tilde p$, where $\tilde p=\sum p_i$. The above scenario has been generalized to
study arbitrary exponential potentials with couplings, $V=\sum^{n} z_s \exp(\sum^{m}\alpha_{sj}\phi_j)$
in Ref.~\cite{Copeland:1999cs}, see also~\cite{Green:1999vv}. Such potentials are expected to
arise in dimensionally reduced SUGRA models~\cite{Bremer:1998zp}.

One particular nice observation for $m$ scalar potentials of chaotic type, $V\sim \sum_{i}f(\phi_{i}^{n}/M_{\rm P}^{n-4})$
(for $n\geq 4$), is that inflation can now be driven at VEVs below the Planck
scale~\cite{Kanti:1999ie,Kanti:1999vt,Kaloper:1999gm,Jokinen:2004bp}~\footnote{The double inflation model has been
studied extensively with two such fields, $V= m_1^2\phi_1^2+m_2\phi_2^2$, in Refs.
\cite{Silk:1986vc,Adams:1991ma,Holman:1991ht,Polarski:1992dq,Peter:1994dx,Langlois:1999dw}. In general one could expect:
$V\sim \sum_{i} m_{i}^2\phi_{i}^2$~\cite{Kanti:1999ie,Kanti:1999vt,Kaloper:1999gm}, or $V\sim \lambda_{i}(\phi^{n}_{i}/M_{\rm P}^{n-4})$, where $n\geq4$~\cite{Jokinen:2004bp}, where $\phi_{i}\ll M_{\rm P}$.}.
The {\it effective} slow-roll parameters
are given by: $\epsilon_{eff}=\epsilon/m \ll 1$ and $|\eta_{eff}|=|\eta|/m\ll 1$, where $\epsilon,~\eta$ are the slow-roll
parameters for the individual fields. Inflation can now occur for field VEVs~\cite{Jokinen:2004bp}:
\begin{equation}
\frac{\Delta \phi}{M_{\rm P}}\sim \left(\frac{600}{m}\right)\left(\frac{N_Q}{60}\right)\left(\frac{\epsilon_{eff}}{2}\right)^{1/2} \ll 1\,,
\end{equation}
where $N_Q$ is the number of e-foldings.
Obviously, all the properties of chaotic inflation can be preserved at VEVs much below the quantum gravity scale,
including the prediction for the tensor to scalar ration for the stochastic gravity waves, i.e. $r=16\epsilon_{eff}$.
For $\epsilon_{eff}\sim 0.01$ and $m\sim 100$, it is possible to realize a sub-Planckian inflation. the spectral
tilt close to the flatness can be arranged in the above example $n_s-1=-6\epsilon_{eff}+2\eta_{eff}$.
Furthermore, realistic {\it assisted inflation} model can be realized with a better UV understanding in string theory
with $m$ complex structure axions arising in type IIB string theory~\cite{Dimopoulos:2005ac,Easther:2005zr}, Kaluza-Klein
scalars~\cite{Kanti:1999ie,Kanti:1999vt,Kaloper:1999gm}, in multi-brane driven inflation~\cite{Mazumdar:2001mm,Cline:2005ty,Becker:2005sg,Piao:2002vf}, and in $SU(N)$ gauge theories~\cite{Jokinen:2004bp}~\footnote{Although it is quite plausible that conditions for late inflation can naturally be created after high scale assisted inflation, where one can have a possible signature for very long wavelength stochastic gravity waves~\cite{Allahverdi:2007ts}.}.

However, the caveat for all these models discussed in Secs.~\ref{sec:originalhybridinflation},~\ref{MSHI},~\ref{SAOVHI},~\ref{AINF}
 is that the connection with the SM physics is still lacking. It is not clear whether
the scalars (i.e. inflaton, waterfall fields, etc.)  can carry the SM charges or not. A partial attempt has been made in 
the context of \emph{assisted inflation} in
Ref.~\cite{Jokinen:2004bp} within $SU(N)$ gauge theories, where the inflatons are gauge invariant under SUSY $SU(N)$. Therefore, all these models bear similar uncertainties for reheating and thermalization as any other gauge singlet models
of inflation, see the discussion in Sec.~\ref{GSICM}.


\subsubsection{Non-Gaussianities from multi-field models}

With several light fields one would expect
isocurvature perturbations~\cite{Gordon:2000hv,Bassett:2005xm,Linde:1996gt,Liddle:1999pr}.
Isocurvature perturbations can also seed second order metric perturbations,
which can yield large non-Gaussianities  (see for
example~\cite{Bernardeau:2002jy,Bernardeau:2002jf,Bartolo:2004if,Jokinen:2005by}).
The generation of non-Gaussianities in hybrid inflation has been studied
in Refs~\cite{Bernardeau:2002jf,Enqvist:2004bk,Alabidi:2006hg,Sasaki:2008uc,Naruko:2008sq,Byrnes:2008zy,Barnaby:2006cq}. It was found that the regular hybrid inflation models do not produce large amount of non-Gaussianities, since inflation is {\it effectively} realized by the slow-roll of one field, while the
fluctuations of the waterfall field are highly suppressed by its super-heavy
mass. Some modifications of the model were proposed
in Refs.~\cite{Bernardeau:2002jf,Alabidi:2006wa,Alabidi:2006hg}, and then generalized to the ``multi-brid inflation'' scenarios~\cite{Sasaki:2008uc,Naruko:2008sq}, where large non-Gaussianities can be generated. 
The model is based on the coupling of several inflaton fields $\phi_i$ to a waterfall field $\chi$~\cite{Sasaki:2008uc}
\begin{equation}
V(\phi_i,\psi) = \frac{1}{2}\sum_i^n g_i^2\phi_i^2\chi^2 + \frac{\lambda}{4}
\left(\chi^2-\frac{\sigma^2}{\lambda}\right)^2~.
\end{equation}
In the two-brid model (case $n=2$), the predictions are found very
different from the original hybrid model. The value of $f_{NL}$ is computed using the
$\delta N$ formalism in \cite{Sasaki:2008uc,Naruko:2008sq,Huang:2009xa}, and
it is found to be~\cite{Sasaki:2008uc}:
\begin{equation}
n_s = 1-(m_1^2+m_2^2)~,\quad r=\frac{8(m_1g_1\cos\gamma+m_2g_2\sin\gamma)^2}{g_1^2\cos^2\gamma+g_2^2\sin^2\gamma}~,\quad
f_{\rm NL}^{\rm local}=\frac{5g_1^2g_2^2}{6\sigma(g_1^2\cos^2\gamma+g_2^2\sin^2\gamma)}~,
\end{equation}
where $m_i$ are the masses of the components $\phi_i$ (in Planck units),
and the VEVs at the end of inflation are parametrized by:
\begin{equation}
\phi_{1,f}=\frac{\sigma}{g_1}\cos \gamma ~,\quad \phi_{2,f}=
\frac{\sigma}{g_2}\sin \gamma~.
\end{equation}
The presence of extra-parameters allows the model to predict
large levels of non-Gaussianities. For example, for $m_1\sim 0.005$,
$m_2\sim 0.035$, $\gamma\ll 1$ and $g_1=g_2\equiv g$, the model predicts
$n_s\simeq 0.96$, $r\simeq 0.04$, and $$f_{\rm NL}^{\rm local}\simeq
\frac{5g\, m_2^2}{6 m_1 \sigma}\sim 40\frac{g}{\lambda^{1/4}}~.$$
The stability of the model requires $\lambda^{1/4}\gg 10^{-3}$ which still
allows $f_{ NL}^{\rm local}\ll 4\times 10^4$.  These results were generalized in
\cite{Naruko:2008sq} to more general potentials and for a larger parameter
space.

A generalized expression for non-Gaussanity for multiple field case has been
derived in~\cite{Lyth:2005du,Seery:2005gb} from the $\delta N$ formalism (see Sec.~\ref{sec:nongaussianities}):
\begin{equation}
-\frac{6}{5}f_{NL}=\frac{r}{16}(1+f)+\frac{\sum_{i,j} N_{,i}N_{,j}N_{,ij}}{\left(\sum_kN^2_{,k}\right)^2}\,,
\end{equation}
where $N_{,i}\equiv \partial N/\partial\phi_i\approx V_{i}/V^{\prime}_{i}$ by using the horizon crossing
approximation~\cite{Lyth:1998xn,Lyth:2005du}, $r$ is the tensor to scalar ratio, and $f$ is a geometrical factor
relating to the triangular bispectrum, lying in the range, $0\leq f\leq 5/6$~\cite{Maldacena:2002vr}. The value of
$r$ for $V=\sum_i\lambda_{i}\phi_i^{\alpha}$ is given by~\cite{Piao:2006nm,Kim:2006ys}:
\begin{equation}
r\approx \frac{8 M_{\rm P}^2}{\sum_i(V_i/V'_i)^2}\approx \frac{4\alpha}{N}\,,
\end{equation}
where $N\approx M_{\rm P}^{-2}\sum_i\int_{\phi_i}^{\phi_i^{end}}(V_i /V_{i}^{\prime})d\phi_i\approx(\sum_{i}\phi_{i}^2/2\alpha M_{\rm P}^2)$. With the help of the above expression, the value of $f_{NL}$ can be given by~\cite{Kim:2006te}
\begin{eqnarray}
\frac{-6}{5}f_{NL} &\approx & M_{\rm P}^2\left(\sum_j\frac{V_j^2}{V^{\prime 2}_{j}}\right)^{-2}
\sum_{i}\frac{V_{i}^2}{V^{\prime 2}_{i}}\left(1-\frac{V_iV^{\prime\prime}_{i}}{V^{\prime 2}_{i}}\right)
\approx  \frac{\alpha M_{\rm P}^2}{\sum_i\phi_i^2}\approx \frac{1}{2N}(2+f)\approx
\frac{r}{8\alpha}(2+f)\,.\nonumber \\
\end{eqnarray}
Unfortunately, in these models the value of $f_{NL}$ is very small and undetectable by the future experiments.
The expression is also a generalization of ~\cite{Vernizzi:2006ve}, derived for the two field case. Similar exercise 
has been performed for models of multi-field inflation with non-canonical kinetic terms in~\cite{Cai:2009hw,Cai:2008if}, 
we will discuss such models of inflation in Sec.~\ref{BI}.

Non-Gaussianities can also be created after inflation through
tachyonic preheating at the end of
inflation~\cite{Enqvist:2005qu,Barnaby:2006cq,Barnaby:2006km}.
Preheating of light fields after inflation has also been
considered in~\cite{Enqvist:2004ey,Enqvist:2005nc,Jokinen:2005by},
and in $\delta N$ formalism~\cite{Kohri:2009ac,Bond:2009xx}. It
was found that the $\lambda\phi^4$ model can lead to levels of non-Gaussianities that
are already on the verge of being excluded by current
observations~\cite{Jokinen:2005by}. Also, very recently, it was
pointed out that the waterfall part of the dynamics in a generic
non-SUSY hybrid type inflation (or hilltop inflation) with a
potential of the generic form
\begin{equation}
V(\phi,\psi)\sim V_0+\frac{\eta V_0}{2M_{\rm P}^2}\phi^2-\lambda
\phi\psi^2 +\mathcal{O}(\phi^3,\dots)~,
\end{equation}
could generate a large amount of non-Gaussian
fluctuations given by~\cite{Mulryne:2009ci}~\footnote{A word of caution when we use
$\delta N$ formalism to estimate non-Gaussianity during parametric resonance or during 
tachyonic prehetaing. Note that prehetaing is a violent and non-adiabatic process, which can 
happen at time scales much shorter than one Hubble time. Especially, during tachyonic prehetaing, the field displacement 
can be very negligible as compared to one Hubble time, during which the $\delta N\approx 0$, therefore 
in a separate universe approach, where the Hubble patches evolve homogeneously and smoothly, one can not trust the $\delta N$ formalism.  }.
\begin{equation}
\frac{f_{NL}}{1.3\times 10^4}\sim
\left(\frac{\gamma}{10^{-2}}\right)^{3/2}
\left(\frac{V_e^{1/4}}{10^{-3}M_{\rm P}}\right)^4
\left(\frac{10^{-2}M_{\rm P}}{\lambda}\right)
\left(\frac{10^{-2}}{\epsilon_e}\right)^{1/2}~,
\end{equation}
where the index $e$ denotes the end of inflation and $\gamma$ is a
semi-analytic parameter in the range $[0.03-0.1]$.


\subsubsection{Challenges for non-SUSY models}\label{sec:radcorrnonSUSY}

For non-SUSY models of inflation there are two more challenges
related at the classical and the quantum level.

\begin{itemize}

\item{Effective couplings and symmetries:\\
At the classical level inflaton can couple to other light scalar fields during inflation. In any of the effective potentials considered so far, there is no symmetry argument which will not allow couplings of type; $\phi^2\sum \chi_i^2$, or some non-renormalizable
couplings to other fields (belonging to other hidden sectors) such as $\sim \phi^{n}\chi^{m}/M_{\rm P}^{n+m-4}$ for $n, m>2$. There may be some discrete symmetries which will forbid some terms or some combinations, but it cannot render all the couplings to be vanishing. Any such light field other than the inflaton would introduce isocurvature perturbations during inflation, which at the classical level leaves such models vulnerable to quantum predictions of the CMB fluctuations. A simple calculation assuming adiabatic nature of density perturbations will not suffice in such cases.

Furthermore, the same couplings can dump almost {\it all the inflaton energy density} into some other non-SM degrees of freedom (or hidden sectors) upon reheating or preheating. One must assume that SM degress of freedom are excited, but such assumptions are always hard to justify if the particle contents are unknown. }

\item{Quantum stability of the potential:\\
The inflaton cannot be considered free from matter couplings, any coupling of the inflaton to fermions and gauge bosons
would introduce loop corrections at the perturbative level~\cite{Coleman:1973jx}. This will spoil the classical flatness of the
inflaton potential even when the scale of inflation is far below the scale of
gravity~\cite{Burgess:2007pt,Burgess:2009ea,Weinberg:2009bg}~\footnote{However, it is interesting to note that the potential of Eq.~(\ref{eq:potentialchaotic}) with $\alpha=2$ or $\alpha=4$ have
some accidental stability with respect to loop corrections due to the
absence of a self-coupling or a mass term respectively.
This is not the case anymore if both the terms are present, if $\alpha=3$,
or in the case of a coupling with other fields, for example in hybrid
inflation and its variants.}. Beyond destabilizing or modifying the shape of the potential,
radiative corrections can substantially alter the CMB predictions of the
models. 

SUSY helps stabilizing the classical potential, as the leading quantum corrections are logarithmic in
nature.  The another classic example is the pseudo Nambu Goldstone Boson (pNGB) as an inflaton, where the potential explicitly breaks the global symmetry with small couplings~\cite{Freese:1990rb,Adams:1992bn,Frieman:1995pm,Kaplan:2003aj}. In Ref.~\cite{Kaplan:2003aj}, it was observed that if the global symmetry is explicitly broken by a combination of couplings, then {\em loop contributions to pNGB masses must involve all of the couplings, and therefore one-loop contribution cannot be quadratically divergent}.  This is due to collective or non-local symmetry breaking as discussed in Ref.~\cite{ArkaniHamed:2001nc}.

Let us follow the discussion of ~\cite{Kaplan:2003aj}, where they begin by considering the simplest model which
involves a pNGB $\theta$ which comes from
the breaking of a global $SO(2)$ symmetry.  After integrating out the
``radial'' degree of freedom and pushing the cutoff of this non-linear sigma
model to the point where the interactions become strongly coupled, namely
$\Lambda\sim 4\pi f$.  The inflaton, $\Phi$, can be parameterized as
\begin{eqnarray}
\Phi = \left(\begin{array}{cc} \cos{(\theta/f)}&\sin{(\theta/f)}\\
-\sin{(\theta /f)}
    & \cos{(\theta /f)}\end{array} \right) \left(\begin{array}{c} -1\\
1\end{array}\right)
    \times {f\over\sqrt{2}}
\end{eqnarray}
Let us consider the tree-level potential to be:
\begin{equation}
V= \lambda \left( \sigma^T \sigma - v^2 \right)^2
+ {g_1\over 4} \,(\sigma^T {\Phi})^2
+ {g_2 \over 4} \,(\sigma^T \tau_1 \Phi)^2
\end{equation}
where $\sigma^T = (\sigma_1 \, \sigma_2)$ and $\tau_1$ is the first Pauli
matrix. Let us consider a simple situation when $g_1, g_2=g \neq 0$.
From expanding out the $\Phi$s in the potential, one finds:
\begin{equation}
V = {g f^2\over 4} (\sigma_1^2 + \sigma_2^2 -
2\sigma_1\sigma_2\cos{(2\theta /f)} )
\end{equation}
Now computing the one-loop corrections to the mass of $\theta$, the authors
of Ref.~\cite{Kaplan:2003aj} obtained that there is no one-loop quadratic divergent
contribution to a $\theta$ mass. This is because $\theta$ only couples to the combination
$\sigma_1\sigma_2$ making it impossible
to close a loop with only one vertex.  There is a logarithmic divergence
at one loop proportional to
$g_1 g_2 = g^2$
\begin{eqnarray}
V_{1-loop} &=&  \frac{g^2}{128\pi^2} \log{\left({\Lambda^2 \over
m_{\theta}^2}\right)} (\Phi^T \tau_1 \Phi )^2 + ...\nonumber\\ \nonumber\\
&=& \frac{g^2 f^4}{128\pi^2} \log{\left({\Lambda^2 \over
m_{\theta}^2}\right)} \cos^2{(2\theta/f)} + ...
\end{eqnarray}
The value of $\Lambda$ could be as large as $M_{\rm P}$ or below,
but the corrections to the potential is only logarithmic
dependent. The pNGB inflaton could also originate in
SUSY inflation models and in extra dimensional
models~\cite{Kaplan:2003aj,ArkaniHamed:2003wu,ArkaniHamed:2003mz}.
}

\end{itemize}

Some of the above mentioned challenges can be addressed if inflation is explicitly embedded within an 
observable sector. One such example of inflaton is the SM Higgs in a non-SUSY context.

\subsection{SM Higgs as the inflaton}\label{sec:SMinflaton}

It is natural to study if the SM Higgs can play the role of the inflaton field. This question has
been discussed long ago \cite{Salopek:1988qh}, and has regained
interest in the last few years~\cite{Bezrukov:2007ep,Bezrukov:2008ej,Barvinsky:2008ia,DeSimone:2008ei,Barbon:2009ya,Burgess:2009ea}.

\subsubsection{Dynamics of the SM Higgs inflation}

It has been proposed long ago to improve this situation by
abandoning the universal coupling to
gravity~\cite{Zee:1978wi,Smolin:1979ca,Salopek:1988qh,Bezrukov:2007ep}~\footnote{Such 
attempts were previously made in connection
with scalar tensor theories of inflation to flatten the inflationary
potential~\cite{Salopek:1988qh,La:1989za,Barrow:1990nv,Barrow:1993nt,Parsons:1995ew,GarciaBellido:1995br,Berkin:1990ju,Berkin:1991nm,Holman:1990hg,Holman:1990wq,DiMarco:2005zn},
or to exit the false vacuum inflation~\cite{La:1989za,La:1989st,La:1989pn,Biswas:2005vz}.
Typically the gravitational part of the action is given by: 
$S =\int d^4 x \sqrt{-g} [ \frac{1}{2}  M^2 f(\phi) R - \frac{1}{2}
\partial_{\mu}\phi \partial^{\mu}\phi]$. The action is dynamically
equivalent to a theory in which the gravitational action is the
usual one, via the conformal transformation:
$\hat{g}_{\mu\nu}=f(\phi)g_{\mu\nu}$, where we use the bar to
indicate a quantity in the new frame. The new action looks like:
$S_{E}= {1\over 2}\int d^4x \sqrt{-\hat{g}}[M^2\hat{R}-K(\phi)(\hat\partial\phi)^2] $, where, 
$K(\phi)\equiv {2f(\phi)+3  M^2 f^{'2}(\phi)\over 2f^2(\phi)}$. }, in order to flatten the Higgs potential at high
energies. The lagrangian
is extended to contain a non-minimal coupling to gravity (only)
for a Higgs field $H$
\begin{equation}\label{eq:lagrangianHiggsInfl}
\mathcal{L}=\mathcal{L}_{\rm SM} - \frac{M^2}{2}R -\xi H^\dagger H R~.
\end{equation}
This non-minimal coupling can be motivated by the renormalizability
of the $\lambda \phi^4$ potential \cite{Birrell:1982ix}. The very form
of this Lagrangian might represent the first challenge
of the model, since the equivalence principal is lost; all particles
are not coupled in the same way to gravity.

The above Lagrangian has been studied in the
past~\cite{Salopek:1988qh,Kaiser:1994vs}, where $H$ is a GUT Higgs field
and was applied in \cite{Bezrukov:2007ep} to the SM Higgs field. If $h$
is the Higgs field in the unitary gauge, the resulting action for $h$,
in the Jordan frame, reads
\begin{equation}
\mathcal{S}_J=\int \mathrm{d}^4x\sqrt{-g}\left\{\frac{M^2+\xi h^2}{2} R
+\frac{1}{2}\partial_\mu h \partial^\mu h -\frac{\lambda}{4}\left(h^2-v^2\right)^2\right\} ~.
\end{equation}
Since the coupling to gravity is not minimal in $S_J$, studying the phenomenology of
the model is simplified once the conformal transformation is
applied~\cite{Salopek:1988qh,Bezrukov:2007ep}
\begin{equation}
\begin{split}
\Omega^2=1+\xi h^2~,\quad  g_{\mu\nu} \rightarrow \hat{g}_{\mu\nu}=\Omega^2g_{\mu\nu} ~, \\
\quad h\rightarrow \chi \;\mathrm{with}\;\; \frac{\mathrm{d}\chi}{\mathrm{d} h}
=\sqrt{\frac{\Omega^2+6\xi^2h^2/M_{\rm P}^2}{\Omega^4}}~,
\end{split}
\end{equation}
such that the action in the Einstein frame reads
\begin{equation}
\mathcal{S}_E=\int \mathrm{d}^4x\sqrt{-\hat{g}}\left\{\frac{M_{\rm P}^2}{2} \hat{R}
+\frac{1}{2}\partial_\mu \chi \partial^\mu \chi - U(\chi)\right\} ~.
\end{equation}
In order to keep canonical kinetic terms once the metric is redefined, the potential $U$
for the new field $\chi$ now reads
\begin{equation}
U(\chi)= \frac{1}{\Omega(\chi)^4}\frac{\lambda}{4}\left[h^2(\chi)-v^2\right]^2~.
\end{equation}
At low energies, that is for small VEVs, $\Omega^2 \simeq 1$, $h\simeq \chi$
and the two frames are indistinguishable. At high energies, $h\propto \exp
\chi$, and the potential tends to
\begin{equation}\label{eq:sminflationpoten}
U(\chi)\simeq \frac{\lambda M_{\rm P}^4}{4\xi^2}\left[1+\exp\left(\frac{-2\chi}{\sqrt{6}M_{\rm P}}\right)\right]~.
\end{equation}
Once the potential is known at high energies, it is straightforward to compute the CMB
predictions for the model. From Eq.~(\ref{eq:sminflationpoten}), the slow-roll
parameters read \cite{Bezrukov:2007ep}
\begin{equation}
\epsilon(\chi)\simeq \frac{4M_{\rm P}^4}{3\xi^2h^4}~,\quad \eta(\chi)\simeq -\frac{4M_{\rm P}^2}{3\xi h^2}~,
\quad \zeta\simeq \frac{16M_{\rm P}^4}{9\xi^2h^4}~,
\end{equation}
which requires inflation to take place in the range $h\in [1.07-9.4]M_{\rm P}/\sqrt{\xi}$.
A correct normalization to the COBE data imposes $\xi\simeq 5\times 10^4
\sqrt{\lambda}$, and the model predicts, at the classical
level~\cite{Bezrukov:2007ep,DeSimone:2008ei},
\begin{eqnarray}
n_s  &\simeq &1-8\frac{(4N_Q+9)}{(4N_Q+3)^2}\simeq 0.97,~~~~~
r\simeq \frac{192}{(4N_Q+3)^2}\simeq 0.0033,\nonumber \\
\alpha_s &\simeq  & -5.2 \times 10^{-4}~,
\end{eqnarray}
which is in agreement with the most recent WMAP data.

The main challenge to this model is to evaluate the quantum
corrections to the inflationary potential at high energyto
evaluate if the flatness of the inflationary potential can be
destabilized by them. Both the quantum gravity corrections and the
quantum corrections due to SM particles can be evaluated, though
the full quantum gravity corrections cannot be rigorously
computed. It has been argued in Ref.~\cite{Bezrukov:2007ep} that
the model is safe since $U(\chi)/M_{\rm P}^4~\lambda/\xi^2\ll 1$
and $\mathrm{d}^2U/\mathrm{d}\chi^2\ll M_{\rm P}^2$.

The leading log method \cite{Bezrukov:2007ep} as well as
renormalization group (RG) methods have been implemented
\cite{Barvinsky:2008ia,DeSimone:2008ei} in both frames  to compute
corrections to $V$, and $\xi$. The corrections from SM particles
to $V$ using RG improved calculation with 2-loop beta-function are
also found in Ref.~\cite{DeSimone:2008ei} to be very small and do
not spoil the model (see Fig.~\ref{fig:potentialSMinflation}).
However, they introduce a dependence of the CMB predictions on the
SM particle masses, in particular on the Higgs mass $m_h$ and the
top quark mass $m_t$ as detailed below.


\begin{figure}[h!] \label{fig:potentialSMinflation}
\scalebox{0.6}{\includegraphics{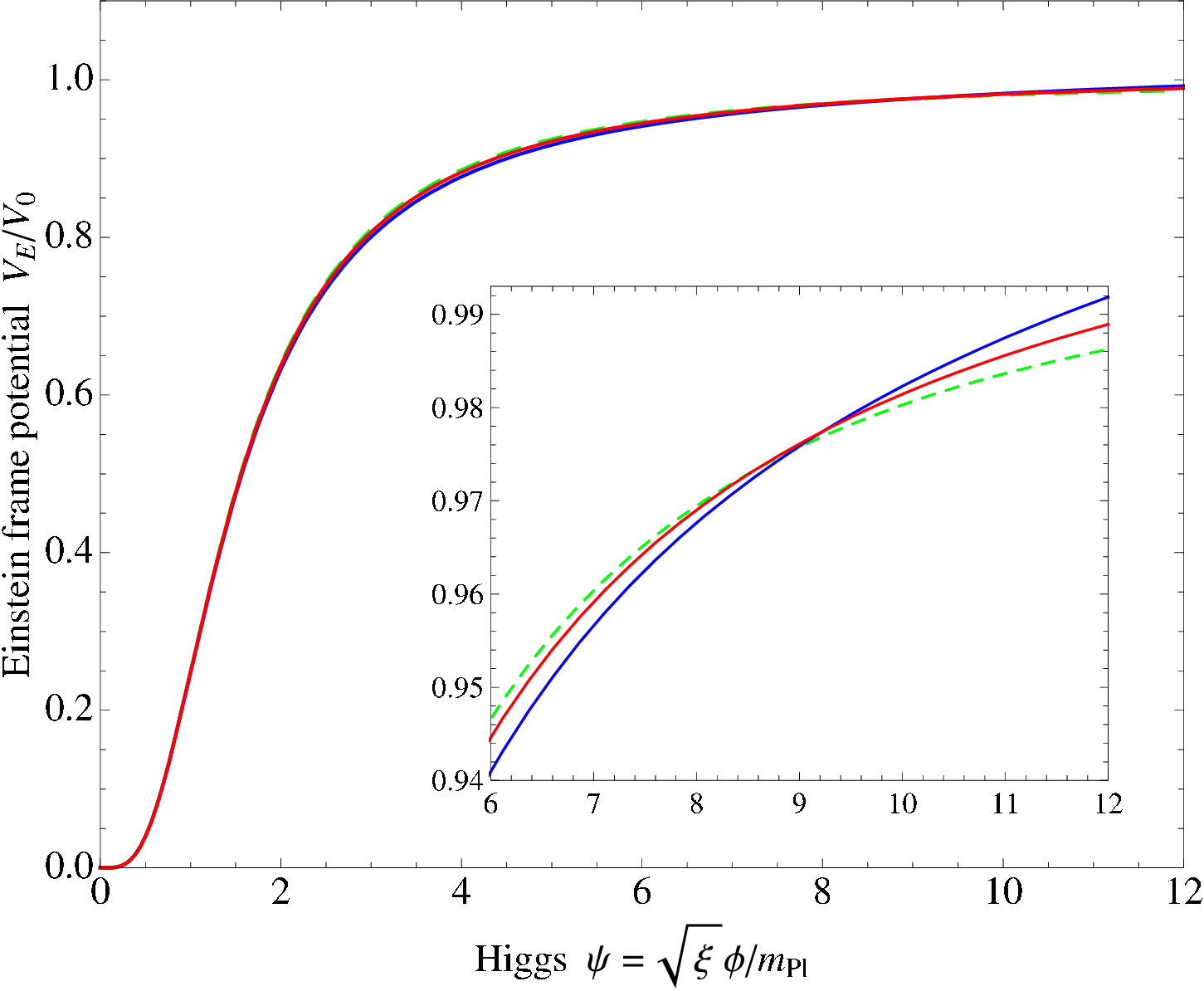}}
\caption{Normalized scalar potential of SM inflation in the Einstein frame, in the classical
approximation (green), or taking into account quantum corrections for $m_h=126.5$ GeV and
$m_h=128$ GeV, in blue and red respectively. Figure is taken from \cite{DeSimone:2008ei}.
\label{fig:potentialSMinflation}}
\end{figure}


The stability of the present classical action could also be affected by the presence
of non-renormalizable operators. This represents another serious challenge to the
model~\cite{Barbon:2009ya,Burgess:2009ea}, though this argument can be applied to
other high scale models of inflation. Indeed, it is shown that the effective
cutoff of the Lagrangian of Eq.~(\ref{eq:lagrangianHiggsInfl}) is $M_{\rm P}/\xi$, whereas
the energy scale of inflation is $M_{\rm P}/\sqrt{\xi}$. One therefore should expect
non-renormalizable contributions to the action to be relevant at energies well below
inflation. This represents a challenge to the model since the implicit assumption is
that the SM of particle physics is the effective theory at least up to the
inflationary scale. In the absence of symmetries to prevent the appearance of
non-renormalizable contributions the model should be considered as fine-tuned, so
that these contributions do not spoil the flatness of the model.


\subsubsection{SM Higgs inflation and implications for collider experiments}

As mentioned above, the important feature of this model of inflation is that masses
of SM particles enter the quantum corrections to the scalar potential, and thus
impact the CMB predictions $n_s$, $r$ and $\alpha_s$. In
particular the most important masses that affect the predictions are the Higgs
mass $m_h$ as it enters the tree level potential, and then the top mass, the
highest mass that enters in loops of SM particle. Independently of inflation,
current accelerator experiments allow these masses to range in \cite{Amsler:2008zzb}
$m_h \in [114.4-182]$ GeV and $m_t \in [169-173]$ GeV. The evolution of the CMB
predictions with respect to these masses is given in details
in~\cite{DeSimone:2008ei}. We will here only mention that compared to the classical
values (equivalent to large $m_h$), $n_s$ $r$ and $|\alpha_s|$ increases when $m_h$
decreases or when $m_t$ increases. In particular, for the spectral index to be
within the $1\sigma$ contour of WMAP ($n_s<0.99$), the mass of the Higgs is
found~\cite{DeSimone:2008ei} to be larger than
\begin{equation}
m_h\gtrsim 125.7 +3.8\left(\frac{m_t-171~\mathrm{GeV}}{2~\mathrm{GeV}}
\right)-1.4\left(\frac{\alpha_{\mathrm{SU(3)}}(M_Z)-0.1176}{0.0020}\right)\pm 2 \mathrm{GeV}~,
\end{equation}
where $\alpha_{\mathrm{SU(3)}}$ is the strong coupling and the $\pm 2$ GeV is
due to theoretical uncertainties from higher order corrections.

Testing this model in the future will therefore require the discovery of the
Higgs particle at the LHC, which is expected if its mass is within the above range,
together with an improvement of the error bar on $n_s$ as given by the PLANCK and
some improvement on the top quark mass measurement at the LHC and the ILC.
From phenomenological point of view SM Higgs inflation is a welcoming news, the Higgs can directly produce
the SM degrees of freedom as shown in Refs.~\cite{GarciaBellido:2008ab,Bezrukov:2008ut}.


\subsection{SUSY models of inflation}\label{sec:models:SUSYmodels}

Historically, SUSY inflation was first introduced to
cure the flatness problem and associated fine tuning of new
inflation \cite{Ellis:1982ed,Ellis:1982ws}, but since then utilizing SUSY
as a tool for inflation has gained in popularity. The fundamental
reason is similar to the resolution of the hierarchy problem for
which SUSY was introduced, i.e. to protect the flatness of the potential.

The scalar sector is now described by a superpotential $W$ and a K\"ahler
potential $K$, instead of just a scalar potential (see
Sec.~\ref{sec:partphys:SUSY}). Another
difference with non-SUSY effective field theory concerns the range
of VEVs allowed, which is now below the Planck scale. Close to the
Planckian VEVs SUGRA corrections become important. One of course recovers
the global SUSY in the limit when $M_{\rm P}\rightarrow \infty$.

In four dimensions, the $N=1$ SUSY potential receives two
contributions; one from the $F$-terms, describing interaction between
chiral superfields, and the second from the $D$-terms, which contains the
gauge interactions.
The scalar potential derived from $W$ and $K$
has to be non-vanishing to support inflation, therefore breaking
(transiently) SUSY. Therefore two classes of models driven
by $F$-terms or by $D$-terms have been proposed. We will discuss first
old and new properties of $F$-and $D$-term hybrid inflation, which is by far
the most popular amongst the model builders. Some of these models have also been reviewed
in Refs.~\cite{Lyth:1998xn,Lazarides:2001zd}.


\subsubsection{Chaotic inflation in SUSY}
\label{sec:models:chaoticSUGRA}

Embedding chaotic inflation within SUSY is a challenging problem, discussed long ago and is still
under investigation by many~\cite{Goncharov:1983mw,Murayama:1993xu,Kawasaki:2000yn,Linde:2005ht,Kadota:2007nc,Kadota:2008pm,Davis:2008fv}. As discussed in Sec.~\ref{sec:models:largefieldmodels}, chaotic inflation in its
non-SUSY description requires super-Planckian VEV to last long
enough. By construction SUSY and SUGRA are only valid when $\phi \ll M_{\rm P}$.  
Another challenge comes from the fact that chaotic inflation requires a UV-complete theory that SUGRA does 
not provide~\cite{Linde:2005ht}.

Even discarding the question of validity, if inflation
was driven by the $F$-terms, the exponential of the K\"ahler potential potential can generically spoil
the flatness of the potential. To tackle this problem, non-minimal K\"ahler
potentials with a logarithmic form have been proposed~\cite{Murayama:1993xu},
which can be combined to a mass-term superpotential, $W=M\Phi^2$, to generate
the right scalar potential, driven by $F$-terms.
It has also been shown that a chaotic potential can also be embedded within
$D$-terms~\cite{Kadota:2007nc,Kadota:2008pm}. They assume 4 chiral superfields
and that the symmetries of the theory are $U(1)_{\rm gauge}\times U(1)_R$~\footnote{A challenging 
problem for these kind of attempts is that the inflaton sector, though it resides in a hidden sector, must couple to
the  observable sector, i.e. MSSM. This modifies the $D$-flatness conditions for any extra $U(1)_{\rm gauge}$ sector
added to the SM gauge group, which opens up new $D$-flat directions and this will eventually modify the inflationary potential by lifting such combinations with the help of $F$-term. Such contributions are often ignored in the literature~\cite{Kadota:2007nc,Kadota:2008pm,Davis:2008fv}.}. A renormalizable superpotential and a minimal K\"ahler 
potential can generate a scalar potential possessing many $F$- flat valleys lifted by $D$-terms required for
generating chaotic inflation. It was pointed out in Ref.~\cite{Davis:2008fv} that the mechanism requires a fine-tuned
moduli sector which is hard to embed within string theory. As a conclusion, the question of managing super-Planckian
VEVs and densities within SUGRA does not make chaotic models appealing from particle physics point of 
view~\footnote{There are attempts to embed natural inflation realizable at low scales within SUGRA, see ~\cite{Adams:1996yd,Adams:1997de,German:1999gi,German:2001tz,German:2001sm}. The generic potential has a form: $V=\Lambda^4(1+\beta|\phi|^2+\gamma |\phi|^3+\delta |\phi|^4+\dots)$, where $\beta,~\gamma,~\delta$ are model dependent coefficients. Challenge for these models is to justify why inflation starts at $\phi\approx 0$ VEV and $\dot\phi\approx 0$. A prior phase of inflation may justify such initial conditions.}.


\subsubsection{Hybrid inflation from $F$-terms}\label{sec:ftermhybridinflation}


The most well-known model of SUSY inflation driven by $F$-terms is of the
hybrid type and based on the superpotential~\cite{Dvali:1994ms}
\begin{equation}
\label{eq:superpotentialFterm}
W^F=\kappa S(\Phi \overline{\Phi}-M^2)~.
\end{equation}
where, $S$ is an absolute gauge singlet while $\Phi$ and
$\overline{\Phi}$ are two distinct superfields belonging to complex
conjugate representation, and $\kappa$ is an arbitrary constant fixed by the
CMB observations~\footnote{It is desirable to obtain an effective singlet $S$ superfield 
arising from a higher gauge theory such as GUT, however to our knowledge 
it has not been possible to implement this idea, see Sect.~\ref{EEMGOIG}. Typically $S$ would 
have other (self)couplings which would effectively ruin the flatness. }.
This form of potential is protected from additional destabilizing
contributions with higher power of $S$, if $S$, $\Phi$ and $\overline{\Phi}$
carrying respectively the charges $+2$, $\alpha$ and $-\alpha$ under
R-parity. Then $W$ carries a charge $+2$ so that the action
$\mathcal{S}=\int \mathrm{d}^2\theta \; W +\dots$ is invariant.

The tree level scalar potential derived from Eq.~(\ref{eq:superpotentialFterm}) reads
\begin{equation}\label{eq:potentialFterm3field}
V_{\rm tree}(S,\phi,\overline{\phi})= \kappa^2|M^2-\overline{\phi}\phi|^2+\kappa^2|S|^2
(|\phi|^2+|\overline{\phi}|^2)^2+\mathrm{D-terms}~,
\end{equation}
where we have denoted by $S,\phi,\overline{\phi}$ the scalar components
of $S,\Phi,\overline{\Phi}$. It has a form similar to the original hybrid
inflation model, though $m=0$, and both $\lambda$ and $\lambda'$ are
replaced by only one coupling constant $\kappa^2$.
In what follows, it is assumed that $\phi^*=\overline{\phi}$ along this direction, the $D$-terms vanish
and that the kinetic terms for the superfields are minimal, which is
equivalent to a minimal K\"ahler potential,
$K=|S|^2+|\Phi|^2+|\overline{\Phi}|^2$.

We can defined two effective
real scalar fields canonically normalized, $\sigma\equiv \sqrt{2}
\mathrm{Re}(S)$, and $\psi\equiv 2\mathrm{Re}(\Phi)=2\mathrm{Re}
(\overline{\Phi})$. The two-field scalar potential can then be put to
the form~\cite{Lazarides:2001zd}
\begin{equation}\label{eq:potentialFterm2field}
V_{\rm tree}(\sigma,\psi)= \kappa^2\left(M^2-\frac{\psi^2}{4}\right)^2+\frac{\kappa^2}{4}
\sigma^2\psi^2~.
\end{equation}
Like in the non-SUSY version, the global minimum of the potential is
located at $S=0$, $\phi\overline{\phi}=M^2$, though at large VEVs,
$S > S_c\equiv M$, the potential also possesses a local valley
of minima (at $\langle\psi\rangle = 0$) in which the field $\sigma$,
identified as the inflaton from now on, lies in a flat direction,
$V_{\rm tree}=V_0\equiv \kappa^2 M^4$. This non-vanishing value of
the potential both sustain the inflationary dynamics and induces a
SUSY breaking. Chaotic initial
conditions are usually assumed \cite{Dvali:1994ms}, which provide for a
large inflaton VEV at the ``beginning'' of inflation (We will
return later to the issue of initial conditions in this model).
The end of inflation is then triggered by slow-roll violation and the
system rapidly settles at the bottom of one of the global minima,
breaking the symmetry $G$, potentially forming topological defects, and
restoring SUSY.

Since, $V(\psi=0)\neq 0$, SUSY is broken. This induces a splitting in the mass of the
fermionic and bosonic components of the superfields $\Phi$ and
$\overline{\Phi}$, with $m^2_B(S)=\kappa^2 |S|^2 \pm \kappa^2 M^2$ and
$m^2_F=\kappa^2 |S|^2$. Note that this description is valid only as
long as $S$ is sufficiently slow-rolling such that $\kappa^2 |S|^2 |\Phi|^2$
can be considered as a mass term.
Therefore radiative corrections do not exactly cancel
out~\cite{Dvali:1994ms,Lazarides:2000ck}, and provide a one-loop potential:
\begin{equation}\label{eq:effectivepotentialFtermvalley}
V^{\rm F}_{\rm 1-loop}(S)=\frac{\kappa^4\mathcal{N}M^4}{32\pi^2}\left[2\ln
\frac{s^2\kappa^2}{\Lambda^2}+(z+1)^2\ln(1+z^{-1})+(z-1)^2\ln(1-z^{-1})
\right]~,
\end{equation}
using the Coleman-Weinberg formula~\cite{Coleman:1973jx}. In this expression
\begin{equation}
z=\frac{|S|^2}{M^2}\equiv x^2~,
\end{equation}
$\Lambda$ represents a non-physical energy scale of renormalization and
$\mathcal{N}$ denotes the dimensionality~\footnote{To be very precise, the
value of $\mathcal{N}$ is less or equal to the dimensionality of $\Phi$ or
$\overline{\Phi}$. This factor should count the number of degrees of freedom
in $\Phi$ that are light enough to be affected by the value of $S$. This can
depend on the precise mass spectrum of the gauge group, for instance a
specific GUT model. See discussion in~\cite{Jeannerot:2006jj}.} of the Higgs
fields $\Phi$ and $\overline{\Phi}$. When discussing the predictions of the
model and the dynamics at the end of inflation, it is important to keep in
mind that the perturbative approach of Coleman and Weinberg breaks down when
close to the inflection point at $z\simeq 1$.


\subsubsection{CMB predictions and constraints}

The predictions of the model differ strongly from the original model, because
the potential is concave down due to the radiative correction is the 
origin for the slope in the potential. For small coupling $\kappa$, the slow-roll conditions
(for $\eta$) are violated infinitely close to the critical point, $z=1$, which
ends inflation. Thus, the quadrupole value for the inflaton, $z_Q
\equiv S^2_Q/M^2$, is obtained by solving
\begin{equation}
\begin{split}
N_Q &=\frac{32\pi^3}{\kappa^2 \mathcal{N}}\frac{M^2}{M_{\rm P}^2}\int_{1}^{z_Q}
\frac{\mathrm{d}z}{z f(z)}~, \\
\mathrm{with} \quad f(z)&=(z+1)\ln(1+z^{-1})+(z-1)\ln(1-z^{-1})~.
\end{split}
\end{equation}
The normalization to COBE allows to fix the scale $M$ as a function of
$\kappa$. If the breaking of $G$ does not produce cosmic strings, the
contribution to the quadrupole anisotropy simply comes from the
inflationary contribution (see Eq.~(\ref{eq:powerspectrum1})) and the
observed value can be obtained even with a coupling $\kappa$ close to
unity \cite{Dvali:1994ms}~\footnote{Small values of $\kappa$ can render
the scale of inflation very low, as low as the TeV
scale~\cite{Randall:1995dj,Randall:1994fr,BasteroGil:1997vn,BasteroGil:2002xs}.}. However it
has been shown that the formation of cosmic strings at the end of
$F$-term inflation is highly probable when the model is embedded in SUSY
GUTs~\cite{Jeannerot:2003qv}. In that case, the normalization to COBE
receives two contributions \cite{Jeannerot:1997is,Rocher:2004my}, one from
inflation $(\delta T/ T)_{\rm infl} \propto V^{3/2}/V'$, and the other from cosmic
strings $(\delta T/ T)_{\rm CS} \propto G\mu$, where
$\mu$ is the mass per unit length of the strings (see
Sec.~\ref{sec:partphys:symmbreak}). The relation between $M$ and
$\kappa$ from the normalization of the power-spectrum is now obtained by
solving~\cite{Rocher:2004my}
\begin{equation}
\left(\frac{\delta T}{T}\right)_{\rm COBE}^2=
\left(\frac{\delta T}{T}\right)_{\rm Infl}^2
+ \left(\frac{\delta T}{T}\right)_{\rm CS}^2~,
\end{equation}
which affects the relation $M(\kappa)$ at large $\kappa$, and imposes
new stringent bounds on $M\lesssim 2\times 10^{15}$ GeV,
and~\cite{Rocher:2004et,Jeannerot:2005mc}
\begin{equation}\label{eq:constraintFtermcoupling}
\kappa \lesssim 7\times 10^{-7}\frac{126}{N_Q}~,
\end{equation}
coming from imposing that the weight of cosmic strings in the WMAP3 data is
less than $\lesssim 10\%$ \cite{Bevis:2007gh}. This constraint can be made
less stringent if some other sector of the Higgs potential imposes that some
components of $\Phi,\overline{\Phi}$ acquire an ultra-large mass before
inflation, during previous symmetry breaking~\cite{Jeannerot:2005mc}.

Once $M$ is fixed, the spectral index $n_s$ can also be computed as a
function of the coupling constant. The range found is $n_s\in [0.98,1]$
whether cosmic strings form or not~\cite{Senoguz:2003zw,Jeannerot:2005mc}, and 
by including the soft-SUSY breaking terms within minimal kinetic terms in the K\"ahler potential, 
the spectral index can be brought down to  $0.928\leq n_s \leq 1.008$~\cite{Rehman:2009nq}.
This represents the most important difference with Linde's original model,
where a blue tilt is generated.
Note however that when cosmic strings form, the coupling constant has to be
suppressed (see Eq.~(\ref{eq:constraintFtermcoupling})) and the predicted value
for $n_s$ is very close to unity, $n_s\simeq 1^-$. This value is precisely in
agreement with the observations when cosmic strings are formed, if they
contribute significantly to the generation of the
anisotropies~\cite{Bevis:2007gh}. 


\subsubsection{SUGRA corrections to $F$-term inflation}


When the VEVs are not negligible compared to the Planck scale,
SUGRA effects become important and may ruin the flatness
of the inflaton potential. Indeed, the potential is now given by
Eq.~(\ref{eq:partphys:sugrapotential}), the $F$-terms containing an
additional exponential factor. The soft breaking mass of the scalar fields are typically
\cite{Dine:1983ys,Goncharov:1983mw,Coughlan:1984yk,Murayama:1993xu,Bertolami:1987xb,Copeland:1994vg,Stewart:1994ts,Dine:1995kz,Dine:1995uk,Dvali:1995mj,Dvali:1995fb,Davis:2008fv}
\begin{equation}
\label{msoftH}
m^2_{soft} \sim \frac{V}{3M_{\rm P}^2} \sim {\cal O}(1)H^2\,.
\end{equation}
Once the inflaton gains a mass $\sim H$, the contribution to the second
slow-roll parameter $\eta$ is of order unity and the field simply rolls
down to the minimum of the potential and inflation stops,
\begin{equation}
|\eta| \equiv {M_{\rm P}^2}\frac{V^{\prime \prime}}{V} \sim
\frac{m^2_{SUGRA}}{H^2} \sim {\cal O}(1) \,,
\end{equation}
where $m^2_{SUGRA}\approx m^2_{susy}+(V_{susy}/3M_{\rm P}^2)\sim m^2_{susy}+{\cal O}(1)H^2$, 
where $m_{susy}\sim {\cal O}(100)$~GeV  contains soft-SUSY breaking mass term for low scale
SUSY breaking scenarios.
For VEVs smaller than the Planck scale it is always possible to
obtain $\epsilon \ll 1$, but in SUGRA $\eta$ can never be made
less than one for a single chiral field with minimal kinetic terms~\footnote{Except when $m_{inf}\gg H_{inf}$. This can be realizable in an inflection point inflation as in the case of MSSM inflation, see Sec.~\ref{See :SUGRA-problem, trans-Planckian, and moduli problems}.}. This is known as the $\eta$ problem in SUGRA models of
inflation~\cite{Copeland:1994vg,Dine:1995kz,Dine:1995uk}.

When there are more than one chiral superfields, as in the $F$-term hybrid
model, it can be possible to cancel the dominant ${\cal O}(1)H$ correction
to the inflaton mass by choosing an appropriate K\"ahler term \cite{Stewart:1994ts,Copeland:1994vg}. 
In hybrid inflation models derived from an $F$-term the dominant ${\cal O}(1)H$ correction
in the mass term can be cancelled if $|S|=0$ exactly, which however seems
to lead to an initial condition problem, as  discussed above.
The fact that the superpotential is linear in $S$ (as in
Eq.~(\ref{eq:superpotentialFterm})) guarantees the cancellation of the
dominant contribution in the mass term for a minimal K\"ahler,
$K_{\rm min}(\Psi_i,\Psi_i^*) =\sum \Psi_i \Psi_i^*= |S|^2
+|\Psi|^2+|\overline{\Psi}|^2$~\cite{Copeland:1994vg},
\begin{equation}\label{eq:potentialFtermSUGRA}
\begin{split}
V_{\rm tree}^{\rm F-SUGRA}&= \kappa ^2 \exp \left(\frac{s^2 + \psi^2}{2 M_{\rm P}^2}\right) \\
& \times \Bigg\{ \left( \frac {\psi^2}{4} - M^2 \right)^2
\left( 1 - \frac{s^2}{2 M_{\rm P}^2} + \frac{s^4}{4 M_{\rm P}^4} \right) \\
&+ \frac{s^2 \psi^2}{4 } \left[ 1+ \frac 1 {M_{\rm P}^2} \left(\frac 1 4 \psi^2
- M^2\right)\right]^2\Bigg\}~.
\end{split}
\end{equation}
Note that these corrections affect the dynamics at large field, though at
small VEVs, the radiative corrections are the
dominant origin of the dynamics. This is generically the case during the
last 60 e-foldings of inflation~\cite{Clesse:2009ur}.


\subsubsection{Non-minimal kinetic terms and the SUGRA $\eta$ problem}

For K\"ahler potentials such as
\begin{equation}\label{eq:models:FtermKahlerNonMin}
K =|S|^2 +|\Phi|^2+|\overline{\Phi}|^2+\kappa_S |S|^4/M_{\rm P}^2+\dots~,
\end{equation}
the kinetic terms $K^{ij}\partial_\mu \Phi_i\partial^\mu \Phi_j^*$ are
non-minimal because $K^{ij}\neq \delta^{ij}$. One obtains in particular
$\left(\partial_{SS^{*}}K\right)^{-1}\sim 1-4\kappa_S |S|^2/M_{\rm P}^2+\dots $
that leads to a problematic $\kappa_S\times {\cal O}(1)H$ contribution to
the inflaton mass, and therefore on the second slow-roll parameter $\eta$,
unless some suppression is invoked. Several mechanisms have been
proposed to tackle this $\eta$-problem.


\begin{itemize}
\item The first one is to constrain the parameter of the leading corrections,
imposing $\kappa_S\sim 10^{-3}$ which is sufficient to keep the model
safe, but without much physical motivation. For a generic inflationary model it is not
possible to compute $\kappa_s$ at all (see the discussion on the footnote below).
In this model $|S|\ll M_{\rm P}$ ensures that higher order terms are
negligible~\cite{Lazarides:2001zd}.


\item Safe non-minimal K\"ahler potentials have also been
proposed~\cite{Brax:2005jv,Antusch:2009ef,Pallis:2009pq,BasteroGil:1998te} making use of
the shift symmetry~\footnote{Under this symmetry, a superfield
$S\rightarrow S+iC$, where $C$ is a constant.}~\cite{Kawasaki:2000yn,Kawasaki:2000ws} to
protect the K\"ahler potential of the form $K(\Phi,\bar{\Phi})
\rightarrow K(\Phi+\bar{\Phi})$. This
symmetry generates an exactly flat direction for an inflaton field and a
non-invariance of the superpotential induces some slope to its potential
to allow slow-roll at the loop level.

Another symmetry - the Heisenberg symmetry - has also been invoked to
protect the form of the K\"ahler potential, see for a recent discussion in \cite{Antusch:2008pn}, generating a
model where the effective K\"ahler is a no-scale potential, that is of the
form $K=\ln (\Phi_i)$. This obviously solves the $\eta$-problem by
canceling the exponential term $\exp (K)$~\footnote{There is a word of caution here, it is
assumed that one can take the inflaton VEV above the
Planck scale,  as in the case of Heisenberg symmetry of the K\"ahler potential, in order to realize chaotic type
models~\cite{Kawasaki:2000yn,Kawasaki:2000ws,Antusch:2009ty}. However,
this assumes that the K\"ahler potential does not obtain any quantum corrections. Further note that
non-renormalization theorem can only protect the superpotential terms~\cite{Grisaru:1979wc}, but the
K\"ahler potential generically obtains quantum corrections, which have been computed explicitly in some
cases~\cite{Berg:2005yu,Berg:2004ek,Berg:2005ja}. In string motivated models it is hard to realize chaotic type
models of inflation with VEVs larger than the $4d$ Planck scale.}.

\end{itemize}

The presence of non-renormalizable correction to the kinetic terms
has important consequences on the CMB predictions of the $F$-term hybrid model. In
particular it can be used to realize a better agreement with the WMAP 5
measurement on the spectral index~\cite{BasteroGil:2006cm,urRehman:2006hu}.
It has been shown that the presence of the $\kappa_s$ term in
Eq.~(\ref{eq:models:FtermKahlerNonMin}) allow the model to; \\
1) lower the value of $M$ responsible for the normalization of the spectrum for a
given $\kappa$ and thus lower the influence of cosmic strings on the
CMB, and \\
2) predict a spectral index lower than $0.98$, easily in the range
$n_s\in [0.9,1]$ as represented in Fig.~(\ref{fig:3rehman}).


\begin{figure}[t]
\scalebox{0.5}{\includegraphics{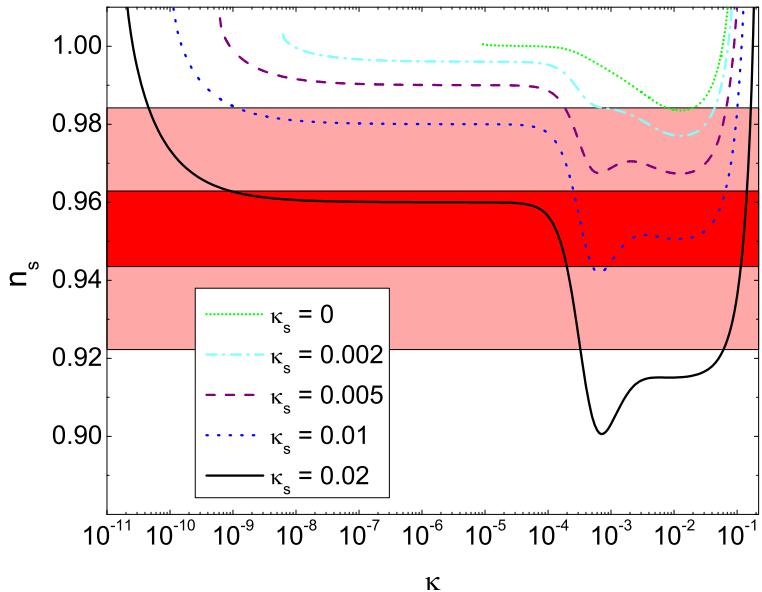}}
\caption{Spectral index for $F$-term hybrid inflation with minimal kinetic
terms (top green curve) and non-renormalizable corrections to the K\"ahler.
Figure is taken from \cite{urRehman:2006hu}.\label{fig:3rehman}}
\end{figure}


\subsubsection{Initial conditions for $F$-term hybrid inflation}

\begin{figure}[t]
\scalebox{0.35}{\includegraphics{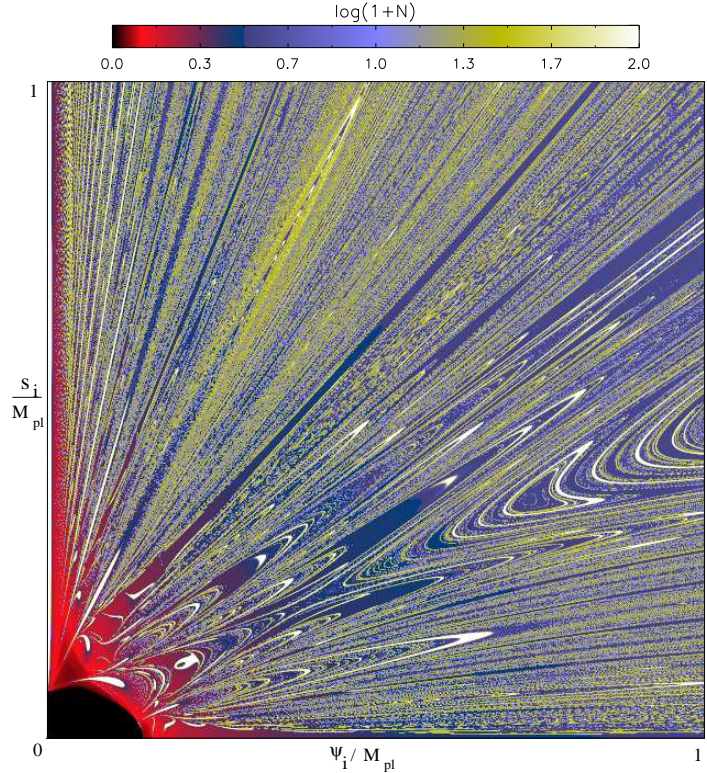}}
\caption{Initial condition space for $F$-term hybrid inflation in minimal
SUGRA, when restricting to the sub-Planckian ($M_{\rm pl}=M_{\rm P}$) initial VEVs and
vanishing velocities. The color code represents the
number of e-foldings generated for each initial VEVs.
This confirms  that hybrid inflation is successful
when the onset of inflation occurs close to the inflationary valley (the white narrow band
along the y-axis), but shows a subdominant space for other trajectories where
inflation is also successful for the sub-Planckian 
VEVs~\cite{Clesse:2009ur}. \label{fig:anamorphosis_SUGRA}}
\end{figure}

We have seen in Sec.~\ref{sec:models:multi-field} that initial
conditions for hybrid inflation was considered one of the challenges for
the hybrid models, and it was considered as a problem for the $F$-term
inflation in~\cite{Lazarides:1997vv}. It was argued that for inflation to be
successful, the initial field value for the waterfall field $\psi_i$ had
to be fine-tuned to an almost vanishing value, so as to start close
enough to the inflationary valley. Note that in this reference, the
radiative corrections to the inflationary potential has been
approximated by a mass term, similarly to the non-SUSY case (see
Sec.~\ref{sec:models:multi-field}). This model has been studied more
recently with a high precision, taking into account of the SUGRA corrections
induced by a minimal K\"ahler potential~\cite{Clesse:2009ur}. 


 In~\cite{Clesse:2009ur}, the space of
successful initial conditions is found composed of a non-fractal
set of successful points but with fractal boundaries. Such a set is
represented in Fig.~(\ref{fig:anamorphosis_SUGRA}), where for given
initial VEVs, the number of e-foldings of inflation realized is
represented. It was pointed out that similar to non-SUSY hybrid inflation,
inflation is realized generically by a first phase of fast roll down the
potential and when the velocity vector is correctly oriented at the bottom
of the potential, the inflaton climbs up and slow-rolls back down the
inflationary valley around $\langle\psi\rangle=0$.



\subsubsection{Other hybrid models and effects of non-renormalizable terms}
\label{sec:models:SUSYmodelswithNR}

The superpotential of $F$-term hybrid inflation given in
Eq.~(\ref{eq:superpotentialFterm}) contains only renormalizable terms.
However VEVs close to the UV cutoff of the theory,
(necessarily smaller than the reduced Planck mass), non-renormalizable
terms play a non-negligible role. Two models have been proposed to
study these effects: the ``smooth'' \cite{Lazarides:1995vr} and
``shifted''~\cite{Jeannerot:2000sv} hybrid inflation models.

The initial motivation for both of these models was to implement hybrid inflation in a GUT model based on the
Pati-Salam gauge group $\gps\equiv SU(4)_{\rm C}\times SU(2)_{\rm L}
\times SU(2)_{\rm R}$ without the formation of
monopoles at the end of inflation. The idea is therefore to break
the symmetry \emph{before} the phase of inflation, which can be achieved by introducing 
one non-renormalizable term in the superpotential.


\begin{itemize}

\item{Smooth hybrid inflation:\\
In this model~\cite{Lazarides:1995vr}, the superpotential has to satisfy a new $Z_2$
symmetry in addition to the R-symmetry and $\ggut$ under which the
pair of Higgs superfields would transform following $\Phi\bar{\Phi}
\rightarrow - \Phi\bar{\Phi}$. This forbids the second term in the
superpotential of standard $F$-term inflation Eq.~(\ref{eq:superpotentialFterm}),
but allows one of the first non-renormalizable term
\begin{equation}\label{eq:superpotentialsmoothhybrid}
W^{\rm smooth}=\kappa S\left[-M^2+\frac{(\bar{\Phi}\Phi)^2}{M_{\rm P}^2}\right]~.
\end{equation}
the scalar potential reads
\begin{equation}
V^{\rm smooth}(S,\Phi)=\kappa^2\left|-M^2+\frac{(\bar{\Phi}\Phi)^2}{M_{\rm P}^2}
\right|^2
+\kappa^2|S|^2\frac{|\Phi|^2|\bar{\Phi}|^2}{M_{\rm P}^4}\left(|\Phi|^2
+|\bar{\Phi}|^2 \right)~,
\end{equation}
where we denote by the same letter the superfield and its scalar
component ($\theta=0$).

We remind the reader that $\bar{\Phi}$, $\Phi$ are 2 fields charged
under $\ggut$. If we follow the original motivation, $\ggut=\gps$ and
we want $\Phi$ to be non-trivially charged under the factors
$SU(4)_{\rm C}\times SU(2)_{\rm R}$ in order to generate the breaking
scheme $\gps\rightarrow \gsm$. The simplest possibility to realize this
is to assign $\Phi$ to the representation ${\bf (4,1,2)}$. It is then
necessary to assign $\bar{\Phi}$ to its complex conjugate
representation ${\bf (\bar{4},1,2)}$ so that the superpotential is invariant
under $G$, $S$ being necessarily an absolute gauge singlet belongs to a hidden sector.

We can define two real scalar fields, $s$ and $\phi$,
as being the relevant component of the representation of the $S$, $\Phi$,
$\bar{\Phi}$ fields such that the potential can be rewritten~\cite{Lazarides:2000ck}
\begin{equation}\label{eq:smoothpotential}
V^{\rm smooth}(s,\phi)=\kappa^2\left(M^2-\frac{\phi^4}{M_{\rm P}^2}\right)^2
+2\kappa^2s^2\frac{\phi^6}{M_{\rm P}^4}~.
\end{equation}
This modifies the picture drastically, since now the valley $\phi=0$
still represents a flat direction for $s$, but is also a local maximum
in the $\phi$ direction. As a consequence, inflation will be realized for non-vanishing values of
$\phi$, which induces the symmetry breaking \emph{during} inflation.
The minimum of the potential at fixed $s$ is indeed reached for
\begin{equation}
\phi^2 =\frac{4}{3}\frac{M^2\mu^2}{s^2}~,\quad\mathrm{for}\quad s\gg \mu M~,
\end{equation}
which correspond to the two symmetric minima of the potential.




Inside the inflationary trajectory described above, the effective one-field
potential is $V(s)=\mu^4(1-(2/27)\mu^2M^2/s^4)$ in the limit $s\gg \mu M$,
a form similar to mutated hybrid inflation.
The predictions of the model have been studied in \cite{Lazarides:1995vr},
assuming an embedding within SUSY GUTs, that is with a unification scale
of $2\times 10^{16}$ GeV and a gauge coupling constant of $\sim 0.7$.
The normalization to the COBE data imposes the mass scales of inflation
is found lower than in the $F$-term case $\mu\simeq 9\times 10^{14}$ GeV and
the cutoff $M$ scale is found close to the reduced Planck mass $M\sim M_{\rm P}$.
The spectral index is then given by~\cite{Lazarides:1995vr}
\begin{equation}
n_s \simeq 1=\frac{5}{3N_Q}\simeq 0.97\quad (\mathrm{for}~~N_Q=60)~,
\end{equation}
and a negligible running spectral index.}


\item{Shifted hybrid inflation:\\
The shifted inflation model is similar to the smooth model except that
the additional $Z_2$ symmetry is not imposed.
As a consequence, the superpotential reads~\cite{Jeannerot:2000sv}
\begin{equation}\label{eq:superpotentialshifted}
W^{\rm shifted}=\kappa S\left[\bar{\Phi}\Phi-M^2\right]
-\beta\frac{S(\bar{\Phi}\Phi)^2}{M_{\rm P}^2}~,
\end{equation}
where $\Phi,\bar{\Phi}$ are two distinct superfields that belong to
non-trivial representations of the Pati-Salam group $\gps=SU(4)_c
\times SU(2)_L\times SU(2)_R$: $\Phi\in \mathbf{(\bar{4},1,2)}$, $\bar{\Phi}\in
\mathbf{(4,1,2)}$. This gives rise to the following $F$-terms (in global
SUSY)
\begin{equation}
V^{\rm shifted}_{\rm F-terms}=\left|-\kappa M^2+\kappa\bar{\Phi}\Phi
-\beta\frac{(\bar{\Phi}\Phi)^2}{M_{\rm P}^2}\right|^2
+\kappa^2|S|^2\left( |\bar{\Phi}|^2+|\Phi|^2\right)\left|1-\frac{2\beta}{\kappa M_{\rm P}^2}
\bar{\Phi}\Phi \right|^2 ~,
\end{equation}
where we have denoted by the same letter the superfield and its $\theta=0$
component in the superspace.

Following \cite{Lazarides:2000ck}, we define the relevant fields $\bar{\phi}$ and
$\phi$ as the component of $\bar{\Phi}$ and $\Phi$ that generates the
symmetry breaking $\gps \rightarrow \gsm $, that is to give a VEV to a
component of $\Phi$ that is charged under $\gps$ but not under $\gsm$.
For the inflaton field, we will define $s\equiv |S|$. We can also choose to
set $\beta >0$ and $\bar{\phi}^*=\phi$, so that the potential is $D$-flat
and becomes
\begin{equation}\label{eq:shiftpotential}
\begin{split}
&V^{\rm shifted}(s,\phi)=\kappa^2\left[- M^2+|\phi|^2
-\frac{\beta}{\kappa M_{\rm P}^2}|\phi|^4\right]^2
+2\kappa^2 s^2 |\phi|^2\left[1-\frac{2\beta}{\kappa M_{\rm P}^2}
|\phi|^2\right]^2 ~,\\
&\frac{V^{\rm shifted}(w,y)}{\kappa^2 M^4}=(y^2-1-\xi y^4)^2+2w^2y^2(1-2\xi y^2)^2 ~.
\end{split}
\end{equation}
The second form is for normalized fields, $w=s/M$, $y=\phi/M$ and $\xi
\equiv \beta M^2/\kappa M_S$. The potential is represented by
Fig.~(\ref{fig:shiftedpoten}).


\begin{figure}[t]
\scalebox{0.5}{\includegraphics{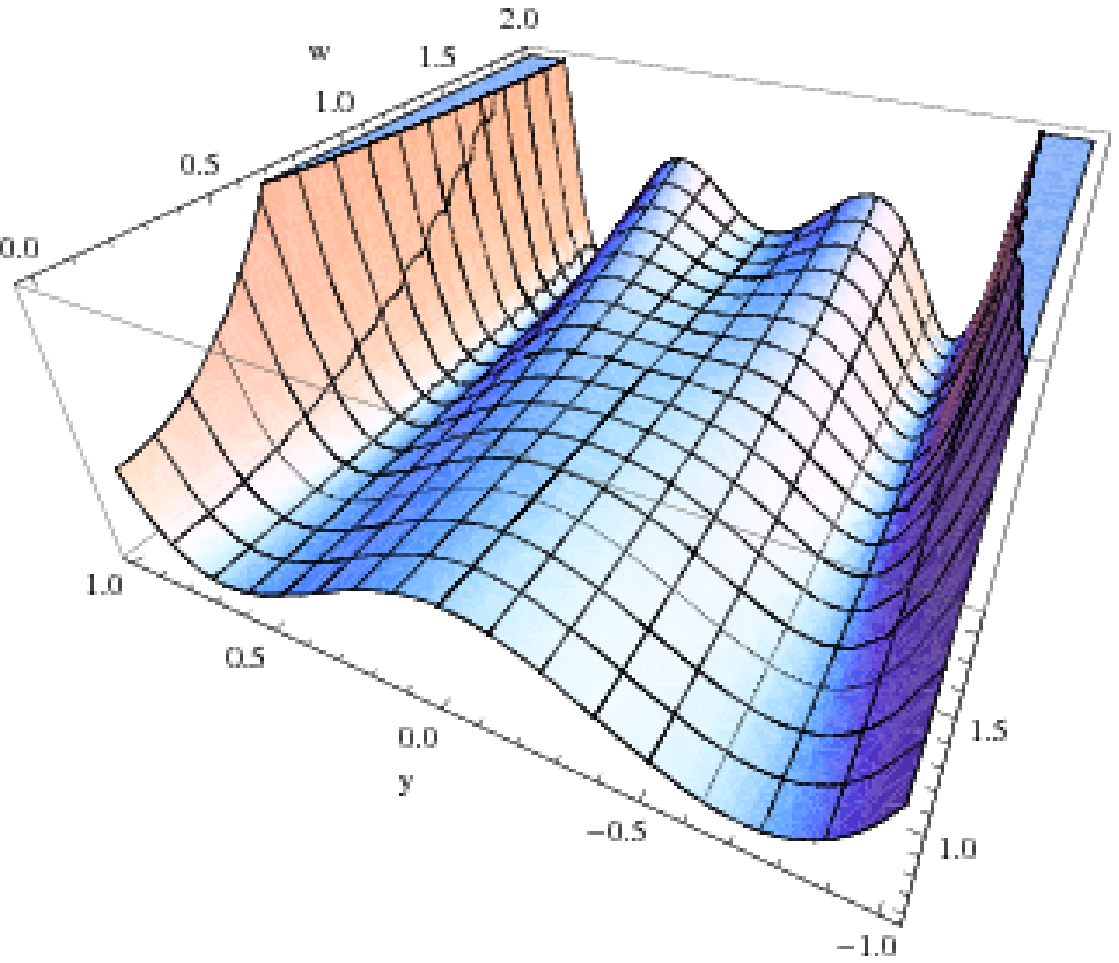}}\quad\quad\quad\quad\quad
\scalebox{0.5}{\includegraphics{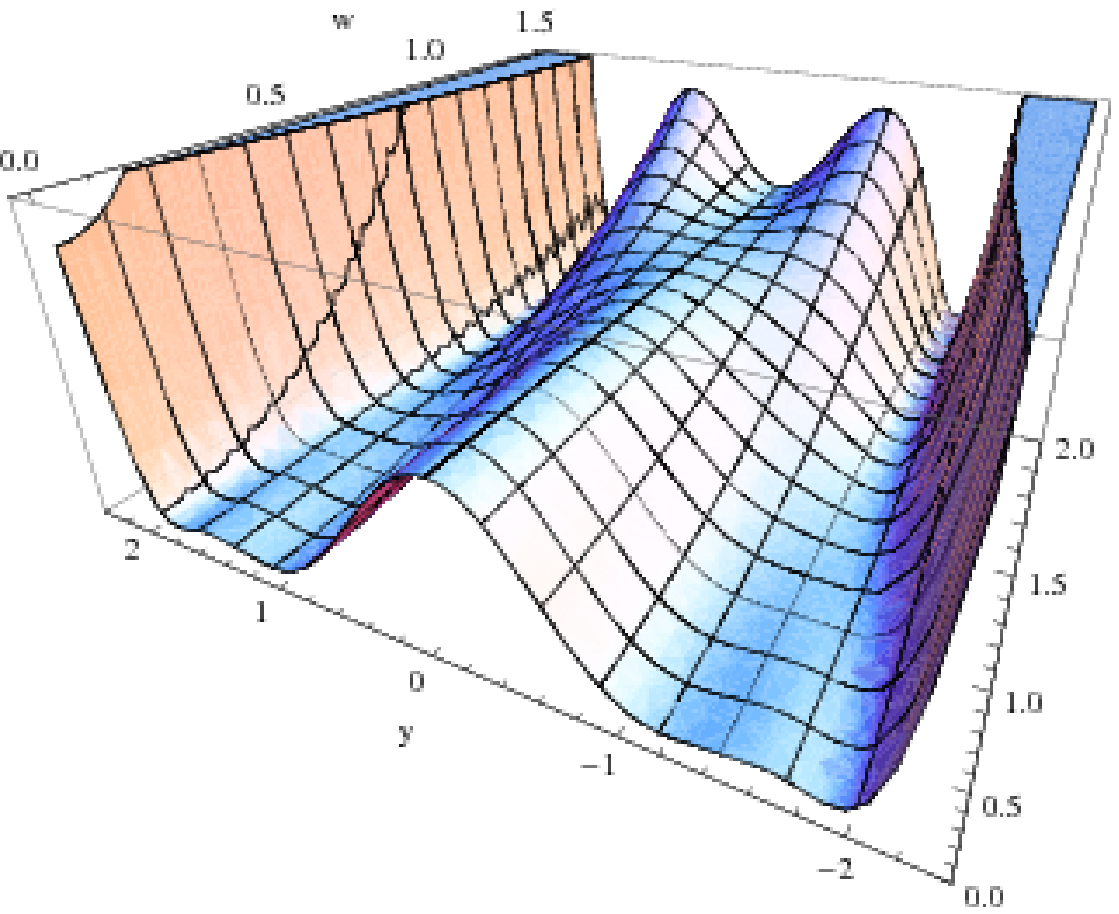}}
\caption{Shifted hybrid inflation potential of Eq.~(\ref{eq:shiftpotential})
for $\xi>1/4$ (left) and $1/6<\xi<1/4$ (right) in the reduced
variable space.\label{fig:shiftedpoten}}
\end{figure}


We can observe that the potential contains three (respectively two) local minima at
high (respectively small ) values of the inflaton field $w$, if $\xi\geq 1/4$ is of
the order of unity (Fig.~(\ref{fig:shiftedpoten}) left). They are located at $y=0$
for the central one while the other valleys are ``shifted'' away from $y=0$,
with a trajectory function of the inflaton field $y=f(w)$.
The number of valleys goes up to four at small inflaton VEVs and 
intermediate values of $1/7.2 <\xi<1/4$ (Fig.~(\ref{fig:shiftedpoten}) right): two
new shifted valley at $y=\pm 1/\sqrt{2\xi}$ appears. At smaller values of
$\xi$, both shifted valleys become indistinguishable.
If inflation is realized in one shifted valley, like for smooth
hybrid inflation, the symmetry $G$ is broken during inflation and no
topological defect can affect the post-inflationary cosmology. This is the
motivation of the model and also an implicit assumption. Note that the dynamics
could impose inflation to take place in the central valley, at vanishing $\psi$.
It would then be possible that the number of e-foldings produced after the symmetry
breaking is small enough that topological defects have some influence on the
CMB. This is still an open question.

If inflation is realized in the $y=\pm 1/\sqrt{2\xi}$ valley, it is possible to
compute the full mass spectrum of the model \cite{Jeannerot:2000sv}. The
classical contribution to the potential is $V_0=\kappa^2\tilde{M}^4$, where
$\tilde{M}^4=M^4(1/4\xi-1)^2$. The one-loop quantum corrections appear like in the 
$F$-term case from the splitting in mass, $2\kappa^2\tilde{M}$, between fermions and bosons of the superfield
$\Phi,\bar{\Phi}$. Consequently, the effective scalar potential in this valley
is identical to Eq.~(\ref{eq:effectivepotentialFtermvalley}) for the original
$F$-term model replacing $M$ by $\tilde{M}$. The CMB predictions are then derived;
the scale of inflation and the spectral index are found as a function of $\kappa$,
for a fixed value of $\beta$ and $M_S$ identified as the string scale. For
$\kappa\in[10^{-2},10^{-3}]$, they are predicted in the
ranges $n_s\in [0.89,0.99]$ and $V_{\rm infl}\in [2,7]\times 10^{14}$ GeV, which is
in agreement with the most recent CMB measurements ~\cite{Jeannerot:2000sv}. The level of predicted
tensor perturbations with these parameters, $r\lesssim 10^{-5}$, is however
beyond the reach of planned experiments.}


\item{Non-renormalizable terms and SUGRA effects:\\
The analysis of the previous models driven by $F$-terms are based on the presence
of non-renormalizable terms (smooth and shifted hybrid inflation), which  can be expected
to become unstable and suffer from the $\eta$-problem when formulated in SUGRA.
It actually turns out not to be a major problem,
the presence of extra non-renormalizable term changes the shape of the tree-level potential and 
avoid the $\eta$-problem as discussed in the previous section. Moreover, we are allowed to assume non-minimal
(non-renormalizable) K\"ahler potentials, which introduce more parameters and more freedom to flatten the
potential~ \cite{BasteroGil:2006cm,urRehman:2006hu}.

For example in the case of a smooth hybrid inflation (with the superpotential given
by Eq.~(\ref{eq:superpotentialsmoothhybrid})) the effective potential in minimal
SUGRA is of the form~\footnote{Note that this expression has a rather small range
of validity, since it is valid only if $s^2\gg \mu M$ and $s\ll M_{\rm P}$.}~\cite{Senoguz:2004vu}
\begin{equation}\label{eq:effective1fieldsmoothSUGRA}
V(s)\simeq \mu^4\left[1-\frac{2\mu^2M^2}{27 s^4}+\frac{s^4}{8 M_{\rm P}^4}\right]~.
\end{equation}
The last term in this expression comes from the SUGRA corrections. Assuming this
expression is valid for relevant scales, it is found that the spectral index is
raised from $n_s\simeq 0.97$ in global SUSY to larger values $n_s\in [0.98,1.03]$ 
depending on the scale $\tilde{M}=\sqrt{\mu M}$~\cite{Senoguz:2004vu}. 

Introducing the non-renormalizable corrections to the
K\"ahler potential, $K=K_{\rm min} + \lambda S^4/(4 M)+\dots$, lead to an
effective potential \cite{Linde:1997sj,urRehman:2006hu}
\begin{equation}
V(s)\simeq \mu^4\left[1-\frac{2\mu^2M^2}{27 s^4}-\lambda\frac{s^2}{2 M}
+\frac{\gamma_S}{8}\frac{s^4}{M_{\rm P}^4}\right]~.
\end{equation}
This allows the model to reduce the tension by increasing the cutoff scale $M$
and reduce of the predicted spectral index to values inside $n_s\in [0.95,
0.99]$.

We conclude this section by pointing out that non-renormalizable corrections
can also affect significantly other predictions such as the running spectral
index, even if inflation takes place well below the scale of new
physics~\cite{Yamaguchi:2004tn,Ballesteros:2005eg}. The reason is that if the
inflationary valley is flat at tree level, all the dynamics is determined by the lifting of a non-renormalizable term.
It was first found in~\cite{Yamaguchi:2004tn} that smooth
hybrid inflation could predict a non-negligible amount of running by balancing
the non-renormalizable contributions to $W$ and $K$ with the SUGRA effects.
This prediction however
requires the modification of the superpotential to include a phase of new
inflation, since the parameter range to generate a large running imposes a
short (less than 60 e-foldings) phase of smooth inflation. A model independent
study of SUSY hybrid type models with potentials flat at tree level, and lifted
by radiative corrections was considered in~\cite{Ballesteros:2005eg},
$$V(\phi)=V_0+\beta \ln\frac{m(\phi)}{\mu}+\phi^4\frac{\phi^{2N}}{M^{2N}}~,$$
where the last term contains an arbitrary non-renormalizable operator. 
It was found that typically running of the spectral tilt is negligible, $\alpha_s \ll 1$, in 
renormalizable models if a large number of e-foldings is realized.
However they also pointed out that if the non-renormalizable mass scale, $M$ is 
larger than the inflationary scale can generate a large
running $\mathcal{O}(-0.05)$.}


\item{ Extensions of hybrid $F$-term inflation:\\
Many models have been proposed, that generalize or extend the idea of
the hybrid inflaton driven by $F$-terms. In Ref.~\cite{BasteroGil:1997vn,BasteroGil:2002xs}, the 
realization of hybrid inflation has been illustrated in conjunction with solving the 
$\mu$-problem of the MSSM within the NMSSM. They proposed that the waterfall field
coupled to the inflaton also induce the mass term for the electroweak Higgs pairs of the MSSM:
\begin{equation}\label{eq:superpotentialnewshifted}
W^{\phi\mathrm{NMSSM}}=\lambda N H_1H_2-\kappa S N^2~.
\end{equation}
At the cost of adding two scalar singlets to the MSSM, this model is found to
solve the $\mu$-problem, prevents from domain walls to form during the
electro-weak symmetry breaking (see~\cite{Abel:1995wk}), and give rise to a phase of
inflation of the (non-SUSY) hybrid type with an effective potential of the form
$V_0+m^2\phi^2$.

Other examples of extensions of the hybrid model via $F$-terms are the smooth
and shifted scalar potentials obtained with only renormalizable
operators~\cite{Lazarides:2007fh,Jeannerot:2002wt}. They
were named ``new smooth'' and ``new shifted'' hybrid inflation, and
require the introduction of additional fields interacting with the inflaton and the waterfall field. 

For example, new shifted hybrid
inflation is still based on a singlet $S$ of the Pati-Salam gauge group $\gps$
coupled to $\Phi$, $\bar{\Phi}$ in ${\bf (\bar{4},1,2)}$ and ${\bf (4,1,2)}$. But the
model also assumes the introduction of new superfields $\Psi$, $\bar{\Psi}$
both in ${\bf (15,1,3)}$ to which the inflaton $S$ would be coupled.
The introduction of these extra-fields has motivations~\cite{Gomez:2002tj}
from the fermion spectrum, in particular the predicted mass of the bottom quark,
that becomes in better agreement with experimental measurements, compared
to the minimal Pati Salam model.

The inflaton sector now relies on the following superpotential~\cite{Jeannerot:2002wt}
\begin{equation}\label{eq:superpotentialnewshifted}
W^{\rm new~shifted}=\kappa S(\Phi\bar{\Phi}-M^2)-\beta S\Psi^2 + m\Psi \bar{\Psi}
+\lambda \bar{\Psi}\Phi\bar{\Phi}~,
\end{equation}
which leads to the following $F$-term contribution to the scalar potential
\begin{equation}
V=\left|\kappa (\Phi\bar{\Phi}-M^2)-\beta \Psi^2\right|^2
+\left|2\beta \Psi -m \bar{\Psi}\right|^2+ \left|m\Psi +\lambda \Phi\bar{\Phi}\right|^2
+\left|\kappa S+\lambda \bar{\Psi}\right|^2\left(\left|\Phi\right|^2
+\left|\bar{\Phi}\right|^2\right)~,
\end{equation}
The flat directions 
of the potential are at $\langle\Phi\rangle=\langle\bar{\Phi}\rangle=\langle\Psi
\rangle=\langle\bar{\Psi}\rangle=0$ for the trivial one, and at
\begin{equation}
\langle \Phi\rangle=\langle\bar{\Phi}\rangle=v~,\quad \langle S\rangle >0~, \quad
\langle \Psi\rangle=-\frac{M}{2}\sqrt{\frac{\kappa}{\beta\xi}}~, \quad
\langle \bar{\Psi}\rangle=-S\frac{\kappa}{\lambda}~,
\end{equation}
where $$v^2\equiv \frac{2\kappa^2(1/4\xi+1)+\lambda^2/\xi}{2(\kappa^2+\lambda^2)}
\quad \xi .$$ Inflation is assumed to take place in this shifted valley,
and thus in this model  breaking of the Pati-Salam group to the SM group (and the formation
of monopoles associated with it) is realized \emph{during} inflation. Indeed,
the VEV of $\Phi$ breaks completely the Pati-Salam gauge group only leaving the SM
group unbroken (the doublet of $SU(2)_R$ breaks it completely and the
$\mathbf{4}$ of $SU(4)_c$ necessarily breaks its $U(1)_{\rm B-L}$
subgroup~\cite{Slansky:1981yr}).

The presence of many free parameters in this model only allows to confirm that
with reasonable parameter values (coupling of order $10^{-2}$, and masses and
unification scales around $10^{15}-10^{16}$ GeV), the normalization to COBE,
around 60 e-foldings of inflation, and a spectral index around $n_s\simeq 0.98$ can be
obtained~\cite{Jeannerot:2002wt}. The  motivations and the mechanism behind
the ``new smooth hybrid inflation''~\cite{Lazarides:2007fh} are identical,
with a similar potential given by Eq.~(\ref{eq:superpotentialnewshifted}). The
predictions of the model are:
$n_s\simeq 0.969~,\quad r\simeq 9.4\times 10^{-7}~,\quad
\alpha_s\simeq -5.8\times 10^{-4}$~\cite{Lazarides:2007fh}.
%

A ``semi-shifted hybrid inflation'' was also proposed~\cite{Lazarides:2008nx}, with a similar framework
of the extended Pati-Salam group, as for new shifted or new smooth hybrid inflation
and the same (super)potential. In this model, however, the chosen inflationary
valley is different, since it takes advantage of a third flat direction
appearing only if $\tilde{M}^2\equiv M^2-m^2/2\kappa^2 >0$
\begin{equation}
\langle \Phi\rangle=\langle\bar{\Phi}\rangle=0~, \quad \langle \Psi\rangle=\pm \tilde{M}~,
\quad \langle \bar{\Psi}\rangle=-\frac{2\kappa \langle \Psi\rangle}{m} S~,\quad
\langle S\rangle >0~.
\end{equation}
This inflation is then called semi-shifted, since only the VEVs of
$\Psi,\bar{\Psi}$ are shifted away from 0. As a consequence the breaking during
inflation is $ \gps \rightarrow SU(3)_c\times SU(2)_L\times U(1)_R \times
U(1)_{\rm B-L}$ leaving the second symmetry breaking $SU(3)_c\times SU(2)_L\times
U(1)_R \times U(1)_{\rm B-L} \rightarrow \gsm$ for the end of inflation. As a
consequence, the model predicts the formation of B-L cosmic strings, which can
have important phenomenological consequences \cite{Jeannerot:2005ah}, including
on the CMB predictions of the model. When taking into account of the SUGRA
corrections, the spectral index \cite{Lazarides:2008nx} is found in the range
$n_s\in [0.98,1.05]$ for $m\in [0.5,2.5]\times 10^{15}$ GeV, in agreement with
WMAP data in presence of cosmic strings. The running and the ratio of
tensor to scalar is again found well below expected detection limits.}

\end{itemize}


\subsection{Inflation from $D$-terms in SUSY and SUGRA}\label{sec:models:Dterminflation}

It was first noticed in Ref.~\cite{Stewart:1994ts} that inflation with
a perfectly flat inflaton potentials in SUSY/SUGRA can be constructed using
a constant contribution to the $D$-term, and a rather complicated
superpotential used to drive the field dependent contributions to the
$D$-terms to 0. In~\cite{Binetruy:1996xj,Halyo:1996pp}, a very simple
superpotential was proposed to achieve the similar result, where it was noticed
that the radiative correction would lift this flat direction and
drive inflation.

In addition, it was noticed that the $\eta$
problem arising in $F$-term models does not appear for $D$-terms driven
inflation even for the non-minimal K\"ahler potential because
the $D$-sector of the potential does not receive exponential contributions
from non-minimal SUGRA. The model 
requires the presence of a Fayet-Iliopoulos (FI) term $\xi$, and 
therefore a $U(1)_\xi$ symmetry that allows or generates it.

This model rapidly became one of the most studied models of inflation
because of its stability in SUGRA, and its stability when embedded in
other high energy frameworks such as SUSY GUTs (see for
example~\cite{Jeannerot:1997is,Dvali:1997mh}), and SUGRA from
superconformal field theory~\cite{Binetruy:2004hh}. The model was also
found to be a good low-energy description of brane
inflation~\cite{Dvali:1998pa,Halyo:2003wd}) proposed in the context of
extra-dimensional cosmology.

Finally, the presence of anomalous $U(1)$
symmetries in weakly coupled string
theories~\cite{AlvarezGaume:1983ig,Dine:1987xk,Dine:1987gj,Atick:1987gy}
generated a lot of hope, though there are problems in fully embedding the model and the
generation of the FI term~\cite{Espinosa:1998ks,Kolda:1998yt}.
Also, the model tends to require large inflaton values (Planckian, even
super-Planckian in part of the parameter space) unlike $F$-term
inflation, which represents another challenge for the model, and thus necessitates 
$D$-term inflationary models to be studied in the context of SUGRA with a VEV below $M_{\rm P}$.


\subsubsection{Minimal hybrid inflation from $D$-terms}

Let us consider both $F$-and $D$-terms contribute to the scalar
potential. Given a K\"ahler potential $K(\Phi_m,\Phi_n)$, the
$D$-terms $$D^{a}=-g_{a} [D_a=\phi_i {(T_a)^i}_j K^j+\xi_a]$$ (where
$K^m\equiv \partial K/\partial \Phi_m$) give rise to a scalar potential within $N=1$ SUGRA:
\begin{equation}
V^{\rm SUGRA}(\phi,\phi^{\ast})=\frac{1}{2}[\mathrm{Re} f(\phi)]^{-1}
\sum D^{a}D_{a}+\mathrm{F-terms}\,
\end{equation}
where $g_{a}$ and $T^{a}$ are respectively the gauge coupling constants and
the generators of each factors of the symmetry of the action, $'a'$ running
over all factors of the symmetry, and $f(\phi)$ is the gauge kinetic
function. If this symmetry contains a factor $U(1)_\xi$, not originating
from a larger non-abelian group, the most general action allows for the
presence of an additional constant contribution $\xi$. Below we will assume
that such an abelian symmetry is the only symmetry of the inflaton sector.

The simplest realization of $D$-term inflation reproduces the hybrid
potential with three chiral superfields, $S$, $\phi_{+}$, and $\phi_{-}$
with non-anomalous $U(1)_\xi$ (an abelian theory is said to be anomalous if
the trace of the generator is non-vanishing $\sum q_n\neq 0$) charges
$q_n=0,+1,-1$~\cite{Binetruy:1996xj,Halyo:1996pp}. The
superpotential can be written as
\begin{equation}\label{eq:dtermsuperpotential}
W^D = \lambda S\phi_{+}\phi_{-}\,.
\end{equation}
In what follows, we assume the
minimal structure for $f(\Phi_i)$ (i.e., $f(\Phi_i)$=1) and take the
minimal K\"ahler potential~\footnote{This is the simplest SUGRA
model and in general the K\"ahler potential can be a more
complicated function of the superfields. }. Then the scalar potential reads
\begin{equation}\label{DpotenSUGRAtot}
\begin{split}
V^{\rm D-SUGRA}_{\rm tree}=
\lambda^2&\exp\left({\frac{|\phi_-|^2+|\phi_+|^2+|S|^2}{M^2_{\rm
P}}}\right)
\left[|\phi_+\phi_-|^2\left(1+\frac{|S|^4}{M^4_{\rm
P}}\right)+|\phi_+S|^2 \left(1+\frac{|\phi_-|^4}{M^4_{\rm
P}}\right) \right.\\
&\left. +|\phi_-S|^2 \left(1+\frac{|\phi_+|^4}{M^4_{\rm P}}\right)
+3\frac{|\phi_-\phi_+S|^2}{M^2_{\rm P}}\right] +
\frac{g_\xi^2}{2}\left(|\phi_+|^2-|\phi_-|^2+\xi\right)^2~,
\end{split}
\end{equation}
where $g_\xi$ is the gauge coupling of $U(1)_\xi$.
Here, we have assumed a minimal K\"ahler potential $K=|\phi_-|^2
+|\phi_+|^2+|S|^2$. The global minimum of the potential is obtained
for $\langle S\rangle =0$ and $\langle\Phi_-\rangle =\sqrt {\xi}$,
which is SUSY preserving but induces the breaking of $U(1)_\xi$.
For $S>S_{inst}\equiv g_\xi\sqrt{\xi}/\lambda$ the potential is
minimized for $|\phi_+|=|\phi_-|=0$ and therefore, at the tree level,
the potential exhibits a flat inflationary valley, with vacuum energy
$V_0=g_\xi^2\xi^2/2$. The radiative corrections depend on the splitting
between the effective masses of the components of the superfields
$\Phi_+$ and $\Phi_-$, because of the transient $D$-term SUSY breaking.
Extracting the quadratic terms from the potential
Eq.~(\ref{DpotenSUGRAtot}), the scalar components $\phi_+$ and
$\phi_-$ get squared masses:
\begin{equation}\label{eq:Dtermmassbosons}
m^2_{\pm}=\lambda^2|S|^2 \exp\left({\frac{|S|^2}{M^2_{\rm
Pl}}}\right)\pm g_\xi^2 \xi~,
\end{equation}
%
while the squared mass of the Dirac fermions reads
\begin{equation}\label{eq:Dtermmassfermions}
m_{\rm f}^2=\lambda^2|S|^2 \exp\left(\frac{|S|^2}{M^2_{\rm Pl}}\right)~.
\end{equation}
The radiative corrections are given by the Coleman-Weinberg
expression~\cite{Coleman:1973jx} and the full potential inside the
inflationary valley reads
\begin{equation}\label{eq:VexactDsugrainsidevalley}
\begin{split}
V^{\rm D-SUGRA}_{\rm eff}=
\frac{g_\xi^2\xi^2}{2} \Bigg\{1+&\frac{g_\xi^2}{16\pi^2}
\Bigg[2\ln \frac{\lambda^2|S|^2}{\Lambda^2} \exp
\left(\frac{|S|^2}{M_{\rm P}^2}\right)+\\
&(z+1)^2\ln(1+z^{-1})+(z-1)^2\ln(1-z^{-1})\Bigg]\Bigg\}~,
\end{split}
\end{equation}
with
\begin{equation}\label{eq:DtermxdeSsugra}
z=\frac{\lambda^2 |S|^2 }{g_\xi^2\xi} \exp\left({\frac{|S|^2}{M^2_{\rm
P}}}\right)~.
\end{equation}
Inflation ends when the slow-roll conditions break down, that is for
$z_{\rm end}\simeq 1$, and the predictions for the inflationary
parameters are very similar to the previous discussion on $F$-term
inflation.
$D$-term inflation based on an anomalous $U(1)_{\rm a}$ symmetry (which
could appear in string theory \cite{Dine:1987xk,Dine:1987gj,Atick:1987gy}) is no different.
More than one anomalous $U(1)_{\rm a}$'s can also give rise to a multiple phase of 
hybrid inflation, see~\cite{Burgess:2005sb}.


\subsubsection{Constraints from CMB and cosmic strings}

The CMB phenomenology is very similar to the $F$-term inflation model
described in the earlier sections. The fitting to the CMB observables can be
done by simply setting the values for energy scale $\sqrt{\xi}$,  superpotential
coupling $\lambda$, and gauge coupling $g_\xi$ of $U(1)_\xi$.


\begin{figure}[h!]
\scalebox{0.6}{\includegraphics{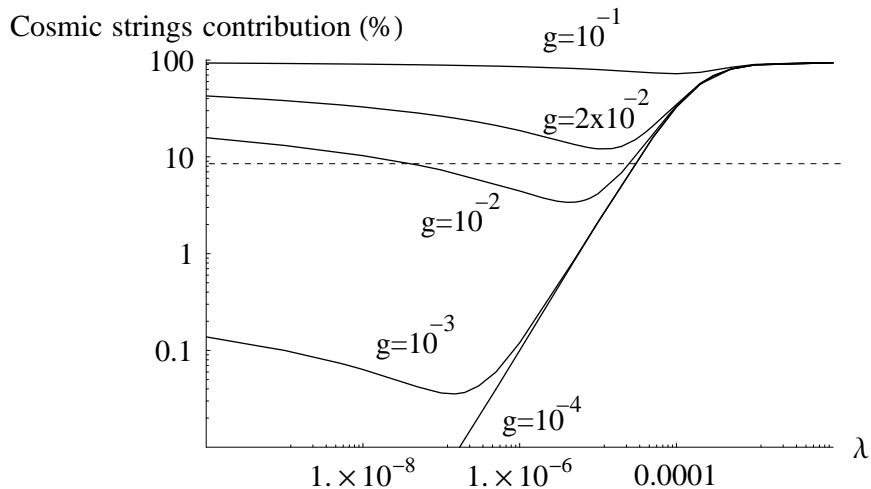}}
\caption{Cosmic strings contribution to the CMB quadrupole
anisotropies as a function of the superpotential coupling $\lambda$
and gauge coupling $g\equiv g_\xi$. Figure is taken from \cite{Rocher:2006nh}.\label{fig:CSvsinflationDterm}}
\end{figure}

By construction, the model leads to the formation of cosmic strings,
since the inflationary phase ends by the breaking of the abelian
symmetry $U(1)_\xi$ \cite{Binetruy:1996xj,Jeannerot:1997is}. Their
formation affects the
normalization of the fluctuation power spectrum by imposing an
additional contribution. This can be described by an additional
contribution to the temperature quadrupole
anisotropy
\begin{equation}
\left.\frac{\delta T}{T}\right|_Q^2=\left.\frac{\delta T}{T}
\right|_{\rm infl}^2+\left.\frac{\delta T}{T}\right|_{\rm CS}^2~,
\end{equation}
where $(\delta T/T)_{\rm CS}= y 2\pi \xi/M_{\rm P}^2$, because the
$D$-term strings are BPS. The contributions from inflation and 
cosmic strings are therefore proportional to the same energy scale
$\sqrt\xi$. When using $y=9$, it was
found~\cite{Kallosh:2003ux,Rocher:2004my,Rocher:2004et,Jeannerot:2006jj}
that the contribution of strings to the anisotropies varies with
the coupling constants; $\lambda$ and $g_\xi$, as represented in
Fig. \ref{fig:CSvsinflationDterm} .

Thus the formation of cosmic strings only imposes a suppressed
superpotential coupling for a successful $D$-term inflation;
\begin{equation}\label{eq:constraintCSDterm}
\lambda \lesssim 10^{-5}~.
\end{equation}
This conclusion was found valid also with next-to-minimal K\"ahler
potentials~\cite{Jeannerot:2006jj,Rocher:2006nh,Battye:2006pk}.
Three mechanisms have been proposed to lift this (slight)
fine-tuning; adding some symmetries to the Higgs $\Phi_\pm$ or
adding more fields to make the strings
unstable~\cite{Binetruy:2004hh,Urrestilla:2004eh}, modifying the
superpotential to produce them during inflation or finally
introducing couplings for the inflaton to flatten its
potential~\cite{Lin:2006xta}. They will be discussed in
Sec.~\ref{sec:dtermwithoutstrings}.

The predicted spectral index of the model can be computed in a similar
way to the $F$-term model. In the inflationary valley, the potential of
Eq.~(\ref{eq:VexactDsugrainsidevalley}) allows to compute $n_s$ at the
quadrupole scale for given values of $\xi$, $z_Q$, $\lambda$ and $g_\xi$.
The general logarithmic slope of the potential being concave down, the
second slow-roll parameter is negative and thus the spectrum turns out to be
red.

More precisely, the normalization of the spectrum and imposing 60
e-foldings of inflation between the field values responsible for the
quadrupole $z_Q$ and $z_{\rm end}$, leaves two parameters
unconstrained (say $\lambda$ and $g_\xi$) out of the four unknowns. 
The model possess two regimes; at large coupling
$\lambda$, $z_Q,z_{\rm end}\gg 1$, and the spectral index can be
approximated by~\cite{Jeannerot:2006jj}
\begin{equation}
n_s=1-\frac{2\lambda}{2 g_\xi N_Q+\lambda}~,
\end{equation}
which can be much smaller than unity.  In the small
coupling limit, $z_Q,z_{\rm end}\simeq 1$, and $n_s-1\simeq 0$, which is
slightly disfavored in the WMAP 5-years data. Note also that in this regime
the computation of loop corrections to the potential using the
Coleman-Weinberg~\cite{Coleman:1973jx}
formula breaks down, since the relevant quantities are computed very
close to the inflection point $z=1$, at which $V''$ and thus $\eta$,~
$n_S$ diverge. A more accurate description of this regime might require
the use of renormalization group improvements (see Sec.~\ref{SPEFT}).

Finally, an additional problem for $D$-term inflation is the
super Planckian VEVs for the relevant parameter
space $S_Q^2/M_{\rm P}^2\gtrsim \mathrm{PLog}[g_\xi^2 \xi/\lambda^2]$~\footnote{$\mathrm{Plog}[x]$ represents the inverse function of $x\rightarrow x e^x$.}. For a gauge coupling, $g_\xi
\gtrsim 10^{-3}$, and a small superpotential coupling in agreement
with a low weight of cosmic strings, the inflaton VEV already
shoots up above the Planck scale for $60$ e-foldings.



The minimal version of $D$-term inflation does not predict a
detectable amount of non-Gaussianity, as only one field
effectively rolls and fluctuates during inflation, at least. The
reason is that in the inflationary valley, the waterfall fields coupled to
the inflaton have mass of the order of the GUT scale, $m_B^2\simeq \lambda^2
S^2\pm g^2_\xi\xi$, much heavier than the Hubble scale during inflation. But in case 
of the presence of a light (super)field $S_2$ with mass $\mu_2$ in the theory, both $S$ and
$S_2$ contribute to the primordial fluctuations, and can create a
large amount of non-Gaussianities, well above the level of one field
inflation~\cite{Bernardeau:2007xi}. Furthermore, this will also
induce large isocurvature perturbations which would modify the
spectral tilt and the primordial spectrum~\footnote{Note that during the tachyonic stage of the waterfall field
the non-Gaussianity was found to be small~\cite{Barnaby:2006km}, but see contradictory results 
found in~\cite{Mulryne:2009ci}.}.



\subsubsection{$D$-term inflation from superconformal field theory}

The standard picture described in the previous section has been modified
in~\cite{Binetruy:2004hh} to take into account of the possibility that SUGRA is
constructed from a superconformal field theory (see~\cite{Kallosh:2000ve}
for a review). In this framework the theory is described by a conformal
K\"ahler $\mathcal{N}$ and a conformal superpotential $\mathcal{W}$. In this
formulation, which allows to embed any SUGRA of $N\leq 4$, the field
content are some chiral superfield $\Phi_i$ and a field $Y$, called the
``conformon'', whose modulus fixes the Planck scale:
\begin{equation}
|Y|^2=M_{\rm P}^2\exp [K(\Phi_i,\Phi_i^*)/3M_{\rm P}^2]~,
\end{equation}
where $K$ is the SUGRA K\"ahler potential, related to the
conformal K\"ahler by
$$-\frac{1}{3}\mathcal{N}(Y,\Phi_i) = |Y|^2\exp \left[-K(\Phi_i,\Phi_i^*)/3M_{\rm P}^2\right]~.$$ The theory is
fully described once a conformal superpotential $\mathcal{W}$ is chosen,
related to the SUGRA superpotential, $W$, by
\begin{equation}\label{eq:superconformalsuperpot}
\begin{split}
\mathcal{W}(Y,\Phi_i)\equiv Y^3 M_{\rm P}^3 W(\Phi_i)~,
\end{split}
\end{equation}
The phase of the conformon, $\Lambda_Y$, is free and should be fixed to break
the invariance of the theory under the K\"ahler transformations that leave
the Lagrangian invariant. $\Lambda_Y$ can be fixed by imposing that the
superconformal
superpotential is real, $\mathcal{W}=\mathcal{W}^*$. This leads to the regular
description of $N=1$~SUGRA, where the Lagrangian is fully described by the function
$G\equiv K/M_{\rm P}^2 + \ln(|W|^2/M_{\rm P}^6)$. But during $D$-term
inflation, this choice is meaningless as the superpotential is
vanishing~\cite{Binetruy:2004hh}. Alternatively, another choice to fix the
gauge is to assume that $Y$ is real~\cite{Kallosh:2000ve}, $Y=Y^*$
(see~\cite{Kallosh:2000ve} for the full description of the Lagrangian with
this choice).

The transformation of $Y$ under $U(1)_\xi$ can be written as an
imaginary constant,
\begin{equation}
\delta_\alpha Y = i \frac{g_\xi  \xi}{3M_{\rm P}^2}~,
\end{equation}
where $\xi$ is the FI term. From Eq.~(\ref{eq:superconformalsuperpot}), it
is clear that imposing the invariance of the conformal superpotential,
$\delta_\alpha \mathcal{W}=0 = 3 \delta_\alpha Y +\delta_\alpha W$, implies that
the existence of the FI term requires the superpotential $W$ not to be
invariant, $\delta_\alpha W = i g_\xi  \xi/M_{\rm P}^2$.

For $W=S\Phi_+\Phi_-$, $\delta_\alpha W = i g_\xi\sum q_i$, and the presence of a FI
term imposes the anomaly (non-vanishing sum of charges) of the $U(1)_\xi$
symmetry, which is given by $\xi/M_{\rm P}^2$. This can be accommodated by modifying 
the charges from $0,+1,-1$, for example to $0$, $(1-\xi/2M_{\rm P}^2)$ and $(-1-\xi/2M_{\rm P}^2)$ for $S$, 
$\Phi_+$ and $\Phi_-$, respectively. This  leads to a tree level potential of the form:
\begin{equation}
V^D_{\rm tree}(\phi_+,\phi_-)=\frac{g^2}{2}\left[\left(1-\frac{\xi}{2M_{\rm P}^2}\right)|\phi_+|^2-\left(1+
\frac{\xi}{2M_{\rm P}^2}\right)|\phi_-|^2+\xi\right]^2~,
\end{equation}
This affects only the $D$-terms of the potential, which modify the effective masses of the components
of $\Phi_\pm$ involving the expression for the one-loop contribution to
$V(S)$, but note that the amplitude, $\xi/M_{\rm P}^2$,  is very
small~\cite{Binetruy:2004hh,Rocher:2006nh}, since $\sqrt\xi$ is found at least
three orders of magnitude below the Planck mass, from the normalization of the
CMB anisotropies and the non-observation of cosmic strings. It
is worth noting  that this construction makes the $D$-term inflation 
more robust against non-renormalizable corrections to $W$, since the new charge
assignment prevents, by symmetry, all terms of the form $S(\Phi_+\Phi_-)^n/
M_{\rm P}^{2n-2}$~\cite{Binetruy:2004hh,Rocher:2006nh}.


\subsubsection{$D$-term inflation without cosmic strings}\label{sec:dtermwithoutstrings}

Several modifications of the original $D$-term model have also
been proposed to avoid the formation of cosmic strings and
therefore lift the constraint on the coupling constant, $\lambda$
given in Eq.~(\ref{eq:constraintCSDterm}). These works were partly
motivated by the fact that the model was thought incompatible with
the observations because the cosmic strings had been
found responsible for most of the
temperature anisotropies ($75\%$), in contradiction with
observations ($\lesssim 10\%$)~\cite{Jeannerot:1997is}. Other motivations include the
embedding of $D$-term inflation in strings theory, where FI term are
generically present, but at a much lower scale than that imposed by
the COBE normalization. They either assume that extra symmetry or
extra fields are present to render the defects unstable, or add
couplings in the superpotential to produce the strings before or
during inflation or add couplings to allow for a normalization to
COBE with lighter cosmic strings (then reducing their impact on
the CMB). 

It is first possible to assume that the Higgs fields are charged not only under
$U(1)_\xi$, but also under some additional non-abelian
local symmetry~\cite{Binetruy:2004hh}. For example, if $\Phi_-$
that takes a non-vanishing VEV, $\sqrt \xi$, is also a doublet under some other
symmetry $SU(2)_a$, the symmetry breaking at the end of inflation follows:
\begin{equation}
SU(2)_a\times U(1)_\xi \rightarrow U(1)'~,
\end{equation}
and no topologically stable cosmic strings form (though embedded strings would
form, like during the electro-weak symmetry breaking). Rendering the cosmic
strings unstable by enlarging the symmetries of the theory can also be achieved
by assuming
two additional fields~\cite{Urrestilla:2004eh,Binetruy:2004hh}. For example,
if a second pair of Higgs fields, $\tilde{\phi}_\pm$, carry some charges identical
to $\phi_\pm$ under $U(1)_\xi$. Thus the theory possesses an extra accidental
global symmetry $SU(2)$, since the potential (written here in global SUSY) is given by~\footnote{In~\cite{Binetruy:2004hh}, the same
idea is proposed, though they employ only 1 extra Higgs fields $\phi'$ with a
charge identical to $\phi_-$.}
\begin{equation}
\begin{split}
V(\phi_\pm,\tilde{\phi}_\pm)=\frac{g^2}{2}&\left(|\phi_+|^2+|\tilde{\phi}_+|^2-
|\phi_-|^2-|\tilde{\phi}_-|^2-\xi\right)^2\\
&+\lambda^2 S^2\left(|\phi_+|^2+|\tilde{\phi}_+|^2
+|\phi_-|^2+|\tilde{\phi}_-|^2\right)
+\left|\phi_+\phi_--\tilde{\phi}_+\tilde{\phi}_-\right|^2~,
\end{split}
\end{equation}
which is now invariant under the exchange $\phi_+\leftrightarrow\tilde{\phi}_+$ and
$\tilde{\phi}_-\leftrightarrow\phi_-$. When the Higgs fields settle in the global minimum,
a global U(1) symmetry is left unbroken and the symmetry breaking at the
end of inflation is of the form:
\begin{equation}
SU(2)_{\rm glob}\times U(1)_\xi \rightarrow U(1)_{\rm glob}~,
\end{equation}
producing again no topologically stable strings, but only embedded (or semi-local)
strings. Note that their stability in this case depends on various parameters;
the ratio of the scalar mass over the $U(1)_\xi$ gauge field mass, the existence
of zero modes and the details of the symmetry breaking~\cite{Urrestilla:2004eh}.
As a conclusion, it is not excluded that these objects have an interesting impact
on cosmology. Both of these ingredients can be found in string theory~\cite{Urrestilla:2004eh,Binetruy:2004hh}.

Other mechanisms have been proposed to eliminate the
impact of cosmic strings by breaking the $U(1)$ symmetry before inflation. For example,
authors in~\cite{Lyth:1997pf} modified the $D$-term superpotential
with the introduction of a pair of vector-like fields [under $U(1)_\xi$]
$\Psi,\bar{\Psi}$ and two singlets $X,\sigma$~\cite{Lyth:1997pf}
\begin{equation}
W=\lambda\Phi_+\Phi_- + X(\Psi\bar{\Psi}-M)+\sigma \Phi_+ \Psi
\end{equation}
Inflation then takes place with $\langle X\rangle=\langle\sigma\rangle=0$,
$\langle\Psi\rangle=\langle\bar{\Psi}\rangle=M$, therefore breaking $U(1)_\xi$
before or during inflation, and diluting the strings.

The Refs.~\cite{Lin:2006xta,Lin:2007va}, explored various ays to reduce the impact of
cosmic strings on the CMB, proposing the ``sneutrino modified
$D$-term inflation''. The superpotential is modified by introducing a
coupling of the inflaton superfield $S$ to the right-handed sneutrino $N_R$
field~\cite{Lin:2006xta,Lin:2007va}
\begin{equation}\label{eq:superpotsneutrinmodifDterm}
W=\lambda S \Phi_+\Phi_-+\lambda_\nu \Phi H_u L + \frac{M_R}{2}N_R^2~,
\end{equation}
where the second and third terms are responsible for generating neutrino masses via
see-saw mechanism. The model also requires a coupling through a non-renormalizable
K\"ahler potential:
\begin{equation}
K_{\rm min} \rightarrow K_{\rm min} +\frac{c\, S^\dagger S N_R^\dagger
N_R }{\Lambda^2}~,
\end{equation}
which affects the main conclusions of the model concerning the impact of cosmic
strings via the predictions for the spectral index. The potential in
Eq.~(\ref{eq:VexactDsugrainsidevalley}) is modified
by an additional term, $-\kappa s^2/2$, where $\kappa=(c-1)M_R^2|N_R|^2/\Lambda^2$, if
the coupling is assumed to be $\lambda\gtrsim 0.1$. For such large couplings,
inflation is realized with $z\gg 1$ and the effective potential reduces to
$V\simeq V_0+\frac{g^4\xi}{16\pi^2}\ln (s^2/\Lambda^2) -\kappa s^2/2$.
For $c>1$, this reduces the spectral index and allows for a better fit to the
observations. In addition, the normalization of the power spectrum of primordial
fluctuations is enhanced, leading to a lowering the
energy scale of inflation, therefore a reducing the energy per unit length for cosmic strings.

Similarly, the ``sneutrino $D$-term inflation'' proposed in~\cite{Kadota:2005mt}
assumed the sneutrino to be the inflaton. The superpotential  is given by:
\begin{equation}
W=\frac{\lambda}{M_{\rm P}} N_R^2\Phi_+\Phi_-+\lambda_\nu N_R H_u L + \frac{M_R}{2}N_R^2~,
\end{equation}
where $N_R$ is assumed to be the lightest right handed (s)neutrino, with $U(1)_R$
charge $+1$ and no $U(1)_\xi$ charge. Therefore, the tree level $D$-terms in the
potential of Eq.~(\ref{DpotenSUGRAtot}) are not affected, though the additional
couplings will affect the radiative correction, and therefore the dynamical
properties of the model. Even with a minimal
K\"ahler potential, inflation is found to be successful in the regime, $M_R^2N_R \ll
g_\xi^2 \xi^2$. The model predicts an almost scale-invariant
power spectrum, $n_s\simeq 1$, and the constraint from the cosmic string tension
is relaxed as compared to the standard case.


\subsubsection{$F_D$-term hybrid inflation}\label{sec:fdtermmodel}

There are also models where both $F$ and $D$-terms are contributing to the
inflationary potential. A mixture of the $F$- and $D$-term inflation was proposed
by~\cite{Garbrecht:2006ft,Garbrecht:2006az}, built as an extension to the
NMSSM. The model is constructed in such a way that the inflaton field $S$
involved in a $F$-term like superpotential also generates the $\mu$ term of the
MSSM, and it is also coupled to the right-handed neutrinos, generating the Majorana
mass scale. 

The symmetries of the model are also extended to $\gsm\times U(1)_\xi$, the
additional abelian factor allowing for the presence of a FI term. This
subdominant contribution to the $D$-terms is employed to control the decay
rate of superheavy fields such as the waterfall fields $X_i$ and the
inflaton field into gravitinos. In this model, the potential is dominated by the $F$-terms. 
The renormalizable superpotential for the $F_D$-terms hybrid  model is given by:
\begin{equation}\label{Wmodel}
W = \kappa\, {S}\, \left( X_1 X_2  - M^2 \right) + \lambda\, S H_u H_d
+ \frac{\rho_{ij}}{2}\, S N_i N_j + h^{\nu}_{ij} L_i H_u N_j+ W_{\rm MSSM}^{(\mu=0)}\; ,
\end{equation}
where $W_{\rm  MSSM}^{(\mu=0)}$ denotes the  MSSM superpotential without the
$\mu$-term, $S$ is  the  SM-singlet inflaton superfield, $N_i$ are the
right-handed Majorana neutrinos and ${X}_{1,2}$ is a chiral multiplet pair
with opposite charges under some $U(1)_\xi$ gauge group. Consequently, the
$D$-term contribution to the scalar potential is given by: $V_D =(g^2/8)(|X_1|^2-|X_2|^2-\xi)^2$.
The soft SUSY-breaking sector can be obtained from Eq.~(\ref{Wmodel}) and
reads:
\begin{equation}\label{Lsoft}
-{\cal L}_{\rm soft} = M^2_S S^*S + \Big(\kappa A_\kappa S X_1X_2 +
\lambda A_\lambda S H_u H_d + \frac{\rho}{2} A_\rho S \widetilde{N}_i\widetilde{N}_i
-\kappa a_S M^2 S  +{\rm  H.c.}\Big)~,
\end{equation}
where   $M_S$,   $A_{\kappa,\lambda,\rho}$    and   $a_S$   are   soft
SUSY-breaking mass parameters of order $M_{\rm SUSY} \sim 1$~TeV.

The second term in Eq.~(\ref{Wmodel}) induces the $\mu$-term when
when the scalar component of $S$ acquires a VEV,
$\mu\ =\  \lambda\, \langle S  \rangle\ \approx\ \frac{\lambda}{2\kappa}\,
|A_\kappa - a_S|$, where the VEVs of  $H_{u,d}$ are neglected
compared to the VEV of $X_{1,2}$. The  third  term in Eq.~(\ref{Wmodel}),
$\frac{1}{2}\,\rho_{ij}\,  {S}\, {N}_i {N}_j$,
gives  rise  to  an  effective lepton-number-violating  Majorana  mass
matrix, i.e.~$M_S  = \rho_{ij}\,  v_S$.  Assuming that  $\rho_{ij}$ is
approximately  SO(3) symmetric,  viz.~$\rho_{ij}  \approx \rho\;  {\bf
1}_3$,  one   obtains  3  nearly   degenerate  right-handed  neutrinos
$\nu_{1,2,3\,R}$, with mass $m_N\ =\ \rho\, v_S$.
If  $\lambda$  and  $\rho$  are  comparable  in  magnitude,  then  the
$\mu$-parameter and  the SO(3)-symmetric Majorana mass  $m_N$ are tied
together, i.e.~$m_N  \sim \mu$, thus  leading to a scenario  where the
singlet   neutrinos  $\nu_{1,2,3\,R}$  can   naturally  have   TeV  or
electroweak-scale masses.

The renormalizable  superpotential Eq.~(\ref{Wmodel}) of the  model may be
uniquely determined by imposing the continuous $R$ symmetry:
${S}\  \to\ e^{i\alpha}\,{S}\, ,{L}\  \to\  e^{i\alpha}\, {L}\,, {Q}\ \to\ e^{i\alpha}\, {Q}$
with $W \to e^{i\alpha} W$,  whereas all other fields remain invariant
under an $R$ transformation.  Notice that the $R$ symmetry
forbids  the  presence of  higher-dimensional  operators  of the  form
${X}_1 {X}_2  {N}_i {N}_j/M_{\rm P}$.

The crucial observation is that  the superpotential  Eq.~(\ref{Wmodel})  is
symmetric   under   the   permutation   of   the   waterfall   fields, ${X}_1
\leftrightarrow  {X}_2$.  This permutation symmetry persists, even after the
spontaneous SUSY breaking of $U(1)_\xi$, since in the ground
state, $\langle X_1  \rangle = \langle X_2 \rangle  = M$. Hence, there is
an exact discrete symmetry acting on the gauged waterfall sector, similar to
the $D$-parity. In
order to  break this unwanted  $D$-parity, a subdominant  FI $D$-term is
required. As a $D$-parity conservation, heavy particles with mass $g M$
are stable and can be considered cosmologically bad, if they are
overproduced after the end of inflation.

Furthermore, in order to avoid the
SUGRA-$\eta$ problem, the K\"ahler potential has to be chosen of the form
$K=S^2+\kappa_s |S|^4/4M_{\rm P}^2$, where the Hubble induced mass correction
to $S$ field turns out to read $\pm 3\kappa_s H^2
S^2$~\cite{Garbrecht:2006ft,Garbrecht:2006az}, and the tree level potential
is similar to $F$-term inflation models, reading in the limit $S\gg M$,
\begin{equation}
V_{\rm infl}\simeq \kappa^2M^4\left[1+\frac{1}{64\pi^2}(4\kappa^2 + 8\lambda^2
+ 6\rho^2)\ln(|S|^2/M^2)\right]+M_S^2S^2-(\kappa a^2M^2S +{\rm h.c.})
+\kappa^2M^4\frac{|S|^4}{2M_{\rm P}}~.
\end{equation}
where the last term corresponds to the SUGRA correction assuming a minimal
K\"ahler potential. The cosmological predictions of the model are typical of any $F$-term inflation
with so many free parameters at disposal, it is always possible to get the
desired spectral tilt and the power spectrum.


\subsubsection{Embedding $D$-term models in  string theory}

Attempts were made to embed $D$-term models within an explicit SUSY
$SU(6)$ model~\cite{Dvali:1997mh}. This will be discussed with other embedding of
inflation in SUSY GUT below at section~\ref{sec:models:SUSYGUTs}.
More recent modifications of the $D$-term models were motivated by the
possibility that the model is generically realized at large field VEV. This opens up the
importance of non-renormalizable corrections to the K\"ahler
potential~\cite{Copeland:1994vg,Seto:2005qg,Rocher:2006nh}, or corrections
which are motivated from string theory~\cite{Hsu:2003cy,Brax:2006yq,Davis:2008sa}.
It was found that the dynamics of the model remains mostly unchanged when
considering any non-renormalizable corrections of order $M_{\rm P}^{-2}$ in
K\"ahler potential~\cite{Seto:2005qg,Rocher:2006nh}. It was observed in
Ref.~\cite{Rocher:2006nh} that despite the new contributions to the potential,
the non-observation of cosmic strings still requires a suppressed superpotential
coupling constant $\lambda\lesssim 10^{-4}$.

The $D$-term inflation models arising from string theories have generated a
lot of activity in the past (within weakly coupled string
theories)~\cite{Espinosa:1998ks,Kolda:1998yt,Halyo:1999bq,Lyth:1998xn}, and more recently
within a brane description or with a more phenomenological approach studying the
moduli corrections to the model. The original interest for $D$-term inflation was in
part due to the observation that anomalous $U(1)$, generically appear in weakly
coupled string theories with $\sum q_i < 0$.
There are many problems to circumvent destabilizing the model by
vacuum shifting due to $\sum q_i < 0$, or to generate the right amplitude for
the FI term and the string coupling, see for discussion in~\cite{Espinosa:1998ks,Lyth:1998xn}.

The $D$-term inflation has also been found to be the low energy description of brane
inflation~\cite{Dvali:1998pa,Halyo:2003wd,Binetruy:2004hh}. The existence of branes
in string theory allows to construct a new class of inflationary models
(see Sec.~\ref{sec:stringstheorymodels}), where the inflaton becomes a modulus describing
the distance between two branes. The most studied examples are the $D3/D7$ and
the $D3-\overline{D3}~$\cite{Dvali:1998pa,Burgess:2001fx,Dvali:2001fw,Jones:2002cv}
brane systems. It has been shown that such a system
give rise to a $D$-term model of inflation, the inter-brane distance possessing
a flat direction at tree level, and the open string degrees of freedom between the
two branes playing the role of the waterfall fields $\Phi_\pm$, as one of them
becomes tachyonic below a certain inter-brane separation, i.e. near the string length scale. The 
formation of cosmic (super)strings occurs at the end, in the form of
$D1$-branes or fundamental ($F$-)strings~\cite{Dvali:2003zh,Copeland:2003bj,BlancoPillado:2005xx}.

A more phenomenological approach is to study the modifications of $D$-term models
when embedded in string inspired SUGRA~\cite{Hsu:2003cy,Brax:2006yq,Davis:2008sa}.
These corrections arise from the coupling (at least gravitationally) to moduli
fields, originating from the string theory compactification.
A volume modulus can be stabilized using the following form of the
potentials~\cite{Burgess:2003ic,Achucarro:2006zf,Brax:2006yq}:
\begin{equation}\label{eq:modulistabilDterm}
W_{\rm mod}(T,\chi)=W_0+\frac{A e^{-aT}}{\chi^b}~,\quad K_{\rm mod}=-3\ln(T+\bar{T}
-|\chi|^2+\delta_{\rm GS}V_2)~,
\end{equation}
where $T$ is a volume modulus, the modulus, $\chi$, is a matter field introduced to allow for an
additional contribution to the superpotential which is charged under a new $U(1)$
abelian symmetry, with $V_2$ its gauge superfield, and
$\delta'_{\rm GS}$ is the Green-Schwarz parameter. The moduli potential derived contains
the $D$-terms of the form:
\begin{equation}
V^{\rm D}_{\rm mod}(T,\chi)=\frac{\{3\delta'_{\rm GS}[1+(a/b)|\chi|^2]\}^2}{8\mathrm{Re}(f(T))
\exp^2(-K/3)}~,
\end{equation}
where $f_{\rm mod}(T)$ is the gauge kinetic function for that sector. Interestingly,
the above stabilization potential has a non-vanishing vacuum which can generate an
effective FI term for inflation, if $T$ and $\chi$ are also charged under the
$U(1)_\xi$, giving $$\xi=\frac{3\delta^\xi_{\rm GS}(1+a/b|\chi|^2)}{4\mathrm{Re}(T)
-2|\chi|^2}~,$$ with $\delta^\xi_{\rm GS}$ the Green-Schwarz parameter of $U(1)_\xi$.

The form of the K\"ahler potential can be invariant under the shift symmetry
of the inflaton field
\begin{equation}
K_{\rm infl}=|S-\bar{S}|^2/2+|\Phi_+|^2+|\Phi_-|^2,
\end{equation}
as the minimal potential would induce a SUGRA-$\eta$ problem, due to non-vanishing
$F$-terms of the moduli sector during inflation. However, if we assume that the total
K\"ahler is simply the sum, $K=K_{\rm mod}+K_{\rm infl}$,  the contribution from that
sector to the (effective mass)$^2$ of the waterfall field would spoil the graceful
waterfall exit from inflation. A total K\"ahler potential of the form:
\begin{equation}
K=-3\ln (T+\bar{T}-|\chi|^2-K_{\rm infl}/3+\delta'_{GS}V_2+\delta^\xi_{\rm GS}V_1)~,
\end{equation}
would preserve the general behavior of the $D$-term inflationary model, as long as
the amplitude for the moduli are small. The current observations constrain
the parameter space of this model, leaving 3 classes of models,
all predicting a spectral index below or close to unity. One last modification due to
the moduli sector is the nature of the cosmic strings formed at the end of
inflation, which are not being BPS anymore and potentially containing massive fermionic
currents.


\subsubsection{Hybrid inflation in $N=2$ SUSY: $P$-term inflation}

There are attempts to embed hybrid inflation in $N=2$
SUSY~\cite{Watari:2000jh,Kallosh:2003ux,Achucarro:2005vz}.
In Ref.~\cite{Kallosh:2003ux}, the authors have unified $F$ and
$D$-terms within $P$-term inflation in the context of a global $SU(2,2|2)$
superconformal gauge theory, which also corresponds to a dual gauge theory
of $D3/D7$ branes~\cite{Dasgupta:2004dw}. The idea is to break the
$SU(2,2|2)$ symmetry down to $N=1$ SUSY by adding the $N=2$ FI
terms. The bosonic part of the superconformal action is given
by~\cite{Kallosh:2003ux}
\begin{eqnarray}
\label{superconformal}
{\cal L}=  D_\mu \Phi_3 D^\mu \Phi_3^*   -{1\over 4}F_{\mu\nu}^2 + {1\over 2} \vec P^2
+  D_\mu \Phi^A D^\mu \Phi_A  +  F^A F_A
+  g   \Phi^A \vec \sigma _A{}^B \vec P \Phi_B - 2g^2 \Phi^A  \Phi_A \Phi_3 \Phi_3\ . \nonumber \\
\end{eqnarray}
where two complex scalar fields forming a doublet under $SU(2)$, $\Phi^A$ and $\Phi_A= (\Phi^A)^*$.
$N=2$ gauge multiplet consists of a complex scalar $\Phi_3$, a vector $A_\mu$, (all singlets in $SU(2)$),  a spin-1/2 doublet $ \lambda^{ A}= \varepsilon^{AB} \gamma_5 C \bar \lambda_B^T$ (gaugino) and an  auxiliary field
$P^{r}$, triplet in $SU(2)$. There is also a doublet of dimension 2 auxiliary fields, $F^A$ with $F_A= (F^A)^*$
The covariant derivatives on the hyperderivatives are given by~\cite{Kallosh:2003ux}:
\begin{eqnarray}
D_\mu \Phi_A = \partial_\mu \Phi_A  +i g A_\mu \Phi_A \ ,~~~~~~~
D_\mu \Phi^A= \partial_\mu \Phi^A -i g A_\mu \Phi^A \ .
\label{covdev}
\end{eqnarray}
After adding the $N=2$ FI terms, the potential is given by:
\begin{eqnarray}
V_{N=2}^P= 2 g^2\left [\Phi^\dagger \Phi |\Phi_3|^2 +{1\over 4} \left(\Phi^\dagger \vec \sigma \Phi -\vec\xi  \right)^2\right]\ .
\end{eqnarray}
where the terms in the potential arises after solving the equations of motion for the auxiliary fields,
$P=- g \Phi^A  (\vec \sigma)_A{}^B \Phi_B $  and $F^A=0$. Note that the FI term proportional to $\xi$
has already been added to the potential now.

Let us isolate the gauge singlet, $S=\Phi_3$, and rest of the charged fields, $\Phi_1=\Phi_+$ ($\Phi_2^*=\Phi_-$) for the positively (negatively) charged scalars, the total potential can be written as~\cite{Kallosh:2003ux}
\begin{eqnarray}
V_{N=2}^P= 2 g^2\left (|S \Phi_+|^2+ |S \Phi_-|^2  +\Bigl|\Phi_+ \Phi_- - {\xi_+ \over 2}\Bigr|^2 \right)+ {g^2\over 2}\Bigl(|\Phi_+|^2 -
|\Phi_-|^2- \xi_3\Bigr)^2\ .
\end{eqnarray}
It was observed that the  $P$-term potential now corresponds to an $N=1$ model,
$V= |\partial W|^2 + {g^2\over 2}D^2$, with an appropriate superpotential and a $D$-term given by:
$W= \sqrt {2} g S(\Phi_+ \Phi_- - \xi_+/2) \ , \qquad D= |\Phi_+|^2 -|\Phi_-|^2- \xi_3 $, where $\xi \equiv \sqrt{|\vec \xi|^2}= \sqrt {\xi_+ \xi_- + (\xi_3)^2}\ , \qquad  \xi_{\pm} \equiv \xi_1\pm i\xi_2$~\footnote{In $D3/D7$ brane construction the three FI terms   $\vec \xi$  are provided by  a magnetic flux  triplet  $ \vec \sigma \, (1+\Gamma_5)\, F_{ab}\Gamma ^{ab} $, where $F_{ab}$ is the field strength of the vector field living on D7 brane in the Euclidean part of the internal space with $a=6,7,8,9$. The spectrum of $D3-D7$ strings depends only on $|\vec \xi|$~\cite{Kallosh:2003ux,Dasgupta:2002ew,Herdeiro:2001zb,Dasgupta:2004dw}.
}.

It was noticed in~\cite{Kallosh:2003ux} that the potential  for $N=2$ SUSY gauge theory at $\xi_+=\xi_-=0$, $\xi_3=|\vec \xi|$ are similar to the case  of $D$-term inflation studied before in \cite{Binetruy:1996xj} with
$W= \lambda S\Phi_+ \Phi_- $ and $ D= |\Phi_+|^2 -|\Phi_-|^2- \xi$, for which the potential is given by:
\begin{equation}
V_{N=2}^D= 2 g^2\Bigl (|S \Phi_+ |^2+ |S \Phi_- |^2  +|\Phi_+ \Phi_-|^2 \Bigr)+ {g^2\over 2}\Bigl(|\Phi_+|^2 -|\Phi_-|^2- \xi\Bigr)^2\ .
\end{equation}
with an assumption, $\lambda = \sqrt{2}g$. If instead, $\xi_+=\xi_-=2M^2= \xi$, one recovers a potential of an $F$-term
inflation model with, $W= \lambda S(\Phi_+'\Phi_-'-M^2) $, and, $  D= |\Phi_+'|^2 -|\Phi_-'|^2$, with a potential:
\begin{equation}
V_{N=2}^F= 2 g^2\Bigl(|S \Phi_+' |^2+ |S \Phi_-' |^2  +|\Phi_+' \Phi_-' - M^2|^2 \Bigr)+ {g^2\over 2}\Bigl(|\Phi_+'|^2 -|\Phi_-'|^2\Bigr)^2\ .
\end{equation}
The first term of the potential now coincides  with the $F$-term inflation potential in $N=1$ theory, proposed by \cite{Dvali:1994ms}, under the assumption that  the gauge coupling $g=\lambda/\sqrt{2}$.

The $P$-term model could be coupled to $N=1$ SUGRA and the inflationary potential is given by~\cite{Kallosh:2003ux,Binetruy:2004hh}
\begin{eqnarray}
V &=&  2g^2 e^{|S|^2\over M_p^2}  \left[
|\Phi_+ \Phi_- - \xi_+/2|^2 \left (1 - {S\bar S \over M_p^2} +\Bigr({S\bar S \over M_p^2}\Bigl)^2\right) +|S \Phi_+|^2+ |S \Phi_-|^2 \right] \nonumber\\
&+& {g^2\over 2}\Bigl(|\Phi_+|^2 -|\Phi_-|^2- \xi_3 \Bigr)^2  \ .
\end{eqnarray}
The inflating trajectory takes place at $\Phi_+=\Phi_-=0$. After
adding the $1$-loop correction from gauge fields, one
obtains~\cite{Kallosh:2003ux}
\begin{equation}
V =  {g^2 \xi^2\over 2}\left(1 + {g^2\over 8\pi^2} \ln {
|S^2|\over {|S_c^2|} } +f   {|S|^4 \over 2 M_p^4} +\dots  \right),
\end{equation}
where $f  = {\xi_1^2+ \xi_2^2\over \xi^2}\ , \qquad 0 \leq f  \leq
1$ and $\dots$ stands for terms ${|S|^6 / 2 M_p^6}$ and higher
order gravitational corrections.  Special case $f =0$ corresponds
to D-term inflation, and  $f =1$  corresponds to F-term inflation.
A general P-term inflation model has an arbitrary $0 \leq f  \leq
1$. The running of the spectral index from blue to red and the
amplitude of the perturbations are similar to what has been
studied in~\cite{Linde:1997sj}.


\subsection{Embedding inflation in SUSY GUTs}\label{sec:models:SUSYGUTs}

Embedding inflation within GUT has a long history. We give here an
overview of some of the old attempts and the current status of some of these models,
and then turn to more recent developments.


\subsubsection{Inflation in non-SUSY GUTs}

The original idea of inflation was built on $SU(5)$ GUT, and the motivation was to
dilute the unwanted relics, i.e. GUT monopoles, besides predicting the universe as
large and as homogeneous on the largest scales as possible. Guth~\cite{Guth:1980zm}
first suggested (see also~\cite{Sato:1980yn,Linde:2005ht}) that the GUT phase
transition is first order, driven by the potential of
the GUT Higgs field $\Phi$ in the adjoint representation $\mathbf{24}$. Once the
finite temperature effects are taken into account,~\cite{Linde:2005ht}
\begin{equation}
V^{\rm old}(\phi,T)=-\frac{N \pi T^4}{90}-\frac{\mu^2-\beta T^2}{2}\phi^2
-\alpha_i T \phi^3+\gamma_i\phi~,
\end{equation}
where $\phi^2 \equiv \mathrm{Tr} \Phi^2$ and $\alpha_i$ and $\gamma_i$ are
constants of the theory.
Assuming an initial temperature much larger than $\mu$ and the GUT scale, the
cooling of the universe induces the GUT phase transition $SU(5)\rightarrow
\gsm$, which could be used to generate a false vacuum inflation if the universe was
trapped in a local minimum of the potential, typically at $\phi=0$.
The false vacuum leads to the so called ``old inflation'', which is terminated by
the formation of bubbles of true vacuum, as the phase transition takes place. This
idea was abandoned because it was plagued with many problems. In particular, inside a 
bubble of new vacuum, the energy density of the false vacuum
is transfered into kinetic energy, inducing the bubble expansion and the collisions
due to bubble walls. This leads to a highly inhomogeneous and anisotropic universe in
strong contradiction with the observations.

In Refs.~\cite{Linde:1981mu,Albrecht:1982wi}, the ``new inflation'' scenario was
proposed to avoid such problems and also implemented within $SU(5)$, as it was
suggested that the field $\phi$ trapped in the false vacuum $\phi=0$ slowly rolls down
its potential described by the Coleman-Weinberg potential~\cite{Coleman:1973jx}
at finite high temperature $T\gg M_X$,~\cite{Linde:2005ht}
\begin{equation}\label{eq:su5newinflation}
\begin{split}
V^{\rm new}(\phi,T)&=\frac{9M_X^4}{32\pi^2}+\frac{5}{8}g^2T^2\phi^2+\frac{25g^4\phi^4}{128\pi^2}
\left(\ln\frac{\phi}{\phi_0}-\frac{1}{4}\right)+c T^4~,\\
&\simeq V(0)-\frac{\lambda}{4}\phi^4~,
\end{split}
\end{equation}
where the second expression describes the effective inflationary potential close
to the origin when the temperature has dropped to $T\sim H$. In this expression,
$V(0)=9M_X^4/32\pi^2$ and $\lambda\simeq 25g^4/32\pi^2\left(\ln H/\phi_0-1/4\right)$.
This model evaded the old inflation problem but predicted the wrong amplitude of
anisotropies $\delta \rho/\rho\sim 110\sqrt{\lambda} \gg 10^{-4}$.

An improved version of this idea followed, the Shafi-Vilenkin model and the Pi 
model~\cite{Shafi:1983bd,Shafi:1984tt,Lazarides:1984pq,Pi:1984pv},
based on an $SU(5)$ theory containing an additional $SU(5)$ singlet $\chi$, which is driving
inflation in order to obtain an effective potential similar to Eq.~(\ref{eq:su5newinflation}) with an 
appropriate level of CMB temperature anisotropies. The theory is based on the following
potential ($\Phi$ still represents the GUT Higgs in the adjoint, and $H_5$ represents
the Higgs in the fundamental representation which is realizing the electro-weak breaking)~\cite{Shafi:1983bd}
(see also \cite{Linde:2005ht,Shafi:2007vq,Rehman:2008qs} for recent reviews)
\begin{equation}\label{eq:su5newinflationshafi}
\begin{split}
V^{\rm new}(\chi,\Phi,H_5)&=\frac{1}{4}a (\tr\Phi^2)^2+\frac{1}{2}b \tr\Phi^4
-\alpha(H_5^\dagger H_5) \tr \Phi^2+\frac{\gamma}{4}(H_5^\dagger H_5)^2
-\beta H_5^\dagger\Phi^2 H_5\\
&+\frac{\lambda_1}{4}\chi^4-\frac{\lambda_2}{2}\chi^2\tr \Phi^2
+\frac{\lambda_3}{2}\chi^2H_5^\dagger H_5~.
\end{split}
\end{equation}
The inflaton develops a Coleman-Weinberg potential due to its coupling to $\Phi$
and $H_5$. Its precise expression
is obtained by minimizing the above potential for $\Phi$ which settles the
system in the inflationary valley. Indeed, the breaking $SU(5)\rightarrow \gsm$
is realized in the usual $T_{24}\propto \mathrm{Diag}(1,1,1,-3/2,-3/2)$ direction,
the VEV of $\Phi$ being a function of that of $\chi$ because of the coupling
$\lambda_2$,
\begin{equation}\label{eq:GUTVEVnewinflation}
\langle\Phi\rangle=\sqrt{\frac{2}{15}}\phi ~\mathrm{Diag}(1,1,1,-3/2,-3/2)~,\quad
\mathrm{with}~~~ \phi^2=(2\lambda_2/\lambda_c) \chi^2~.
\end{equation}
($\lambda_c\equiv a+7b/15$ represents the mixture of the $\Phi^4$ terms in $V$.)
Discarding the pure $H_5$ sector (relevant at the EW scale) and computing the
masses of the triplet and doublet in $H_5$ that enter the Coleman-Weinberg formula,
one can reduce the potential to $V(\phi,\chi)$ and then to the effective inflationary
potential using Eq.~(\ref{eq:GUTVEVnewinflation})~\cite{Shafi:1983bd}
\begin{equation}
V^{\rm new}_{\rm eff}(\chi)=A\chi^4\left[\ln \left(\frac{\chi}{\chi_0}\right)-\frac{1}{4}\right]
+\frac{A\chi_0^4}{4}~,
\end{equation}
where $\chi_0$ is the position of the minimum of $V^{\rm new}_{\rm eff}(\chi)$ and
$A$ is a function of the couplings $\lambda_2$, $\lambda_c$ and the gauge coupling
$g_5$. The system after inflation is trapped in the global minimum at $\chi=\chi_0$
and $\phi=\phi_0=\sqrt{2\lambda_2/\lambda_c}$. The mass of the superheavy gauge
bosons inducing the proton decay is proportional to $\phi_0$,
\begin{equation}
M_X = \sqrt{\frac{5}{3}}\frac{g\phi_0}{2}~,
\end{equation}
Thus the phase of inflation take place at an energy close to the mass scale
involved in the proton decay, $M_X\sim 2 V_0^{1/4}$, and its stability constrains
the inflationary scale.

The predictions for $SU(5)$ singlet 
inflation~\cite{Linde:1981mu,Albrecht:1982wi,Shafi:1983bd,Shafi:1984tt,Pi:1984pv} is
similar to that of the potential; $V=V_0[1-\lambda_\chi(\chi/\mu)^4]$, with
$\lambda_\chi=A\ln{\chi/\chi_0}$. The predictions depend on $A$ or
alternatively on $V_0$, and for $V_0^{1/4}\in [2\times 10^{15},4\times 10^{16}]$~GeV,
they are found in the range:
$n_s\in [0.93,0.96]~, \quad r\in [10^{-5}-10^{-1}]~, \quad \alpha_s\in [0.6,1.3]
\times 10^{-3} $.
These predictions are found within the $2\sigma$ limit of the WMAP
data provided that $V_0$ is large,~\cite{Shafi:2006cs,Rehman:2008qs}, though this
requires that inflation taking place for super Planckian VEVs~\cite{Rehman:2008qs}. A large 
scale $V_0\sim 10^{16}$ GeV also implies a
proton lifetime roughly estimated in the range $\tau(p\rightarrow \pi^0e^+)\in
[10^{34},10^{38}]$ years~\cite{Rehman:2008qs}.
It was also proposed in Ref.~\cite{Rehman:2008qs} to realize
a chaotic-like inflation with the same potential but at large field $\chi>\chi_0$,
but this also imposes super Planckian VEVs. Taking the VEV above the
Planck scale does not make any
physical meaning as it would be hard to rely on the Coleman-Weinberg one-loop corrected
potential away from the renormalization scale.

Other challenges for this model makes it unappealing or unrealistic. The analysis
of the parameter space $(a,b,g)$ in~\cite{Breit:1983pp} revealed that the potential
possess a local minimum with symmetries $SU(4)\times U(1)$ in which the system gets
trapped, even when starting close to the SM minimum (with symmetries
$SU(3)\times SU(2)\times U(1)$). Obtaining the standard model at low energy is
therefore only possible after another first order transition that leads to the
problems of old inflation. Other problems were found related to the reheating and
the generation of baryon asymmetry (see for e.g.~\cite{Linde:2005ht})~\footnote{One 
can also think of realizing a brief period of
thermal inflation close to the electroweak scale with the help of the GUT
singlet~\cite{Lyth:1995hj,Lyth:1995ka}. The idea is that thermal effects will keep the
singlet scalar field close to the symmetric phase, but once inflation begins, the
universe would cool down and the field would roll down the potential and settle in its
minimum, which is close to the GUT scale.  The simplest potential will be like: $V\sim
V_0-m^2\phi^2$. The idea is appealing, but challenging to realize. The model requires a
very light GUT singlet, i.e. $m\sim 100$~GeV to be kept in thermal equilibrium with
other light degrees of freedom to get  $+T^2\phi^2$ correction. However the singlet
should break the GUT group down to the SM when it develops a VEV and excite the SM
degrees of freedom alone at the electroweak scale.  If $V_0^{1/4}\sim 10^{16}$~GeV,
then the initial temperature of the universe prior to such a phase transition ought to
be very high fairly close to the GUT temperatures and final temperature should be close
to the electroweak scale to get sufficient inflation, i.e. roughly $7-10$ e-foldings of
inflation, which is sufficient to dilute the unwanted relics such as excess gravitons or
damping the moduli oscillations, etc. To our knowledge there is no explicit light GUT
singlet which has been constructed to execute this idea.}.


\subsubsection{Hybrid inflation within SUSY GUTs and topological defects}

The embedding of inflation in SUSY GUTs has mostly been studied for hybrid
inflation as those models consider a coupling between the inflaton sector
and the GUT Higgs sector. It was therefore natural to first inquire if some
GUT Higgs field already present in theories which can play the role
of the waterfall fields $\Phi_\pm$. For the $D$-term model, the presence of
a constant FI term would require that the group $U(1)_\xi$ is not the
subset of a non-abelian group~\cite{Jeannerot:1997is}. As a consequence,
if the SM is embedded in a model based on the group $\ggut$, the whole
theory is based on $\ggut\times U(1)_\xi$ and inflation takes place in the
chain~\cite{Jeannerot:1997is},
\begin{equation}
\ggut\times U(1)_\xi \rightarrow \dots H \times U(1)_\xi \overset{\rm Infl}{\rightarrow} H
\rightarrow \dots\rightarrow \gsm~,
\end{equation}
where ``Infl'' identify the symmetry breaking that triggers the waterfall at the
end of inflation. This represents the lowest possible level of embedding of
inflation within SUSY GUTs, since both the inflaton field and the Higgs
fields are introduced in addition to the field content motivated by the particle
physics.

The $F$-term inflation model does not contain the restriction due to the presence
of a FI term, therefore its embedding withing SUSY GUT has a richer
phenomenology. The superfields $\Phi$ and $\overline{\Phi}$ are assumed to
belong to a non-trivial representation and the complex conjugate representation
of some group $G_{\rm infl}$, whereas $S$ is a singlet of $G_{\rm infl}$, in such
a way that $W=\kappa S(\Phi \overline{\Phi})$ is invariant under $G_{\rm infl}$.

The general picture is when inflation takes place, within the cascade of
spontaneous symmetry breaking induced by the Higgs sector is
therefore~\cite{Jeannerot:1997is},
\begin{equation}
\ggut \rightarrow \dots H \times G_{\rm infl} \overset{\rm Infl}{\rightarrow} H
\rightarrow \dots \rightarrow \gsm~.
\end{equation}
Cosmological observations can constrain what $G_{\rm infl}$ can and cannot be, i.e.
one of the motivation to introduce inflation in GUT was to solve the monopole
problem.
Therefore, the breaking of $G_{\rm infl}$, and all subsequent SUSY
breaking cannot give rise to the formation of monopoles. Following the argument
of Ref.~\cite{Davis:1995bx} for SO(10), a
systematic study has been done in Ref.~\cite{Jeannerot:2003qv} for all possible
models that can be constructed using SUSY GUTs based on all possible
GUT group of rank lower than 8 (including $SO(10)$, $SU(n)$, $E_6$). It was
assumed that a discrete $Z_2$
symmetry is left unbroken at low energies to protect the proton from a too rapid
decay. It was shown that generically the waterfall at the end of
inflation gives rise to the formation of cosmic strings, since almost all
possible ways to break $\ggut$ down to $\gsm$ involve the generation of a $U(1)$
symmetry (leading to monopoles) or its breaking (leading to cosmic strings).

In addition, for an $SO(10)$ or $E_6$ based models, the waterfall accompanies
generically the breaking of the $U(1)_{B-L}$ symmetry~\cite{Jeannerot:2003qv}.
Consequently, the Higgs coupled to the inflaton can also be involved in the
see-saw mechanism. Assuming an $SO(10)$ model that preserves the R-parity, and
using a minimum number of fields to realize this breaking, this is realized
generically employing a pair $\Phi=\mathbf{126},\overline{\Phi}=
\mathbf{\overline{126}}$~\cite{Martin:1992mq}, though $\Phi=\mathbf{16},
\overline{\Phi}=\mathbf{\overline{16}}$ is also possible, and does not preserve
the R-parity. Since the $U(1)_{B-L}$ symmetry breaking scale is also commonly
used to generate the Majorana mass for right-handed neutrinos, this opens up
the possibility to combine constraints from neutrino measurement and CMB
constraints on the parameter space in specific models with these ingredients.


\subsubsection{Embedding inflation within GUT }
\label{EEMGOIG}

Several embeddings of $F$-term models of inflation in a specific
and realistic SUSY GUT model have been proposed, see
Refs~\cite{Jeannerot:1995yn,Kyae:2005vg,Jeannerot:2006jj,bajc:2009fe}.
Most of these
references~\cite{Jeannerot:1995yn,Kyae:2005vg,Jeannerot:2006jj}
have considered SUSY $SO(10)$ models that are capable of accounting
for enough proton stability, a (semi-)realistic mass matrix for
fermions, and a doublet-triplet (D-T) splitting usually through the
Dimopoulos-Wilczek mechanism. We will discuss such models in this
section and also discuss a more minimal model of
$SO(10)$~\cite{bajc:2009fe} in the next subsection.

Imposing that the R-parity is unbroken at low energy is usually assumed in
order to protect the proton from a too rapid decay through dimension 4
operators involving sparticles (see Sec.~\ref{sec:partphys:SUSYGUTs}).
$SO(10)$ contains a $Z_2$ symmetry subgroup of $U(1)_{B-L}\subset SU(4)_C
\subset SO(10)$ that can play this role provided only ``safe'' Higgs
representations are employed to realize $SO(10)\rightarrow \gsm$, namely via
$\mathbf{10},~\mathbf{45},~\mathbf{54},~\mathbf{126},~\mathbf{\overline{126}},~
\mathbf{210}$, etc. Therefore, in Ref.~\cite{Jeannerot:1995yn}, the model
has the following field content; two pairs of adjoint
$\mathbf{45}$ and $\mathbf{54}$ denoted $A_{45},A_{45}',S_{54},S_{54}'$ are
assumed to break $SO(10)$ down to $3_c2_L1_R1_{B-L}$ while a pair
$\mathbf{126}+\mathbf{\overline{126}}$ denoted $\Phi+\bar{\Phi}$ is
used to break the ${B-L}$ symmetry and obtain the MSSM $\gsm\times Z_2^R$.
An additional pair $(H,H')$ of fundamental $\mathbf{10}$ are assumed for the
electroweak symmetry breaking and one last $\mathbf{45}$ denoted $A''$ is
also required to avoid dangerous light degrees of freedom.

Finally $F$-term hybrid inflation is realized adding an $SO(10)$ singlet $\mathcal{S}$
to this field content. The superpotential then contains $5$ sectors, the first two implementing the
breaking of $SO(10)$ and the electroweak symmetry breaking respectively, and
the D-T splitting at the same time,~\cite{Jeannerot:1995yn}
\begin{equation}
W_1=m_A A^2+m_S S^2 + \lambda_S S^3+\lambda_A A^2S~,\quad W_2=HAH+
m_H' H'^2~.
\end{equation}
The third sector is a replica of the first sector for the field $S'$ and
$A'$. If each sector 1 and 3 can break individually $SO(10)$ down to
$3_c2_L2_R1_{B-L}$, their combination can break $SU(2)_R$ further down to
$U(1)_R$. The fourth sector
breaks the $B-L$ symmetry dynamically, realizing the $F$-term hybrid inflation
at the same time using
\begin{equation}
W_4=\kappa \mathcal{S}(\Phi\bar{\Phi}-M^2)~.
\end{equation}
It is argued that though the inflaton was not required to break the B-L
symmetry, its presence can be motivated by the fact that then the symmetry
breaking is realized dynamically and at a scale close to the GUT scale.
One last sector has to be introduced to avoid massless goldstone bosons
$W_5= AA'A''$. The VEV in the appropriate direction for all those fields
can break $SO(10)$ to the MSSM$\times Z_2$ and it is clear that at least
one of the global minima of the potential emerging from, $W_1+W_2+W_3+W_4+W_5$, possess 
the right symmetries. However it seems unclear from what part of the initial conditions in field
space the dynamical evolution of the system would indeed end up on the
inflationary valley and give rise to $F$-term inflation. Indeed the risk is
the presence of tachyonic instabilities in this multi-dimensional potential
that would destabilize the dynamics and avoid the inflationary valley.

Another problem that may arise is the stability of the
superpotential, since many other terms are allowed by the symmetries of the
model,  for example coupling the inflaton to itself or with other
fields would have potential to ruin the inflationary success, as it would
introduce mass terms or quartic terms for the inflaton potential destroying
the flat direction of the potential ensured by the linearity of $W$ in
$\mathcal{S}$. Note however that the non-renormalizability theorem protects
the form of a given superpotential against the generation of other terms
from radiative corrections.

A similar attempt to embed $F$-term hybrid inflation in the Barr-Raby model
has also been described in the appendix of~\cite{Jeannerot:2006jj} with a
different field content; 1 adjoint, $A=\mathbf{45}$, instead of 2, the same pair
of fundamental $(H,H')$, the breaking of $B-L$ symmetry employing one of the
two pairs of $\mathbf{16},\mathbf{\overline{16}}$ (therefore breaking the
R-parity), and 4 singlets instead of 1 are assumed. The D-T splitting is
then realized giving a VEV along B-L to the adjoint and the presence of the
singlets is required to avoid the presence of massless pseudo-goldstone
bosons. For the symmetry breaking, like in the previous model, the VEVs of
the adjoint and one pair of spinors is enough to break $SO(10)$ down to the
MSSM, the inflaton $S$ being again coupled to the pair of spinors
$\mathbf{16},\mathbf{\overline{16}}$ that breaks $B-L$. The
superpotential of the theory contains non-renormalizable terms and reads:
\begin{eqnarray}
W=&&S(\Phi\bar{\Phi}-M^2)+\frac{\alpha}{4M_S}A^4+\frac{1}{2}M_A A^2+T_1AT_2
+M_T T_2^2+\bar{\Phi}'\left[\zeta\frac{PA}{M_S}+\zeta_Z Z_1\right]\Phi \,\nonumber\\
&&+\bar{\Phi}\left[\xi\frac{PA}{M_S}+\xi_Z Z_1\right]\Phi'
+M_\Phi\bar{\Phi}'\Phi'~.
\end{eqnarray}
where the spinors are denoted by: $\Phi,\bar{\Phi},\Phi',\bar{\Phi}'$ and
$S,P,Z_1,Z_2$ being the 4 gauge singlets.
Plugging in the VEV required to achieve the symmetry breaking of SO(10)
allows to identify which of the components of $\mathbf{16}
+\mathbf{\overline{16}}$ stay light during inflation, and which are coupled
to the GUT Higgs fields and thus acquire a GUT scale mass. This is important
for the predictions of the $F$-term hybrid inflation model, since only the
light states will contribute to the radiative corrections and to the
effective $\mathcal{N}$ in front of the $\ln$ term (see
Sec.~\ref{sec:ftermhybridinflation} and
Eq.~(\ref{eq:effectivepotentialFtermvalley})). In this particular model,
$\Phi$ is a 16-dimensional spinor, only two of its components will
remain light, although having such light states is rather model dependent issue.

Finally, an embedding of shifted hybrid inflation in a very
elaborated $SO(10)$ model has been proposed in \cite{Kyae:2005vg}, and an
embedding of $D$-term hybrid model has been proposed in~\cite{Dvali:1997mh}.


\subsubsection{Origin of a gauge singlet inflaton within SUSY GUTs}

It is desirable to seek an answer to the question, whether $F$-term hybrid 
inflation \emph{coupled} or \emph{embedded} within SUSY GUTs?
Independently of the assumed  GUT group, because of the form of the superpotential
Eq.~(\ref{eq:superpotentialFterm}), the inflaton superfield, $S$, is
necessarily an {\it absolute gauge singlet}, since $S M^2$ cannot be a gauge invariant
if $S$ is not an singlet superfield~\cite{bajc:2009fe}.

To fully embed the inflation model within SUSY GUTs, one could generate $M$ by the
VEV of some other field, though this has not yet been done so far,
and attempts in this direction within the minimal $SO(10)$ model
of~\cite{Bajc:2004xe} are unsuccessful~\cite{bajc:2009fe}.

These models are based on a minimal number of superfields, and a
protected R-parity at low energies. It contains only a
$\Sigma\equiv \mathbf{210}$, a pair $\Phi\equiv
\mathbf{126},\overline{\Phi}\equiv \mathbf{\overline{126}}$ and a
fundamental $H=\mathbf{10}$, and the most general superpotential
with this field content given by~\cite{Bajc:2004xe}:
\begin{equation}
W_{\rm min}^{SO(10)}=m_\Sigma\Sigma^2+m_\Phi\Phi\overline{\Phi}
+\lambda_\Sigma\Sigma^3+\eta\Sigma\Phi\overline{\Phi} + m_H
H^2+\Sigma H(\alpha\Phi+\bar{\alpha} \overline{\Phi})~.
\end{equation}
One can show that any VEV that would generate the mass term
$M^2$ in the inflaton superpotential would also generate a mass for
the inflaton field~\cite{bajc:2009fe}. As a conclusion it is
not possible to generate \emph{exactly} the $F$-term model, at least
within this minimal version of $SO(10)$.

Another concerning issue is the stability of the
superpotential~\cite{bajc:2009fe}. The form of the superpotential
of Eq.~(\ref{eq:superpotentialFterm}) is supposedly protected by a
$U(1)_R$ symmetry, under which $S$ is doubly charged, while $\Phi$
and $\overline{\Phi}$ have opposite charges. This property is
important for example to prevent contributions to $W$ such as
quadratic or cubic terms in the inflaton superfield, that would
spoil the flatness of the inflationary potential. Furthermore,
assigning $R$-charges to obtain flatness of the inflaton potential
in presence of the Higgs sector is also challenging. For example,
the minimal $SO(10)$ model of \cite{Bajc:2004xe}, whose most
general superpotential cannot be invariant under such an
$R$-symmetry. Trying to accommodate a phase of $F$-term inflation
in such model would receive destabilization from various sources.

\begin{itemize}
\item Quadratic and cubic terms in the inflaton potential, which
would a-priori ruin the flatness of the potential. It is possible
to arrange the parameters to realize a saddle point or an
inflection point inflation. More work is required in this
direction.

\item Additional couplings between the inflaton, $S$, and other
singlets (including $S \Phi_i^2$ for all the superfields $\Phi_i$
in the theory), which would demand additional assumptions such
that these $\Phi_i$ superfields remain heavy during inflation via
their couplings.

\item Like the original hybrid model, those additional couplings
would generate dangerous quadratic and cubic terms at one-loop
level, even if they were assumed vanishing at the tree level. One
would have to ensure that extra terms do not spoil the flatness of
the potential.
\end{itemize}

Note that these stability problems are also present in earlier
attempts to embed $F$-term hybrid inflation in SUSY
GUTs. If these stability issues are left aside, it is interesting
to note that among all possible global minima of the minimal
$SO(10)$ theory, one and only one of them can accommodate a phase
of $F$-term inflation, when the stability of the potential against
VEV shifting and the formation of topological defect formation are
taken into account~\cite{bajc:2009fe}.


\subsubsection{Other inflationary models within SUSY GUTs}

Inspite of the challenge to seek an origin of a  gauge singlet inflaton within the GUT group, 
there have been many examples to embed the waterfall field within SUSY GUTs.
Here we briefly sketch some examples of the embedding.

\begin{itemize}
\item{Shifted and smooth hybrid inflation:\\
As we discussed in Sec.~\ref{sec:models:SUSYmodelswithNR}, the shifted
hybrid inflation model was built embedded in the Pati-Salam subgroup $\gps\equiv
SU(4)_C\times SU(2)_L\times SU(2)_R$ of $SO(10)$~\cite{Jeannerot:2000sv,Jeannerot:2001qu}. The idea is to
break the gauge group during inflation in a single step which also
avoids the monopole problem arising in the breaking of the
Pati-Salam group down to the MSSM group, breaking that can be done
using a pair of Higgs superfields, $H^{c}=({\bf \bar 4, 1, 2})$
and $\bar H^{c}=({\bf 4, 1, 2})$. These Higgs fields also give
through their VEVs a Majorana mass to the right-handed neutrino
directions, inducing the see-saw mechanism to account for the tiny
mass of the left-handed neutrinos. This pair of Higgs fields is
also assumed to be the one coupled to the inflaton field,
realizing the symmetry breaking during inflation, and supporting
inflation through their $F$-terms. The vanishing of the $D$-term
part of the potential constraints, $\vert \bar{H}^c \vert = \vert
H^c\vert $. The inflationary potential is embedded in a potential
where there are many more fields, which are relevant for the
computation of one-loop corrected inflationary potential. The
actual potential also is invariant under the global $U(1)_R$ of
conventional SUSY and under a Peccei-Quinn symmetry, $U(1)_{PQ}$.
Like for the original $F$-term inflation, the $R$-symmetry prevents
from non-linear terms for $S$, such as $S^2, S^3$ terms in the
superpotential, which would ruin the flat direction for the
inflaton at tree level.\\

In Ref.~\cite{Kyae:2005vg}, the authors embedded this shifted
model within an explicit model based on SO(10)and also implemented
the Dimopoulos-Wilczek mechanism~\cite{Dimopoulos:1981xm}, and
addressed the MSSM $\mu$-problem. The inflationary superpotential
is given by~\cite{Jeannerot:2000sv,Kyae:2005vg}
\begin{eqnarray}
W_{\rm infl}&\approx& -\kappa S\bigg[ M_{B-L}^2- {\bf
16}_H{\bf\overline{16}}_H+\frac{\rho}{\kappa M_*^2 }({\bf
16}_H{\bf\overline{16}}_H)^2
\nonumber\\
&&-\frac{\kappa_1}{\kappa}P\overline{P}+\frac{\rho_1}{\kappa
M_*^2}(P\overline{P})^2
-\frac{\kappa_2}{\kappa}Q\overline{Q}+\frac{\rho_2}{\kappa M_*^2}
(Q\overline{Q})^2 \bigg]
\\
&\equiv& -\kappa SM_{\rm eff}^2 ~, \nonumber
\end{eqnarray}
where a gauge singlet $S$ is the inflaton, ${\bf 16}_H$, ${\bf
\overline{16}}_H$, additional singlets are $P$, $\overline{P}$,
$Q$, and $\overline{Q}$ are displaced from the present values, for
which two eigenvalues can potentially act like a moduli, while two
of them are heavy as the GUT scale with a large VEV. The
dimensionless constants are $\kappa,~\kappa_i,~\rho$~\cite{Kyae:2005vg}.

It is assumed that initially, $|\langle S\rangle|^2 \approx
M_{B-L}^2[1/(4\zeta)-1]/2$, with $1/4<\zeta<1/7.2$, and
$\langle{\bf 16}_H\rangle$, $\langle{\bf\overline{16}}_H\rangle$,
$\langle P\rangle$, $\langle\overline{P}\rangle$, $\langle
Q\rangle$, $\langle \overline{Q}\rangle \neq 0$, where $M_{\rm
eff}^2\sim M_{B-L}^2$. With $D_SW\approx -\kappa M_{\rm eff}^2(1+|S|^2/M_P^2)$, the
$F$-term part of the potential is given by:
\begin{eqnarray}
V_{F}\approx\bigg(1+\sum_{k}\frac{|\phi_k|^2}{M_{\rm
P}^2}+\cdots\bigg)\bigg[ \kappa^2M_{\rm
eff}^4\bigg(1+\frac{|S|^4}{2M_{\rm P}^4}\bigg)
+\bigg(1+\frac{|S|^2}{M_P^2}+\frac{|S|^4}{2M_{\rm P}^4}\bigg)
\sum_{k}|D_{\phi_k}W|^2\bigg]\,, \nonumber\\
\end{eqnarray}
where all scalar fields except $S$ contribute to $\phi_k$. The
factor $(1+\sum_{k}|\phi_k|^2/M_P^2+\cdots)$ in front originates
from $e^{K/M_P^2}$. Provided, $|D_{\phi_k}W|/M_{\rm P}$ are much
smaller than the Hubble scale ($\sim \kappa M_{\rm eff}^2/M_{\rm
P} $), the flatness of $S$ will be guaranteed even after including
SUGRA corrections. This can be realized by the choice $\kappa\ll
1$, and $\kappa/\rho \ll 1$. The $U(1)$ $R$-symmetry ensures the
absence of terms proportional to $S^2$, $S^3$, etc. in the
superpotential, which otherwise could spoil the slow-roll
conditions. The spectral tilt arising from this model is very
close to one, i.e. $n_s=0.99\pm 0.01$ and $|dn_s/d\ln k|\ll
0.001$.\\

The smooth hybrid inflation model was also originally built within
$SO(10)$~\cite{Lazarides:1995vr,Jeannerot:2001qu} though to our
knowledge this model has not be embedded into a specific SUSY GUTs
model. Its proximity to the shifted model allows us to guess
that there should not be any  problem as long as an additional
singlet is assumed to play the role of the inflaton. 

As mentioned earlier in this chapter, these two models can also be realized
without the non-renormalizable superpotential terms leading to the
new shifted~\cite{Jeannerot:2002wt}, and the new smooth hybrid
inflation~\cite{Lazarides:2007fh} models, at the cost of adding
more fields in the picture, see Sec.\ref{sec:models:SUSYmodelswithNR}.}

\item{Flipped $SU(5)$:\\
Inflation in flipped $SU(5)~(=SU(5)\times U(1)_X$), which is a
maximal subgroup of $SO(10)$ with a chiral superfield in the
spinorial representation $\mathbf{16}$ per family, was also
considered in~\cite{Kyae:2005nv,Rehman:2009yj}. The breaking of $SU(5)\times
U(1)_X$ to the MSSM gauge group happens when ${\bf 10_H}$ and
$\overline{\bf 10_H}$ develops a VEV. The relevant superpotential
is given by~\cite{Kyae:2005nv}
\begin{eqnarray} \label{superpot}
W &=& \kappa S\left[{\bf
10}_H{\bf\overline{10}}_H-M^2\right]
+\lambda_1{\bf 10}_H{\bf 10}_H{\bf
5}_h+\lambda_2\overline{{\bf 10}}_H\overline{{\bf
10}}_H\overline{{\bf 5}}_h
\\
&&+y_{ij}^{(d)}{\bf 10}_i{\bf 10}_j{\bf 5}_h +y_{ij}^{(u,\nu)}{\bf
10}_i\overline{{\bf 5}}_j\overline{{\bf 5}}_h + y_{ij}^{(e)}{\bf
1}_i\overline{{\bf 5}}_j{\bf 5}_h \nonumber \,,
\end{eqnarray}
where $\kappa, \lambda_1, \lambda_2$ are constants and the
appropriate Yukawas are given by $Y_{ij}$, i.e.
$Y_{ij}^{(\mu,\nu)}$, provide masses to up-type quarks and
neutrino Dirac masses. The $U(1)_R$ symmetry eliminates terms such
as $S^2$ and $S^3$ from the superpotential. Higher dimensional
baryon number violating operators such as ${\bf 10}_i{\bf
10}_j{\bf 10}_k{\bf\overline{5}}_l\langle S\rangle/M_P^2$, ${\bf
10}_i{\bf\overline{5}}_j{\bf \overline{5}}_k{\bf 1}_l\langle
S\rangle/M_P^2$, etc. are suppressed as a consequence of $U(1)_R$.
The proton decay proceeds via dimension six operators mediated by
the superheavy gauge bosons. The dominant decay mode is
$p\rightarrow e^+/\mu^+,\pi^0$ and the estimated lifetime is of
order $10^{36}$ yrs.~\cite{Ellis:2002vk,Dorsner:2004xx}. This $Z_2$
ensures that the LSP is absolutely stable. Inflation happens for
$\kappa\leq 10^{-2}$ and matches the standard predictions, i.e.
$n_s\approx 1$ with negligible gravity waves and running of the
spectral tilt~\cite{Kyae:2003vn,Kyae:2005nv} for $M\sim
10^{16}$~GeV. By introducing soft-SUSY breaking mass terms with minimal
K\"ahler potential it is possible to bring down the spectral tilt, $n_s\sim 0.96-0.97$ for
$M_s\sim 8\times 10^{15}-2\times 10^{16}$~GeV~\cite{Rehman:2009yj}. }

\item{5D SO(10):\\ 
Inflationary models from 5D $SO(10)$ were also
constructed in Refs.~\cite{Kyae:2002hu,Kyae:2002ss}. There are
certain advantages of orbifold constructions of five dimensional
(5D) SUSY GUTs, in which $SO(10)$ can be readily broken
to its maximal subgroup
$H$~\cite{Dermisek:2001hp,Asaka:2001eh,Hall:2001xr}, with the
doublet-triplet splitting problem addressed due to construction
where $SO(10)$ is compactified on $S^{1}/(Z_2\times Z_2^{\prime})$
(where $Z_2$ reflects, $y\rightarrow -y$, and $Z_2^{\prime}$
reflects $y^{\prime}\rightarrow -y^{\prime}$ with
$y^{\prime}=y+y_c/2$. The two orbifold points are $y=0$ and
$y=y_c/2$). In order to realize inflation in $4D$, the $N=2$
SUSY in the 5D bulk is broken down to $N=1$ SUSY
on the orbifold fixed points, below the compactification scale
$\pi/y_c$, where the branes are located.

The $F$-term inflation potential can be constructed on the branes,
assuming that the inter brane separation is fixed. The two branes
preserve different symmetries, on one the full $SO(10)$ is
preserved while in the other $SU(4)_c\times SU(2)_L\times
SU(2)_R$~\cite{Hebecker:2001jb}. There also exists a bulk scalar
field, $S$, which couples to the singlets of $SO(10)$ on the
branes. For instance, on the brane where $SO(10)$ is preserved,
the superpotential would be given by: $W=\kappa S(Z\overline
Z-M_1^2)$, while in the second brane $W=\kappa S(\phi\overline
\phi-M_2^2)$, where $\phi,\overline\phi$ belong to $({\bf \bar 4,
1, 2})$ and $({\bf 4, 1, 2})$~\cite{Kyae:2002hu}. Similar
constructions were made in Ref.~\cite{Kyae:2002ss} where on one
brane $SU(5)\times U(1)_X$ and on the other $SU(5)^{\prime}\times
U(1)^{\prime}_X$ were preserved, inflationary potential arises to
the breaking of $U(1)$ at a scale close to the GUT scale.}
\end{itemize}

In all the above
examples, see Refs.~\cite{Jeannerot:2000sv,Jeannerot:2001qu,Jeannerot:2002wt,Kyae:2005vg,Lazarides:2007fh,Kyae:2002hu,Kyae:2002ss,Lazarides:2007ii},
it is possible to excite non-thermal leptogenesis either from the
direct decay of the inflaton or the Higgs coupled to it or from
non-perturbative excitation from the coherent oscillations of the
inflaton.


\subsubsection{Inflation, neutrino sector and family replication}

The right handed Majorana sneutrino as an inflaton has been
proposed as a particle physics candidate for
inflation~\cite{Murayama:1992ua,Murayama:1993xu}. These initial
models were based on a chaotic type potential for the inflaton
with a superpotential~\footnote{We are using the superfield and
the field notation to be the same for the right handed neutrinos,
i.e. $N$.}
\begin{equation}
W=\frac{1}{2}MN_{i}N_{i}+\mu H_{u}H_d+h^{ij}N_iL_jH_u+k^{ij}e_iL_jH_d\,,
\end{equation}
where the right handed Majorana neutrino superfield, $N$, has been
treated as a gauge singlet. In order to avoid higher order
contributions such as $N^3$ term in the superpotential, the right
handed neutrino can be assigned odd under R-parity. The lightest
right handed electron sneutrino acts as an inflaton with an
initial VEV larger than $M_{\rm P}$. After the end of inflation
the coherent oscillations of the sneutrino field generates lepton
asymmetry with the interference between tree-level and one-loop
diagrams. The largest lepton asymmetry is proportional to the
reheat temperature, $n_L/s\sim \epsilon (3T_{\rm R}/4M_{\rm P})$.
The CP asymmetry, $\epsilon \sim (\ln 2/8\pi){\rm Im}h^{\ast
2}_{33}$. Models of inflation and non-thermal leptogenesis were
also considered in
Refs.~\cite{Allahverdi:2002gz,Mazumdar:2003va,Mazumdar:2003wm}.

Sneutrino hybrid inflation was constructed in
Refs.~\cite{Antusch:2004hd,Antusch:2005kf}.
In~\cite{Antusch:2004hd} the following superpotential has been
used to generate inflation and the masses for the right handed
neutrinos:
 \begin{eqnarray}
{W} =
\kappa S \left(\frac{{\Phi}^4}{{M'}^2} -
M^2\right) \!+
\frac{(\lambda_{N})_{ij}}{M_*}N_iN_j\Phi\Phi\,
+ \dots \,,
\end{eqnarray}
where $\kappa, (\lambda_{N})_{ij}$ are dimensionless Yukawa couplings,
$M, M^{\prime}, M_{\ast}$ are three independent mass scales. $\Phi, S, N$ are
 all gauge singlets. The waterfield is $\Phi$ generates masses for the right handed
 sneutrinos.  During inflation $N$ can take large VEVs as $\Phi$ is stuck at the zero VEV.
The K\"ahler potential was given by~\cite{Antusch:2004hd}:
\begin{eqnarray}
K & =&
|{S}|^2 \!+ |{\phi}|^2 \!+ |{N}|^2
+\kappa_S \frac{|{S}|^4}{4 M_{\rm P}^2}
+\kappa_N \frac{|N|^4}{4 M_{\rm P}^2}
+\kappa_\phi \frac{|\phi|^4}{4 M_{\rm P}^2} \, \nonumber \\
&+& \kappa_{S\phi} \frac{|{S}|^2 |{\phi}|^2}{M^2_{\rm P}}
+ \kappa_{S N} \frac{|{S}|^2 |N|^2}{M^2_{\rm P}}
+\kappa_{N \phi} \frac{|N|^2 |{\phi}|^2}{M^2_{\rm P}}
+ \dots \, ,
\end{eqnarray}
and the $F$-term potential is given by:
\begin{eqnarray}
V &=&
\kappa^2 \left(\frac{|\phi|^4}{{M'}^2} - M^2 \right)^2
\left(1 +(1-\kappa_{S\phi}) \frac{|\phi|^2}{M_{\rm P}^2}
+(1-\kappa_{S N}) \frac{|N|^2}{M_{\rm P}^2}
- \kappa_S \frac{ |S|^2 }{M_{\rm P}^2}\right)
\nonumber \\
&&+ \frac{4 \lambda_N^2}{M_*^2} \,(|{N}|^4 |\phi|^2 +
|N|^2 |\phi|^4) +\dots \,,
\end{eqnarray}
Inflation is driven by the singlet field, $S$, where $\phi$ has a zero VEV.
By virtue of the coupling between $\phi$ and $N$, the $N$ field remains massless
during inflation and therefore subject to random fluctuations during inflation. The authors
assumed that the last $60$ e-foldings of inflation arises when the sneutrino field is rolling
slowly down the potential given by:
\begin{equation}
V\approx \kappa^2M^4\left(1+(1-\kappa_{SN})\frac{|N|^2}{2M_{\rm P}^2}+\delta\frac{|N|^4}{4M_{\rm P}^4}
\right)+\dots\,,
\end{equation}
where
$\delta=0.5+\kappa_{SN}^2-\kappa_{SN}\kappa_N+1.25\kappa_{N}+\dots$.
The predictions of the model is typical of a hybrid inflation with
a nearly flat spectrum. The WMAP data constraints parameters such
as $|(1-\kappa_{SN})|\leq 0.02$, also the running in the spectral
tilt and tensor to scalar ratio are negligible. The model also
generates isocurvature fluctuations, since both $S$ and $N$ in
principle can be light during inflation, but they have assumed
that $S$ obtains a heavy mass and therefore settled down to its
minimum in one Hubble time or so during
inflation~\cite{Antusch:2004hd}.

Hybrid inflation model with Dirac neutrinos was constructed in
Refs.~\cite{Antusch:2005kf,EytonWilliams:2004bm,EytonWilliams:2005bg},
in a specific Type-I string theory with the help of anisotropic
compactification~\cite{EytonWilliams:2005bg}. The model explains
the smallness of the $\mu$-term, strong-CP  and the Dirac nature
of neutrinos. The relevant part of the superpotential for
inflation is given by:
\begin{equation}
W=\lambda\phi H_uH_d+\kappa \phi N^2\,,
\end{equation}
where $\lambda\sim \kappa\sim 10^{-10}$. Including the soft
SUSY breaking terms, the $F$-term potential is given
by~\cite{Antusch:2005kf,EytonWilliams:2004bm}
\begin{equation}
V=V_0+\lambda A_{\lambda}\phi H_u H_d+\kappa A_\kappa \phi N^2+{\rm h.c.}+m_0^2(|H_u|^2+|H_d|^2+
|N|^2)+m_{\phi}\phi^2\,,
\end{equation}
where the origin of $V_0$ arises from the Peccei-Quinn breaking
scale, $\Lambda=2\pi f_a$, where $10^{10}\leq f\leq 10^{13}$~GeV.

The phase transition associated with the spontaneous breaking of
family symmetry, by flavons, which is responsible for the
generation of the effective quark and lepton Yukawa couplings
could also be responsible for inflation~\cite{Antusch:2008gw}.

In order to understand the origin of fermion masses and mixing,
one can extend the  SM by some horizontal family symmetry $G_F$,
which may be continuous or discrete, and gauged or global. It must
be broken at high scales with the help of flavons, $\phi$, whose
VEV will break the family symmetry. The Yukawa couplings are
generically forbidden by the family symmetry $G_F$, but once it is
broken, effective Yukawa couplings can be generated by
non-renormalizable operators, i.e. $(\phi /M_c)^n\psi \psi^c H$. This gives rise to 
an effective Yukawa coupling $\varepsilon^n\psi \psi^c H$, where 
$\varepsilon=\langle\phi \rangle /M_c\sim {\cal O}(0.1)$
and $\psi, \psi^c$ are SM fermion fields, $H$ is a SM Higgs field,
and $M_c$ is the messenger scale.

The relevant superpotential for inflation can be given by~\cite{Senoguz:2004ky,Antusch:2008gw}:
\begin{eqnarray}
W = \kappa S \left[ \frac{(\phi_1 \phi_2 \phi_3)^n}{M_*^{3n-2}} - \mu^2  \right]
\end{eqnarray}
for $n \ge 1$ and $\kappa\sim 1$. The fields, $\phi =
(\phi_1,\phi_2,\phi_3)$ is in the fundamental triplet
$\underline{3}$ representation of $A_4$ or $\Delta_{27}$. At the
global minimum of the potential, the $\phi_i$ components get VEVs
of order $M=M_*\left(\frac{\mu}{M_*}\right)^{2/3n}$ ~\footnote{The
initial motivations were proposed
in~\cite{Senoguz:2004ky,Asaka:1999jb} in order to obtain a
phenomenologically viable new inflationary potential from the
superpotential: $W=S(-\mu^2+(\bar\Phi \Phi)^{m}/M_{\ast}^{2m-2})$.
Here $\bar\Phi(\Phi)$ denotes a conjugate pair of superfields
charged under some gauge group, and $S$ is a gauge singlet. There
is an $U(1)_R$ symmetry under which $\Phi\rightarrow \Phi$,
$\bar\Phi\rightarrow \bar\Phi$, $S\rightarrow e^{i2\alpha}S$ and
$W\rightarrow e^{i2\alpha}W$. Under these symmetries $m=n$ for odd
$n$ and $m=n/2$ otherwise. Along the $D$-flat direction, the
k\"ahler potential is given by; $
K=|S|^2+(2|\Phi|^2+\kappa_1|\Phi|^4/4M_{\rm
P}^2+\kappa_2|S|^2|\Phi|^2/M_{\rm P}^2)+ \kappa_3|S|^4/4M_{\rm
P}^4\dots$ . The resultant potential is given by: $V\approx
\mu^4(1-\kappa_3 |S|^2/M_{\rm P}^2 +2(1- \kappa_2)|\phi|^2/M_{\rm
P}^2-2|\phi|^{2m}/M_{\ast}^{2m}+\dots)$. For $\kappa_3<-1/3$, the
S field obtains heavy mass compared to the Hubble expansion rate
and the field rolls to its minimum in one e-foldings leaving
behind the dynamics of $\phi$ field to slowly roll over the
potential; $V\approx \mu^4(1-0.5(\kappa_2-1)\phi^2/M_{\rm
P}^2-2\phi^{2m})$~\cite{Izawa:1996dv}. The spectral can match the
CMB data for values of $m=2- 5$ with a negligible running of the
spectral index $dn_s/d\ln k\leq 10^{-3}$. One particular issue  is that
the initial condition for $\phi$ field which needs close to zero
VEV, this is a nontrivial initial condition, one proposal is to
have a pre-inflationary phase.}.

The K\"ahler potential is of the non-minimal
form~\cite{Senoguz:2004ky,Antusch:2008gw}
\begin{eqnarray}
 K &=& |S|^2 + |\phi|^2 +  \kappa_2 \frac{|S|^2 |\phi|^2}{M_P^2} +
 \kappa_1\frac{|\phi|^4}{4 M_{Pl}^2} + \kappa_3\frac{|S|^4}{4 M_{Pl}^2} + ...
\label{eq:inf_flav_A4}
\end{eqnarray}
Along the $D$-flat direction, the VEVs are
$\phi_1=\phi_2=\phi_3\ll M\ll  M_{\rm P}$. One can further make an assumption such 
that $\kappa_3 <-1/3$, so that the $S$ field becomes heavy compared to the
Hubble expansion rate, and therefore the field relaxes to its
minimum VEV, $S=0$, within one e-folding. By assuming
$|\phi_i|=\varphi/\sqrt{2}$, $\beta = (\kappa_2 -1)$, $\lambda =
(\beta(\beta+1)+1/2+\kappa_1/12)$ and $\gamma
=2/(6)^{3n/2}\lesssim 0.14$, the potential along $\varphi$ is
given by~\cite{Antusch:2008gw}
\begin{equation}
V \simeq \mu^4 \left[ 1 - \frac{\beta}{2} \frac{\varphi^2}{ M_{P}^2} +
\frac{\lambda}{4}\frac{\varphi^4}{ M_{P}^4} - \gamma
\frac{\varphi^{3n}}{M^{3n}} + \cdots\right]\,.
\end{equation}
One needs to suppress the quartic term in the potential, i.e.
$\varphi^4$ term, otherwise for $|\gamma{\varphi^{3n}}/{M^{3n}}| \ll
|({\lambda}/{4})({\varphi^4}/{ M_{\rm P}^4})|$,  it turns out
that $M\geq M_{\rm P}$. For a specific parameter space, $M\approx
10^{15}-10^{16}$~GeV and $\mu\approx 10^{13}-10^{14}$~GeV and
$n=2,3,4$ and $\beta \leq 0.03$, it is possible to match the
amplitude of the perturbations and the spectral tilt within
$n_s=0.96\pm0.014$ for $N=60$ e-foldings of inflation. The model
requires pre-inflationary phase to set the initial conditions for
$\varphi\approx 0$ to be realizable.

Another example of flavon  is  to consider a vacuum alignment
potential as studied in the SU(3) family symmetry model of
\cite{Antusch:2008gw,deMedeirosVarzielas:2005ax}. It was assumed
that $<\phi_{23} > \propto (0,1,1)^T$ and $<\Sigma >=
\mbox{diag}(a,a,-2a)$ are already at their minima, and that the
relevant part of the superpotential which governs the final step
of family symmetry breaking is given
by~\cite{deMedeirosVarzielas:2005ax}
\begin{eqnarray}
W = \kappa S (\bar\phi_{123} \phi_{123} - M^2) + \kappa' Y_{123} \bar\phi_{23} \phi_{123}
+ \kappa'' Z_{123} \bar\phi_{123} \Sigma \phi_{123} + ... \,.
\end{eqnarray}
where a singlet, $S$, is the driving superfield for the flavon
$\phi_{123}$ and the non-minimal K\"ahler potential is given by:
\begin{eqnarray}
K&=&
|S|^2 + |\phi_{123}|^2 + |\bar\phi_{123}|^2 + |Y_{123}|^2 + |\bar\phi_{23}|^2 + |\phi_{23}|^2 + |Z_{123}|^2 + |\Sigma|^2 \nonumber \\
&&
+ \kappa_S\frac{|S|^4}{4 M_{\rm P}^2}
+ \kappa_{SZ}\frac{|S|^2 |Z_{123}|^2}{4 M_{\rm P}^2}
+ ...\,.
\end{eqnarray}
During inflation the fields with larger VEVs
$Y_{123},\,\phi_{123},\,\bar\phi_{123}$ do not evolve, the
inflationary potential is dominated by
\begin{equation}
V =  \kappa^2 M^4\,   \left[ 1 -  \gamma \frac{\xi^2}{2M_{\rm P}^2}
- 2\kappa_{S} \frac{\sigma^2}{2 M_{\rm P}^2} + ...\right],
\end{equation}
where $|S| = \sigma/\sqrt{2}$, $|Z_{123}| = \xi/\sqrt{2}$ and $\gamma = \kappa_{SZ}-1$.
Inflation can happen if the coefficients in front of the mass terms for $\sigma$
and/or $\xi$ are sufficiently small.

The inflaton could be $\sigma$ if $\gamma < -1/3$, the mass of
$\xi$ becomes heavy compared to the Hubble scale during inflation.
For $\kappa_{S} \approx (0.005 - 0.01)$ and $\kappa \approx (0.001
- 0.05)$, the spectral index is consistent with the current data,
$n_s = 0.96 \pm 0.014$~\cite{BasteroGil:2006cm}. Finally, the
scale, $M\sim 10^{15}$~GeV, of family symmetry breaking along the
$\langle\phi_{123}\rangle$-direction will be  determined by the temperature anisotropy in the CMB.

So far, as shown in details regarding $F$ and $D$-term hybrid inflation models, all of them 
have to assume an existence of a hidden sector physics, an extra gauge singlet playing the role of 
the inflaton field, whose mass and couplings are constrained only from the CMB data. There are  of course 
sneutrino driven models of inflation which employ a SM gauge singlet with an additional motivation of
connecting inflationary sector to the neutrino physics. However, it is desirable to seek models of inflation
which are truly embedded within an observable sector particle physics. Such models are based on
beyond the SM physics within a robust framework, where the shape of the potential, interactions and various parameters are well motivated from the low energy particle physics point of view. Furthermore, such an observable sector
model of inflation can be directly put to the test by both LHC and CMB data from PLANCK.

\section{MSSM gauged inflatons}\label{sec:mssminflation}
%
%

\subsection{Inflation due to MSSM flat directions}
\label{MI}

Since MSSM introduces so many squarks and sleptons, one obvious question is why can't they be an inflaton?
Indeed the squarks and sleptons, being light compared to the high scale model of inflation, can also be displaced away from their 
minimum. However, since they do not minimize $F$-and $D$-terms of the total potential, they cost more energetically as compared 
to a {\it gauge invariant} combination of squarks and sleptons. In the SUSY limit both $F$ and $D$-contributions
are vanishing for {\it gauge invariant} flat directions, which maintain their $D$-flatness, but they can be lifted by
the $F$-term contribution away from the point of {\it enhanced symmetry}.

A simple observation has been made in \cite{Allahverdi:2006iq,Allahverdi:2006cx,Allahverdi:2006we},
where the inflaton properties are directly related to the soft SUSY breaking mass term and the A-term.
In the limit of unbroken SUSY the flat directions have exactly vanishing potential. This situation changes
when soft SUSY breaking and non-renormalizable superpotential terms of
the type~\cite{Enqvist:2003gh,Dine:2003ax}
\begin{equation} \label{supot}
W_{non} = \sum_{n>3}{\lambda_n \over n}{\Phi^n \over M^{n-3}}\,,
\end{equation}
are included. Here $\Phi$ is a {\it gauge invariant} superfield which
contains the flat direction.  Within MSSM all the flat directions are
lifted by non-renormalizable operators with $4\le n\le 9$~\cite{Gherghetta:1995dv},
where $n$ depends on the flat direction. 
Let us focus on the lowest order superpotential term in
Eq.~(\ref{supot}) which lifts the flat direction. Softly broken SUSY
induces a mass term for $\phi$ and an $A$-term so that the scalar
potential along the flat direction reads~\cite{Allahverdi:2006iq,Allahverdi:2006cx}
\begin{equation} \label{scpot}
V = {1\over2} m^2_\phi\,\phi^2 + A\cos(n \theta  + \theta_A)
{\lambda_{n}\phi^n \over n\,M^{n-3}_{\rm P}} + \lambda^2_n
{{\phi}^{2(n-1)} \over M^{2(n-3)}_{\rm P}}\,,
\end{equation}
Here $\phi$ and $\theta$ denote respectively the radial and the
angular coordinates of the complex scalar field
$\Phi=\phi\,\exp[i\theta]$, while $\theta_A$ is the phase of the
$A$-term (thus $A$ is a positive quantity with dimension of mass).
Note that the first and third terms in Eq.~(\ref{scpot}) are positive
definite, while the $A$-term leads to a negative contribution along
the directions whenever $\cos(n \theta + \theta_A) < 0$. 


\subsubsection{Inflaton candidates}
\label{INFCA}

As discussed in~\cite{Allahverdi:2006iq,Allahverdi:2006cx,Allahverdi:2006we,Allahverdi:2008bt}, among nearly 
300 flat directions~\cite{Gherghetta:1995dv}, there are two that can lead to a successful inflation along the lines
discussed above.

One is $udd$ , up to an overall phase factor, which is parameterized by:
\begin{eqnarray}
\label{example}
u^{\alpha}_i=\frac1{\sqrt{3}}\phi\,,~
d^{\beta}_j=\frac1{\sqrt{3}}\phi\,,~
d^{\gamma}_k=\frac{1}{\sqrt{3}}\phi\,.
\end{eqnarray}
Here $1 \leq \alpha,\beta,\gamma \leq 3$ are color indices, and $1
\leq i,j,k \leq 3$ denote the quark families. The flatness constraints
require that $\alpha \neq \beta \neq \gamma$ and $j \neq k$.

The other direction is $LLe$, parameterized by (again up to
an overall phase factor)
\begin{eqnarray}
L^a_i=\frac1{\sqrt{3}}\left(\begin{array}{l}0\\ \phi\end{array}\right)\,,~
L^b_j=\frac1{\sqrt{3}}\left(\begin{array}{l}\phi\\ 0\end{array}\right)\,,~
e_k=\frac{1}{\sqrt{3}}\phi\,,
\end{eqnarray}
where $1 \leq a,b \leq 2$ are the weak isospin indices and $1 \leq
i,j,k \leq 3$ denote the lepton families. The flatness constraints
require that $a \neq b$ and $i \neq j \neq k$.  Both these flat
directions are lifted by $n=6$ non-renormalizable
operators~\cite{Gherghetta:1995dv},
\begin{eqnarray}\label{suppot-1}
W_6\supset\frac{1}{M_{\rm P}^3}(LLe)(LLe)\,,\hspace{1cm}
W_6\supset \frac{1}{M_{\rm P}^3}(udd)(udd)\,.
\end{eqnarray}
The reason for choosing either of these two flat
directions is twofold: 

\begin{enumerate}
\item{\it within MSSM, a non-trivial $A$-term arises, \underline{at the lowest order}, only at $n=6$, and}
\item {\it we wish to obtain the correct COBE normalization of the CMB
spectrum.}

\end{enumerate}

Since $LLe$ and $udd$ are independently $D$- and
$F$-flat, inflation could take place along any of them but also, at
least in principle, simultaneously. The dynamics of multiple flat
directions are however quite involved~\cite{Enqvist:2003pb,Jokinen:2004bp}. 

Those MSSM flat directions which are lifted by operators with
dimension $n=7,9$ are such that the superpotential term contains at
least two monomials, i.e. is of the type~\cite{Dine:1995kz,Dine:1995uk,Gherghetta:1995dv}
\begin{eqnarray}\label{doesnotcontri}
W \sim \frac{1}{M_{\rm P}^{n-3}}\Psi\Phi^{n-1}\,.
\end{eqnarray}
If $\phi$ represents the flat direction, then its VEV induces a large
effective mass term for $\psi$, through Yukawa couplings, so that
$\langle \psi \rangle =0$. Hence Eq. (\ref{doesnotcontri}) does not
contribute to the $A$-term.

The scalar component of $\Phi$ superfield, denoted by $\phi$, is given by
\begin{equation} 
\phi = {{u} + {d} + {d} \over \sqrt{3}} ~ ~ ~ , ~ ~ ~ \phi = {{L} + {L} + {e} \over \sqrt{3}},
\end{equation}
for the $udd$ and $LLe$ flat directions respectively.

After minimizing the potential along the angular direction $\theta$, we can situate the real 
part of $\phi$ by rotating it to the corresponding angles $\theta_{\rm min}$. The scalar 
potential is then found to be~\cite{Allahverdi:2006iq,Allahverdi:2006cx,Allahverdi:2008bt}
\begin{equation} 
\label{scalar-pot}
V(\phi) = {1\over2} m^2_\phi\, \phi^2 - A {\lambda\phi^6 \over 6\,M^{6}_{\rm P}} + \lambda^2
{{\phi}^{10} \over M^{6}_{\rm P}}\,,
\end{equation}
where $m_\phi$ and $A$ are the soft breaking mass and the $A$-term respectively 
($A$ is a positive quantity since its phase is absorbed by a 
redefinition of $\theta$ during the process).


\subsubsection{Inflection point inflation}
\label{INPI}

Provided that
\begin{equation} \label{dev}
{A^2 \over 40 m^2_{\phi}} \equiv 1 + 4 \alpha^2\, ,
\end{equation}
where $\alpha^2 \ll 1$, there exists a point of inflection in $V(\phi)$
\begin{eqnarray}
&&\phi_0 = \left({m_\phi M^{3}_{\rm P}\over \lambda \sqrt{10}}\right)^{1/4} + {\cal O}(\alpha^2) \, , \label{infvev} \\
&&\, \nonumber \\
&&V^{\prime \prime}(\phi_0) = 0 \, , \label{2nd}
\end{eqnarray}
at which
\begin{eqnarray}
\label{pot}
&&V(\phi_0) = \frac{4}{15}m_{\phi}^2\phi_0^2 + {\cal O}(\alpha^2) \, , \\
\label{1st}
&&V'(\phi_0) = 4 \alpha^2 m^2_{\phi} \phi_0 \, + {\cal O}(\alpha^4) \, , \\
\label{3rd}
&&V^{\prime \prime \prime}(\phi_0) = 32\frac{m_{\phi}^2}{\phi_0} + {\cal O}(\alpha^2) \, .
\end{eqnarray}
From now on we only keep the leading order terms in all expressions. Note that in gravity-mediated SUSY breaking, the $A$-term and the soft SUSY breaking mass are of the same order of magnitude as the gravitino mass, i.e. $m_{\phi} \sim A \sim m_{3/2} \sim (100~{\rm GeV}-1~{\rm TeV})$. Therefore the condition in Eq.~(\ref{dev}) can indeed be satisfied. We then have $\phi_0 \sim {\cal O}(10^{14}~{\rm GeV})$. Inflation occurs within an interval
\begin{equation} \label{plateau}
\vert \phi - \phi_0 \vert \sim {\phi^3_0 \over 60 M^2_{\rm P}} ,
\end{equation}
in the vicinity of the point of inflection, within which the slow-roll parameters $\epsilon \equiv (M^2_{\rm P}/2)(V^{\prime}/V)^2$ and $\eta \equiv M^2_{\rm P}(V^{\prime \prime}/V)$  are smaller than $1$. The Hubble expansion rate during inflation is given by
\begin{equation} \label{hubble}
H_{\rm MSSM} \simeq \frac{1}{\sqrt{45}}\frac{m_{\phi}\phi_0}{M_{\rm P}}
\sim (100~{\rm MeV}-1~{\rm GeV})\,.
\end{equation}
The amplitude of density perturbations $\delta_H$ and the scalar spectral index $n_s$ are given 
by~\cite{Allahverdi:2006iq,Allahverdi:2006cx,BuenoSanchez:2006xk,Allahverdi:2006wt}:
\begin{equation} \label{ampl}
\delta_H = {8 \over \sqrt{5} \pi} {m_{\phi} M_{\rm P} \over \phi^2_0}{1 \over \Delta^2}
~ {\rm sin}^2 [{N}_{Q}\sqrt{\Delta^2}]\,, \end{equation}
and
\begin{equation} \label{tilt}
n_s = 1 - 4 \sqrt{\Delta^2} ~ {\rm cot} [{N}_{Q}\sqrt{\Delta^2}], 
\end{equation}
respectively where
\begin{equation} \label{Delta}
\Delta^2 \equiv 900 \alpha^2 {\cal
N}^{-2}_{Q} \Big({M_{\rm P} \over \phi_0}\Big)^4\,. \end{equation}
${ N}_{Q}$ is the number of e-foldings between the time when the observationally 
relevant perturbations are generated till the end of inflation and follows: ${N}_{Q} \simeq 66.9 + (1/4) {\rm ln}({V(\phi_0)/ M^4_{\rm P}}) \sim 50$, provided that the universe is immediately thermalized after the end of inflation~\cite{Burgess:2005sb,Liddle:2003as}.
We note that reheating after MSSM inflation is very fast, due to gauge couplings of the inflaton to gauge/gaugino fields, and results in a radiation-dominated universe within few Hubble times after the end of inflation~\cite{Allahverdi:2006iq,Allahverdi:2006cx}.


\subsubsection{Parameter space for MSSM inflation}

\begin{figure}
\vspace*{-0.0cm}
\begin{center}
\includegraphics[width=9.0cm]{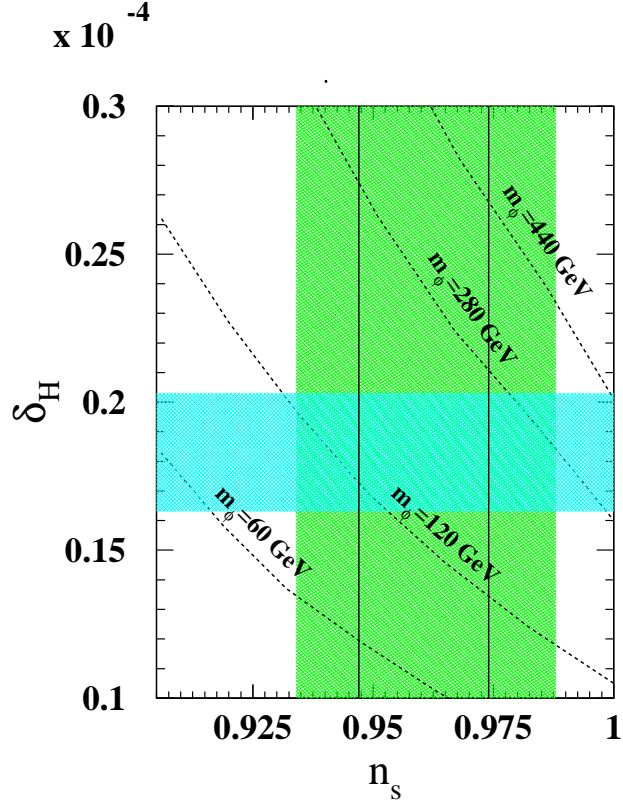}
\caption{$n_s$ is plotted as a function of $\delta_H$ for different
values of $m_{\phi}$. The $2\sigma$ region for $\delta_H$ is shown by the blue horizontal band and the $2\sigma$ 
allowed region of $n_s$ is shown by the vertical green band. The $1\sigma$ allowed region of $n_s$ is within the 
solid vertical lines. We choose $\lambda =1$~\cite{Allahverdi:2008bt}.} \label{nsdel0}
\end{center}
\end{figure}


A remarkable property of MSSM inflation, which is due to inflation occurring near a point of inflection, is that it can give rise to a wide range of scalar spectral index. This is in clear distinction with other models (for example, chaotic inflation, hybrid inflation, natural inflation, etc.) and makes the model very robust. Indeed it can yield a spectral index within the whole $2 \sigma$ allowed range by 5-year WMAP data $0.934 \leq n_s \leq 0.988$. Note that for  $\alpha^2 = 0$, Eqs.~(\ref{ampl},\ref{tilt}) are
reduced to the case of a saddle point inflation, for which the spectral index
is strictly $0.92$, for details see~\cite{Allahverdi:2006we,Allahverdi:2008bt,BuenoSanchez:2006xk}~\footnote{Varying range of spectral tilt is in concordance with the statistical nature of the vacua at low energies below the cut-off. MSSM harbors a 
mini-landscape~\cite{Allahverdi:2008bt} with a moduli space of $37$ complex dimensions~\cite{Gherghetta:1995dv}, which has more than $700$ gauge invariant monomials~\cite{Basboll:2009tz}. Although, its much smaller compared to the string landscape, but one would naturally expect a distribution of discrete values of non-renormalizable $A$-terms along with the soft breaking terms. This would  
inevitably give rise to many realizations of our universe with varied range of spectral tilts. }.

For $\alpha^2 < 0$, the spectral index will be smaller than the $0.92$, which is more than $3 \sigma$ away
from observations. The more interesting case, as pointed out
in~\cite{Allahverdi:2008bt,BuenoSanchez:2006xk}, happens for $\alpha^2 > 0$. This happens for
\begin{equation} \label{Delta2}
2 \times 10^{-6} \leq \Delta^2 \leq 5.2 \times 10^{-6}\,. \end{equation}
In Fig.~(\ref{nsdel0}), we have shown $\delta_H$ as a function of $n_s$ for
different values of $m_{\phi}$. The horizontal blue band shows the $2 \sigma$ experimental band for $\delta_H$. The vertical green shaded region is the 2$\sigma$ experimental band for $n_s$. The region enclosed by  solid lines shows the 1$\sigma$ experimental allowed region. Smaller values of $m_{\phi}$ are preferred for smaller values of $n_s$. Note
that the allowed range of $m_{\phi}$ is $90-330$~GeV for the
experimental ranges of $n_s$ and $\delta_H$. This figure is drawn for $\lambda \simeq
1$, which is natural in the context of effective field theory (unless it is suppressed due to some symmetry). 
Smaller values of $\lambda$ will lead to an increase in $m_{\phi}$, see Fig.~(\ref{lamcon}), where we have plotted
$(n_s,m_{\phi})$ for different values of $\lambda$ for the case of $udd$ direction with different masses of 
$m_{u},~m_{d}$, and fixed value of $\tan(\beta)$. We will explain the blue bands when we discuss the parameter space of MSSM inflation at low energies close to the dark matter production scale~\cite{Allahverdi:2007vy,Allahverdi:2008bt}.

\begin{figure}[t]
\vspace{1cm}
\includegraphics[width=8.0cm]{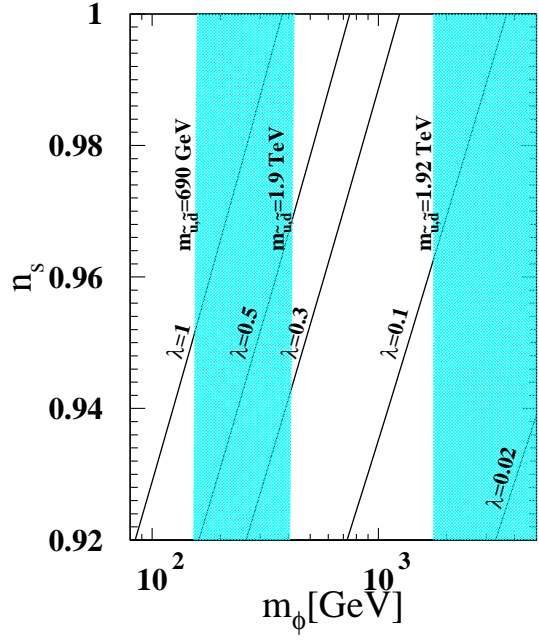}
\caption{Contours of $\lambda$  for $\delta_H=1.91\times 10^{-5}$ in
the $n_s$-$m_{\phi}$ plane. The blue band on the left is due to the
stau-neutralino coannihilation region for $\tan\beta=10$ and the
blue band on the right (which continues beyond the plotting range)
denotes the focus point region~\cite{Allahverdi:2007vy}. } \label{lamcon}
\end{figure}



\subsubsection{Embedding MSSM inflation in $SU(5)$ or $SO(10)$ GUT}

By embedding MSSM inflation in GUT makes a mild assumption that there
exists new physics which encompasses MSSM beyond the unification
scale $M_{\rm G}$. We remind the readers that inflation
occurs around a flat direction VEV $\phi_0 \sim 10^{14}$ GeV. Since
$\phi_0 \ll M_{\rm G}$, heavy GUT degrees of freedom play no role in
the dynamics of MSSM inflation, and hence they can be ignored. Here
we wish to understand how such embedding would affect inflationary
scenario.

It is generically believed that gravity breaks global symmetries~\cite{Kallosh:1995hi}.
Then all {\it gauge invariant} terms which are $M_{\rm P}$
suppressed should appear with $\lambda \sim {\cal O}(1)$. Obviously
the above terms in Eq.~(\ref{suppot-1}) are invariant under the SM.
Once the SM is embedded within a GUT at the scale $M_{\rm G}$, where
gauge couplings are unified, the gauge group will be enlarged. Then
the question arises whether such terms in Eq.~(\ref{suppot-1}) are
invariant under the GUT gauge group or not. Note that a GUT singlet
is also a singlet under the SM, however, the vice versa is not
correct. 

\begin{itemize}

\item{\underline{\bf $SU(5)$}:\\
We briefly recollect representations of matter fields in this case:
$L$ and $d$ belong to ${\bf {\bar 5}}$, while $e$ and $u$ belong to
${\bf 10}$ of $SU(5)$ group. Thus under $SU(5)$ the superpotential
terms in Eq.~(\ref{suppot-1}) read~\cite{Allahverdi:2007vy}
\begin{equation} 
\label{su5}
{{\bf \bar {5}} \times {\bf \bar {5}}
\times {\bf 10} \times {\bf \bar {5}} \times {\bf \bar {5}} \times {\bf 10}
\over M^3_{\rm P}}.
\end{equation}
This product clearly includes a $SU(5)$ singlet. Therefore in the case
of $SU(5)$, we expect that $M_{\rm P}$ suppressed terms as in
Eq.~(\ref{supot}) appear with $\lambda \sim {\cal O}(1)$. If
we were to obtain the $(LLe)^2$ term by integrating out the heavy
fields of the $SU(5)$ GUT, then $\lambda=0$. This is due to the fact
that $SU(5)$ preserves $B-L$.}\\

\item{\underline{\bf $SO(10)$}:\\
In this case all matter fields of one generation are included in the
spinorial representation ${\bf 16}$ of $SO(10)$. Hence the
superpotential terms in Eq.~(\ref{supot}) are $[{\bf 16}]^6$ under
$SO(10)$, which does not provide a singlet. A {\it gauge invariant}
operator will be obtained by multiplying with a $126$-plet Higgs.
This implies that in $SO(10)$ the lowest order {\it gauge invariant}
superpotential term with $6$ matter fields arises at $n=7$ level:
\begin{equation}
\label{so10}
{{\bf 16} \times {\bf 16} \times {\bf 16} \times {\bf 16} \times {\bf 16}
\times {\bf 16} \times {\bf 126}_H \over M^4_{\rm P}}\,.
\end{equation}
Once ${\bf 126}_H$ acquires a VEV, $S0(10)$ can break down to a lower
ranked subgroup, for instance $SU(5)$. This will induce an effective
$n=6$ non-renormalizable term as in Eq.~(\ref{supot}) with
\begin{equation}
\label{solam}
\lambda \sim \frac{\langle {\bf 126}_H
\rangle}{M_{\rm P}} \sim \frac{{\cal O}(M_{\rm GUT})}{M_{\rm P}}\,.
\end{equation}
Hence, in the case of $SO(10)$, we can expect $\lambda \sim {\cal
O}(10^{-2}- 10^{-1})$ depending on the scale where SO(10) gets
broken.}\\

\end{itemize}


By embedding MSSM in $SO(10)$ naturally implies
that $\lambda \ll 1$. Smaller values of $n_s$ (within the range
$0.92 \leq n_s \leq 1$) point to smaller $\lambda$, as can be seen
from  figure 6. This, according to Eq.~(\ref{solam}), implies a
scale of $SO(10)$ breaking, i.e. $\langle {\bf 126}_H \rangle$,
which is closer to the GUT scale. Further note that embedding the MSSM 
within $SO(10)$ also provides an advantage for obtaining a right handed neutrino.
It was concluded in Ref.~\cite{Allahverdi:2007vy} that if we include the RH neutrinos, then 
$udd$ direction is preferred over $LLe$.


\subsubsection{Gauged inflaton in $SM\times U(1)_{B-L}$}

If we augment MSSM with three right-handed (RH) neutrino multiplets, then
it is possible to realize neutrinos of Dirac type with an appropriate Yukawa 
coupling. Whether the nature of neutrino is
Dirac or Majorana can be determined in the future neutrinoless
double beta decay experiment. 

For various reasons, which will become clear, the inclusion of a
gauge symmetry under which the RH (s)neutrinos are not singlet is
crucial. As far as inflation is concerned, a singlet RH sneutrino
would not form a gauge-invariant inflaton along with the Higgs and
slepton fields. The simplest extension of the SM
symmetry, $SU(3)_C\times SU(2)_L\times U(1)_Y\times U(1)_{B-L}$, 
which is also well motivated: anomaly cancelation requires
that three RH neutrinos exist, so that they pair with LH neutrinos
to form three Dirac fermions.

The relevant superpotential term is
\begin{equation}\label{supotN} 
W  \supset h {N} {H}_u { L} . 
\end{equation}
Here ${ N}$, ${ L}$ and ${H}_u$ are superfields containing
the RH neutrinos, left-handed (LH) leptons and the Higgs which gives
mass to the up-type quarks, respectively.

In this model there is an extra $Z$ boson ($Z^{\prime}$) and one
extra gaugino ($\tilde Z'$). The $U(1)_{B - L}$ gets broken around
TeV by new Higgs fields with ${B - L} = \pm 1$.  This also prohibits
a Majorana mass for the RH neutrinos at the renormalizable level
(note that ${N N}$ has ${B - L} = 2$). The Majorana mass can be
induced by a non-renormalizable operator, but it will be very small.

\begin{figure}
\includegraphics[width=9cm]{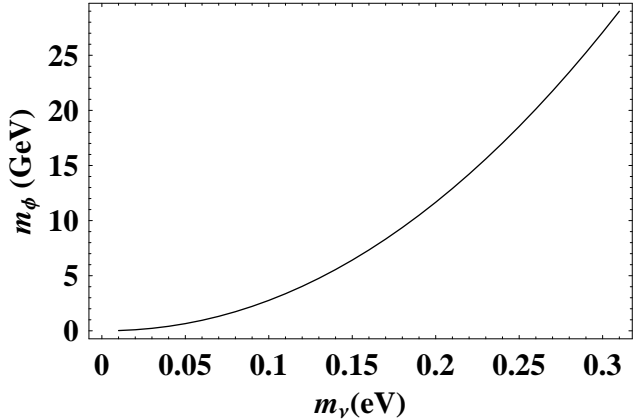}
\caption{The inflaton mass $m_{\phi}$ is plotted as a function of
the neutrino mass $m_{\nu}$~\cite{Allahverdi:2007wt}.} \label{lneutrinoinf}
\end{figure}


The value of $h$ needs to be small, i.e. $h \leq 10^{-12}$, in order
to explain the light neutrino mass, $\sim {\cal O}(0.1~{\rm eV})$
corresponding to the atmospheric neutrino oscillations detected by
Super-Kamiokande experiment. Note that the ${N} {H}_u {L}$ monomial represents a
$D$-flat direction under the $U(1)_{B-L}$, as well as the SM gauge
group~\cite{Allahverdi:2006cx,Allahverdi:2007wt}.
\begin{equation} \label{flat} {\phi} = {{\tilde N} + {H}_u + {\tilde L} \over
\sqrt{3}} , 
\end{equation}
where ${\tilde N}$, ${\tilde L}$, $H_u$ are the scalar components of
corresponding superfields. Since the RH sneutrino ${\tilde N}$ is a
singlet under the SM gauge group, its mass receives the smallest
contribution from quantum corrections due to SM gauge interactions,
and hence it can be set to be the lightest SUSY particle
(LSP). Therefore the dark matter candidate arises from the RH
sneutrino component of the inflaton, see Eq.~(\ref{flat}). The
potential along the flat direction, after the minimization along the
angular direction, is found to be~\cite{Allahverdi:2006cx,Allahverdi:2007wt},
\begin{eqnarray} \label{flatpot}
V (\vert \phi \vert) = \frac{m^2_{\phi}}{2} \vert \phi \vert ^2 +
\frac{h^2}{12} \vert \phi \vert^4 \, - \frac{A h}{6\sqrt{3}}
\vert \phi \vert^3 \,.
\end{eqnarray}
The flat direction mass $m_{\phi}$ is
given in terms of ${\tilde N},~H_u,~{\tilde L}$ masses:
$m^2_{\phi} = {m^2_{\tilde N} + m^2_{H_u} +
m^2_{\tilde L} \over 3}$.
For $A \approx 4 m_{\phi}$, there exists an {\it inflection point} for
which $V^{\prime}(\phi_0) \neq 0, V^{\prime \prime}(\phi_0) = 0$, where
inflation takes place
\begin{eqnarray} \label{sad} \phi_0 = \sqrt{3}\frac{m_{\phi}}{h}=
6 \times 10^{12} ~ m_{\phi} ~ \Big({0.05 ~
{\rm eV} \over m_\nu} \Big)\,,~~
V(\phi_0) = \frac{m_{\phi}^4}{4h^2}=3 \times 10^{24} ~ m^4_{\phi} ~
\Big({0.05 ~ {\rm eV} \over m_\nu} \Big)^2 \,.\nonumber\\
\label{sadpot}
\end{eqnarray}
Here $m_\nu$ denotes the neutrino mass which is given by $m_\nu = h
\langle H_u \rangle$, with $\langle H_u \rangle \simeq 174$ GeV. For
neutrino masses with a hierarchical pattern, the largest neutrino
mass is $m_\nu \simeq 0.05$ eV in order to explain the atmospheric
neutrino oscillations~\cite{Strumia:2006db}, while the current upper bound on
the sum of the neutrino masses from cosmology, using WMAP and SDSS
data alone, is $0.94$ eV~\cite{Tegmark:2006az}.

The amplitude of density perturbations 
follows~\cite{Allahverdi:2006cx,Allahverdi:2007wt}.
\begin{equation} \label{amp} \delta_{H} \simeq
\frac{1}{5\pi}\frac{H^2_{inf}}{\dot\phi} \simeq 3.5 \times 10^{-27}
~ \Big( {m_\nu \over 0.05 ~ {\rm eV}} \Big)^2 ~  \Big({M_{\rm P}
\over m_{\phi}} \Big) ~ {N}_{Q}^2\,. \end{equation}
Here $m_{\phi}$ denotes the loop-corrected value of the inflaton
mass at the scale $\phi_0$ in
Eqs.~(\ref{sad},\ref{amp}). In Fig.~(\ref{lneutrinoinf}), we 
have shown the neutrino mass as a function of the inflaton mass
for $\delta_H =1.91 \times 10^{-5}$. We see that the neutrino mass
in the range $0$ to $0.30$~eV corresponds to the inflaton mass of
$0$ to $30$~ GeV. The spectral tilt as usual has a range of values 
$0.90 \leq n_s\leq 1.0$ ~\cite{Allahverdi:2006cx,Allahverdi:2007wt}.


\subsubsection{Inflection point inflation in gauge mediation}

In GMSB the two-loop correction to the flat direction potential
results in a logarithmic term above the messenger scale, i.e. $\phi >
M_S$~\cite{deGouvea:1997tn,Giudice:1998bp,Enqvist:2003gh,Dine:2003ax}. Together 
with the $A$-term this leads to the scalar potential~\cite{Allahverdi:2006ng}
\begin{eqnarray} \label{GMSBscpot}
V = M_{F}^4\ln\left(\frac{\phi^2}{M_{S}^2}\right) + 
A\cos(n \theta  + \theta_A)
\frac{\lambda_{n}\phi^n}{n\,M^{n-3}_{\rm GUT}} + \lambda^2_n 
\frac{{\phi}^{2(n-1)}}{M^{2(n-3)}_{\rm GUT}}\,,
\end{eqnarray}
where $M_F \sim (m_{SUSY} \times M_S)^{1/2}$ and $m_{SUSY} \sim 1$~TeV
is the soft SUSY breaking mass at the weak scale. For $\phi >
M^2_F/m_{3/2}$, usually the gravity mediated contribution, $m^2_{3/2}
\phi^2$, dominates the potential where $m_{3/2}$ is the gravitino
mass.  Here we will concentrate on the VEVs  such that 
$M_s \ll \phi \leq  M^2_F/m_{3/2}$. A
successful inflation can be obtained near the point of inflection;
\begin{eqnarray}
\phi_0 &\approx& \left( \frac{M^{n-3}_{\rm GUT} M_F^2}{\lambda_n}
\sqrt{\frac{n}{(n-1)(n-2)}} \right)^{1/(n-1)}\,, \\ 
A &\approx & \frac{4(n-1)^2
\lambda_n}{n M^{n-3}_{\rm GUT}} \phi_0^{n-2}\,. \label{A2}
\end{eqnarray}
In the vicinity of the inlection point, the total energy
density is given by
\begin{eqnarray}
V(\phi_0) \approx M_F^4 \left[ \ln\left(\frac{\phi_0^2}{M_S^2} \right) -
  \frac{3n-2}{n(n-1)} \right]\,, 
\end{eqnarray}
There are couple of interesting points, first of all note that the
scale of inflation is extremely low, the Hubble scale during inflation is given
by: $H_{\rm inf}\sim {M_{F}^2}/{M_{\rm P}} \sim 10^{-3}-10^{-1}~{\rm eV} $
for $M_F \sim 1-10$ TeV.  The relevant number of e-foldings is 
${N}_{Q}\sim 40$~\cite{Allahverdi:2006ng}. For $M_F \sim 10$~TeV, and 
 $\phi_0\sim 10^{11}$~GeV it is possible to match the CMB temperature anisotropy and 
 the required tilt in the spectrum~\cite{Allahverdi:2006ng}.

The validity of Eq.~(\ref{GMSBscpot}) for such a large VEV
requires that $M^2_F > (10^{11}~{\rm GeV}) \times m_{3/2}$. For 
$M_F\sim 10$ TeV this yields a bound on the gravitino mass, $m_{3/2} < 1-10$ MeV, which 
is compatible with the warm dark matter constraints.


\subsection{Quantum stability}

\subsubsection{Radiative correction}\label{Rad-correct-MSSM-pot}

Since the MSSM inflaton candidates are represented by 
{\it gauge invariant} combinations of squarks and sleptons, the inflaton parameters 
receive corrections from gauge interactions which can be computed in a straightforward way.
Quantum corrections result in a logarithmic running of the soft
SUSY breaking parameters $m_\phi$ and $A$. The effective potential
at phase minimum $n\theta_{\min} = \pi$ is then given by~\cite{Allahverdi:2006we}:
\begin{eqnarray}
V_{eff}(\phi, \theta_{min}) &=& \frac{1}{2} m_0^2 \phi^2 \left[
  1+ K_1 \log \left(\frac{\phi^2}{\mu_0^2}\right) \right] -
  \frac{\lambda_{n,0} A_0}{nM^{n-3}} \phi^n \left[ 1+ K_2
  \log\left(\frac{\phi^2}{\mu_0^2}\right) \right] \nonumber \\ & & +
  \frac{\lambda_{n,0}^2}{M^{2(n-3)}} \phi^{2(n-1)} \left[ 1 + K_3 \log
  \left(\frac{\phi^2}{\mu_0^2}\right) \right]\,.
\end{eqnarray}
where $m_0$, $A_0$, and $\lambda_{n,0}$ are the values of $m_{\phi}$,
$A$ and $\lambda_n$ given at a scale $\mu_0$. Here $A_0$ is chosen to
be real and positive (this can always be done by re-parameterizing the
phase of the complex scalar field $\phi$), and $|K_i|<1$ are
coefficients determined by the one-loop renormalization group
equations.

In the limit when $|K_i| \ll 1$, one finds a simple relationship~\cite{Allahverdi:2006we}
\begin{eqnarray} 
A^2 = 8(n-1) m_{\phi}^2(\phi_0)
\left( 1+ K_1 - \frac{4}{n} K_2 + \frac{1}{n-1} K_3 \right)\,,\\
\phi_0^{n-2} = \frac{M^{n-3} m_{\phi}(\phi_0)}{\lambda_n \sqrt{2(n-1)}}
\left( 1+\frac{1}{2} K_1 - \frac{1}{2(n-1)} K_3 \right)\,.
\end{eqnarray}
For $n=6$, the coefficient is $A^2\sim 40 m_{\phi}^2$, where $\phi_0$
denotes the point of inflection. The coefficients $K_i$ need to be solved from
the renormalization group equations at the inflationary scale $\mu=\phi_0$. 
Since $K_i$ are already one loop corrections, taking the tree-level 
value as the renormalization scale is sufficient~\footnote{In general there is 
no prospect of measuring the non-renormalizable $A_6$ term, because 
interactions are suppressed by $M_{\rm P}$. However, a knowledge of SUSY breaking sector
and its communication with the observable sector may help to link the
non-renormalizable $A$-term under consideration to the renormalizable
ones. In the case of a Polonyi model where a general
$A$-term at a tree level is given by: $m_{3/2}[(a-3)W+\phi (dW/d\phi)],$
with $a=3 - \sqrt{3}$~\cite{Nilles:1983ge,Haber:1984rc}. One then finds a relationship
between $A$-terms corresponding to $n=6$ and $n=3$ superpotential
terms, denoted by $A_6$ and $A_3$ respectively, at high scales~\cite{Allahverdi:2006we}:
$A_6={3 - \sqrt{3} \over 6 - \sqrt{3}} A_3$.}.

{\it The radiative corrections do
not remove the inflection point nor shift it to unreasonable values. The
existence of an inflection point is thus insensitive to radiative
corrections. }


\begin{figure}
\vspace*{-0.0cm}
\begin{center}
\epsfig{figure=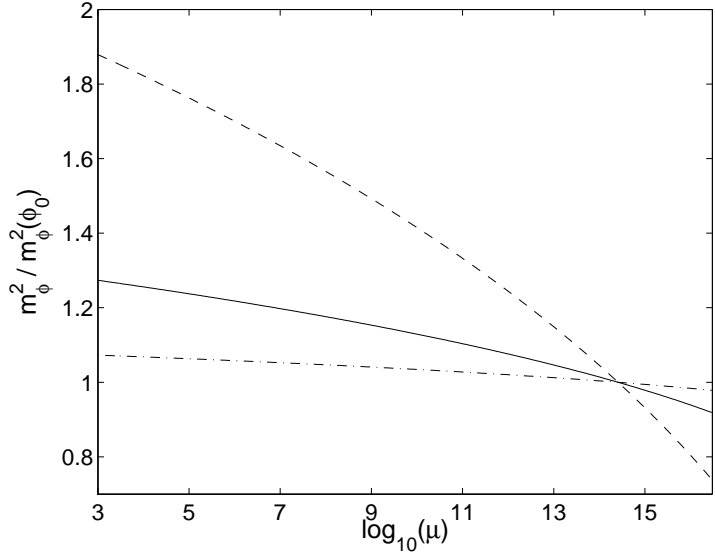,width=.64\textwidth,clip=}
\vspace*{-0.0cm}
\end{center}
\caption{The running of $m^2_{\phi}$ for the $LLe$ inflaton when the
saddle point is at $\phi_0 = 2.6 \times 10^{14}$GeV (corresponding to
$n=6$, $m_{\phi}=1$ TeV and $\lambda = 1$). The three curves
correspond to different values of the ratio of gaugino mass to flat
direction mass at the GUT scale: $\xi = 2$ (dashed), $\xi = 1$ (solid)
and $\xi = 0.5$ (dash-dot)~\cite{Allahverdi:2006we}.  }\label{LLE-RUN0}
\end{figure}


One can analytically obtain the values of $K_i$ for the $LLe$ flat direction.
For ${LLe}$ the one-loop RG equations governing the running of 
$m^2_{\phi}$, $A$, and $\lambda$ with the scale $\mu$ are given 
by~\cite{Nilles:1983ge,Haber:1984rc,Allahverdi:2006we}
\begin{eqnarray} \label{RGE}
\mu {d m^2_{\phi} \over d \mu} &=& - {1 \over 6 \pi^2} \left({3 \over 2}
{\tilde m_2}^2 g^2_2 + {3 \over 2} {\tilde m_1}^2 g^2_1 \right) \, , \nonumber
\\
\mu {d A \over d \mu} &=& - {1 \over 2 \pi^2} \left({3 \over 2} {\tilde m_2}
g^2_2 + {3 \over 2} {\tilde m_1} g^2_1 \right) \, , \nonumber \\
\mu {d \lambda \over d \mu} &=& - {1 \over 4 \pi^2} \lambda \left({3 \over 2}
g^2_2 + {3 \over 2} g^2_1 \right) \, .
\end{eqnarray}
Here ${\tilde m_1}$, ${\tilde m_2}$ denote the mass of the $U(1)_Y$
and $SU(2)_W$ gauginos respectively and $g_1,~g_2$ are the associated
gauge couplings. The running of gauge couplings and gaugino masses obey the usual
equations~\cite{Nilles:1983ge,Haber:1984rc,Allahverdi:2006we}:
\begin{eqnarray}
\mu {d g_1 \over d \mu} &=& {11 \over 16 \pi^2} g^3_1 \, , \nonumber \\
\mu {d g_2 \over d \mu} &=& {1 \over 16 \pi^2} g^3_2 \, , \nonumber \\
{d \over d \mu}\left({{\tilde m_1} \over g_1^2}\right) &=&
{d \over d \mu}\left({{\tilde m_2} \over g_2^2}\right) = 0 \, .
\end{eqnarray}
By solving the above equations, one finds:
\begin{eqnarray}
K_1 &\approx& - {1 \over 4 \pi^2} \left[\left({{\tilde m_2}
\over m_{\phi_0}}\right)^2 g^2_2 + \left({{\tilde m_1}
\over m_{\phi_0}}\right)^2 g^2_1 \right] \, , \nonumber \\
K_2 &\approx& - {3 \over 4 \pi^2} \left[\left({{\tilde m_2}
\over A_0}\right) g^2_2 +
\left({{\tilde m_1} \over A_0}\right) g^2_1 \right] \, , \nonumber \\
K_3 &\approx& - {3 \over 8 \pi^2} \lambda_0 \left[g^2_2 + g^2_1 \right] \, ,
\end{eqnarray}
where the subscript $0$ denotes the values of parameters at the high scale
$\mu_0$.

For universal boundary conditions, as in minimal grand unified
SUGRA, the high scale is the GUT scale $\mu_X \approx 3 \times
10^{16}$~GeV, ${\tilde m_1}(\mu_X) = {\tilde m_2}(\mu_X) = {\tilde m}$
and $g_1 = \sqrt{\pi/10} \approx 0.56$, $g_2 = \sqrt{\pi/6} \approx
0.72$. With the help of RG equations to run the coupling constants and masses
to the scale of the saddle point $\mu_0 = \phi_0 \approx 2.6 \times
10^{14}$~GeV for $M_{\rm P} = 2.4 \times 10^{18}$~GeV, $m_{\phi_0}=
1$~TeV, $\lambda_0=1$. With these values one obtains~\cite{Allahverdi:2006we}
\begin{equation}
K_1 \approx  -0.017 \xi^2,~~ 
K_2 \approx  -0.0085 \xi,~~
K_3 \approx  -0.029\,.
\end{equation}
where $\xi = {\tilde m}/ m_{\phi}$ is calculated at the GUT scale. Similar calculation 
can be performed for the $NH_uL$ flat direction also~\cite{Allahverdi:2006cx}.



\subsubsection{SUGRA $\eta$ problem, trans-Planckian, and moduli problems}
\label{See :SUGRA-problem, trans-Planckian, and moduli problems}

{\bf SUGRA} corrections often destroy the slow-roll predictions of
inflationary potentials. In general, the effective potential depends on the
K\"ahler potential $K$ as $ V\sim
\left(e^{K(\varphi^{\ast},\varphi)/M_{\rm P}^2} V(\phi)\right) $ so
that there is a generic SUGRA contribution to the flat direction
potential of the type for minimal choice of $K$,
\begin{equation}
\label{mflat}
V(\phi)=H^2M_{\rm P}^2 f\left(\frac{\phi}{M_{\rm P}}\right)\,,
\end{equation}
where $f$ is some function (typically a polynomial).  Such a
contribution usually gives rise to a Hubble induced correction to the
mass of the flat direction with an unknown coefficient, which depends
on the nature of the K\"ahler potential. If the K\"ahler
potential has a shift symmetry, i.e. no scale type, then at tree level there is no Hubble
induced correction. 

Let us compare the non-gravitational contribution, Eq.~(\ref{scpot}),
to that of Hubble induced contribution, Eq.~(\ref{mflat}). Writing
$f\sim \left( \phi/M_{\rm P}\right)^p$ where $p\ge 1$ is some power,
we see that non-gravitational part dominates whenever~\cite{Allahverdi:2006we}
\begin{equation}
H_{\rm inf}^2M_{\rm P}^2\left(\frac{\phi}{M_{\rm P}}\right)^p \ll
m_{\phi}^2\phi_0^2\,,
\end{equation}
so that the SUGRA corrections are negligible as long as $\phi_0 <<
M_{\rm P}$, as is the case here (note that $H_{\rm inf} M_{\rm P} \sim
m_{\phi} \phi_0$).  The absence of SUGRA corrections is a generic
property of this model. Note also that although non-trivial K\"ahler
potentials give rise to non-canonical kinetic terms of squarks and
sleptons, it is a trivial exercise to show that at sufficiently low
scales, $H_{\rm inf}<< m_{\phi}$, and small VEVs, they can be rotated
to a canonical form without affecting the potential. 

The same reason, i.e. $H_{\rm inf}<< m_{\phi}$ also precludes any large
{\bf trans-Planckian} correction. Any such correction would generically go
as $(H_{\rm inf}/M_{\ast})^2\sim (m_{\phi}/M_{\ast})\ll 1$, where $M_{\ast}$ is the scale at which
one would expect trans-Planckian effects to kick in~\cite{Brandenberger:2000wr,Danielsson:2002kx,Burgess:2002ub}.
Note that in our case the initial vacuum is the Bunch-Davis and the evolution of the modes is adiabatic.
The latter condition is important to make sure that unknown physics at the high scale is less and less sensitive to
the low energy world~\cite{Burgess:2002ub,Burgess:2003zw}.

Finally, we also make a comment on the cosmological 
{\bf moduli problem}~\cite{deCarlos:1993jw,Dine:2000ds,Banks:1995dt,Kofman:2004yc}.
The moduli are generically displaced from their true
minimum if their mass is less than the expansion rate during
inflation. In our case $H_{\rm inf} \ll m_{moduli}\sim {\cal O}({\rm TeV})$ . This
implies that quantum fluctuations cannot displace the moduli from
their true minima during the inflationary epoch driven by MSSM flat
directions. Moreover, any oscillations of the moduli will be
exponentially damped during the inflationary epoch. Therefore, MSSM inflation
can naturally address the infamous moduli problem~\cite{Allahverdi:2006we}.


\subsection{Exciting SM baryons and cold dark matter}

Interesting aspect of MSSM inflaton is that inflation
takes place away from the point of {\it enhanced gauge symmetry}.
Keep in mind that the VEV of the MSSM flat direction inflaton breaks 
the gauge symmetry spontaneously, for
instance ${udd}$ breaks $SU(3)_C \times U(1)_Y$ while ${LLe}$
breaks $SU(2)_{W}\times U(1)_{Y}$, therefore, induces a SUSY
conserving mass $\sim g \langle \phi(t) \rangle$ to the gauge/gaugino
fields in a similar way as the Higgs mechanism, where $g$ is a gauge
coupling. When the flat direction goes to its minimum, $\langle \phi(t) \rangle = 0$, 
the gauge symmetry is restored. In this respect
the origin is a point of enhanced symmetry.

After the end of inflation, the flat direction starts rolling towards
its global minimum. At this stage the dominant term in the scalar
potential will be: $m_\phi \phi^2/2$. Since the frequency of
oscillations is $\omega \sim m_{\phi} \sim 10^3 H_{\rm inf}$, the flat
direction oscillates a large number of times within the first Hubble
time after the end of inflation. Hence the effect of expansion is
negligible. Further note that the motion is strictly along the radial direction,
i.e. one dimensional. 

In all the examples inflaton has gauge couplings to the
gauge/gaugino fields and Yukawa couplings to the Higgs/Higgsino
fields. As we will see particles with a larger couplings are produced
more copiously during inflaton oscillations~\footnote{The flat direction is coupled to the scalar fields through gauge and Yukawa interactions. For instance, in the $LLe$ case since the lepton Yukawas are $\leq 10^{-2}$ we can safely neglect them. The gauge couplings arise from the $D$-term part of the scalar potential. The $D$-terms corresponding to $SU(2)_W$ and $U(1)_Y$ symmetries follow: $V_D \supset  {1 \over 2} g^2 [ \sum_{i=1}^{3}{({\tilde L}^{\dagger}_1 
T^i {\tilde L}_1 + {\tilde L}^{\dagger}_2 T^i {\tilde L}_2 )^2} ] + {1 \over 2} g^{\prime 2} [ ({\vert {\tilde e}^{*}_3 \vert}^2 - {1 \over 2} {\vert {\tilde L}_1 \vert}^2 - {1 \over 2} {\vert {\tilde L}_2 \vert}^2) ] $. Here $T^1,~T^2,~T^3$ are the $SU(2)$ generators (i.e. $1/2$ times the Pauli matrices) and $g,~g^{\prime}$ are the gauge couplings of $SU(2)_W,~U(1)_Y$, respectively.\\
Similarly, couplings of the flat direction to the gauge fields are obtained from the flat direction kinetic terms: ${\cal L} \supset (D^\mu {\tilde L}_1)^{\dagger} (D_\mu {\tilde L}_1) +  (D^\mu {\tilde L}_2)^{\dagger} (D_\mu {\tilde L}_2) + (D^\mu {\tilde e}_3)^{\dagger} (D_\mu {\tilde e}_3)$,
where $D_\mu {\tilde L}_1  =  (\partial_\mu + {i \over 2} g^{\prime} B_\mu - i g W_{1,\mu} T^1 - i g W_{2,\mu} T^2 - i g W_{3,\mu} T^3) {\tilde L}_1 $,~ $D_\mu {\tilde L}_2 =  (\partial_\mu + {i \over 2} g^{\prime} B_\mu - i g W_{1,\mu} T^1 - i g W_{2,\mu} T^2 - i g W_{3,\mu} T^3) {\tilde L}_2 $, and $D_\mu {\tilde e}_3  =  (\partial_\mu - i g^{\prime} B_\mu ) {\tilde e}_3 $, 
where $W_{1,\mu},~W_{2,\mu},~W_{3,\mu}$ and $B_\mu$ are the gauge fields of $SU(2)_W$ and $U(1)_Y$, respectively.\\
The flat direction couplings to the fermions are found from the following part of the Lagrangian: ${\cal L} \supset \sqrt{2} g \sum_{i=1}^{3}{[ {\tilde L}^{\dagger}_1 {\tilde W}_i^t T^i (i \sigma_2 L_1) + {\tilde L}^{\dagger}_2 {\tilde W}_i^t T^i (i \sigma_2 L_2) ] } + \sqrt{2} g^{\prime} [ {\tilde e}^{\dagger}_3 {\tilde B}^t (i \sigma_2 e_3) - {1 \over 2} {\tilde L}^{\dagger}_1 {\tilde B}^t  (i \sigma_2 L_1) - {1 \over 2} {\tilde L}^{\dagger}_2 {\tilde B}^t (i \sigma_2 L_2)] +  {\rm h.c.} $,
where ${\tilde W}_1,~{\tilde W}_2,~{\tilde W}_3$ and ${\tilde B}$ are the gauginos of $SU(2)_W$ and $U(1)_Y$ respectively. Superscript $t$ denotes transposition, and $\sigma_2$ is the second Pauli matrix. The field content of $L_1,~L_2,~e_3$ are given by:
${L}_1 = \left(\begin{array}{ll} \psi_1 \\ \psi_2 \end{array}\right)$, 
${L}_2 = \left(\begin{array}{ll} \psi_3 \\ \psi_4 \end{array}\right)$,~${e}_3 = \psi_5$, where $\psi_i$ are left-handed Wyle spinors. Similarly, one can also work out all the relevant couplings of $udd$.}.


\subsection{Particle creation and thermalization}
\label{RTCSB}

There are distinct phases of particle creation in this model, here
we briefly summarize them below~\footnote{Readers may wish to revisit this section after reading the following 
Secs.~\ref{See: Tachyonic preheating},~\ref{See: Instant preheating},~\ref{SGR/P},~\ref{FDAT}, and \ref{RTOTU}.}.  

\begin{itemize}

\item{\underline{Tachyonic preheating}:\\ 
Right after the end of inflation,
the second derivative is negative in the case of inflection point inflation. The inflaton fluctuations with a 
physical momentum $k \ls m_{\phi}$ will have a tachyonic instability, see Sec.~\ref{See: Tachyonic preheating}.  
Moreover $V^{\prime \prime} < 0$ only at field values which are 
$\sim \phi_0\sim 10^{14}$~GeV.  Tachyonic effects are
therefore expected to be negligible since, unlike the case
in~\cite{Felder:2000hj}, the homogeneous mode has a VEV which is
hierarchically larger than $m_{\phi}$, and oscillates at a frequency $\omega \sim
m_{\phi}$. Further note fields which are coupled to the inflaton
acquire a very large mass $\sim h \phi_0$ from the homogeneous piece
which suppresses non-perturbative production of their quanta at large
inflaton VEVs. Therefore tachyonic effects, although genuinely
present, do not lead to significant particle production in this case~\footnote{One interesting 
observation for the $LLe$ direction is that its VEV naturally
gives masses to the Hypercharged fields, thus breaking the conformal invariance required for the 
photons~\cite{Enqvist:2004yy,Mazumdar:2008up}. The excited hypercharge can be converted into 
normal electromagnetism after the electroweak phase transition. This would seed vector perturbations 
for the observed large scale magnetic field in the intergalactic medium~\cite{Brandenburg:2004jv}.}.    }\\

\item{\underline{Instant preheating}:\\ 
An efficient bout of particle creation
occurs when the inflaton crosses the origin, which happens twice in
every oscillation. The reason is that fields which are coupled to the
inflaton are massless near the {\it point of enhanced symmetry} (see~Sec.~\ref{See: Instant preheating}). Mainly
electroweak gauge fields and gauginos are then created as they have
the largest coupling to the flat direction. The production takes place
in a short interval, $\Delta t \sim \left(g m_{\phi} \phi_0\right)^{-1/2}$, 
where $\phi_0\sim 10^{14}$~GeV is the initial
amplitude of the inflaton oscillation, during which quanta with a
physical momentum $k \leq \left(g m_{\phi} \phi_0 \right)^{1/2}$ are
produced. The number density of gauge/gaugino degrees of freedom is given
by~\cite{Felder:1998vq,Allahverdi:2006we}
\begin{equation} \label{chiden}
n_{g} \approx {\left(g m_{\phi} \phi_0
\right)^{3/2} \over 8 \pi^3}\,.
\end{equation}
As the inflaton VEV is rolling back to its maximum value $\phi_0$, the
mass of the produced quanta $g \langle \phi(t) \rangle$ increases. The
gauge and gaugino fields can (perturbatively) decay to the fields
which are not coupled to the inflaton, for instance to (s)quarks. Note
that (s)quarks are not coupled to the flat direction, hence they
remain massless throughout the oscillations. The total decay rate of
the gauge/gaugino fields is then given by 
$\Gamma = C \left(g^2/48\pi\right) g\phi $, where $C \sim {\cal O}(10)$ is a numerical factor
counting for the multiplicity of final states.

The decay of the gauge/gauginos become efficient when
$\langle \phi \rangle \simeq \left({48 \pi m_{\phi} \phi_0/C g^3}\right)^{1/2}$, 
where we have used $\langle \phi(t) \rangle \approx \phi_0 m_{\phi} t$,
which is valid when $m_{\phi} t \ll 1$, and $\Gamma \simeq t^{-1}$,
where $t$ represents the time that has elapsed from the moment that
the inflaton crossed the origin. Note that the decay is very quick
compared with the frequency of inflaton oscillations, i.e. $\Gamma \gg
m_{\phi}$. It produces relativistic (s)quarks with an energy~\cite{Allahverdi:2006we}:
\begin{equation} \label{energy}
E =\frac{1}{2}g\phi(t)
\simeq \left({48 \pi m_{\phi} \phi_0 \over C g}\right)^{1/2}\,.
\end{equation}
The ratio of energy density in relativistic particles thus produced
$\rho_{rel}$with respect to the total energy density $\rho_0$ follows
from Eqs.~(\ref{chiden}),~(\ref{energy}):
\begin{equation}
\label{ratio}
{\rho_{rel} \over \rho_0} \sim 10^{-1} g\,,
\end{equation}
where we have used $C \sim {\cal O}(10)$.  This implies that a
fraction $\sim {\cal O}(10^{-1})$ of the inflaton energy density is
transferred into relativistic (s)quarks every time that the inflaton
passes through the origin. Within $10-50$ oscillations the inflaton would 
loose its energy into relativistic MSSM degrees of freedom.}\\

\item{\underline{Reheating and thermalization}:\\
A full thermal equilibrium is reached when ${\it (a)~kinetic }$ and
${\it (b)~chemical~equilibrium }$ are established. The maximum
(hypothetical) temperature attained by the plasma would be given by:
\begin{equation}
\label{tmax}
T_{max} \sim V^{1/4} \sim \left(m_{\phi}\phi_0\right)^{1/2}
\approx 10^{9}~{\rm GeV}\,.
\end{equation}
However, not all the MSSM degrees of freedom can be in thermal equilibrium
at such a high temperature. Depending on the nature of a flat direction inflaton,
the final reheat temperature can be quite low. 

For instance, if $LLe$ is the inflaton then $udd$ can acquire a large VEV independently.
The VEV of $udd$ will spontaneously generate masses to the gluons and gluinos, i.e. 
\begin{equation}
m_{G} \sim  g \langle \varphi(t) \rangle < g \phi_0\,.
\end{equation}
To develop and maintain such a large VEV for $udd$, it is
not necessary that ${udd}$ potential has a saddle point, or point of inflection as
well. It can be trapped in a false minimum during inflation, which
will then be lifted by thermal corrections when the inflaton decays~\cite{Allahverdi:2006we,Allahverdi:2006dr}.
The above inequality arises due to the fact that the VEV of $\varphi$ cannot
exceed that of the inflaton $\phi$, since its energy density should be
subdominant to the inflaton energy density.

If $g \varphi_0 \gg T_{max}$ the gluon/gluino fields will be too heavy
and not kinematically accessible to the reheated plasma. Here
$\varphi_0$ is the VEV of ${ udd}$ at the beginning of inflaton
oscillations.  In a radiation-dominated Universe the Hubble expansion
redshifts the flat direction VEV as $\langle \varphi \rangle \propto
H^{3/4}$, which is a faster rate than the change in the temperature $T
\propto H^{1/2}$. Once $g \langle \varphi \rangle \simeq T$,
gluon/gluino fields come into equilibrium with the thermal bath. When 
this happens the final reheat temperature is generically small, i.e.
$T_{\rm R} \leq 10^{7}$~GeV~\cite{Allahverdi:2006we,Allahverdi:2005mz}. 
See Secs.~\ref{FDAT}, \ref{RTOTU}, and \ref{deviation QATEOTU}.}

\end{itemize}


\subsubsection{Benchmark points for MSSM inflation and dark matter abundance}
\label{BPMC}

In this section we
explore the available parameter space for inflation in conjunction
with a thermal cold dark matter abundance within the minimal
SUGRA model. Remarkably for the inflaton, which is a
combination of squarks and sleptons, there is a stau-neutralino
coannihilation region below the inflaton mass $500$~GeV for the
observed density perturbations and the tilt of the spectrum.  For such
a low mass of the inflaton the LHC is capable of discovering the
inflaton candidates within a short period of its operation.  Inflation
is also compatible with the focus point region which opens up for the
inflaton masses above TeV.

Since $m_{\phi}$ is related to the scalar masses, sleptons ($LLe$
direction) and squarks ($udd$ direction), the bound on $m_{\phi}$
will be translated into the bounds on these scalar masses which are
expressed in terms of the model parameters~\cite{Allahverdi:2006we}. 

Note that CMB constraints $m_{\phi}$ at $\phi_0\sim 10^{14}$ GeV
which is two orders of magnitude below the GUT scale. From this
$m_{\phi}$,  $m_0$ and $m_{1/2}$ are determined at the GUT scale by 
solving the RGEs for fixed values of $A_0$ and $\tan\beta$. The RGEs 
for $m_{\phi}$ are
\begin{eqnarray}
\mu{dm_{\phi}^2\over{d\mu}}&=&{-1\over{6\pi^2}}({3\over
2}{M_2^2}g_2^2+{9\over {10}}{M_1^2}g_1^2)\,,\quad\, ({\rm for\,
LLe})\, \nonumber \\
\mu{dm_{\phi}^2\over{d\mu}}&=&{-1\over{6\pi^2}}({4}{M_3^2}g_3^2+
{2\over {5}}{M_1^2}g_1^2)\,, \quad\, ({\rm for\, udd})\,.
\end{eqnarray}
$M_1,~M_{2}$ and $M_3$ are $U(1),~SU(2)$ and $SU(3)$ gaugino masses
respectively. After determine $m_0$ and $m_{1/2}$ from $m_{\phi}$, one can
determine the allowed values of $m_{\phi}$ from the experimental
bounds on the mSUGRA parameters space~\cite{Freedman:1976xh,Deser:1976eh,Barbieri:1982eh,Hall:1983iz}.

The models of mSUGRA depend only on four parameters and one sign. These are
$m_0$ (the universal scalar soft breaking mass at the GUT scale
$M_{\rm G}$); $m_{1/2}$ (the universal gaugino soft breaking mass at
$M_{\rm G}$); $A_0$ (the universal trilinear soft breaking mass at
$M_{\rm G}$); $\tan\beta =\langle H_u \rangle \langle H_d \rangle$ at the electroweak scale
(where $H_u$ gives rise to $u$ quark masses and $H_d$ to $d$ quark
and lepton masses); and the sign of $\mu$, the Higgs mixing
parameter in the superpotential ($W_{\mu} = \mu H_u H_d$).

Unification of gauge couplings within SUSY suggests that
$M_{\rm G} \simeq 2 \times 10^{16}$ GeV. The model parameters are
already significantly constrained by different experimental results.
Most important constraints are: (1) The light Higgs mass bound of $M_{h^0} > 114.0$~GeV 
from LEP \cite{Barate:2003sz}. (2)~The $b \rightarrow s \gamma$ branching ratio~\cite{Alam:1994aw}:
$2.2\times10^{-4} < {\cal B}(B \rightarrow X_s \gamma) <4.5\times10^{-4}$.
(3)~In mSUGRA the $\tilde\chi^0_1$ is the candidate for CDM. (4)~The
$2\sigma$ bound from the WMAP \cite{Komatsu:2008hk} gives a relic density bound
for CDM to be $0.095 < \Omega_{\rm CDM} h^2 < 0.129 $. (5)~The bound on the lightest chargino mass of
$M_{\tilde\chi^{\pm}_1} > 104$~GeV from LEP~\cite{Eidelman:2004wy}. (6)~The possible $3.3~\sigma$ deviation
(using $e^+e^-$ data to calculate the leading order hadronic contribution)from the SM expectation of the
anomalous muon magnetic moment from the muon $g-2$ collaboration~\cite{Bennett:2004pv}.


\begin{figure}[t]
\vspace{1cm} \center
\includegraphics[width=5.8cm]{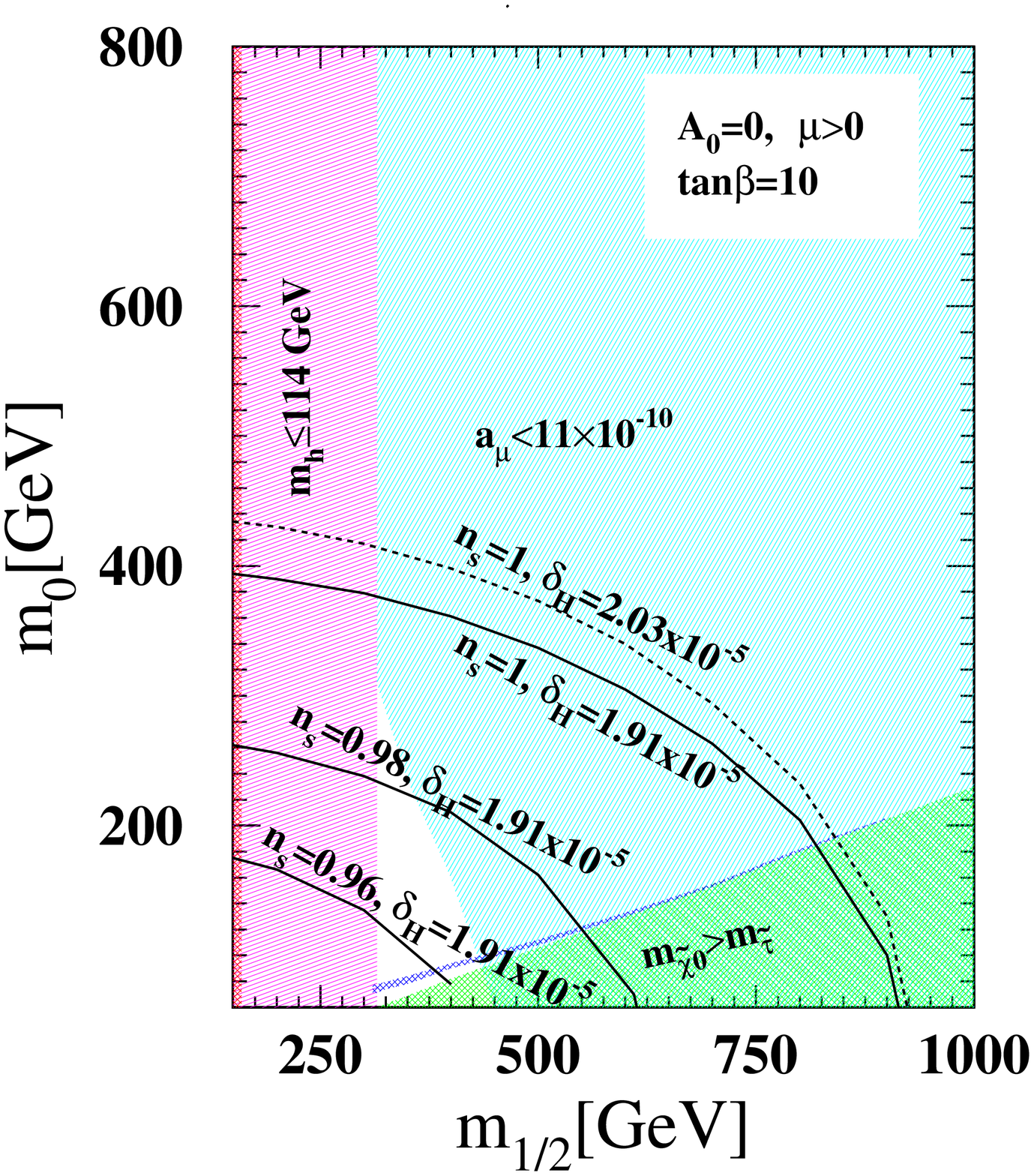}~
\includegraphics[width=5.8cm]{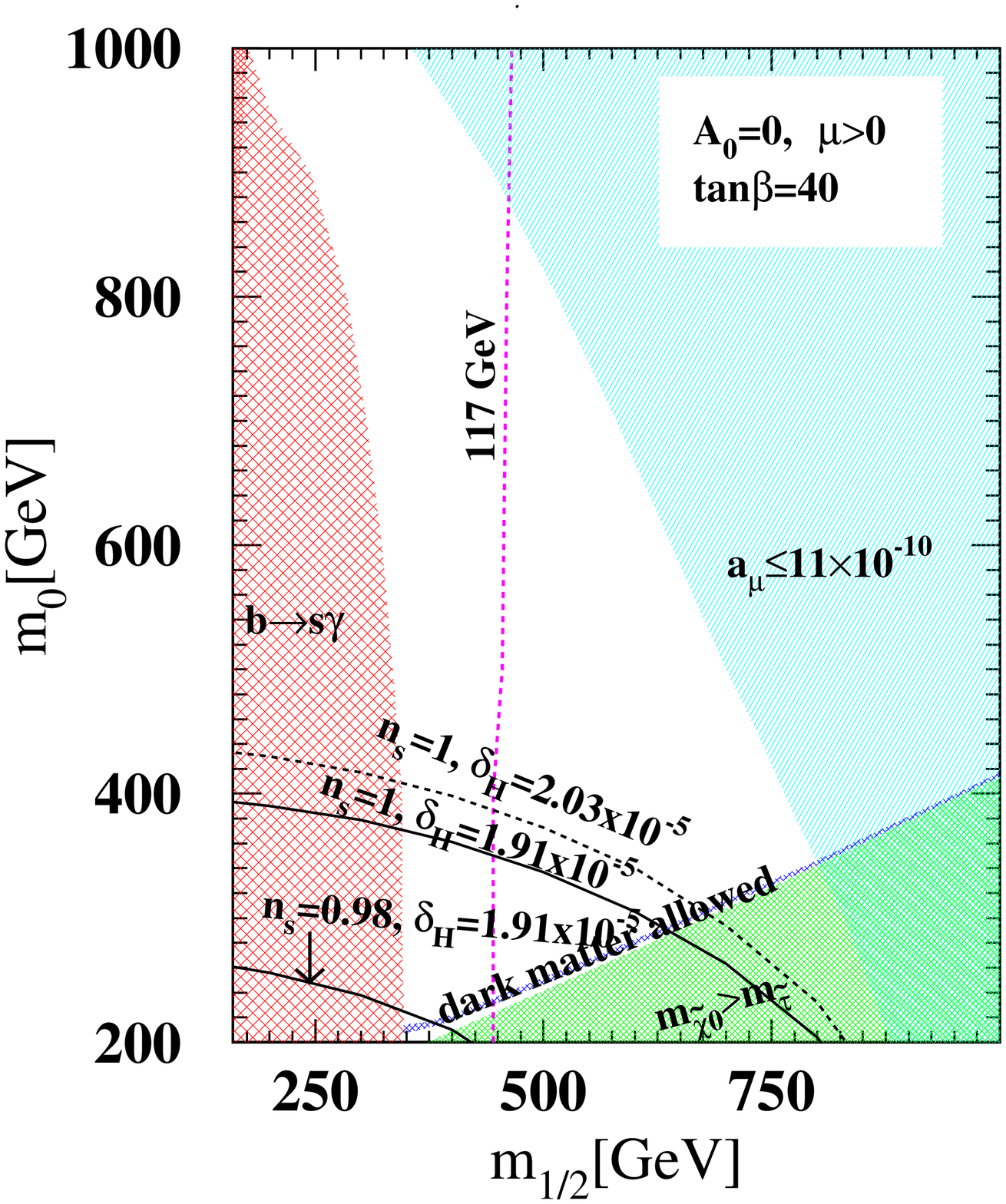}~
\includegraphics[width=5.8cm]{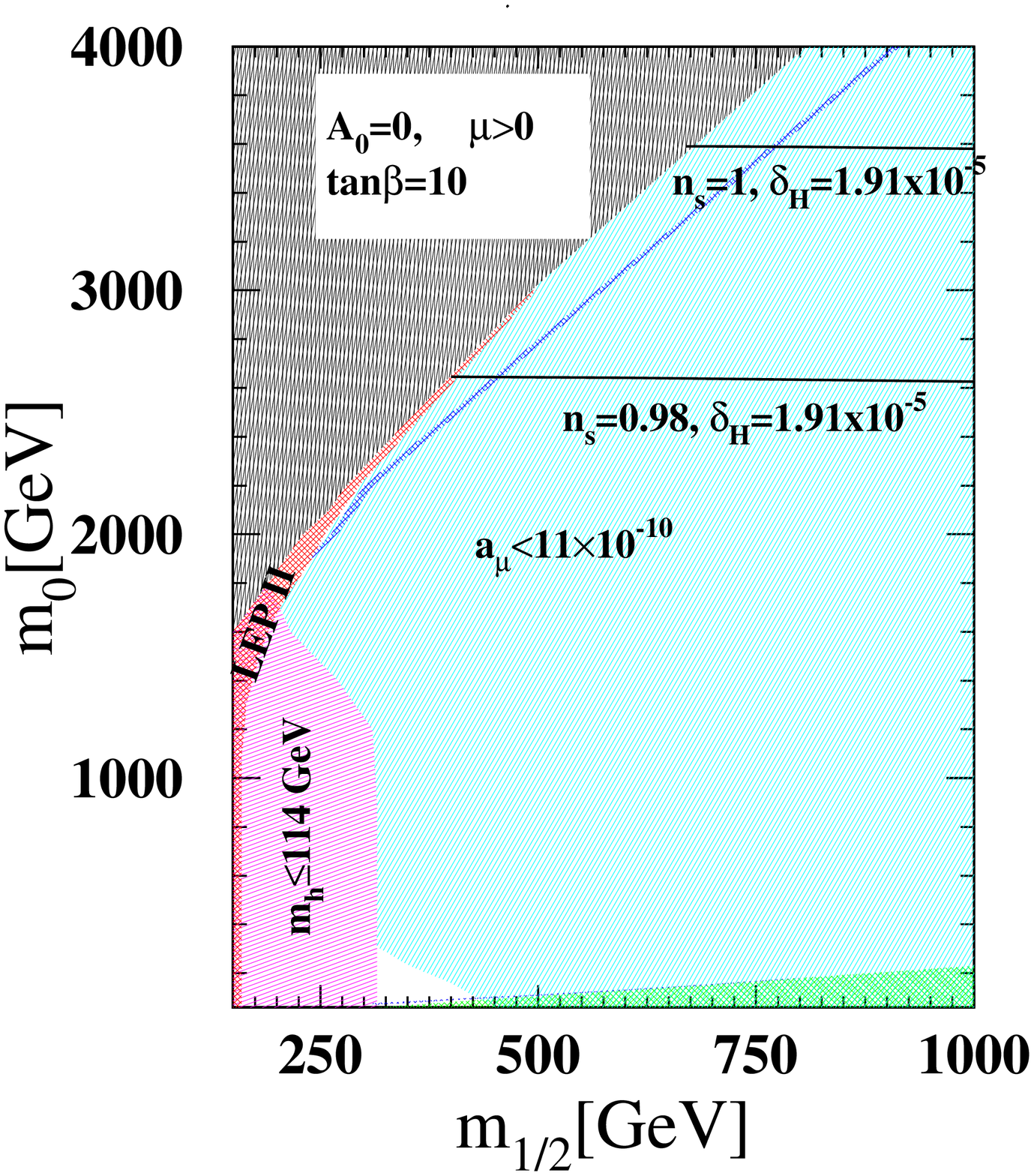}
\caption{The
contours for different values of $n_s$ and $\delta_H$ are shown in
the $m_0-m_{1/2}$ plane for $\tan\beta=10, 40$, and $\lambda=1$ for
the contours. We show the dark matter allowed region {narrow blue
corridor}, (g-2)$_\mu$ region (light blue) for $a_{\mu}\leq
11\times10^{-8}$, Higgs mass $\leq 114$ GeV (pink region) and LEPII
bounds on SUSY masses (red).
In the third panel $\lambda=0.1$ were chosen, The black region is
not allowed by radiative electroweak symmetry breaking, and
$m_t=172.7$~GeV for this graph. Note that black curved lines denote the 
cosmological parameters, $(\delta_H, n_s)$, within $95\%$c.l. Note that
smaller values of $n_s<1$ is preferred by the dark matter abundance in
this scheme of parameters. The plots are taken from~\cite{Allahverdi:2007vy}.} 
\label{10flat}
\end{figure}


The allowed mSUGRA parameter space has mostly three
distinct regions: (i)~the stau-neutralino
($\tilde\tau_1~-~\tilde\chi^1_0$), coannihilation region where
$\tilde\chi^1_0$ is the LSP, (ii)~the
$\tilde\chi^1_0$ having a dominant Higgsino component (focus point)
and (iii)~the scalar Higgs ($A^0$, $H^0$) annihilation funnel
(2$M_{\tilde\chi^1_0}\simeq M_{A^0,H^0}$). 

The mSUGRA parameter space in Figs.~(\ref{10flat}),
 for $\tan\beta=10$ and $40$ with the $udd$ flat
direction using $\lambda=1$. In the
figures, we show contours correspond to $n_s=1$ for the maximum
value of $\delta_H=2.03\times 10^{-5}$ (at $2\sigma$ level) and
$n_s=1.0,~0.98,~0.96$ for $\delta_H=1.91\times 10^{-5}$. It is also
interesting to note that the allowed region of $m_{\phi}$, as
required by the inflation data for $\lambda=1$ lies in the
stau-neutralino coannihilation region which requires smaller values
of the SUSY particle masses.  See Fig.~(\ref{lamcon}), where
both co-annihilation and focus point regions have been illustrated for 
$\lambda\sim 1-0.02$. 

The SUSY particles in this parameter
space are, therefore, within the reach of the LHC very quickly. The
detection of the region at the LHC has been considered in~\cite{Arnowitt:2006jq}. 
From the figures, one can also find that as
$\tan\beta$ increases, the inflation data along with the dark
matter, rare decay and Higgs mass constraint  allow smaller ranges
of $m_{1/2}$. For example, the allowed ranges of  gluino masses are
765 GeV-2.1 TeV and 900 GeV-1.7 TeV for $\tan\beta=10$ and $40$
respectively. Now if $\lambda$ is small, i.e.
$\lambda\ls 10^{-1}$, the allowed values of $m_{\phi}$
would  be large. In this case the dark matter allowed region requires
the lightest neutralino to have larger Higgsino component in the
mSUGRA model. 


\subsubsection{Can dark matter be the inflaton?}
\label{SNDM}

If the reheat temperature of the
universe is higher than the mass of the inflaton, then the plasma
upon reheating will, in addition, have a thermal distribution of
the inflaton quanta. If the inflaton is absolutely stable, due to
some symmetry, then it can also serve as
the cold dark matter. One such example, $NH_uL$ as an observable sector inflaton is an 
interesting scenario, as it can explain successful inflation,  exciting 
SM quarks and leptons, the observed neutrino masses via Dirac coupling 
and the right handed sneutrino, $\widetilde N$, as a dark matter candidate~\cite{Allahverdi:2007wt}.

It is well known that for particles with gauge interactions, the unitarity bound puts an
absolute upper bound  on the dark matter mass to be 
less than $\sim 100$ TeV~\cite{Jungman:1995df}. 
Once the temperature drops below the inflaton
mass, its quanta undergo thermal freeze-out and yield the
correct dark matter abundance. Furthermore, scatterings via the new $U(1)_{\rm B-L}$ 
gauge interactions also bring the right handed
sneutrino into thermal equilibrium. Note that part of the inflaton,
i.e. its ${\widetilde N}$ component see Eq.~(\ref{flat}), has never
decayed; only the coherence in the original condensate that drives
inflation is lost. However, the neutrino Yukawa, $h$, is way too small to
allow acceptable thermal dark matter. Note that ${\widetilde N}$ would dominate
the universe right after the end of inflation if it had no gauge
interactions. 

In order to calculate the relic abundance of the RH sneutrino, it is necessary
to know the masses of the additional gauge boson $Z^{\prime}$
and its SUSY partner ${\tilde Z}^{\prime}$, the new Higgsino masses,
Higgs VEVs which break the new $U(1)$ gauge symmetry, the RH
sneutrino mass, the new gauge coupling, and the charge assignments
for the additional $U(1)$.  Assuming that the new gauge symmetry is
broken around 2 TeV, and the existence of
two new Higgs superfields to maintain the theory anomaly free.

The primary diagrams responsible to provide the right amount of relic
density are mediated by ${\tilde Z}^{\prime}$ in the $t$-channel.
 In Fig.~(\ref{sneutrinoominf}), we show the relic density values for smaller masses
of sneutrino.  In the case of a larger
sneutrino mass in this model, the correct dark matter abundance can
be obtained by annihilation via $Z^{\prime}$ pole~\cite{Lee:2007mt,Allahverdi:2007wt}.
A sneutrino mass in the $1-2$~TeV range provides a good fit to the PAMELA data 
and a reasonable fit to the ATIC data~\cite{Allahverdi:2009ae}. 


\begin{figure}[t]
\includegraphics[width=9cm]{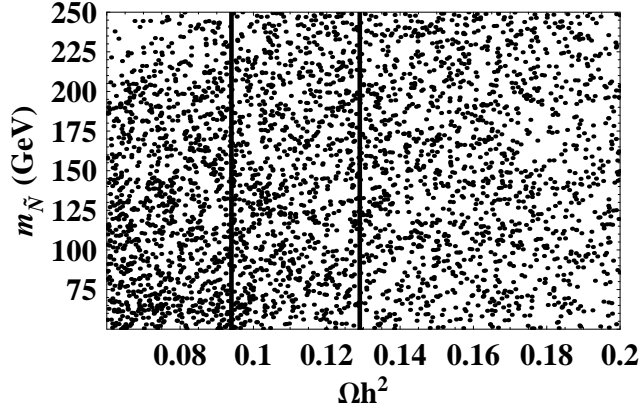}
\caption{$\Omega h^2$ vs $m_{\tilde N}$. The solid lines from left
to right are for $\Omega h^2 =$ 0.094 and 0.129 respectively. The
$Z^{\prime}$-ino mass is equal to the Bino mass since the new $U(1)$
gauge coupling is the same as the hypercharge gauge coupling. 
}\label{sneutrinoominf}
\end{figure}


Since the dark matter candidate, the RH sneutrino, interacts with
quarks via the $Z^{\prime}$ boson, it is possible to see it via the
direct detection experiments. The detection cross sections are not
small as the interaction diagram involves $Z^{\prime}$ in the
$t$-channel. The typical cross section is about 2$\times 10^{-8}$ pb
for a $Z^{\prime}$ mass around 2 TeV, makes it possible
to probe the dark matter candidate in direct detection~\cite{Baudis:2007dq}.


\subsection{Stochastic initial conditions for low scale inflation}
\label{ICMI}

It is conceivable that the universe gets trapped in a metastable vacuum at earlier stages, irrespective of how it began.  
Metastable vacua are ubiquitous in string theory, see the discussion in section~\ref{STLAGE}. Inflation can be driven from 
one vacuum to another, either via quantum tunneling or via transient phase of non-inflationary 
dynamics~\cite{Burgess:2005sb}. In any case, it is important to note that these high scale inflation provides a natural 
initial condition for MSSM inflation.

Let us imagine that there are almost degenerate metastable vacua. Once the energy density of the false vacuum dominates, inflation begins and the universe undergoes accelerated expansion with a constant Hubble rate $H_{\rm false}$. False vacuum is not stable and decays via bubble nucleation. The rate per volume for the decay of a metastable vacuum to the true vacuum is given by:
\begin{equation} \label{tunnel} 
\Gamma/V = C \exp\left(-\Delta S_E \right)\ ,
\end{equation}
where $C$ is a one-loop determinant and $\Delta S_E$ is the difference in Euclidean actions between the instanton and the
background with larger cosmological constant. The determinant $C$ can at most be $C\lesssim M^4_{\rm P}$, simply because 
$M_{\rm P}$ is the largest scale available, and estimates (ignoring metric fluctuations) give a value as small as 
$C\sim r^{-4}_0$, with $r_0$ the instanton bubble radius \cite{Coleman:1980aw,Garriga:1993fh}. Therefore a typical decay rate in a (comoving) Hubble volume is given by
\begin{equation} \label{tunnel2}
\Gamma \lesssim \frac{M^4_{\rm P}}{H^3_{\rm false}}\exp\left(-\Delta S_E \right)\,. 
\end{equation}
Especially with a large Hubble scale $H_{\rm false}$, the associated
decay time is much longer than $H^{-1}_{\rm false}$, given that
typically $\Delta S_E \gg 1$. This implies that most of the space is
locked in the false vacuum and inflate forever, while bubble
nucleation creates pockets of true vacuum whose size grow only linearly
in time. Here we will concentrate on one such true vacuum, where MSSM flat directions 
can be excited.


\subsubsection{Quantum fluctuations of MSSM flat directions}
\label{QFMFD}

During false vacuum inflation quantum fluctuations displace any scalar field whose mass is 
smaller than $H_{\rm false}$. The question is whether these
fluctuations can push the MSSM inflaton sufficiently to the plateau
of its potential around the point of inflection $\phi_0$, see Eqs.~(\ref{infvev},\ref{plateau}).
If MSSM inflaton begins with a small VEV $\phi < \phi_0$, the mass term in Eq.~(\ref{scpot}) dominates.
Hence, for any $H_{\rm false} > {\cal O}({\rm TeV})$, it
obtains a quantum jump, induced by the false vacuum inflation, of
length $H_{\rm false}/2\pi$, within each Hubble
time~\cite{Linde:2005ht}. Typically the quantum
fluctuations have a Gaussian distribution, and the r.m.s
(root mean square) value of the modes which exit the inflationary
Hubble patch within one Hubble time is given by $H_{\rm
false}/2\pi$. These jumps superimpose in a random walk fashion,
which is eventually counterbalanced by the classical slow-roll due to the mass term, 
resulting in~\cite{Linde:1982uu,Vilenkin:1982wt,Vilenkin:1983xq,Vilenkin:1983xp,Enqvist:1987au,Linde:2005ht}
\begin{equation} \label{fluct} \langle \phi^2 \rangle = {3 H^4_{\rm false} \over
8 \pi^2 m^2_{\phi}} \Big[1 - {\rm exp}\Big(-{2 m^2_{\phi} \over 3
H_{\rm false}} t\Big)\Big], \end{equation}
which for $t \to \infty$ yields
\begin{equation} \label{rms}
\phi_{r.m.s} = \sqrt{3 \over 8 \pi^2} {H^2_{\rm
false} \over m_{\phi}}\,.
\end{equation}
If $\phi_{r.m.s} \geq \phi_0$, then $\phi$ will lie near
$\phi_0$, see Eq.~(\ref{infvev}),in many regions of space. This requires that
\begin{equation} \label{cond1}
H_{\rm false} \geq \Big({8 \pi^2 \over 3}\Big)^{1/4} (m_{\phi} \phi_0)^{1/2} \gs 10^8~{\rm GeV},
\end{equation}
where $m_{\phi}\sim 100$~GeV and $\phi_0\sim 3\times 10^{14}$~GeV.
$\phi_{r.m.s}$ settles at its final value when $t > 3 H_{\rm false}/2m_{\phi}^2$, which
amounts to
\begin{equation} \label{flucefold}
N_{\rm false} > {3 \over 2} \Big({H_{\rm false} \over m_{\phi}}\Big)^2 \gs 10^{16},
\end{equation}
e-foldings of false vacuum inflation. The large number of e-foldings required is
not problematic as inflation in the false vacuum is eternal in nature.


\subsubsection{Inflection point as a dynamical attractor}
\label{IPDA}

In general the MSSM inflaton can have a VEV above the point of inflection $\phi > \phi_0$ and/or a large velocity ${\dot \phi}$ at the beginning of false vacuum inflation~\cite{Allahverdi:2006iq,Allahverdi:2006cx,Allahverdi:2006we,Allahverdi:2008bt,Allahverdi:2007wh}. In this case the classical motion of the field becomes important. There are typically three regimes in the evolution of $\phi$ field~\cite{Allahverdi:2008bt}; (1)~Oscillatory regime: If the initial VEV of $\phi$, denoted by $\phi_i$, is such that 
$V^{\prime \prime}(\phi_i) > H^2_{\rm false}$, then it starts in the oscillatory regime, (2)~Kinetic energy dominance regime: If 
$\phi_i < \phi_{\rm tr}$, then $V^{\prime \prime}(\phi) < H^2_{\rm false}$, and the potential is flat during false vacuum inflation,
then  the dynamics of $\phi$ in this case depends on its initial velocity denoted by ${\dot \phi}_i$. If ${\dot \phi}^2_i > 2 V(\phi_i)$, the kinetic dominance prevails. In principle the inflaton will overshoot the point of inflection only if it begins very close to $\phi_0$ and has a large negative velocity initially. It is interesting to note that $\phi$ can end up above the point of inflection even if $\phi_i < \phi_0$, provided that ${\dot \phi}_i > 3 H_{\rm false} (\phi_0 -\phi_i)$. The most important regime is (3)~the slow-roll regime: 
which we will discuss below.

Once an initial phase of oscillations or kinetic energy dominance ends, $\phi$ starts a slow-roll motion towards  the inflection point, $\phi_0$. The equation of motion in this regime is $3H_{\rm false} \dot \phi + V^{\prime}(\phi)\approx 0$. Initially the field is under the influence of the non-renormalizable potential, i.e. $V(\phi) \propto \phi^{10}$, see Eq.~(\ref{scalar-pot}), however, as $\phi$ moves toward 
$\phi_0$, the $\phi^{10}$ term becomes increasingly smaller. Eventually, for $\phi \approx \phi_0$, we have $V^{\prime}(\phi) = V^{\prime}(\phi_0) + V^{\prime \prime \prime}(\phi_0) (\phi - \phi_0)^2/2$. It happens that this is the longest part of 
$\phi$ journey. It is desirable to find how long does it take for $\phi$ to reach the edge of the plateau in Eq.~(\ref{plateau}). Outside the plateau, $V^{\prime}(\phi_0)$ is subdominant, see Eqs.~(\ref{1st},~\ref{3rd}), and hence:
\begin{equation} \label{slow}
{\dot \phi} = - {32 m^2_{\phi} (\phi - \phi_0)^2 \over 3 H_{\rm false} \phi_0}.
\end{equation}
This results in
\begin{equation} \label{att}
(\phi - \phi_0) \approx {3 H_{\rm false} \phi_0 \over 32 m^2_{\phi} t},
\end{equation}
for large $t$. Therefore the inflection point acts as an attractor for the classical equation of motion. After using Eq.~(\ref{plateau}),  $t_{\rm slow}\gs 10^{10}$ e-foldings of false vacuum inflation (note that $\phi_0 \sim 10^{14}$ GeV, and $H_{\rm false} > m_{\phi}\sim {\cal O}(100)$~GeV). After this time $\phi$ is settled within the plateau in the bulk of the inflating space. This implies that the bubbles which nucleate henceforth have the right initial conditions for a subsequent stage of MSSM inflation.

\begin{figure}
\vspace*{-0.0cm}
\begin{center}
\includegraphics[width=8.0cm]{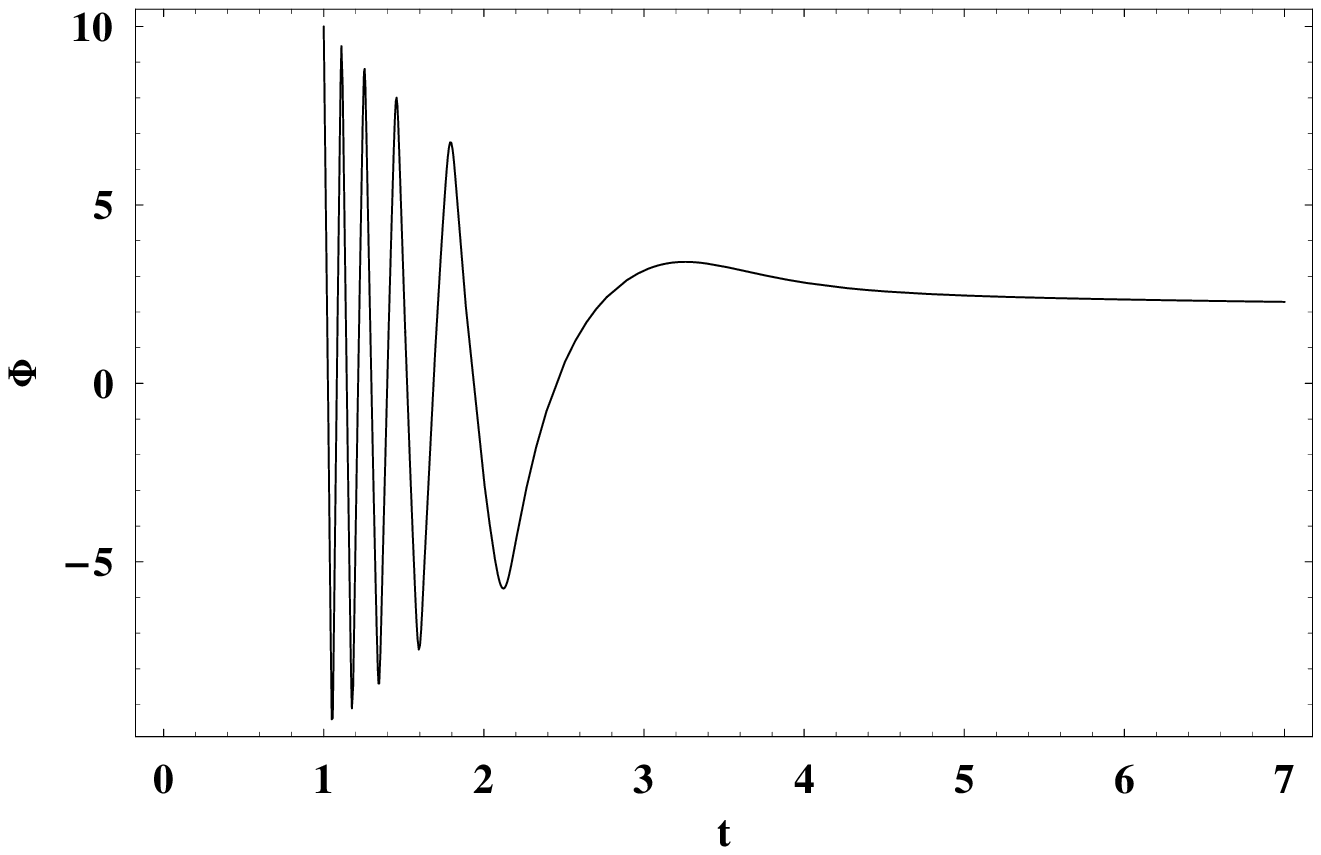}~~~~~~
\includegraphics[width=8.0cm]{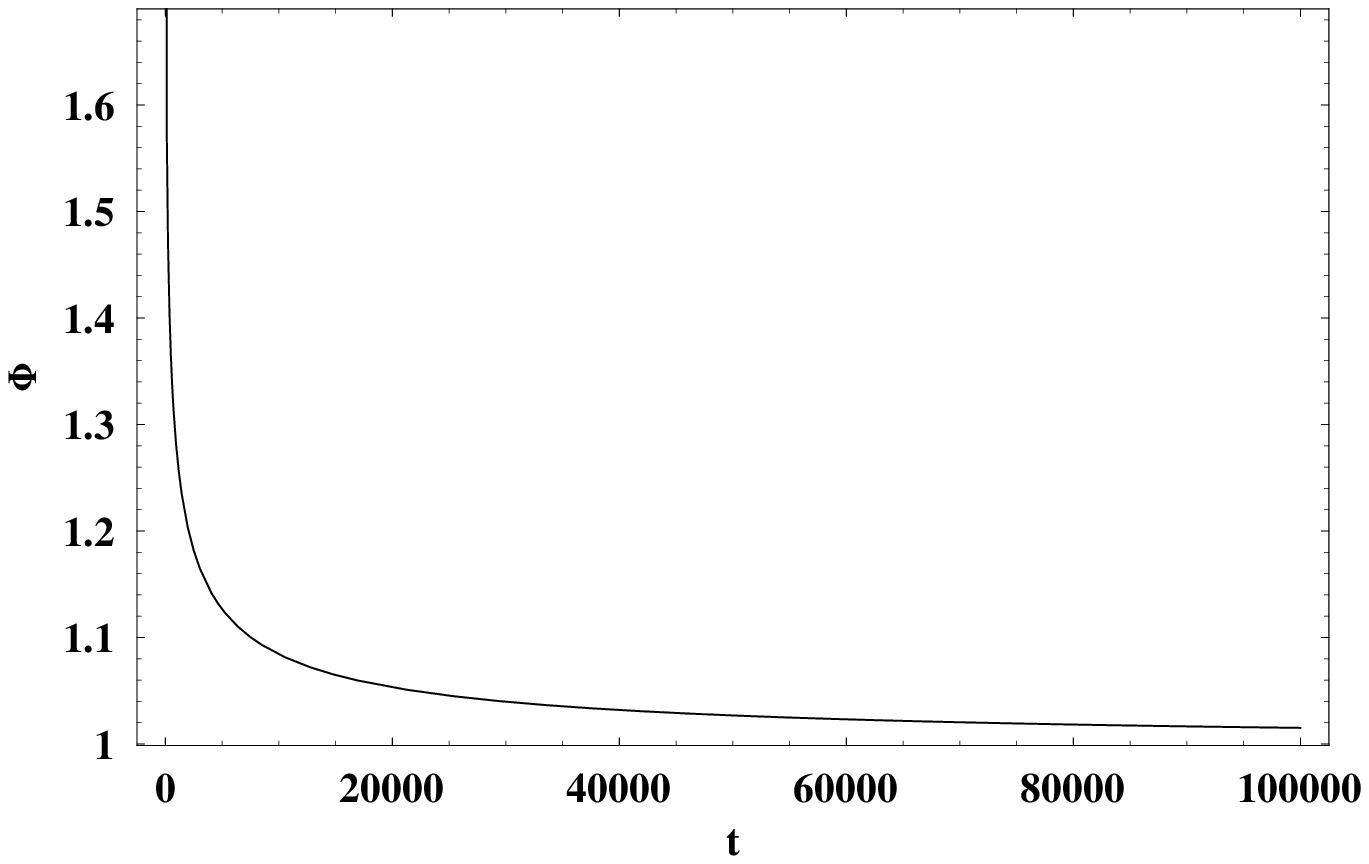}
\end{center}
\caption{The plot shows $\phi$ (scaled by $\phi_0$) as a function of time for $\phi_i = 10$ and ${\dot \phi}_i = 100$, with $H_{\rm false} = 10^3 m_{\phi}$. There is an initial oscillatory phase since $V^{\prime\prime}(\phi_i) \geq H^2_{\rm false}$. It ends quickly as the amplitude of oscillations decreases fast due to Hubble expansion. Then the slow-roll motion begins which lasts much longer.  In the second plot the slow-roll phase for the same initial conditions is depicted. It lasts very long but the field asymptotes to the {\it point of inflection} $\phi_0$ with $\dot \phi\rightarrow 0$. 
The plots are take from~\cite{Allahverdi:2008bt}.
}
\label{oscillations}
\end{figure}

\begin{figure}
\vspace*{-0.0cm}
\begin{center}
\includegraphics[width=8.0cm]{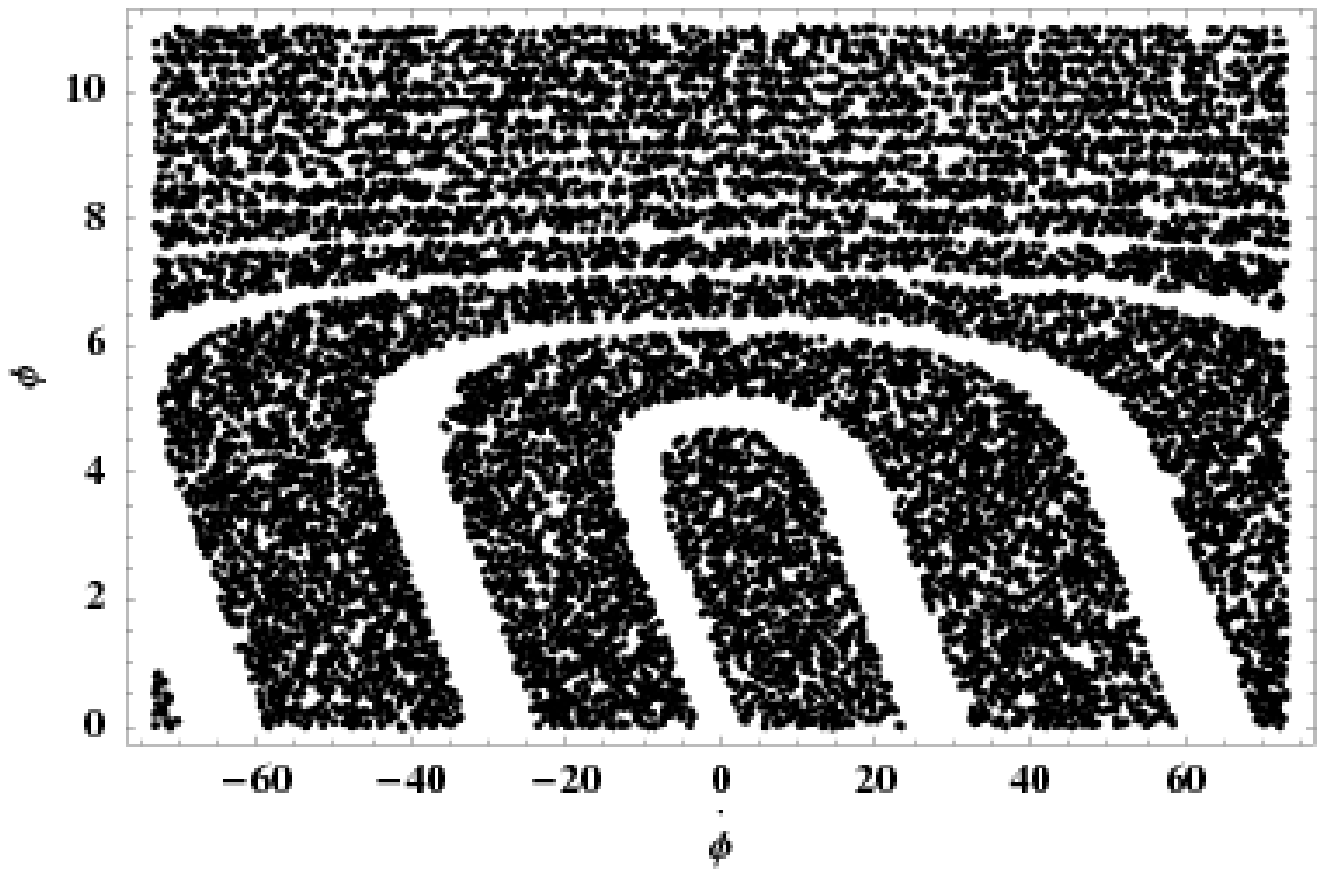}~~~~~~
\includegraphics[width=8.0cm]{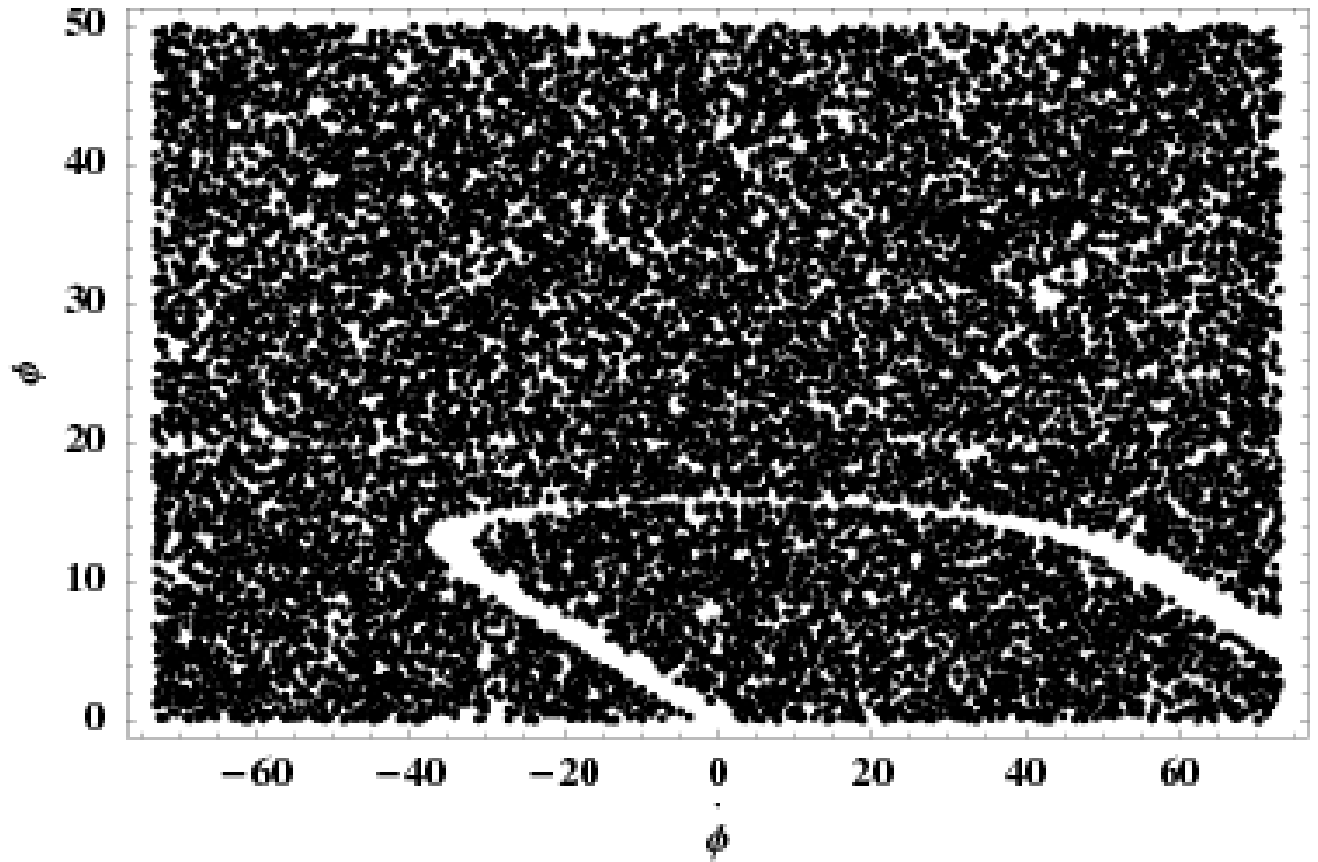}
\end{center}
\caption{The attractor behavior holds for a wide range of initial values of $\phi$ and ${\dot \phi}$. In the left panel the initial values of $\phi$ versus $\dot\phi$ for $H_{\rm false}=10^2 m_{\phi}$ is shown. The dots show the initial values for which 
$\phi$ settles to $\pm\phi_0$ and the white bands $\cap$ show the critically damped regions where $\phi$ settles to zero at late times. The situation improves a lot for a larger $H_{\rm false}=10^4 m_{\phi}$ as shown in the right hand panel. The plots are symmetric under $\phi \rightarrow -\phi$. Here we have shown the upper half of $\phi-{\dot \phi}$ plane where $\phi \geq 0$. The plots are take from~\cite{Allahverdi:2008bt}.}
\label{phase1}
\end{figure}


\subsubsection{Inflating the MSSM bubble}
\label{See: Inflating the bubble}

Let us consider the bubbles that have the right initial conditions for MSSM inflation, i.e. $\phi$ has settled in the plateau around a point of inflection according to Eq.~(\ref{plateau}). The initial size of the bubble is $r_0 < H^{-1}_{\rm false}$. Inside of an expanding bubble has the same geometry as an open FRW universe. The Hubble rate inside the bubble is therefore given by~\cite{Allahverdi:2007wh}
\begin{equation} \label{bubhub} H^2 = {V_{\phi} + V_{\varphi} \over 3 M^2_{\rm P}} + {1
\over a^2}, \end{equation}
where $a$ is the scale factor of the universe and $V_{\varphi}$ is the energy density in the $\varphi$ field.
Note that $\varphi$ is the field responsible for forming the false vacuum, which could either arise within MSSM or from some other sector.There are examples of $\varphi$ field as an MSSM flat direction.

Since $\phi$ is inside the plateau of its potential, its dynamics is frozen, hence $V(\phi)\sim V(\phi_0)$ as long as $H > H_{\rm MSSM}$. Right after tunneling, $H \equiv {\dot a}/a = r^{-1}_0 > H_{\rm false}$. This implies that the last term on the right-hand side of Eq.~(\ref{bubhub}) dominates over the first two terms, and hence the
universe is curvature dominated. The $\varphi$ field oscillates around the
true vacuum of its potential at the origin, and quickly decays to radiation whose energy density is redshifted $\propto a^{-4}$. On the other hand the curvature term is redshifted $\propto a^{-2}$, while $V(\phi)$ remains essentially constant (due to extreme flatness of the inflaton potential) for $H > H_{\rm MSSM}$. As a result, the universe inside the bubble will remain curvature dominated until $H \simeq H_{\rm MSSM}$.

At this point the inflaton field $\phi$ dominates the energy density and a phase of MSSM inflation begins. This blows the open universe inside the bubble and inflates away the curvature term. As long as the total number of e-foldings is $N_{Q}$ plus few, the observable part of the universe looks like flat today (within the limits of 5 year WMAP data)~\cite{Komatsu:2008hk}. Perturbations of the correct size with acceptable spectral index will be generated during the slow-roll phase, and the SM degrees of freedom will be created from the decay of $\phi$ field in the post-inflationary phase.


\subsection{Other examples of gauge invariant inflatons}

\noindent
{\it Within MSSM}:\\
So far we have studied the flat direction inflaton represented by a monomial superfield, $\Phi$, instead one
can also imagine  a polynomial $I$ spanned by the Higgses and the sleptons as an example~\cite{Enqvist:2003pb},
\begin{equation}
\label{polyn1}
I = \nu_1 H_u L_1 + \nu_2 H_u L_2 + \nu_3 H_u L_3
\end{equation}
where $\nu_i$ are complex coefficients, $H_u$ is the up-type Higgs and
$L_i$ are the sleptons. For a following field configuration, the polynomial
$I$ has a vanishing matter current and vanishing gauge fields
\cite{Enqvist:2003pb},
\begin{equation}
\label{struct321}
L_i = e^{-i\chi/2} \phi_i \left( \ba{ll} 0 \\ 1 \ea \right), \quad H_u =
e^{i\chi/2} \sqrt{\sum_i |\phi_i|^2} \left( \ba{ll} 1 \\ 0 \ea \right)\,,
\end{equation}
where $\phi_i$ are complex scalar fields and the phase $\chi$ is a
real field constrained by
\begin{equation}
\label{constr320}
\partial_{\mu} \chi = \frac{\sum_j J_j^\phi}{2i \sum_k |\phi_k|^2}\,, \qquad
J_i^{\phi} = \phi_i^* \partial_{\mu} \phi_i - \phi_i \partial_{\mu} \phi_i^*
\,.
\end{equation}
The field configuration in Eq.~(\ref{struct321}) leads to an effective
Lagrangian for the flat direction fields $\phi_i$,
\begin{equation}
\label{lagr324}
{\mathcal{L}} = \left |D_{\mu} H_u \right |^2 + \sum_{i=1}^3|D_{\mu} L_i|^2 -V =
\frac{1}{2} \partial_{\mu} \Phi^{\dagger} \left(1 + P_1 - \frac{1}{2}P_2
\right) \partial^{\mu} \Phi - V\,,
\end{equation}
where $D_{\mu}$ is a gauge covariant derivative that reduces to the
partial derivative when the gauge fields vanish, $P_1$ is the
projection operator along $\Phi$ and $P_2$ along $\Psi$, where
\begin{eqnarray}
\bar{\phi} = \left( \phi_1 \quad \phi_2 \quad \phi_3 \right)^T \,, \qquad 
\Phi = \left( \begin{array}{ll} \bar{\phi} \\ \bar{\phi}^*
  \end{array} \right) \,, \qquad \Psi = \left( \begin{array}{ll}
    \,\,\bar{\phi} \\ -\bar{\phi}^* \end{array} \right) \,,
\end{eqnarray}
and the corresponding equation of motion
\begin{eqnarray}
\label{eqm327}
\partial_{\mu}\partial^{\mu}\Phi + 3H\dot\Phi +
\left(1-\frac{1}{2}P_1+P_2\right) \dd{V}{\Phi^{\dagger}}
- R^{-2} \left[\partial_{\mu}\Psi\,(\Psi^{\dagger}
  \partial^{\mu}\Phi)\right. & & \nonumber \\ + \left.
  \Psi\,(\partial_{\mu}\Psi^{\dagger} P_2 \partial^{\mu}\Phi) +
  \frac{1}{2}\Phi\,\partial_{\mu} \Phi^{\dagger}
  \left(1-P_1-\frac{3}{2}P_2\right) \partial^{\mu}\Phi \right] &=& 0,
\end{eqnarray}
where $R=\sqrt{\Phi^{\dagger}\Phi}$. We are interested in the
background dynamics where all the fields are homogeneous in time, and
for simplicity we study only the radial motion, such that $\Phi = R
\hat{e}_{\Phi}$, where $\dot{\hat{e}}_{\Phi}=0$ (the dot denotes 
derivative w.r.t time). Then the equation of motion simplifies to
\begin{equation}
\ddot R + 3H\dot R + \frac{1}{2} \frac{\partial V}{\partial R} = 0\,,
\end{equation}
A notable feature is that the
fields have non-minimal kinetic terms, since the field manifold
defined by the flat direction is curved, actually a hyperbolic
manifold. This results into the usual equation of motion for one
scalar field with a potential for the radial mode except for the
factor $1/2$, which makes the potential effectively flatter in this
direction. This can be traced to the square root nature of $H_u$ in
Eq.~(\ref{struct321}).

As far as our
example of $LH_{u}$ is concerned there are only three families which
we can account for. The flattest MSSM direction, $QuQue$, is lifted by
$n=9$ superpotential operator, $QuQuQuH_{d}ee$. The flat direction
$QuQue$ is an 18 complex dimensional manifold, Ref.~\cite{Gherghetta:1995dv}. The largest D-flat direction
is only 37 complex dimensional \cite{Gherghetta:1995dv}. One can imagine
a larger representation which will have larger number of $D$-flat directions which 
can mimic {\it assisted inflation}~\cite{Jokinen:2004bp}~\footnote{Let us consider $M$ fields $H_i$ and $N-1$
fields $G_j$ in the fundamental representation $N$ of the gauge
group $SU(N)$. Note that the matter content is also enhanced, which has a total
$N-1+M$ degrees of freedom.  Then there exists a $D$-flat direction
described by a gauge invariant polynomial
$I=\sum_{j=1}^M \alpha_j \eps_{d_1\cdots d_{N-1} e} H_1^{d_1}\cdots H_{N-1}^{d_{N-1}} G_j^e$,
which after solving the constraint equations
${\partial I}/{\partial H_j^a} = C H_j^{a*},~ {\partial I}/{\partial G_i^a} = C G_i^{a*}$
produces a vacuum configuration: $H_j^a = \delta_N^a \phi_j$, for $ j=1,\cdots,M$, and 
$G_i^a = \delta_i^a \sqrt{\sum_{j=1}^M |\phi_j|^2}$, for  $i=1,\cdots,N-1$.
When one substitutes these into $D$-terms, one finds that all $D$-terms vanish.
The Lagrangian for the flat direction is given by ${\cal L} = c\sum_{j=1}^{M} |D_{\mu} H_j|^2 + c\sum_{i=1}^{N-1}
|D_{\mu} G_i|^2 - V(\{H_i,G_j\})$, where $c=1/2$ for the real fields, and $c=1$ for the complex fields~\cite{Jokinen:2004bp,Brandenberger:2003zk}. 
The lagrangian reduces to: ${\cal L} = \frac{1}{2} \partial_{\mu}\Phi^{\dagger} \left[1+(N-1)P\right]
\partial^{\mu}\Phi - V(\Phi)$ where $P = \Phi \Phi^{\dagger} / (\Phi^{\dagger}\Phi)$ is the
projection operator, and the field configurations of the real fields are:
$\Phi =  (\phi_1,\ldots,\phi_M)^T$, for $ \phi_i \in {\cal R}$.
Similar generalizations can be made for $N\times N$ non-commutative hermitian matrices, 
see~\cite{Ashoorioon:2009wa,Ashoorioon:2009sr}.}.\\

\noindent
{\it Beyond MSSM}:\\
A gauge invariant inflationary model has been proposed sometime
ago in in Ref.~\cite{Lazarides:1993fi}. The idea is that $(SU(3))^3$
gauge group is spontaneously broken down to $SU(3)_c\times SU(2)_L\times SU(2)_R\times U(1)_{B-L}$.
The quartic contribution to the superpotential is given by a $D$-flat direction of $(SU(3))^3$,
 $\Phi$, which is invariant under $27$,
 \begin{equation}
 W\sim \frac{\lambda}{M_{\rm P}}(27 \overline{27})^2\,,
 \end{equation}
where $\lambda$ is a small number determined by matching the amplitude of the CMB observations,
and $\Phi$ is the monomial representing, $N, \bar N$ or $ \nu^{c}, \bar\nu^{c}$.
The potential along such a $D$-flat direction is given by~\cite{Lazarides:1993fi}:
\begin{equation}
V(\phi) \approx - M_S^2 \mid \phi\mid^2 + \frac{\lambda^2}{3}
\frac{\mid\phi\mid^6}{M_P^2}\,,
\end{equation}
where $M_S \sim 10^{3}$~GeV, denotes the soft SUSY breaking scale. It was argued that 
the negative mass squared term would appear due to running in presence of strong 
dynamics~\cite{Lazarides:1993fi}. Inflation happens near $\phi\sim 0$, and ends with a VEV,
$\phi \sim M\sim \lambda^{-1/2}\sqrt{M_{\rm P}M_S}$~GeV. In order to match the CMB
temperature anisotropy,  $\Delta T/T\approx 0.023\lambda N_Q^2$, where $N_Q=(2\pi/3)(\phi/M_{\rm P})^2$
is the number of e-foldings before the end of inflation. We require $M\sim 10^{15}$~GeV 
and $\lambda\sim 10^{-7}$~\cite{Lazarides:1993fi}. The spectral index for the scalar perturbations 
tend to be small $n \simeq 0.92 - 0.88$, while the ratio of the tensor to the scalar 
ratio is given by; $r \approx 0.4 - 0.7$.


\section{Inflaton decay, reheating and thermalization}\label{sec:reheating}
%
%

\subsection{Perturbative decay and thermalization}
\label{PDT}

For a plasma which is in full thermal equilibrium, the energy density,
$\rho$, and the number density, $n$, of  relativistic particles are
given by~\cite{Kolb:1988aj,Sarkar:1995dd}
\begin{eqnarray} \label{full}
\rho &=& \left({\pi^2/30}\right) T^4\,, ~~~~~~~~~~n = 
\left({\zeta(3)/\pi^2}\right)T^3\,, ~~~~~~~~~~~~ ({\rm Boson}) \, , 
\nonumber \\ 
\rho &=& \left({7/8}\right) \left({\pi^2/30}\right) T^4\,, ~~~ 
n = \left({3/4}\right) \left({\zeta(3)/\pi^2}\right) T^3\,, ~~~~ 
({\rm Fermion}) \, ,
\end{eqnarray}
where $T$ is the temperature of a thermal bath. Note that in a full
equilibrium the relationships, $\langle E \rangle \sim \rho^{1/4}$,
and $n \sim \rho^{3/4}$ hold, with $\langle E \rangle = \left(\rho/n
\right) \simeq 3 T$ being the average particle energy. On the other
hand, right after the inflaton decay has completed, the energy density
of the universe is given by: $\rho \approx 3 \left(\Gamma_{\rm d}
M_{\rm P}\right)^2$. For a perturbative decay, which generates
entropy, we have $\langle E \rangle \approx m_{\phi} \gg \rho^{1/4}$. Then, from
the conservation of energy, the total number density is found to be,
$n \approx \left(\rho/m_{\phi}\right) \ll \rho^{3/4}$. Hence the {\it
complete inflaton decay} results in a dilute plasma which contains a
small number of very energetic particles. This implies that the
universe is far from full thermal equilibrium initially~\cite{Albrecht:1982mp,Turner:1983he,Davidson:2000er,Allahverdi:2005mz,Kofman:1997yn,Kofman:1994rk,Shtanov:1994ce,Traschen:1990sw,Abbott:1982hn}.

Reaching full equilibrium requires re-distribution of the energy among
different particles, {\it kinetic equilibrium}, as well as increasing
the total number of particles, {\it chemical equilibrium}. Therefore
both the number-conserving and the number-violating reactions must be involved. 

\vspace*{16mm}
%
\SetScale{0.6} \SetOffset(50,40)
\begin{picture}(350,100)(0,0)
\ArrowLine(75,90)(175,90) \Text(30,65)[l]{$f_1$}
\Vertex(175,90){3}
\ArrowLine(75,10)(175,10) \Text(30,0)[l]{$f_2$}
\Gluon(175,90)(175,10){5}{4} 
\ArrowLine(175,90)(275,90) \Text(175,65)[r]{$f_1$}
\Vertex(175,10){3}
\ArrowLine(175,10)(275,10) \Text(175,0)[r]{$f_2$}~~~~~~~~~~~~~~~~~~~~~~~~~~~~~~~~~~~~~~~~~~~~~~~~~~~
\SetScale{0.6} \SetOffset(50,40)
\ArrowLine(75,90)(175,90) \Text(30,65)[l]{$f_1$}
\Vertex(175,90){3}
\ArrowLine(75,10)(175,10) \Text(30,0)[l]{$f_2$}
\Gluon(175,90)(175,10){5}{4} 
\ArrowLine(175,90)(275,90) \Text(180,65)[r]{$f_1$}
\Vertex(175,10){3}
\Vertex(215,90){3}
\Gluon(215,90)(275,150){5}{4}
\ArrowLine(175,10)(275,10) \Text(180,0)[r]{$f_2$}
\end{picture}
\vspace*{-13mm}

\vspace*{-0mm}
\noindent
{ Fig.~A}~Typical scattering diagram which builds kinetic
equilibrium in the reheat plasma. Note that the $t-$channel
singularity which results in a cross-section $\propto \vert t \vert^{-1}$. 
The second panel shows a typical scattering diagram which increase the number of 
particles.

\vspace*{16mm}

\begin{itemize}

\item{Kinetic equilibrium among SM fermions:\\
The most important processes are $2 \rightarrow 2$ scatterings with gauge boson exchange
in the $t$-channel, shown in Fig.~(A). The cross-section for these
scatterings is $\sim \alpha \vert t \vert^{-1}$.  Here $''t''$ is
related to the exchanged energy, $\Delta E$, and the momentum,
$\overset {\longrightarrow}{\Delta p}$, through 
$t = {\Delta E}^2 -{\vert \overset {\longrightarrow}{\Delta p} \vert}^2$. The fine
structure constant is denoted by $\alpha$ (note that 
$\alpha \geq 10^{-2}$ in the SM/MSSM).  This cross section can be understood as
follows: the gauge boson propagator introduces a factor of ${\vert t\vert}^{-2}$, 
while phase space integration results in an extra factor
of $\vert t \vert$. Scalar exchange in $t$-channel diagrams are usually 
suppressed, similarly a fermion-fermion-scalar vertex, which arises from a
Yukawa coupling, flips the chirality of the scattered fermion, are also suppressed.
Due to an infrared singularity, these scatterings are very efficient even in a dilute 
plasma~\cite{Davidson:2000er,Allahverdi:2005mz}.
}

\item{Chemical equilibrium:\\
In addition one also needs to achieve chemical equilibrium by changing
the number of particles in the reheat plasma. The {\it relative}
chemical equilibrium among different degrees of freedom is built
through $2 \rightarrow 2$ annihilation processes, occurring through
$s-$channel diagrams. Hence they have a much smaller cross-section
$\sim \alpha s^{-1}$.  More importantly the total number of particles
in the plasma must also change.  It turns out from Eq.~(\ref{full})
that in order to reach full equilibrium, the total number of particles
must {\it increase} by a factor of: $n_{\rm eq}/n$, where $n \approx
\rho/m_{\phi}$ and the equilibrium value is: $n_{\rm eq} \sim
\rho^{3/4}$. This can be a very large number, i.e. $n_{\rm eq}/n\sim {\cal O}(10^3)$. 
It was recognized in~\cite{Davidson:2000er,Allahverdi:2002pu}, see 
also~\cite{Jaikumar:2002iq,Allahverdi:2005fq,Allahverdi:2000ss}, that the most
relevant processes are $2 \rightarrow 3$ scatterings with gauge-boson
exchange in the $t-$channel. Again the key issue is the infrared
singularity of such diagrams shown in Fig.~(A). The cross-section for
emitting a gauge boson, whose energy is ${\vert t \vert}^{1/2} \ll E$,
from the scattering of two fermions is $\sim \alpha^3 {\vert t
\vert}^{-1}$. When these inelastic scatterings become efficient, i.e.,
their rate exceeds the Hubble expansion rate, the number of particles
increases very rapidly~\cite{Enqvist:1993fm}, because the produced gauge bosons
subsequently participate in similar $2 \rightarrow 3$ scatterings.
Decays (which have been considered in~\cite{Allahverdi:2000ss}) are helpful, but in 
general they cannot increase the number of particles to the required level. }

\end{itemize}

The full thermal equilibrium will be established
shortly after the $2 \rightarrow 3$ scatterings become efficient. For
this reason, to a very good approximation, one can use the rate for
inelastic scatterings as a thermalization rate of the universe
\begin{equation}
\label{pert-dec-time}
\Gamma_{\rm th}\sim \alpha^3\left(\frac{M_{\rm P}}{m_{\phi}}\right)\Gamma_{\rm d}\,,
\end{equation}
Since the inflaton decay products have SM
gauge interactions, the universe reaches full thermal equilibrium
immediately after the inflaton decay. The reason is that the 
$2 \rightarrow 3$ scatterings with gauge boson exchange in the
$t-$channel are very efficient, see~\cite{Davidson:2000er,Allahverdi:2002pu,Allahverdi:2005mz}.


Even before all inflatons decay, the decay products form a plasma
can very quick thermalize, and the plasma has the instantaneous
temperature given by Eq.~(\ref{instT}). The plasma can reach its 
maximum $T_{\rm max}$ soon after
the inflaton field starts to oscillate around the minimum of its
potential, which happens for a Hubble parameter $H_I \leq m_\phi$. 
During this era the
energy density of the universe is still dominated by the
(non--relativistic) inflatons that haven't decayed yet. The scale
factor of the universe $a$ then varies as $a \propto T^{-8/3}$
\cite{Kolb:1988aj}. The universe remains in this phase as long as $H >
\Gamma_{\rm d}$. During this phase one can produce massive
long--lived or stable weakly interacting massive particles (WIMPS) \cite{Chung:1998zb,Chung:1998ua,Chung:1998rq,Giudice:2000dp,Kudo:2001ie,McDonald:1999hd,Allahverdi:2002nb,Allahverdi:2002pu,Giudice:2000ex}, see also gravitational production of 
particles~\cite{Chung:2001cb,Chung:1999ve}. There
are three possible scatterings which have been discussed in the 
literature.

Particle creation via Soft-soft scatterings were
investigated in detail in Refs.~\cite{Chung:1998zb,Chung:1998ua,Chung:1998rq,Giudice:2000dp,Kudo:2001ie,Giudice:2000ex},
where the relevant Boltzmann equations governing the production and annihilation
of stable particles, $\chi$'s, are solved both numerically and analytically. In
Refs.~\cite{Chung:1998zb,Chung:1998ua,Chung:1998rq,Giudice:2000ex}, out of equilibrium production of $\chi$ from
scatterings in the thermal bath were studied and the final result is
found to be (the superscript ``ss'' stands for $\chi$ production from
``soft--soft'' scatterings)~\footnote{Particle creation via hard--soft scatterings, and 
hard--hard scatterings were also considered in~\cite{Allahverdi:2002nb,Allahverdi:2002pu}.
It was found that the $\chi$ production through hard--hard scattering is most efficient {\em before} thermalization is completed.}:
\begin{eqnarray}  \label{ssrate1}
\Omega^{\rm ss}_\chi h^2 &\sim& \left ({200 \over g_*} \right )^{3/2}
\alpha_\chi^2 {\left ({2000 T_{\rm R} \over m_\chi} \right )}^7 \,.
\end{eqnarray}
Here $\Omega_\chi$ is the $\chi$ mass density in units of the critical
density, and $h$ is the Hubble constant in units of $100$
km$/$(s$\cdot$Mpc). The cross section for $\chi$ pair
production or annihilation is given by: $\sigma \simeq \alpha_\chi^2 /
m^{2}_{\chi}$. Most $\chi$ particles are produced at $T \simeq
m_\chi/4$. The density of earlier produced particles is
strongly red--shifted, while $\chi$ production at later times is
suppressed by the Boltzmann factor.


\subsection{Non-perturbative inflaton scatterings}
\label{NPID}

Many studies have  been devoted to understand non-perturbative effects during reheating. 
Various non-thermal and non-perturbative effects may lead to a rapid transfer of the
inflaton energy to other degrees of freedom by the process known as
{\it preheating}.  The requirement is that  the inflaton quanta  couple to other
(essentially massless) field $\chi$ through, i.e. terms like $\phi^2\chi^2$.
The quantum modes of $\chi$ may then be excited during the inflaton
oscillations via a {\it parametric resonance}. Preheating has been treated
both analytically
\cite{Traschen:1990sw,Shtanov:1994ce,Kofman:1994rk,Cooper:1994hr,Boyanovsky:1996sq,Kofman:1997yn,Boyanovsky:1997xt,Boyanovsky:1996rw,Boyanovsky:1997cr,Cooper:1996ii,Kofman:1995fi,Allahverdi:1996xc,GarciaBellido:1997wm, Greene:1997fu,BasteroGil:1999fz,Baacke:1996se,Baacke:1996kj,Baacke:1997rs,Greene:1997ge,Cormier:2001iw,GarciaBellido:2003wd,Felder:1998vq,Felder:1999pv,GarciaBellido:2008ab} (for an elaborate discussion on reheating and preheating, 
see~\cite{Kofman:1997yn}), and numerically on lattice simulations 
\cite{Khlebnikov:1996mc,Khlebnikov:1996wr,Khlebnikov:1996zt,Micha:2002ey,Micha:2004bv,Prokopec:1996rr,Tranberg:2003gi,Arrizabalaga:2004iw,Aarts:1998td,Aarts:1999zn,Podolsky:2005bw,Felder:2006cc}. 
Like bosons, fermions can also be excited during preheating
\cite{Dolgov:1989us,Baacke:1998di,Greene:1998nh,Giudice:1999fb,Greene:2000ew,Peloso:2000hy}. In fact, it has been argued that fermionic preheating is more efficient than
bosonic preheating, however these fermions can not be related to the chiral fermions of the SM. The SM 
fermions can only couple to a gauge singlet inflaton via non-renormalizable dimensional $5$ operators, therefore the effective
couplings are very small. During preheating it is possible to excite gravity waves~\cite{Khlebnikov:1997di,GarciaBellido:2007dg,Mazumdar:2008up,GarciaBellido:2007af,Fenu:2009qf,Easther:1999ws,Easther:2007vj,Kusenko:2008zm,Kusenko:2009cv,Gumrukcuoglu:2008gi,Dufaux:2008dn}, magnetic field~\cite{DiazGil:2007dy,DiazGil:2008tf}, gravitino abundance with spin 
$\pm 1/2,~\pm 3/2$~\cite{Maroto:1999ch,Kallosh:1999jj,Giudice:1999am,Kallosh:2000ve,Allahverdi:2000fz,Nilles:2001ry,Nilles:2001fg}, moduli and non-thermal stringy relics~\cite{Giudice:2001ep}, phase transitions~\cite{Khlebnikov:1998sz,Felder:2000hj}. A successful cold electroweak baryogenesis were also studied in the context of preheating~\cite{GarciaBellido:1999sv,Cornwall:2000eu,Cornwall:2001hq}.  For a recent review on reheating and 
preheating, see~\cite{Allahverdi:2010xz}.

During preheating, it is also possible to excite the perturbed FRW metric potential, see~\cite{Bassett:1998wg,Maartens:1997fg,Bassett:1999mt,Gordon:2000hv,Finelli:2000ya,Easther:1999ws}, however as shown in Ref.~\cite{Jedamzik:1999um}, it is hard to excite large metric  perturbations in $g^2\phi^2\chi^2$ theory. For large $g$ and large inflaton VEV, $\phi$, the initial $\chi$ perturbations in the vacuum are very much suppressed, see also~\cite{Bassett:1999ta}. The second order metric perturbations can also leave non-Gaussian 
signature during preheating~\cite{Enqvist:2004ey,Enqvist:2005qu,Enqvist:2005nc,Jokinen:2005by,Bond:2009xx}, which may put severe
constraint on a simple $\lambda\phi^4$ inflation model~\cite{Jokinen:2005by}.\\


\subsubsection{Parametric Resonance}

Let us briefly review the initial stages of a  gauge singlet inflaton decay,
which happens typically non-perturbatively, i.e. preheating, in most of the 
non-SUSY cases. In SUSY, there are complications with the potential itself, 
as well as the presence of MSSM flat directions.

Our focus is on bosonic preheating which acts most
efficiently in transferring the energy density from the inflaton
oscillations. We consider models of large field inflation, such as
chaotic inflation or hybrid models, for which bosonic preheating is most pronounced.
The relevant renormalizable couplings between the inflaton $\phi$
and a scalar field $\chi$ will read from the following potential:
\begin{equation} \label{nonpot}
V = {1 \over 2} m^2_{\phi} \phi^2 + \frac{1}{2}m_{\chi}^2\chi^2+\sigma \phi \chi^2 + h^2 \phi^2 \chi^2 +
\lambda \chi^4\,, \end{equation}
where we have considered $\phi$ and $\chi$ to be real. Here $\sigma$
is a coupling which has a [mass] dimension. The only scalar field in
the SM is the Higgs doublet. Therefore in a realistic case $\chi$
denotes the real and imaginary parts of the Higgs
components. The cubic interaction term is
required for a complete inflaton decay.  The quartic self-coupling of
$\chi$ is required to bound the potential from below along the $\chi$
direction. The dimensionless couplings $\sigma/m_{\phi}$ and $h$ (as
well as $\lambda$) are not related to each other, hence either of the
cubic or the quartic terms can dominate at the beginning of inflaton
oscillations (i.e. when the Hubble expansion rate is $H(t) \simeq
m_{\phi}$ and the amplitude of oscillations is ${\hat \phi} \sim {\cal
O}(M_{\rm P})$). 

In a non-SUSY case efficient
preheating happens over a narrow window $3 \times 10^{-4} \leq h \leq
10^{-3}$. The reason is that the $h^2 \phi^2 \chi^2$ term yields a
quartic self-coupling for the inflaton at a one-loop level which is
constrained by the CMB normalization of the density
perturbations, i.e. $\lambda\leq 10^{-12}$. However, in SUSY this
correction is canceled out by that from fermionic partner of $\chi$,
so in principle one could expect a rather broader range of parameter
space. Neglecting the self interaction for $\chi$ field, the 
equation of motion for $\chi_{k}$ quanta is given by:
\begin{equation}
\ddot\chi_{k}+3\frac{\dot a}{a}\dot\chi_{k}+\left(\frac{k^2}{a^2}+m^2_{\chi}+2(\sigma\phi+h^2\phi^2)\right)\chi_{k}=0\,.
\end{equation}
It is assumed that the inflaton oscillations are homogeneous, $\phi(t)=\hat\phi(t)\sin(m_{\phi}t)$, where 
$\hat\phi(t)\approx (M_{\rm P}/\sqrt{3\pi}m_{\phi}t)$, for chaotic inflation with mass $m_{\phi}$. 
The occupation number for 
the excited $\chi_k$ is given by:
\begin{equation}
n_{k}=\frac{\omega_{k}}{2}\left(\frac{|\dot\chi_k|^2}{\omega_k^2}+|\chi_k|^2\right)-\frac{1}{2}\,,
\end{equation}
It was observed in Refs.~\cite{Traschen:1990sw,Shtanov:1994ce,Kofman:1994rk,Kofman:1997yn}, that 
in general one can have a {\it narrow resonance}, when expansion of the universe and the trilinear interaction
are neglected, then the evolution for $\chi_k$ yields a Mathieu 
equation, which has well known instability bands, during which the mode 
grows exponentially, $\chi_k\propto \exp(\mu^{n}_k z)$, where  $\mu^n_k$ is set by the instability band 
$\Delta_k^{n}$ labeled by an integer $n$, and $z=m_{\phi}t$. The 
resonance occurs for $k=0.5 m_{\phi}(1\pm q/2)$, where $\mu_k$ vanishes at the edges and takes the maximum value 
$\mu_k=q/2$, where $q=g^2(\hat\phi^2/4m_{\phi}^2)$. Thus the occupation number grows exponentially. The situation 
changes quite dramatically when one switches the expansion rate of the universe, the evolution of the scalar field during 
the first $10-50$ oscillations modifies to:
\begin{equation}\label{inf-osc}
\phi (t) \simeq \frac{M_{\rm P}}{\sqrt{3\pi}}
\frac{\cos\left( m_\phi  t \right)}{m_\phi  t} \,,
\end{equation}
where $t$ is the physical time. The presence of the $t$ at the denominator shows the damping of the oscillations due to the
expansion of the universe. During this period the {\it stochastic resonance }
come into the picture~\cite{Kofman:1997yn}, where there are resonance bands as well as decrease in the particle number due to quantum effects. 

In either case (expanding or non-expanding background), based on initial VEV of $\sigma$ there would be two distinct cases.

\begin{itemize}

\item{$\underline{\sigma \ll h^2 M_{\rm P}}$:\\
In this regime the $h^2 \phi^2
\chi^2$ term is dominant at the beginning of the inflaton
oscillations.  This case has been studied in detail in first two
references of~\cite{Kofman:1994rk,Kofman:1997yn}.  For a nominal value of the inflaton
mass, i.e. $m_{\phi} = 10^{13}$ GeV in chaotic inflation case, non-perturbative $\chi$ production
during every oscillations of $\phi$ field, with a physical momentum, $k \ls \left( h m_{\phi} {\hat
\phi}\right)^{1/2}$ (where $\hat\phi\sim M_{\rm P}$), takes place if $h > 10^{-6}$.  Particle
production is particularly efficient if $h > 3 \times 10^{-4}$, and
results in an explosive transfer of energy to $\chi$ quanta. The number
density of $\chi_{k}$ quanta increases exponentially. The parametric resonance
ends when re-scatterings destroy the inflaton condensate. The whole process happens over a time scale
$\sim 150 m^{-1}_{\phi}$, which depends logarithmically on $h$~\cite{Kofman:1997yn,Micha:2004bv}.
}

\item{$\underline{\sigma \gg h^2 M_{\rm P}}$:\\
In this regime the cubic term
$\sigma \phi \chi^2$ dominates.  This case was recently considered in
Refs.~\cite{Shuhmaher:2005mf,Podolsky:2005bw}, where the the $\chi$ field becomes
tachyonic during half of each oscillation.  For $\sigma >
m^2_{\phi}/M_{\rm P}$ (which amounts to $\sigma > 10^7$ GeV for
$m_{\phi} = 10^{13}$ GeV) this tachyonic instability transfers energy
from the oscillating condensate very efficiently to the $\chi$ quanta
with a physical momentum $k \ls \left(\sigma {\hat \phi}
\right)^{1/2}$. Particle production ceases when the back-reaction from
$\chi$ self-coupling induces a mass-squared $\gs \sigma {\hat
\phi}$. Depending on the size of $\lambda$, most of the energy density
may or may not be in $\chi$ quanta by the time back reaction becomes
important~\cite{Dufaux:2006ee}.}

\end{itemize}

Couple of points to note here. In the borderline regime $ \sigma
\sim h^2 M_{\rm P}$, the cubic and quartic interaction terms are
comparable.  The inflaton decay happens due to a combination of
resonant and tachyonic instabilities. If $h \ll m_{\phi}/M_{\rm P}$
and $ \sigma \ll m^2_{\phi}/M_{\rm P}$, the inflaton decays
perturbatively via the cubic interaction term. However this requires
very small couplings; $h,~\left(\sigma/m_{\phi}\right) < 10^{-6}$.
Therefore, unless the inflaton is only gravitationally coupled to
other fields, the initial stage of its decay will be generically
non-perturbative. 

Resonant particle production and re-scatterings lead to the formation
of a plasma consisting of $\phi$ and $\chi$ quanta with typical
energies $\sim 10^{-1} \left(h m_{\phi} M_{\rm
P}\right)^{1/2}$, see~\cite{Kofman:1997yn,Greene:1997ge,Khlebnikov:1998sz,Felder:2000hj}. This plasma is in kinetic
equilibrium but full thermal equilibrium is established over a much
longer time scale than preheating~\cite{Felder:2006cc,Micha:2004bv}.

The occupation number of particles in the preheat plasma is $\gg 1$
(which is opposite to the situation after the perturbative
decay). This implies that the number density of particles is larger
than its value in full equilibrium, while the average energy of
particles is smaller than the equilibrium value.  It gives rise to
large effective masses for particles which, right after preheating, is
similar to their typical momenta~\cite{Kofman:1997yn,Greene:1997ge,Khlebnikov:1998sz}.  Large occupation
numbers also lead to important quantum effects due to identical
particles and significant off-shell effects in the preheat plasma.
Because of all these, a field theoretical study of thermalization is
considerably more complicated in case of preheating. Due to the large
occupation numbers, one can consider the problem as thermalization of
classical fields at early stages~\cite{Micha:2002ey,Felder:2006cc,Micha:2004bv,Podolsky:2005bw}.  In the course of
evolution towards full equilibrium, however, the occupation numbers
decrease. Therefore a proper (non-equilibrium) quantum field theory
treatment will be inevitably required at late stages when
occupation numbers are close to one.

Preheating ends due to back reaction as well as the expansion of the universe. Preheating
does not destroy the zero mode of the inflaton condensate completely,
although the amplitude of the inflaton oscillations diminish, but the
inflaton decay is completed when the zero mode perturbatively decays
into the SM or some other degrees of freedom,
see~\cite{Traschen:1990sw,Shtanov:1994ce,Kofman:1994rk,Kofman:1997yn}.
It was found that the relevant time scale for thermalization after preheating is given by~\cite{Micha:2002ey,Micha:2004bv}:
\begin{equation}
\Gamma_{\rm th}\sim \left(\frac{m_{\phi}}{(c_r \rho_{inf})^{1/4}}\right)^{1/p}\,,
\end{equation}
where $c_r$ is the fraction of the inflaton energy density stored in the coherent oscillations at the onset 
of the turbulent phase and $p\sim 1/7$ obtained from the numerical simulations. Typically this time scale can be, 
$t_{\rm th}\sim c_r^{7/4}10^{21}$ for $m_{\phi}\sim 10^{13}$~GeV and $\rho_{inf}\sim 10^{64}~({\rm GeV})^4$,
comparable to the perturbative inflaton decay rate, see Eq.~(\ref{pert-dec-time}).

One of the most interesting effects of preheating is the copious production 
of particles which have a mass greater than
the inflaton mass $m_{\phi}$.  Such processes are impossible in perturbation 
theory and in the theory of narrow parametric resonance. However, 
superheavy $\chi$-particles with mass $M \gg m_{\phi}$ can be produced in the regime
of a broad parametric resonance. For very small $\phi(t)$ the change in the
frequency of oscillations $\omega (t)$ ceases to be
adiabatic when the adiabaticity condition is violated~\cite{Kofman:1997yn}
\begin{equation}
\label{adiab}
\frac{d\omega(t)}{d t} \geq \omega^2(t)\,.
\end{equation}
The momentum dependent frequency, $\omega_k(t)$ violates
the above condition when
\begin{equation}\label{adiabAAA}
{k^2 + m^2_{\chi}} \lesssim (h^2\phi m_\phi
\hat\phi)^{2/3} - h^2\hat\phi^2 \ .
\end{equation}
The maximal range of momenta for
which particle production occurs corresponds to $\phi(t) = \phi_*$, where
$\phi_* \approx {\textstyle {1 \over 2}} \sqrt {m_\phi\hat\phi\over h}$. The
maximal value of momentum for particles produced at that epoch can be
estimated by $k^2_{\rm max} + m^2_{\chi} = {h m_\phi \hat\phi \over 2}$.  The resonance becomes efficient for
$ h m_\phi \hat\phi \gtrsim 4 m^2_{\chi} $.
Thus, the inflaton oscillations may lead to a copious production of
superheavy particles with $m_{\chi} \gg m$ if the amplitude of the field $\phi$ is
large enough, $h\hat\phi \gtrsim 4m^2_{\chi}/m_{\phi}$.

During the second stage of preheating both $m_\phi$ and $\hat\phi$ change very
rapidly, but their product remains almost constant because the energy density
of the field $\phi$, which is proportional to $m^2_\phi \hat\phi^2/2$, practically
does not change until the very end of preheating. Therefore it is sufficient to
check that $hm\hat\phi \gtrsim 4m^2_{\chi} $ at the end of the first stage of preheating.
One can represent this criterion in a simple form~\cite{Kofman:1997yn}:
\begin{equation}\label{QQQ}
m_{\chi} \lesssim {m_{\phi}\over \sqrt 2}\, q^{1/4} \approx  {m_{\phi} } \left({ hM_{\rm p}\over 
3m_{\phi}}\ln^{-1} {10^{12} m_{\phi}
\over h^5 M_{\rm p} }\right)^{1/2} .
\end{equation}
For example, one may take $ m_{\chi}= 2m_{\phi}$ and $h \approx 0.007$, which corresponds to
$q_0 = h^2\hat\phi^2/m^2_{\phi}=10^6$ . The production of $\chi$-particles with $m_{\chi} = 10 m_{\phi}$ is possible for   
$h \gtrsim 10^{-2}$. Anyway, as we shall see in a realistic SUSY case, the existence of heavy mass of $\chi$
induced by the flat direction VEV of MSSM can kinematically block resonant preheating altogether~\cite{Allahverdi:2007zz},
see Sec.~\ref{See: kinematical blocking of preheating}.


\subsubsection{Instant preheating}
\label{See: Instant preheating}

Let us focus solely on the interaction $h^2\phi^2\chi^2$.
In instant preheating the particle production occurs
during one oscillation of the inflaton~\cite{Felder:1998vq}. The particle production occurs
when the inflaton passes through the minimum of the potential
$\phi=0$. In this case the process can be approximated by writing
$\phi = \dot\phi_0 (t-t_0)$, where $\dot\phi_0$ is the velocity of the field when it passes
through the minimum of the potential at time $t_0$. The time interval
within which the production of $\sigma$ quanta occurs
is~\cite{Felder:1998vq} $\Delta t_* = ( g|\dot\phi_0|)^{-1/2}$,
which is much smaller than the Hubble expansion rate; thus expansion
can be neglected. The occupation number of produced particles jumps from its initial
value zero to a non-zero value during $-\phi_{\ast}\leq
\phi\leq \phi_{\ast}$. In the momentum space the occupation
number is given by~\cite{Felder:1998vq} $n_k = \exp \left( - \frac{\pi k^2}{g |\dot\phi_0|} \right)$,
and the largest number density of produced particles in $x$-space
reads~\cite{Felder:1998vq}
\begin{equation}
\label{number}
n_{\chi} \approx \frac{(h|\dot\phi_0|)^{3/2}}{8\pi^3}\,,
\end{equation}
with the particles having a typical energy of
$(g|\dot\phi_0|/\pi)^{1/2}$, so that their total energy density is
given by
\begin{equation}
\label{energy}
\rho_{\chi}\sim \frac{1}{2} (\delta^{(1)}\dot\chi)^2
 \sim \frac{(g|\dot\phi_0|)^2}{8\pi^{7/2}} \,.
\end{equation}
These expressions are valid if $m_{\chi}^2 < g|\dot\varphi_0|$. Instant preheating has applications especially when 
gauge fields and fermions are involved~\cite{Kofman:2004yc,Allahverdi:2006we}. One particular interesting point is when
the modulus is carrying the SM gauge charges and passing through the point of {\it enhanced gauge symmetry}, i.e. $\langle \phi\rangle \approx 0$ (see Sec.~\ref{RTCSB}). Where the gluons are nearly massless, then they can be excited with a similar abundance given by Eq.~(\ref{number}). When the modulus is displaced away from the
point of enhanced gauge symmetry, the gluons become heavy due to modulus induced VEV dependent mass $\sim g\langle \phi\rangle$,
where $g$ is the gauge coupling. As the gluons become heavy they rapidly 
decay into fermions to reheat the plasma~\cite{Allahverdi:2005mz}. Transferring the inflaton energy through this mechanism is quite 
fast and efficient, see the discussion on reheating due to MSSM inflaton, Sec.~\ref{RTCSB}.


\subsubsection{Tachyonic preheating}
\label{See: Tachyonic preheating}

The second scenario is known as {\it tachyonic preheating}. Let us 
consider a simple example of tachyonic potential 
\begin{equation}
V=V_0-\frac{1}{2}m^2\chi^2+\frac{\lambda}{4}\chi^4
\end{equation}
The rolling of a tachyon in itself results in an exponential
instability in the perturbations of $\chi$ with physical momenta
smaller than the mass. The tachyonic growth takes place within a short
time interval, $t_{\ast}\sim (1/2m)\ln(\pi^2/\lambda)$
(see~\cite{Felder:2000hj}).  During this short period the occupation
number of $\chi$ quanta grows exponentially for modes $k<m$ up to
$n_{k}\sim \exp(2mt_{\ast})\sim \exp(\ln(\pi^2/\lambda))\sim
\pi^2/\lambda$.  For very small self-coupling, which is required for a
successful inflation, the occupation number, which depends inversely
on the coupling constant, can become much larger than one.
First, the number density of the produced particles in $x$-space is
given by $n_\chi\sim m^3/(8\pi\lambda)$. Hence the total energy
density stored in produced $\chi$ quanta is given by~\cite{Felder:2000hj}
\begin{equation}
\label{energy2}
\rho_{\chi}\sim \frac{1}{2} (\delta^{(1)}\dot\chi)^2
 \sim m n_{\chi} \sim \frac{1}{8 \pi} \frac{m^4}{\lambda}\,.
\end{equation}
The plasma from the non-perturbative inflaton decay eventually reaches
full thermal equilibrium, though, at time scales much longer than that
of preheating itself~\cite{Felder:2006cc,Micha:2004bv}. The occupation number of
particles is $f_{k}\gg 1$ in the meantime. This implies that dangerous
relics (such as gravitino and moduli) can be produced much more
copiously in the aftermath of preheating than in full thermal
equilibrium. This is a negative aspect of an
initial stage of preheating. One usually seeks a late stage of entropy
release, in order to dilute the excess of relics. As we shall show,
SUSY naturally provides us a tool to undo preheating~\cite{Allahverdi:2007zz}, see Sec.~\ref{See: kinematical blocking of preheating}


\subsubsection{Fermionic preheating}

The resonant and instant preheating calculations can be reanalyzed for a fermionic coupling,
$h\phi \bar\psi\psi$. In both the cases one would expect a large exponential growth in particle creation. 
It is also possible to excite superheavy fermions from resonant preheating~\cite{Greene:1998nh,Giudice:1999fb,Greene:2000ew,Peloso:2000hy}. However note that both $\phi$ and $\psi$ are SM gauge singlets. 

For an inflaton field coherently oscillating about the minimum
of the potential $V= \frac{1}{2} m_\phi^2  \phi^2$,
If one neglects the back reaction of the created particles, then
after few oscillations, the inflaton evolves according to the
formula Eq.~(\ref{inf-osc}). Thus, there exists a final time
after which $ | \phi | < m_\psi / h$, and the total mass no longer
vanishes, then the resonant production of fermion ends.

The Dirac equation (in conformal time $\eta$) for a fermionic field is given by:
\begin{equation} \label{dir1}
\left( \frac{i}{a}\,\gamma^\mu \,\partial_\mu + i\, \frac{3}{2}
H \gamma^0 - m(\eta) \right) \psi = 0\,,
\end{equation}
where $a$ is the scale factor of the universe, $H=a'/a^2$ the
Hubble rate and $\prime$ denotes derivative w.r.t. $\eta$, and 
$m (\eta)=m_\psi + h \phi( \eta)$, where $m_\psi $ is the bare mass of the
fermion. The particle density per physical volume $V=a^3$ at time
$\eta$ is given by:
\begin{eqnarray}
n ( \eta ) \equiv \langle 0 | \frac{N}{V} | 0 \rangle =\frac{1}{\pi^2 \, a^3} \int\, dk \, k^2
\left| \beta_k\right|^2 ,
\end{eqnarray}
where $\alpha_k, \beta_k$ are the Bugolyubov's coefficients satisfying: $|\alpha_k|^2+|\beta_k|^2=1$.
The occupation number of created
fermions is thus given by $n_k = | \beta_k |^2$, and the above
condition ensures that the Pauli limit $ n_k < 1$ is
respected.  One important physical quantity is the scaling of the total energy
\begin{equation}
\rho_\psi \propto m_\psi  N_\psi \propto q  m_\psi^{1/2}
\end{equation}
which is linear in $q=h^2\hat\phi^2/m_\phi^2$, as generally 
expected~\cite{Greene:1998nh,Giudice:1999fb,Greene:2000ew,Peloso:2000hy}, but 
also note that $m_\psi (t)\propto q^{1/2}$.

In a realistic case, since the SM fermions are chiral, if the inflaton is a 
SM gauge singlet, then it can only couple via dimension-$5$ operators, i.e.
\begin{equation}\label{inf-ferm-coup}
\frac{\lambda}{M_{\rm P}}\phi(H\bar q_l)q_R\,,
\end{equation}
where $\lambda\sim {\cal O}(1)$, $H$ is the SM Higgs doublet and $q_l, q_R$ are the $SU(2)_l$ doublet and the
right handed SM fermions, respectively. As a result preheating of SM fermions from a gauge singlet inflaton becomes less
important due to weak coupling. 

In Ref.~\cite{Giudice:1999fb}, it was argued that an inflaton coupling to right 
handed neutrino, $h\phi \bar N N$, where $N$ is right handed neutrino, will induce non-thermal leptogenesis, 
where the right handed neutrinos were treated gauge singlets. Anyway, if
we embed the right handed neutrinos in a gauge sector, where they get their masses via some Higgs mechanism, 
then one requires non-renormalizable couplings like Eq.~(\ref{inf-ferm-coup}).

Similar argument holds for coupling to the SM gauge bosons, where the inflaton can only couple 
via non-renormalizable operator, i.e. 
\begin{equation}
\frac{\lambda}{M_{\rm P}}\phi F_{\mu\nu}F^{\mu\nu}\,, 
\end{equation}
where $\lambda\sim {\cal O}(1)$. Therefore, exciting the SM gauge bosons and the SM fermions through parametric 
resonance of a gauge singlet inflaton is a daunting task. Inflaton would rather prefer perturbative decay~\footnote{The only
way one can excite SM fermions and gauge fields copiously, if they are directly excited by the oscillations of the SM Higgs boson.
This can happen in low scale electroweak baryogenesis~\cite{GarciaBellido:1999sv,Cornwall:2000eu,Cornwall:2001hq}, 
or in the context of SM Higgs inflation~\cite{GarciaBellido:2008ab}. During the Higgs oscillations the SM degrees of freedom
can be excited via parametric resonance, instant preheating and also via tachyonic preheating. All three phases of 
preheating are present. The other notable example is the MSSM inflation discussed in Sec.~\ref{RTCSB}, where gluons and MSSM fermions were excited via instant preheating. }.


\subsubsection{Fragmentation of the inflaton}

One very curious aspect of fermionic coupling to the inflaton is fragmentation 
of the inflaton to form an inflating non-topological solitons, known as Q-balls~\footnote{The Q-balls 
known to evaporate from their surface, see for a review~\cite{Enqvist:2003gh}, therefore suppressing 
the reheating and thermalization time scale.}. Let us illustrate this idea by studying a chaotic inflation model
where the inflaton field is not real but complex. Provided the fermions live in a 
larger representation than the bosons, the inflaton mass obtains a Logarithmic
correction~\footnote{Similar corrections to the potential arises for the MSSM flat directions in a gravity mediated
scenarios, where $m\sim {\cal O}(\rm TeV)$ and $K\sim \frac{\alpha}{8\pi}\frac{m^2_{1/2}}{m^2_{\tilde l}}$, where 
$m_{1/2}$ is the gaugino mass and $m_{\tilde l}$ is the slepton mass. Fragmentation of such flat direction can excite Q-balls and also gravity waves, see~\cite{Kusenko:2009cv,Kusenko:2008zm}.}:
\begin{equation}
    \label{qpotr}
    V = m^2 |\Phi|^2
    \left[ 1 - K\log\left(\frac{|\Phi|^2}{M^2}\right) \right]\,, \end{equation}
where the value of $K$ is determined by the
Yukawa coupling $h$ with $ K =-C({h^2}/{16\pi^2})$, where $C$ is
some number. If $K < 0$,
the inflaton condensate feels a negative pressure 
for field values $\phi \ll M$, we find:
\begin{equation}
    \label{pot000}
    V(\phi) \simeq \frac{1}{2}m_{3/2}^2\phi^2
    \left(\frac{\phi^2}{2M^2}\right)^K \propto \phi^{2+2K}\,.
\end{equation}
where we assume $|K|\ll 1$.  The equation of state for a field rotating in such a potential is 
\begin{equation}
    \label{state}
    p\simeq \frac{K}{2+K} \rho \simeq -\frac{|K|}{2}\rho \,,
\end{equation}
where $p$ and $\rho$ is a pressure and energy density of the scalar
field, respectively.  Evidently, the negative value of $K$ corresponds to the 
negative pressure, which signals the instability of the condensate. A linear perturbation
analysis~\cite{Enqvist:2002si} shows that the fluctuations grow exponentially if the following condition
is satisfied:
\begin{equation}
    \frac{k^2}{a^2}\left( \frac{k^2}{a^2}+2m_{3/2}^2K \right) < 0.
\end{equation}
Clearly, the instability band exists for negative $K$, as expected from the negative pressure arguments~\cite{Enqvist:2003gh}. The 
instability band, $k$,  is in the range~\cite{Enqvist:2002si}
$  0 < \frac{k^2}{a^2} < \frac{k_{max}^2}{a^2} \equiv  2m_{3/2}^2|K|$,
where $a$ is the expansion factor of the universe. 
The most amplified mode lies in the middle of the band, and the maximum growth rate  of the perturbations is determined by  
$\dot{\alpha} \sim ~|K|m_{3/2}/2$~\cite{Enqvist:2003gh}. When $\delta \phi/\phi_0\sim {\cal O}(1)$, the  fluctuations become nonlinear.  
This is the time when the homogeneous condensate breaks down into Q-balls and anti-Q-balls~\footnote{In general $K$ and $h$ are not independent quantities but are related to each other by $|K| \sim C(h^2/16\pi^2)$.  In this regime the evaporation rate is
saturated by: $  \Gamma_Q = \frac{1}{Q}\frac{dQ}{dt}\simeq \frac{1}{|K|^{3/2}}\left(\frac{m}{M_{\rm P}}\right)^2 m$~\cite{Enqvist:2002si}. Even
though  coupling is large, i.e. $h\sim {\cal O}(0.1)$, the decay rate
mimics that of a Planck suppressed interaction of the inflating
$Q$-ball with matter fields. }.


\begin{figure}[t!]
\centering
\hspace*{-7mm}
\leavevmode\epsfysize=4cm \epsfbox{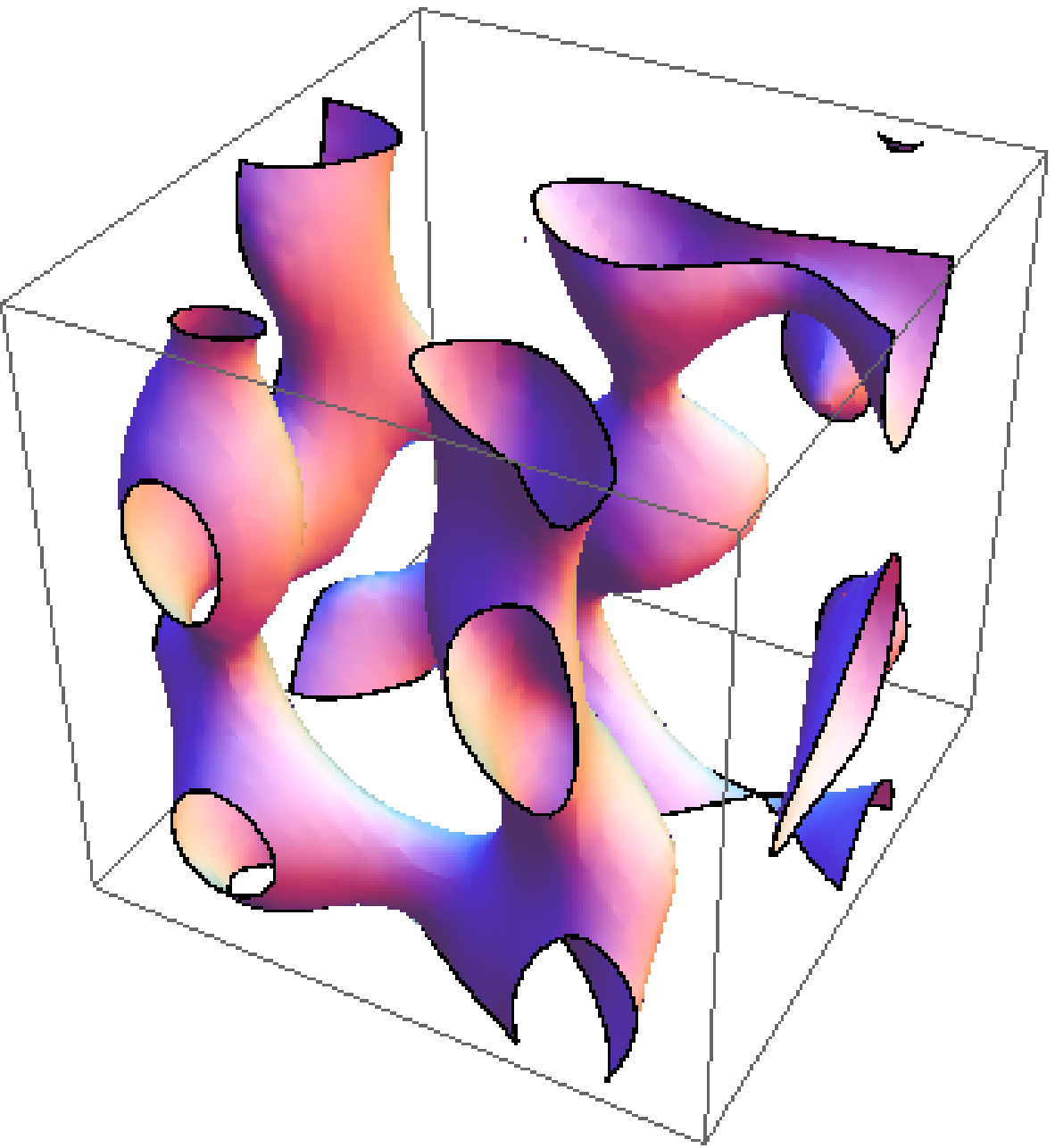}
\leavevmode\epsfysize=4cm \epsfbox{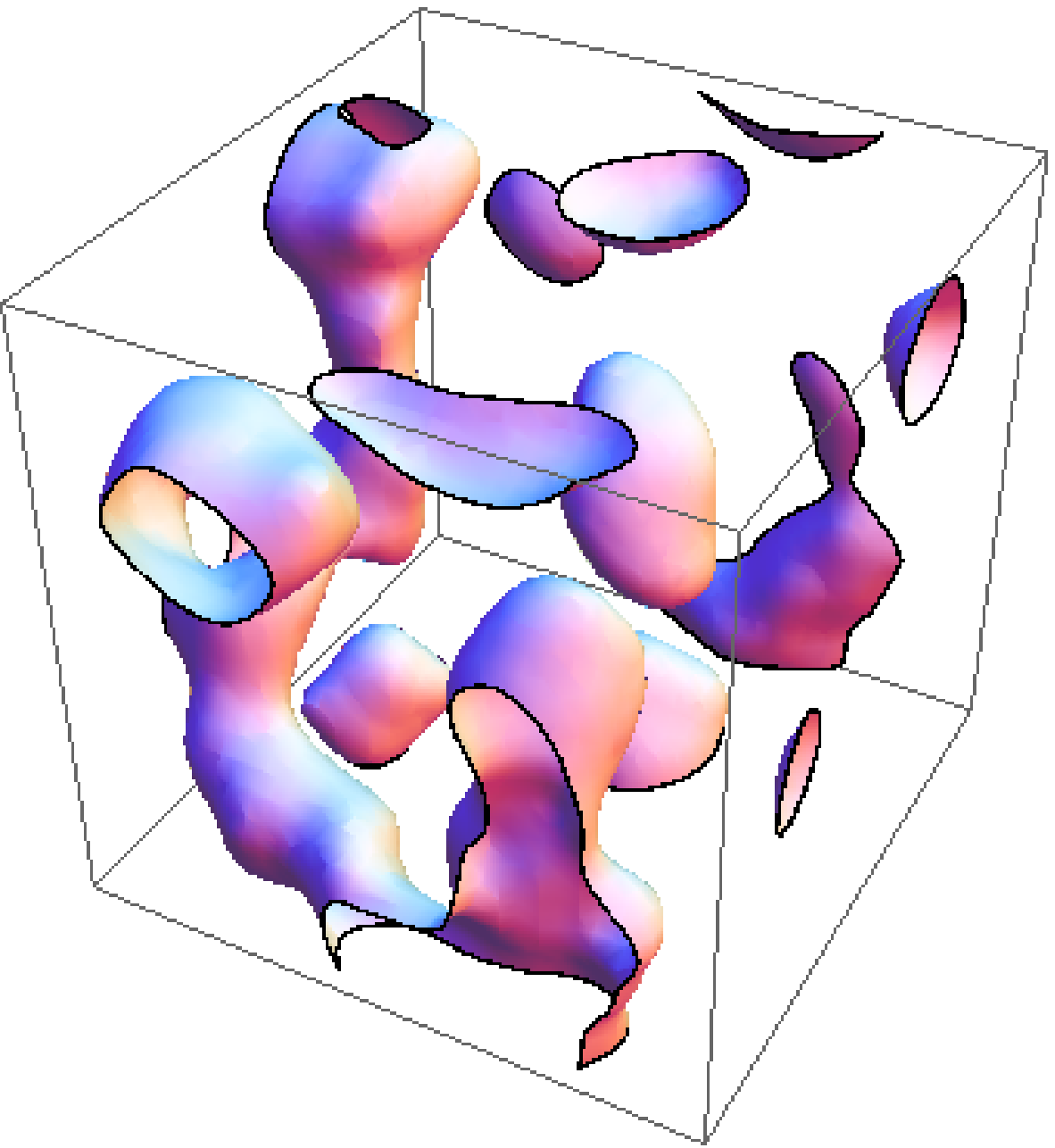}
\leavevmode\epsfysize=4cm \epsfbox{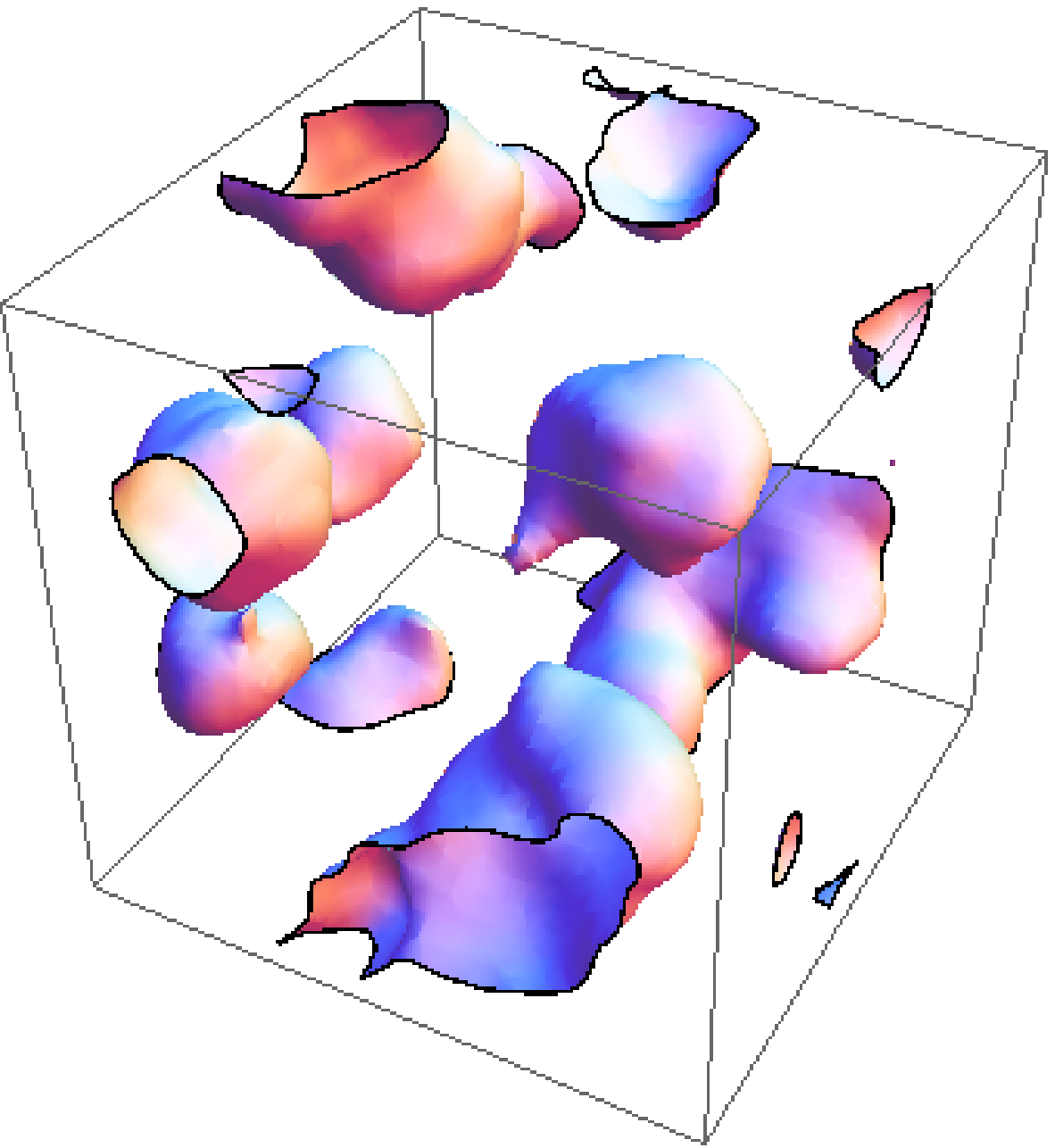}
\leavevmode\epsfysize=4cm \epsfbox{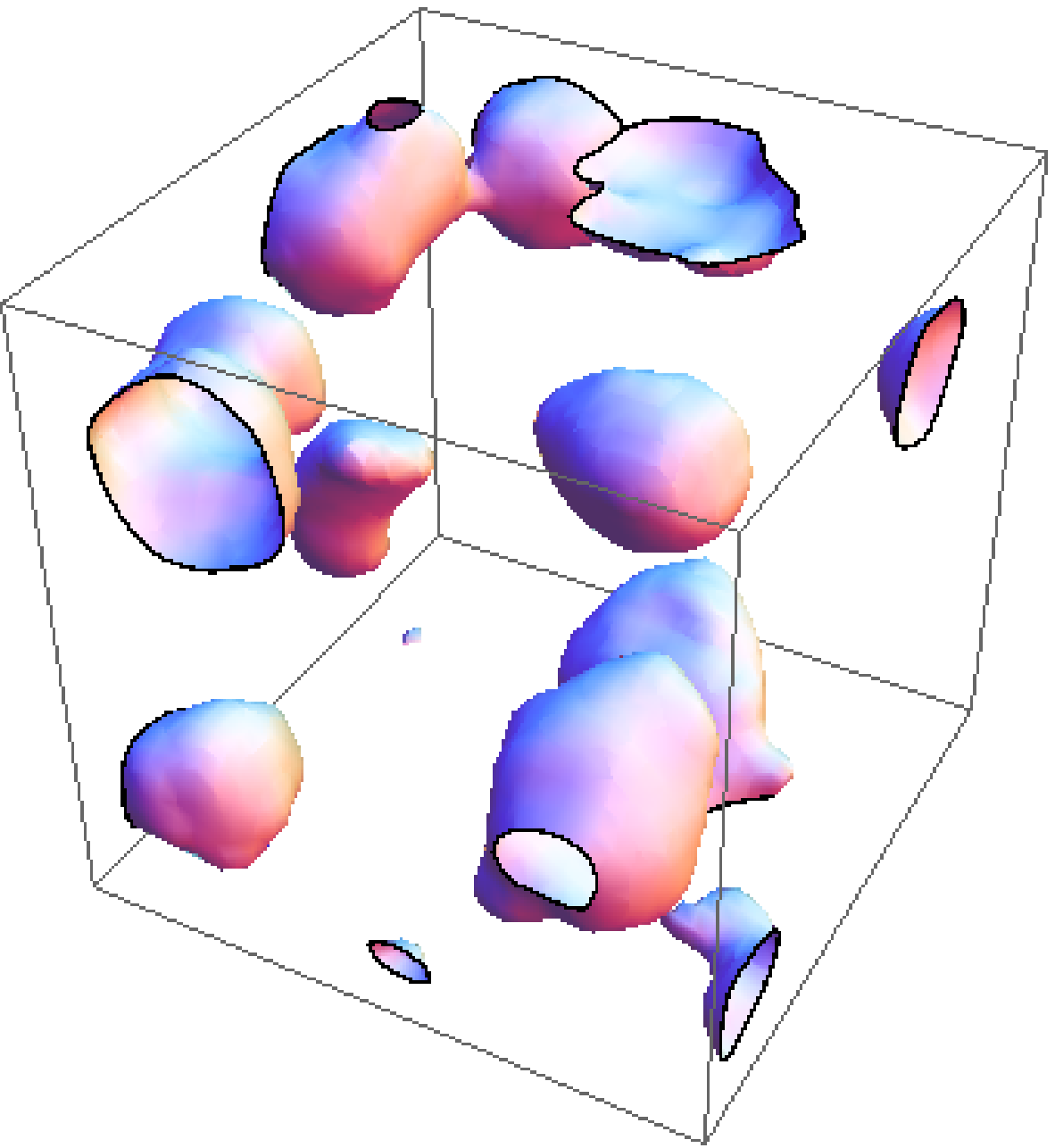}\\[2mm]
\caption{\label{3Dinf}
\small Lumps of Q-matter are formed during the fragmentation of a condensate with a potential given by~
Eq.~(\ref{qpotr}).}
\end{figure}


\subsubsection{Non-perturbative creations of gravity waves}
\label{EGWGF}

The  gravity waves are generated during preheating, as the excitations involve inhomogeneous, non-spherical, 
anisotropic motions of the excited scalar modes. As a result, the stress energy tensor receives anisotropic 
stress-energy contribution. The generation of gravity waves were studied in
Refs.~\cite{Khlebnikov:1997di,GarciaBellido:2007dg,Mazumdar:2008up,GarciaBellido:2007af,Fenu:2009qf,Easther:1999ws,Easther:2007vj,Kusenko:2008zm,Kusenko:2009cv,Gumrukcuoglu:2008gi,Dufaux:2008dn}. Typically, the peak frequency of the gravity 
waves is such that they correspond to the sub-Hubble wavelengths at the time of production. Gravity waves excitations can be 
studied numerically by following the transverse-traceless (TT) components of the 
stress-energy momentum tensor. By perturbing the  Einstein's equation, we obtain the evolution of 
the tensor perturbations, see~\cite{GarciaBellido:2007af}:
\begin{equation}
\ddot h_{ij}+3H\dot h_{ij}-\frac{\nabla^2}{a^2}h_{ij}=16\pi G \Pi_{ij}\,,
\end{equation}
where $\partial_{i}\Pi_{ij}=\Pi_{ii}=0$ and 
$\partial_{i}h_{ij}=h_{ii}=0$.  The TT part of the spatial
components of a symmetric anisotropic stress-tensor
$T_{\mu\nu}$ can be found by using the spatial projection operators,
$P_{ij}=\delta_{ij}-\hat{k}_i\hat{k}_j$, with $\hat k_{i}=k_{i}/k$:
\begin{equation}
\Pi_{ij}(k)=\Lambda_{ij,mn}(\hat k) T_{mn}(k)\,,
\end{equation}
where $\Lambda_{ij,mn}(\hat k)\equiv \left(P_{im}(\hat k)P_{jn}(\hat k)-(1/2)P_{ij}(\hat k)P_{mn}(\hat k)\right)$.
The TT perturbation is written as $h_{ij}(t,{\hat{k}})=\Lambda_{ij,lm}(\hat{k})u_{ij}(t,{k})$, where
\begin{equation}
\ddot u_{ij}+3H\dot u_{ij}-\frac{1}{a^2}\nabla^2 u_{ij}=16\pi GT_{ij}\,.\label{uequ}
\end{equation}
The source terms for the energy momentum tensor in our case are just the gradient terms of the scalar field
$\chi$ involved during preheating.
\begin{equation}
\label{ttsource}
T_{ij}=\frac{1}{a^2}(\nabla_i\chi_1\nabla_j\chi_1+\nabla_{i}\chi_2\nabla_{j}\chi_2)\,, 
\end{equation}
where $ \chi_1$ and $ \chi_2$ represent the real and imaginary parts of $\phi$, respectively. 
The gravitational wave (GW) energy density is given by:
\begin{equation}
\label{rhogw}
\rho_{GW}=\frac{1}{32\pi G}\frac{1}{V}\int d^3\vec{k}\, \dot{h}_{ij}\dot{h}^*_{ij}\approx
 \frac{1}{32\pi G V}\int d^3\vec{x}\, \dot{u}_{ij}\dot{u}^*_{ij}.
\end{equation}
where $V$ is the volume of the lattice. As an application, let us consider exciting gravity waves
in hybrid inflation model of inflaton, with inflaton, $\phi$, and the Higgs field, $\chi$, see~\cite{GarciaBellido:2007af}.

\begin{figure}[t]
\begin{center}
\includegraphics[width=5.5cm,height=7.5cm,angle=-90]{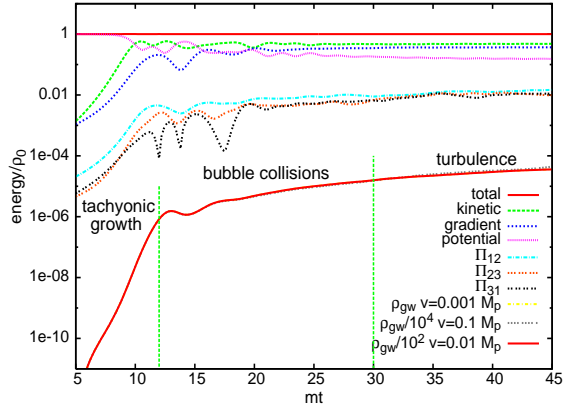}
\end{center}
\vspace*{-5mm}
\caption{The time evolution of the different types of energy (kinetic,
gradient, potential, anisotropic components and gravitational waves
for different lattices), normalized to the initial vacuum energy,
after hybrid inflation, for a model with $v=10^{-3}\,M_P$. One can
clearly distinguish here three stages: tachyonic growth, bubble
collisions and turbulence. The plot is taken from~\cite{GarciaBellido:2007af}.  }
\label{fig3}
\vspace*{-3mm}
\end{figure}

The coupled evolution equations that have to be solved numerically on a lattice for the hybrid model of inflation
are given by~\cite{GarciaBellido:2007af}~\footnote{Note that the weakness of gravity renders negligible  on scalar fields.}:
\begin{eqnarray}\label{GWeqHybrid}
\ddot\chi - \nabla^2\chi + \left(g^2|\phi|^2 +
\mu^2\right)\chi = 0 \,,\\
\ddot\phi_a - \nabla^2\phi_a +
\left(g^2\chi^2 + \lambda|\phi|^2 - m^2\right)\phi_a = 0\,,
\end{eqnarray}
where $\chi$ is the inflaton and the $\phi_a$ are the complex water-fields.

The initial energy density at the end of hybrid inflation is given by:
$\rho_0 = m^2v^2/4$, with $m^2 = \lambda v^2$, where $v$ is the VEV of the Higgs field (water-field), so the fractional
energy density in gravitational waves is r is roughly given by~\cite{GarciaBellido:2007af}:
\begin{equation}
\indent {\rho_{_{\rm GW}}\over\rho_0} = {4t_{00}\over v^2\,m^2} =
\frac{1}{8\pi\,G v^2 m^2}\left\langle\dot h_{ij}\dot  h^{ij}\right\rangle_{\rm V}\,,
\end{equation}
where $\left\langle\dot h_{ij}\dot h^{ij}\right\rangle_V$, defined as a 
volume average like 
${1\over {\rm V}}\int d^3{ x}\dot h_{ij}\dot h^{ij}$, is extracted from 
the numerical simulations. 

There are three stages of preheating which contribute to gravity waves. 
First, an exponential growth driven by the tachyonic instability
of the long-wave modes of the Higgs field. Second, the Higgs field
oscillates around the true vacuum, as the Higgs' bubbles collide and
scatter off each other. Third, a period of turbulence is reached,
during which the inflaton oscillates around its minimum and the Higgs
is already settled in its vacuum~\cite{GarciaBellido:2007af,Khlebnikov:1998sz,Felder:2000hj}.
In Ref.~\cite{Dufaux:2007pt}, it was pointed out that the turbulence phase does not contribute
to any enhancement in gravity waves, but if one includes the gauge fields in the stress-energy tensor
it is possible to mimic the results of~\cite{GarciaBellido:2007af}.

Gravity wave production has also been studied in a potential given by Eq.~(\ref{qpotr}).
The fractional energy density can be estimated by following Eq.~(\ref{rhogw}), given 
by~\cite{Kusenko:2008zm,Kusenko:2009cv}~\footnote{This is one of the realistic cases of gravity 
wave production from the fragmentation of a SUSY flat direction due to running of the soft scalar masses within MSSM. 
In Refs.~\cite{Kusenko:2008zm,Kusenko:2009cv} it was argued that {\it only} $B-L=0$ flat direction can dominate 
the energy density at the time of the fragmentation. Rest of the other MSSM flat directions will generate possibly too 
large baryon asymmetry. There is also a study of gravity wave production in a simple toy model with a 
complex scalar field in Ref.~\cite{Dufaux:2009wn}.}:
\begin{equation}
\label{omest}
\Omega_{GW}=\frac{\rho_{GW}}{m^2\phi(t)^2}\sim |K|^2\left(\frac{\phi(t)}{M_{Pl}}\right)^2\,.
\end{equation}
The scalar field  is dominating the energy density of the universe at the time of fragmentation.  For physically motivating
parameters, we have chosen; $m \sim 100$~GeV, $\phi(t)\sim 10^{16}$~GeV, and $H(t)\sim 1$~GeV, therefore,
for a reasonable value of $K\sim 0.1$, we obtain, $\Omega_{GW}\sim 10^{-6}$. Including the cosmological expansion,
the current abundance of gravity waves will be: $\Omega_{GW} h^2 \sim 10^{-11}$, where $h\sim 0.7$ is the Hubble constant, 
with a peak frequency ranging from mHz --$10$~Hz. Note that $\Omega_{GW}$ depends 
on the value of $K$, for $K=0$ there are no excitations of gravity waves, see Fig.~(\ref{RSphi}).


\begin{figure}[t!]
\centering
\hspace*{-7mm}
\leavevmode\epsfysize=5.5cm \epsfbox{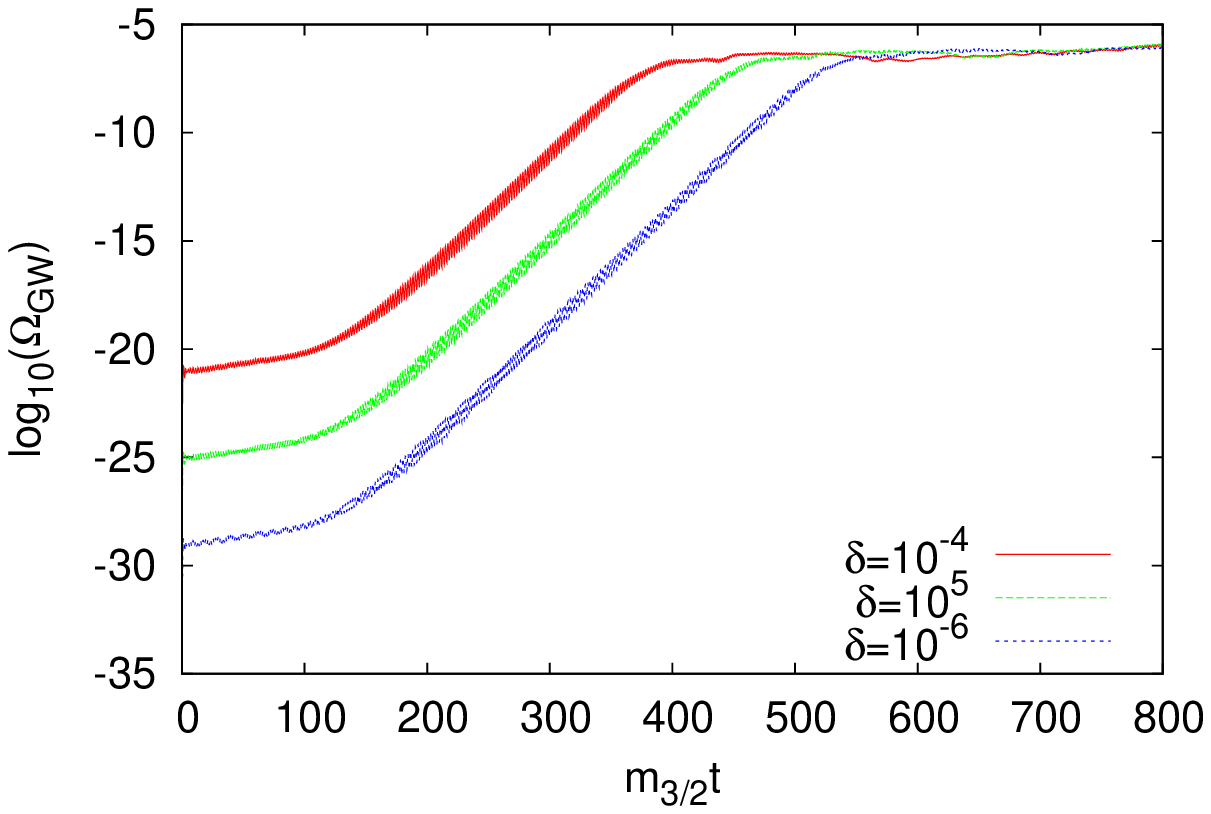}
\leavevmode\epsfysize=5.5cm \epsfbox{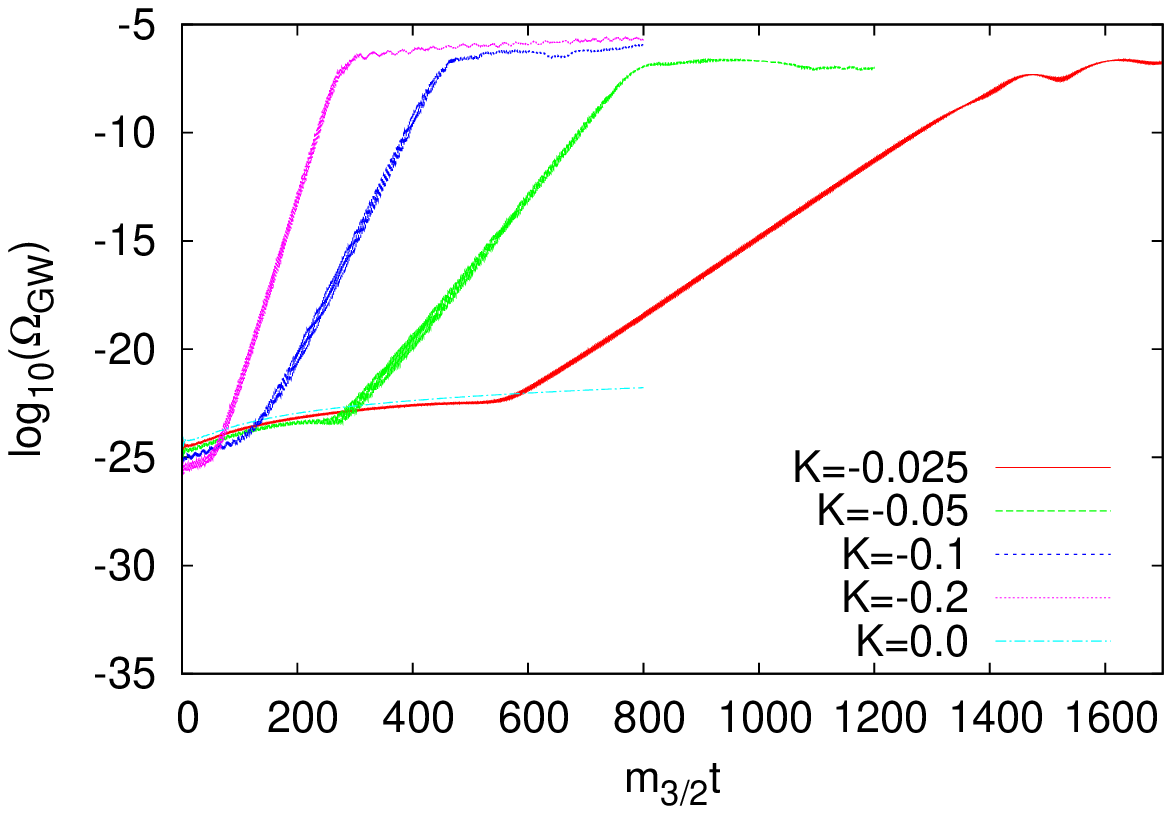}\\[2mm]
\caption{\label{RSphi}
\small The final amplitude of the gravity waves does not depend on the initial perturbations, $\delta$, and The final amplitude of gravity waves saturates for different values of $K$. Note for $K=0$ there is no fragmentation at all, therefore there is no gravity waves.The field value is: $\phi_0=10^{16}$~GeV and $m\sim m_{3/2}=100$~GeV.}
\end{figure}


\subsubsection{Non-perturbative production of gauge fields}

Let us now consider non-perturbative effects on gauge fields. For simplicity, let us consider
an example where the tachyonic field, $\chi$, is charged under $U(1)\times U(1)$, which arises 
quite naturally in a brane-anti-brane inflation~\cite{Mazumdar:2008up}. Here $F^{+}$ and $F^{-}$ are 
the gauge fields that live in the world volume of the brane and anti-brane respectively.
The tachyon, $\chi$, is a bi-fundamental field that couples only 
to a linear combination of the two gauge fields:
$D_{\mu}\chi = \partial_{\mu}\chi - \left(A_{\mu}^{+}-A_{\mu}^{-}\right)\chi$, and $\phi$ is the inflaton 
field~\footnote{The physical situation is quite similar to that of a hybrid inflation. The tachyon field in brane-anti-brane
inflation plays a dynamical role of a water-fall field or the Higgs field.}. The Lagrangian is given by:
\begin{eqnarray}
{\mathcal L} = \frac{1}{2}\partial_{\mu}\phi\partial^{\mu}\phi +
\frac{1}{2}D_{\mu}\chi D^{\mu}\chi^{*} +
\frac{1}{2}m_{\phi}^{2}\phi^{2}
+\frac{1}{2}m_{T}^{2}\left|\chi\right|^{2}\left(\phi^{2}-\phi_{0}^{2}\right)+
\frac{\lambda}{4}\left|\chi\right|^{4}
-\frac{1}{4}F^{2}\nonumber \\
\end{eqnarray}
The tachyon mass is fixed in terms of the string scale, $m_{T}=M_s$, 
while the mass of the inflaton is determined by the amplitude of the temperature anisotropy, which turns out to be: 
$m_{\phi}=0.01M_s$. The stress-energy tensor now gets contributions from  both the charged tachyon, 
$\chi$ and uncharged inflaton, $\phi$, as well as the gauge fields, which is given by~\cite{Mazumdar:2008up}:
\begin{eqnarray}
\Pi_{ij} = F_{iC}F_{j}^{\;\;C} - 
\frac{1}{3}\delta_{ij}F_{kC}F^{kC} 
- D_{i}\chi D_{j}\chi^{*} 
+ \frac{1}{3}\delta_{ij}D_{k}\chi D^{k}\chi^{*}
 -\partial_{i}\phi \partial_{j}\phi^{*} +  
\frac{1}{3}\delta_{ij}\partial_{k}\phi \partial^{k}\phi^{*}\nonumber \\
\end{eqnarray}
The energy density of the gravitational 
waves is simply given by the $`t_{00}'$ component of the 
energy-momentum tensor of the gravitational 
waves calculated in a synchronous gauge, i.e. $h_{0i}=0,~h_{00}=0$. In Fig.~(\ref{GG}), the 
energy pumped into the gauge fields and the gravity waves are shown for 
$\lambda=1$, $m_{\phi}=0.01M_S$ and the tachyon mass is given by $M_S=10^{-4}M_{\rm P}$.. 
Of course, larger the value of a tachyon mass, greater is the instability, and the growth in the respective
perturbations, but compare the energy densities in gauge fields and gravity waves.

\begin{figure}[tbph]
\begin{center}
\includegraphics[width=0.50\textwidth,angle=0]{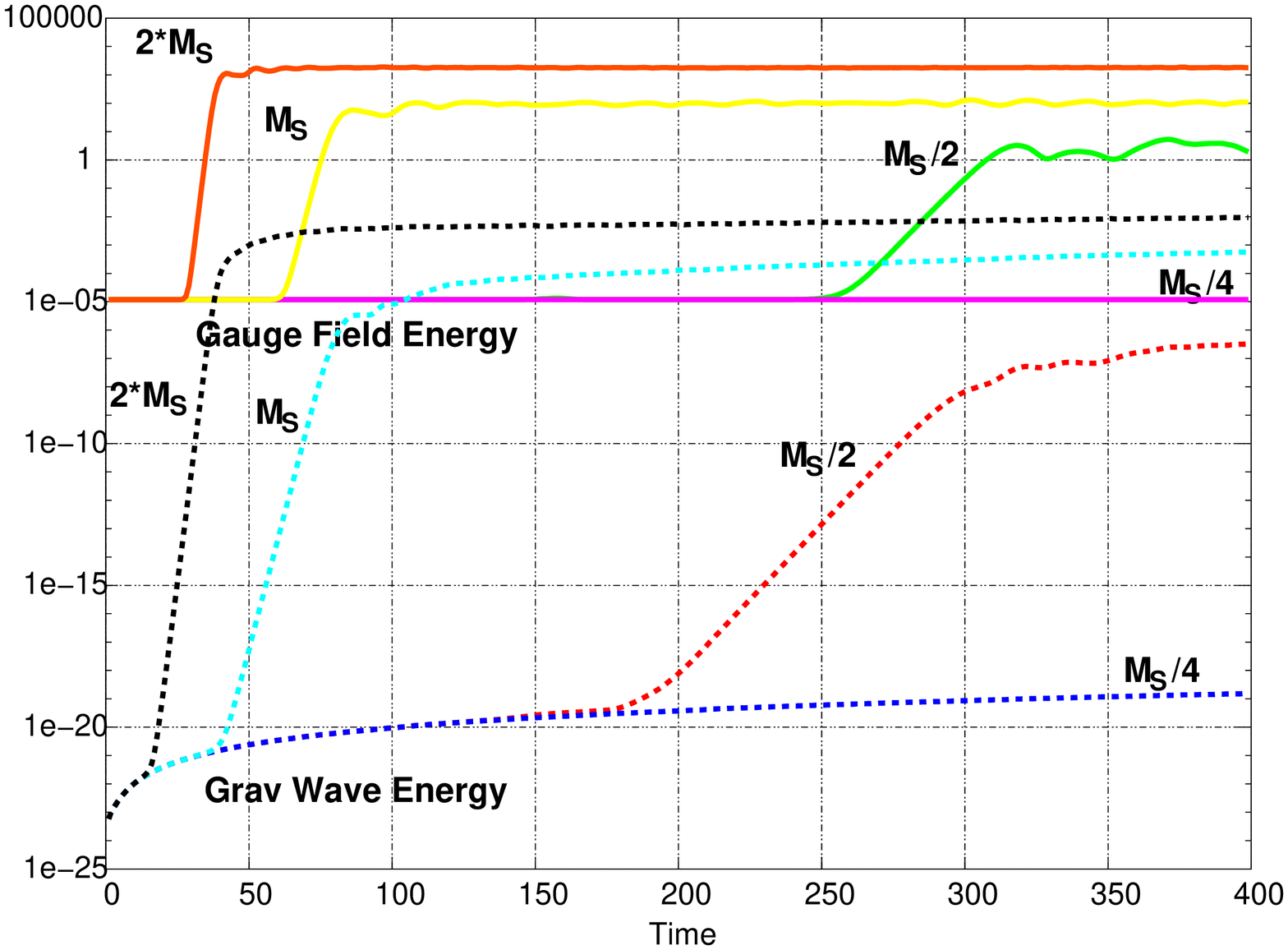}
\end{center}
\caption{Different colors show how the gravitational wave energy and gauge field energy grow for different values of the 
tachyonic mass. }
\label{GG}
\end{figure}
\begin{figure}[tbph]
\includegraphics[width=0.24\textwidth,angle=0]{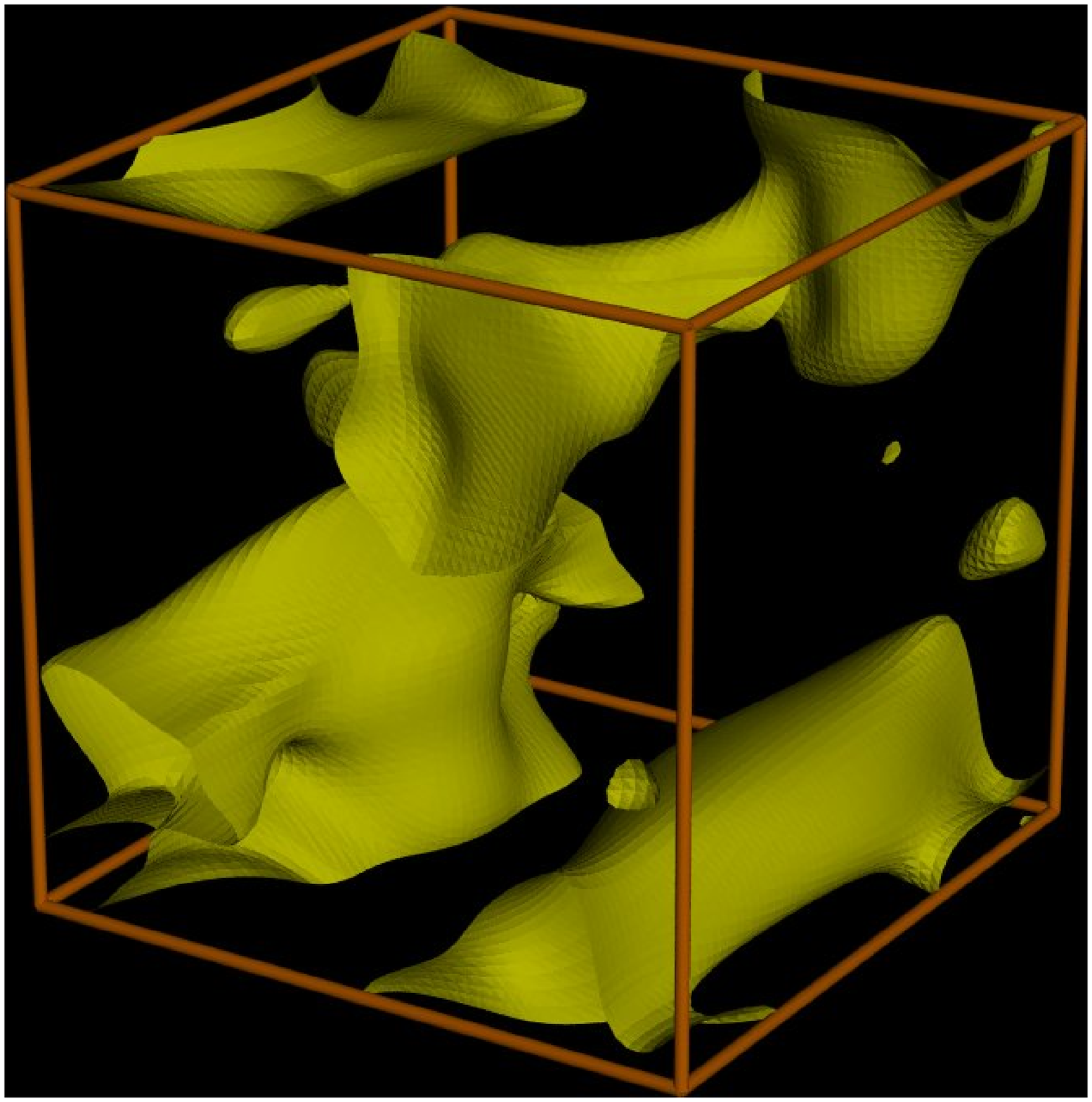}
\includegraphics[width=0.24\textwidth,angle=0]{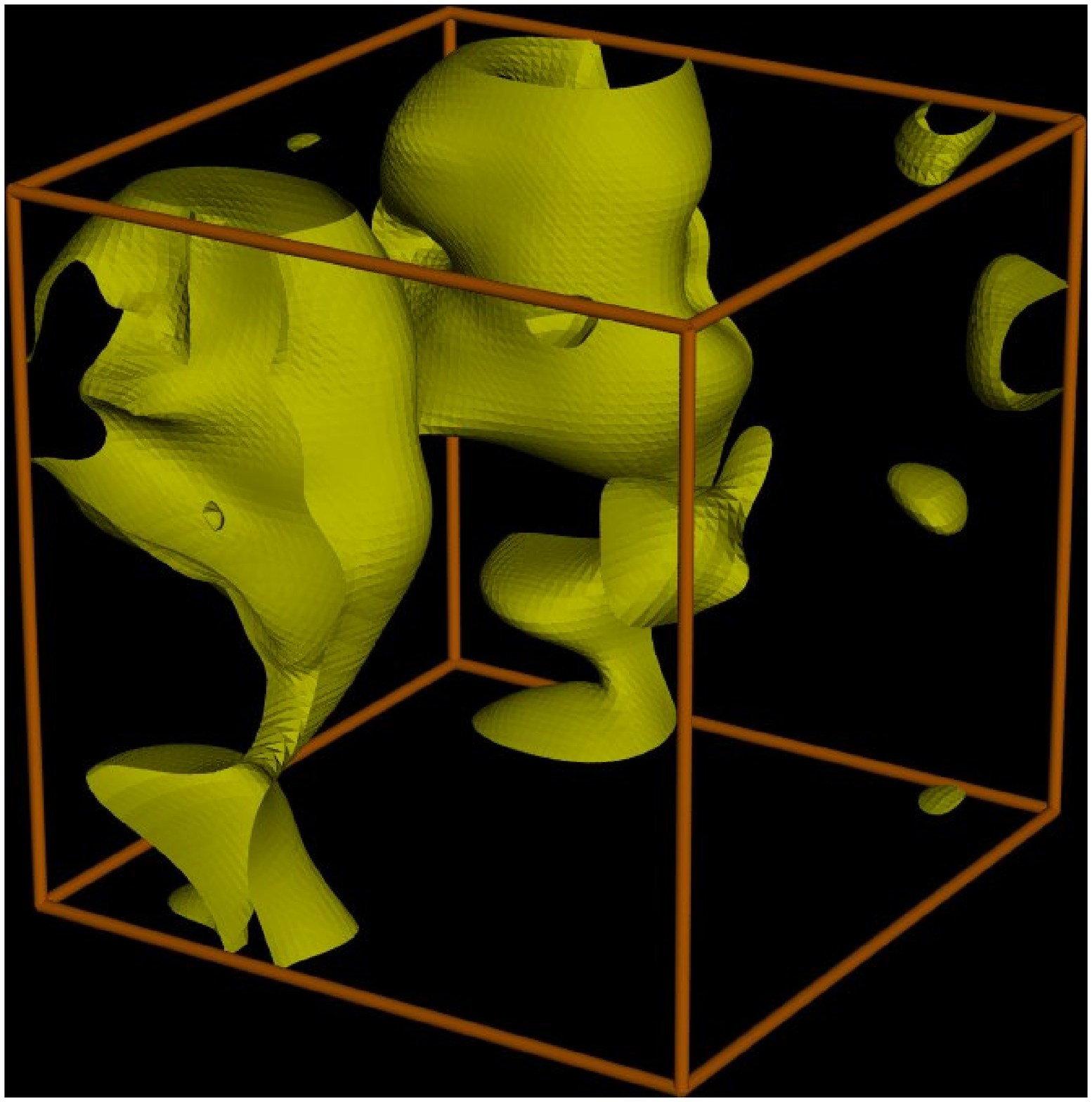}
\includegraphics[width=0.24\textwidth,angle=0]{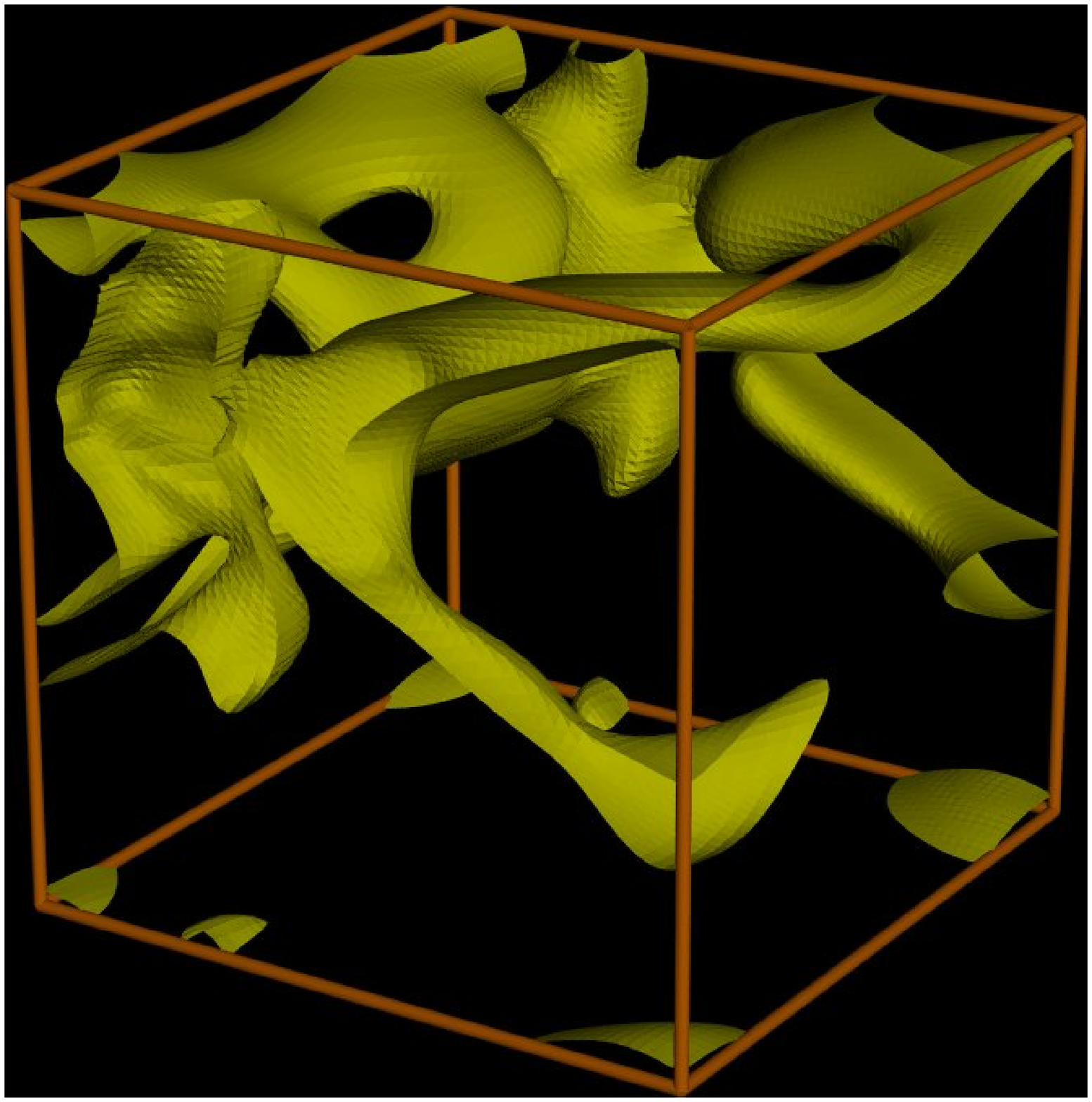}
\includegraphics[width=0.24\textwidth,angle=0]{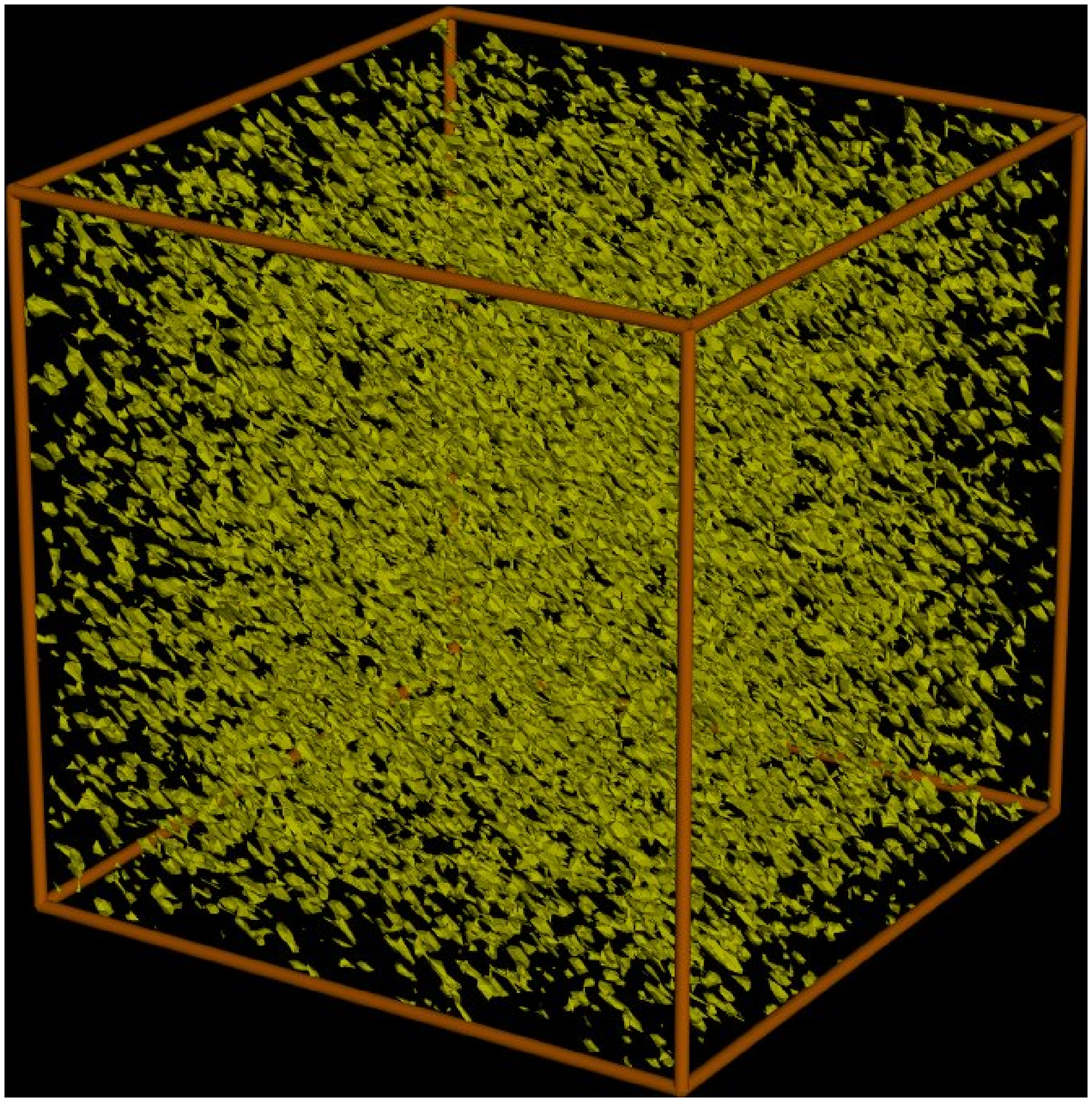}
\vspace{6pt}
\caption{Snap-shots of iso-surface of the energy density for gauge field, gravity 
waves, tachyon  and the inflaton (from left-to-right) at a particular instant of time, $t=300$. The plots are taken 
from~\cite{Mazumdar:2008up}.}
\label{tachyonic-1}
\end{figure}

In Figs.~(\ref{tachyonic-1}), 4 snap-shots of the iso-surface of the constant 
energy density of gauge field, gravity waves, tachyon and inflaton are depicted. All the fields 
except the inflaton show a remarkable departure in the 
homogeneity. Except the inflaton, all fields undergo long wavelength excitations (they all 
look relatively smooth on small scales), while the inflaton obtains the largest inhomogeneity 
on the smaller scales. This is due to the fact that the there is no long wavelength amplification 
for the inflaton in this case~\cite{Mazumdar:2008up}. 


\subsubsection{SM Higgs preheating}

On particular realistic example of preheating~\cite{Bezrukov:2008ut,GarciaBellido:2008ab} can be illustrated by the 
SM Higgs inflation~\cite{Bezrukov:2007ep}. After inflation the Higgs field evolves in time, $h = h(\chi(t))$, see Eq.~(\ref{eq:sminflationpoten}), so the effective masses of the fermions and of the gauge bosons also obtain time
dependence.
\begin{equation}\label{massesWZ}
m_W = m_Z\,{\cos\theta_W} = \frac{1}{2}g_2h(\chi(t))\,, ~~~~m_{f} = \frac{1}{2}y_f\,h(\chi(t))\,,
\end{equation}
where $\theta_W$ is the Weinberg angle $\theta_W=\tan^{-1}(g_1/g_2)$, and $y_f$, $g_1$ and $g_2$ are the Yukawa and the $U(1)_Y$ and $SU(2)_L$ couplings, respectively. The time dependent Higgs VEV spontaneously breaks the gauge symmetry, and 
give masses to $W$ and $Z$ gauge bosons masses. The relevant interactions are given by the charged and neutral Currents, coupling the SM fermions to gauge bosons through the $J_\mu^\pm$, $J^Z_\mu$ currents, and the Yukawa sector, and coupling the SM fermions with the Higgs:
\begin{eqnarray}
&&S_{SSB} = \int d^4x \sqrt{-g} \Big\{ m_W^2 W_{\mu}^+ W^{\mu -}+\frac{1}{2} m_Z^2 Z_\mu Z^\mu  \Big\}\,,~~
S_{Y} = \int d^4x \sqrt{-g} \Big\{ m_d {\bar\psi_{d}}{\psi_{d}} + m_u { \bar\psi_{u}}{ \psi_u}\Big\}, \nonumber \\
&&S_{CC} + S_{NC} =  \int d^4x \sqrt{-g} \left\{\frac{g_2}{\sqrt{2}}W_\mu^+J^-_\mu +\frac{g_2}{\sqrt{2}}W_\mu^-J^+_\mu+ 
\frac{g_2}{\cos\theta_W} Z_\mu J_Z^\mu\right\}\,.
\end{eqnarray}
One can redefine the fields and masses with a specific conformal weight as to keep the canonical kinetic terms for
the gauge fields and fermions, provided:
$\tilde{W}_\mu^{\pm}\equiv {{W}_\mu^{\pm}}/{\Omega}$, $\tilde{Z}_\mu\equiv{{Z}_\mu}/{\Omega}$,~
$\tilde{\psi}_{d}\equiv {\psi_{d}}/{\Omega^{3/2}}$, $ \tilde{\psi}_{u}\equiv {\psi_{u}}/{\Omega^{3/2}}$,~
$\tilde{m}^2_W = \tilde{m}^2_Z\,\cos^2\theta_W = {m^2_W}/{\Omega^2}=(g^2_2 M_{\rm P}^2 /4\xi)
(1-e^{-\alpha\kappa\vert\chi\vert})$,~$\tilde{m}_f \equiv {m_f}/{\Omega} = (y_f M_P/\sqrt{2\xi})\left(1-e^{-\alpha\kappa\vert\chi\vert}\right)^{1/2}$, where $\Omega^2\approx \exp(\alpha\kappa\chi)$ with $\alpha=\sqrt{2/3}$ and $\kappa=1/M_{\rm P}$.
The oscillations of the Higgs field can be approximated by a simple quadratic potential $U(\chi)\approx(1/2)M^2\chi^2$,
where $M=\sqrt{\lambda/3}M_{\rm P}/\xi\sim 10^{-5}M_{\rm P}$. The oscillations evolve with $\chi(t)\approx X(t)\sin(Mt)$, where
$X(t)\propto (Mt)^{-1}$, like in the matter-dominated case. There is a small departure from a quadratic potential, which
is given by Eq.~(\ref{eq:sminflationpoten}),  $U(\chi)=(1/2)M^2\chi^2+\Delta U(\chi)$, but during reheating this correction is 
negligible. As $\chi$ passes through the minimum, by virtue of its couplings, the gauge fields and fermions become massless, therefore they can be excited through parametric resonance or via instant preheating during every oscillations~\footnote{This feature was first discussed in the context of MSSM inflation, where instant preheating excites the gluons abundantly near the point of enhanced gauge symmetry~\cite{Allahverdi:2006we}. }. However during one cycle of oscillation, the gauge bosons and fermions become heavy, as $\chi$ goes away from the minimum and obtains a time dependent VEV. During this epoch the gauge fields can decay into lighter fermions, but there is a kinematical blocking. The process of preheating is not as efficient as one would have thought. Nevertheless, the Higgs energy can be transferred at a faster rate compared to that of the perturbative decay of the 
Higgs, as shown in the Fig.~(\ref{EnergyEvolution})~\cite{GarciaBellido:2008ab}. Furthermore, in order to obtain a full thermalization, Higgs must decay completely, which happens in a time scale similar to that of a perturbative decay rate.


\begin{figure}
\centering
\includegraphics[width=5cm,height=8cm,angle= -90]{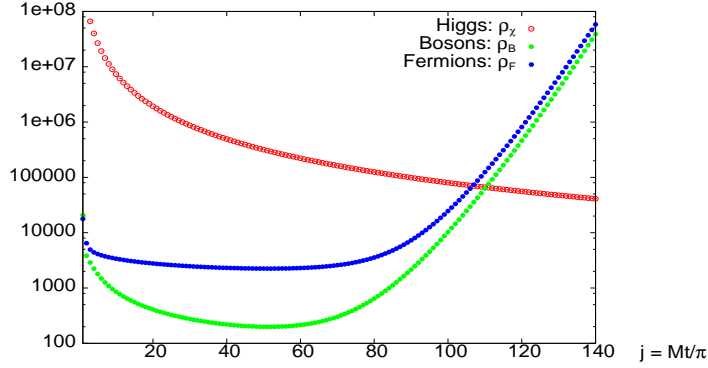}
\caption{Evolution of the energy density transferred into the gauge bosons and into the fermions as a function of $j$, for $\lambda = 0.2$ and $\xi = 44700\sqrt{\lambda}$. All densities are in units of $M^4$~\cite{GarciaBellido:2008ab}.}
 \label{EnergyEvolution}
\end{figure}


\subsection{SUSY generalization of reheating and preheating}
\label{SGR/P}

Reheating, preheating and thermalization issues are quite 
different once SUSY is introduced. 
SUSY introduces new degrees of freedom and
new parameters. Cosmology also acts as a test bed where some of
the SUSY particles can be tested from the success of BBN,  a well
known example is the {\it gravitino problem} in the context of a 
SUSY cosmology.  


\subsubsection{Gravitino problem}
\label{GP}

The gravitino is a spin-$3/2$ partner of a graviton,
which is coupled to the SM particles with the gravitational strength.
Gravitinos with both the helicities can be produced from a thermal
bath. There are many scattering channels which include fermion,
sfermion, gauge and gaugino quanta all of which have a cross-section
$\propto 1/M^2_{\rm P}$~\cite{Ellis:1984eq,Moroi:1993mb,Moroi:1995fs}, and \cite{Bolz:2000fu}, which results in a gravitino
abundance (up to a logarithmic correction):
\begin{eqnarray} \label{gravtherm}
{\rm Helicity}~ \pm {3 \over 2}: {n_{3/2} \over s} 
&\simeq & \left({T_{\rm R} \over 10^{10}~{\rm GeV}}\right) ~ 
10^{-12}\,, \nonumber \\
& & \hspace{7cm} ({\rm full ~ equilibrium}) \, \nonumber \\
{\rm Helicity} ~ \pm {1 \over 2}: {n_{3/2} \over s}
&\simeq & \left(1 + {M^2_{\widetilde g} \over 12 m^2_{3/2}}\right) 
\left({T_{\rm R} \over 10^{10}~{\rm GeV}}\right) ~ 10^{-12}\,;
\end{eqnarray}
where $M_{\widetilde g}$ is the gluino mass. Note that for
$M_{\widetilde g} \leq m_{3/2}$ both the helicity states have
essentially the same abundance, while for $M_{\widetilde g} \gg
m_{3/2}$ production of helicity $\pm 1/2$ states is enhanced due to
their Goldstino nature~\footnote{ Since the
cross-section for the gravitino production is $\propto M^{-2}_{\rm
P}$, the production rate at a temperature, $T$, and the abundance of
the gravitinos produced within one Hubble time will be $\propto T^3$
and $\propto T$ respectively. This implies that the gravitino
production is efficient at the highest temperature of the
radiation-dominated phase of the universe, i.e. $T_{\rm R}$.}.

An unstable gravitino decays to particle-sparticle pairs, and its
decay rate is given by $\Gamma_{3/2} \simeq m^{3}_{3/2}/4 M^2_{\rm
P}$, see~\cite{Moroi:1995fs}. If $m_{3/2} < 50$ TeV, the gravitinos decay during or after BBN
~\cite{Sarkar:1995dd,Olive:1999ij}, which can ruin its successful
predictions for the primordial abundance of light
elements. If the gravitinos decay radiatively, the most
stringent bound, $\left(n_{3/2}/s\right) \leq 10^{-14}-10^{-12}$,
arises for $m_{3/2} \simeq 100~{\rm GeV}-1$ TeV~\cite{Cyburt:2002uv}.
On the other hand, much stronger bounds are derived if the gravitinos
mainly decay through the hadronic modes. In particular, for a hadronic
branching ratio $\simeq 1$, and in the same mass range,
$\left(n_{3/2}/s\right) \leq 10^{-16}-10^{-15}$ will be
required~\cite{Kawasaki:2004yh,Kawasaki:2004qu}.

For a radiatively decaying gravitino the tightest bound
$\left(n_{3/2}/s \right) \leq 10^{-14}$ arises when $m_{3/2} \simeq
100$ GeV~\cite{Cyburt:2002uv}.  Following Eq.~(\ref{gravtherm}) the bound on
reheat temperature becomes: $T_{\rm R} \leq 10^{10}$ GeV. 
For a TeV gravitino which mainly decays into
gluon-gluino pairs (allowed when $m_{3/2} > M_{\widetilde g}$) a much
tighter bound $\left(n_{3/2}/s \right) \leq 10^{-16}$ is
obtained~\cite{Kawasaki:2004yh,Kawasaki:2004qu}, which requires quite a low reheat temperature:
$T_{\rm R} \leq 10^6$~GeV.

The gravitino will be stable if it is the LSP, where $R$-parity is conserved. The gravitino abundance
will in this case be constrained by the dark matter
limit, $\Omega_{3/2} h^2 \leq 0.12$, leading to
\begin{equation} \label{dmlimit}
{n_{3/2} \over s} \leq 
5 \times 10^{-10} ~ \left({1~{\rm GeV} \over m_{3/2}}\right)\,.
\end{equation}
For $m_{3/2} < M_{\widetilde g}$, the helicity $\pm 1/2$ states
dominate the total gravitino abundance.  As an example, consider the
case with a light gravitino, $m_{3/2} = 100$~KeV, which can arise very
naturally in gauge-mediated models~\cite{deGouvea:1997tn}.  If $M_{\widetilde g}
\simeq 500$ GeV, see Eq.~(\ref{gravtherm}), a very severe constraint,
$T_{\rm R} \leq 10^4$ GeV, will be obtained on the reheat
temperature~\footnote{Gravitinos could also be produced by non-perturbative processes,
as was first described in \cite{Maroto:1999ch} for helicity $\pm 3/2$ component of gravitino.
Later the production of the helicity $\pm 1/2$ state were also studied
\cite{Kallosh:1999jj,Giudice:1999am,Kallosh:2000ve,Allahverdi:2000fz,Nilles:2001ry,Nilles:2001fg,Nilles:2001my}.
The helicity $\pm 1/2$ component obtains a major contribution from the inflatino (superpartner of inflaton),
however the inflatino decays with the same rate as that of the inflaton, therefore the $\pm 1/2$ abundance 
does not any major role~\cite{Allahverdi:2000fz,Nilles:2001ry,Nilles:2001fg,Nilles:2001my}.
A very late decay of inflatino could however be possible, as argued
in \cite{Nilles:2001ry,Nilles:2001fg,Allahverdi:2000fz}. In \cite{Allahverdi:2001ux}, it was argued that
if the inflatino and gravitino were not LSP, then late off-shell
inflatino and gravitino mediated decays of heavy relics could be
significant.}~\footnote{Recently, non-thermal abundance of gravitino from a modulus decay  has been 
revisited~\cite{Endo:2006zj,Nakamura:2006uc,Dine:2006ii,Endo:2006tf}. The 
gravitino abundance is given by: $Y_{3/2}  \sim B_{3/2} \frac{3T_{\rm R}}{4m_{\phi}}$,
where $B_{3/2}$ is the branching ratio into  gravitino and would be
$B_{3/2} = 10^{-2} - 1$~with the mixing between modulus and the 
SUSY-breaking field, where we have used an
approximation ${n_{\phi}}/{s} \sim({3T_{\rm R}}/{4m_{\phi}})$, essentially
the moduli decay is creating all the entropy of the universe. The branching ratio of the gravitino
production from a modulus  decay is little more contentious
than one would naively expect. For $B_{3/2}\sim 1$, there is a possibility of 
overproducing gravitinos. However, as it was pointed out in Ref.~\cite{Dine:2006ii},
the decay rate generically obtains a helicity suppression. The precise value 
depends on the details of the SUSY  breaking
hidden sector. There are well known examples of hidden
sectors, where one naturally obtains the expected value of $B_{3/2}\sim 10^{-2}$~\cite{Dine:2006ii}.}.

Note that the above discussions assume that all the MSSM degrees of freedom are in thermal equilibrium instantly
right after inflation, however this basic assumption is in contradiction in presence of MSSM flat 
directions~\cite{Allahverdi:2005mz}, which alters the thermal history of the universe and also affects the gravitino abundance.


\subsubsection{Gauge singlet inflaton couplings to MSSM}
\label{GSICM}

Within MSSM there exists two gauge-invariant combinations of only two superfields:
\begin{equation} \label{two}
{H_u} {H_d}, ~ ~ { H_u} { L}.
\end{equation}
The combinations which include three superfields are:\\
\begin{equation}
{H_u} {Q} {u},~~~ {H_d} {Q} {d},~~~ {H_d}{L}{e},~~~ {Q} {L}{d}, ~~~{u}{d}{d},~~{\rm  and}~~~{L}{L}{e}\,.
\end{equation}
SUSY together with gauge symmetry requires that the inflaton (a SM gauge singlet)
superfield be coupled to these combinations. The superpotential terms ${ \Phi} { H_u} { H_d}$ and 
${\Phi} { H_u} { L}$ have dimension four, and hence are
renormalizable. On the other hand, the interaction terms that couple
the inflaton to the combinations with three superfields have dimension
five and are non-renormalizable. In following we focus on
renormalizable interactions of the inflaton with matter which play the
dominant role in its decay. Further note that terms representing
gauge-invariant coupling of the inflaton to the gauge fields and
gauginos are also of dimension five, and hence preheating into
them will be suppressed~\footnote{It is possible that the inflaton mainly decays to 
another singlet (for example, the RH neutrinos) superfield, however, 
the underlying interactions of a gauge singlet to the MSSM superfields do not change. One must 
transfer the inflaton energy into the MSSM sector at any cost.}.

Preserving $R$-parity at the renormalizable level further constrains
inflaton couplings to matter. Note that ${ H}_u { H}_d$ is
assigned $+1$ under $R$-parity, while ${ H}_u { L}$ has the
opposite assignment $-1$. Therefore only one of the couplings 
preserves $R$-parity: 
${ \Phi} { H}_u {H}_d$ if $R_{ \Phi} = +1$,
and ${ \Phi} { H}_u { L}$. If $R_{ \Phi} = -1$ (such as models where the RH 
sneutrino plays the role of the inflaton~\cite{Kusenko:2008zm}). Therefore the 
renormalizable inflaton coupling to matter can be represented as~\cite{Allahverdi:2007zz}:
\begin{equation} \label{infsup}
2 g { \Phi} { H}_u { \Psi} \,
\end{equation}
where ${ \Psi} = {H}_u$ {\it if} $R_{ \Phi} = +1$ and
${ \Psi} = { L} $ {\it if}  $R_{ \Phi} = -1 $. Taking into account 
of the inflaton superpotential mass term:
$\left(m_{\phi}/2 \right){ \Phi} { \Phi}$, and defining
${ X}_{1,2} = ({ H}_u \pm { \Psi} )/ \sqrt{2}$, the {\it renormalizable part of the
potential}, which is relevant for the inflaton decay into MSSM scalars is given by:
\begin{eqnarray}
\label{infpot} V \supset {1 \over 2} m^2_{\phi} {\phi}^2 + g^2
{\phi}^2 \chi^2 \pm
{1 \over \sqrt{2}} g m_{\phi} \phi \chi^2 \,,
\end{eqnarray}
where $\chi$ denotes the scalar component of ${ X}_{1,2}$
superfields, and we have only considered the real parts of the
inflaton, $\phi$, and $\chi$ field. Further note that the cubic
interaction term appears with different signs for $\chi_1$ and
$\chi_2$, but this is irrelevant during inflaton oscillations.

In addition to the terms in Eq.~(\ref{infpot}), there are also the
self- and-cross-couplings,
\begin{equation}
V_D\supset \left(\frac{g^2}{4}\right) {\left(\chi^2_1 -\chi^2_2 \right)}^2 + \alpha \chi^2_1 \chi^2_2\,,
\end{equation}
arising from the superpotential and $D$-terms respectively ($\alpha$ is a gauge fine
structure constant). Therefore even in the simplest SUSY set up the
scalar potential is more involved than the non-SUSY case given in
Eq.~(\ref{nonpot}), which can alter the picture of preheating
presented in the literature, see for the detailed discussion in Refs.~\cite{Allahverdi:2007zz,Allahverdi:2005mz}~\footnote{A 
remarkable feature in Eq.~(\ref{infpot}) is that SUSY naturally
relates the strength of cubic $\phi \chi^2$ and quartic $\phi^2 \chi^2$
interactions, which is required for complete decay of the inflaton field.
One can also include couplings of the inflaton to fermionic partners of $\chi$.
Regarding the prospects for fermionic preheating  the same conclusions hold
as that of a bosonic case.}.


\subsection{MSSM flat directions, reheating and thermalization}
\label{FDCIDP}

The MSSM flat directions have important role to play in SUSY reheating and 
thermalization~\cite{Allahverdi:2005mz,Allahverdi:2007zz}. 
Consider a MSSM flat direction, $\varphi$, with the corresponding
superfield denoted by $\varphi$ (only for flat directions we are
denoting the superfield and the field with the same notation in this section). Note that since ${\varphi}$ and 
${X}$ superfields are linear combinations of the MSSM superfields (defined in the earlier subsection after Eq.~(\ref{infsup})), and hence are coupled through the MSSM
superpotential in Eq.~(\ref{mssm}).  
\begin{equation}
W \supset \lambda_1 { H}_u {\varphi}
{\Sigma}_1 + \lambda_2 { \Psi} {\varphi} {\Sigma}_2 + ... ,
\end{equation}
where ${\Sigma}_{1,2}$ are some MSSM superfields such  that
${\Sigma}_1 \neq { \Psi}$ and ${ \Sigma}_2 \neq { H}_u$, since
${\varphi}$ is a non-gauge-singlet. 

For example consider the case
where $\varphi$ is a flat direction classified by the ${u} {d}
{ d}$ monomial, and ${ \Psi} = { H}_d$. In this case 
${\Sigma}_{1,2}$ are ${ Q}$ superfields and $\lambda_{1,2}$
correspond to ${\lambda}_u$ and $\lambda_d$ respectively.
Then with the help of ${ X}_{1,2} = ({ H}_u \pm { \Psi} )/ \sqrt{2}$, 
part of  MSSM  superpotential can be written as:  $W \supset {\lambda_1 \over \sqrt{2}} { X}
{\varphi} {\Sigma_1} + {\lambda_2 \over \sqrt{2}} { X} {\varphi}
{\Sigma_2}$. This results in~\cite{Allahverdi:2007zz}:
\begin{equation} \label{chimass}
V \supset \lambda^2 {\vert \varphi \vert}^2 {\chi^2} ~ ~ , ~ ~
\lambda \equiv \left({\lambda^2_1 + \lambda^2_2 \over 8}\right)^{1/2}\,,
\end{equation}
where we have again considered the real part of $\chi$. Note that the
first generation of (s)leptons and (s)quarks have a Yukawa coupling
$\sim {\cal O} (10^{-6}-10^{-5})$, while the rest of the SM Yukawa
couplings are: $\lambda \geq 3\times10^{-4}$.

The most important point is to note that $60-70$ e-foldings of inflation is 
sufficient for the MSSM flat directions to take large VEVs
during and after inflation by virtue of stochastic jumps during inflation, see the discussion 
in Sec.~\ref{QFMFD} and the Refs.~\cite{Enqvist:2003gh,Dine:2003ax}. This however requires
that the MSSM flat directions do not obtain positive Hubble induced corrections during inflation.


\subsubsection{kinematical blocking of preheating}
\label{See: kinematical blocking of preheating}

In order to understand the preheating dynamics it is important to take
into account of $\chi$ coupling to the inflaton $\phi$, as well as to
the MSSM flat direction, $\varphi$, which is displaced away from its
minimum (towards large VEVs) during inflation. The governing potential
can be obtained from Eqs.~(\ref{infpot},\ref{chimass}):
\begin{equation} \label{mssm1}
V = {1 \over 2} m^2_{\phi} {\phi}^2 + g^ 2 {\phi}^2 {\chi}^2 +
{g \over \sqrt{2}} m_{\phi} \phi {\chi}^2 + \lambda^2 {\varphi}^2 {\chi}^2.
\end{equation}
As mentioned in the previous section, we generically have $\lambda
\geq 3 \times 10^{-4}$, and $g$ can be as large as $\sim {\cal O}(1)$.
After mode decomposition of the field $\chi$, the energy of the mode
with momentum $k$, denoted by $\chi_{k}$, is given by:
\begin{equation} \label{energy}
\omega_k = {\left(k^2 + 2 g^2 {\langle \phi \rangle}^2 +
\sqrt{2} g  m_{\phi}
\langle \phi \rangle +
2 \lambda^2 {\langle \varphi \rangle}^2 \right)}^{1/2}.
\end{equation}
Let us freeze the expansion of the universe first. Including the expansion
will not change our conclusions anyway.  Let us even consider the most 
opportunistic case for preheating with a large inflaton VEV, i.e. 
$\langle \phi \rangle > M_{\rm P}$.  Therefore if $g > 10^{-6}$, the inflaton induces a large
mass $g \langle \phi \rangle > H_{inf}$ for $\chi$ during inflation.
As a result, $\chi$, quickly settles down to the minimum, i.e.
$\langle \chi \rangle =0$, even if it is initially displaced, and
remains there. Therefore, $\varphi$, does not receive any mass
corrections from its coupling to $\chi$ during inflation.  Note that
the VEV of the flat direction, $\varphi$, induces a large mass,
$\lambda \varphi_0$, to the $\chi$ field during inflation.

In the interval $m_0 \leq H (t) \leq m_{\phi}$, where $m_0\sim {\cal O}(100)$~GeV is the mass of the 
MSSM flat direction,  the flat direction
VEV slides very slowly because of the under damped motion due to
large Hubble friction term, the flat direction effectively slow
rolls. Non-perturbative production of $\chi$ quanta will occur if
there is a non-adiabatic time-variation in the energy, i.e. that ${d
\omega_k/dt} \gs \omega^2_k$. The inflaton oscillations result in a
time-varying contribution to $\omega_k$, while the flat direction
coupling to $\chi$ yields a virtually {\it constant} piece. The piece induced by 
the flat direction VEV weakens the non-adiabaticity condition. Indeed time-variation
of $\omega_k$ will be adiabatic at all times: 
${d \omega_k / dt} < \omega_k^2$,  provided
$\lambda^2 \langle \varphi \rangle^2 > g {\hat \phi}m_{\phi}$,
where ${\hat \phi}$ is the amplitude of the inflaton oscillations.  
There will be no resonant production of $\chi$ quanta, provided that
\begin{equation} \label{cond}
\varphi_0 >  \lambda^{-1} \left(g M_{\rm P} m_{\phi}\right)^{1/2}\,,~~{\rm Typically}~~~\lambda\geq 3\times 10^{-4}\,,
\end{equation}
except the first generation of (s)leptons and (s)quarks which have a Yukawa coupling $\sim {\cal O}(10^{-6}-10^{-5})$,
and $g\geq 10^{-6}$, in order to have large inflaton couplings to matter, see Eq.~(\ref{infsup}).
This surmounts to a kinematical blocking of preheating by inducing a
piece (which is virtually constant at time scales of interest) to the
mass of inflaton decay products due to their couplings to a flat
direction which has a large VEV. Similar argument holds for kinematical blocking 
for fermionic preheating, as the symmetry between bosons and fermions implies similar
equations for the momentum excitations, see Eq.~(\ref{energy}).


\subsubsection{Late inflaton decay in SUSY}
\label{FDID}

The inflaton decay at the leading order will be kinematically forbidden if 
$y \vert \varphi \vert \geq m_{\phi}/2$ ($y$ is a SM gauge or Yukawa 
coupling)~\cite{Allahverdi:2007zz,Allahverdi:2005mz}. 
One should then wait until the Hubble expansion has redshifted, $\vert \varphi \vert$, down to
$(m_{\phi}/2y)$. The decay happens when (note that $\vert \varphi \vert \propto H$, after 
the flat direction starts oscillating and before the inflaton decays):
\begin{equation} \label{dec1}
H_1 = {\rm min} ~ \left[\left({m_{\phi} \over y \varphi_0} \right) m_0 ,
\Gamma_{\rm d} \right]\,,~~(\Gamma_d\equiv~{\rm total~inflaton~decay~width})\,,
\end{equation}
where $m_0\sim {\cal O}(100)$~GeV is the MSSM flat direction mass.
The inflaton also decays at higher orders of perturbation theory to
particles which are not directly coupled to it~\cite{Allahverdi:2002nb,Allahverdi:2005mz}. This mode
is kinematically allowed at all times, but the rate is suppressed by a
factor of $\sim \left(m_{\phi}/ y \vert \varphi \vert \right)^2
\Gamma_{\rm d}$. It becomes efficient at:
\begin{equation} \label{dec2}
H_2 \sim \left({m_{\phi} m_0 \over \varphi_0}\right)^{2/3} 
\Gamma^{1/3}_{\rm d}.
\end{equation}
Therefore, if the decay products are coupled to a flat direction
with a non-zero VEV, the inflaton will actually decay at a time when
the expansion rate of the universe is given by~\cite{Allahverdi:2005mz}
\begin{equation} \label{infdec}
H_{\rm d} = {\rm max} ~ \left[H_1, H_2 \right].
\end{equation}
In general it is possible to have, $H_{\rm d} \ll \Gamma_{\rm d}$,
particularly for large values of $\varphi_0$. Flat
directions can therefore significantly delay inflaton decay on purely
kinematical grounds.

 
 \subsubsection{Decay of a flat direction}

There is a crucial difference between a rotating and an oscillating flat direction
when it comes to decay. It is well known that an oscillating condensate can decay non-perturbatively 
via preheating, as we discussed in the earlier subsections.

In the case of a rotating condensate, there is the conservation of 
global charges associated with the net particle number density in 
fields~\cite{Affleck:1984fy,Dine:1995kz,Dine:1995uk}~\footnote{Charges identified by the net particle number
in fields, which are included in a flat direction are most notably baryon and lepton number, which
are preserved by the $D$-terms}. This ensures that the total 
number density of quanta will not decrease and, consequently, the average  energy of quanta will not increase.
The actual decay of a rotating flat direction into other fields happens perturbatively, through the $F$-term 
couplings, as originally envisaged by Affleck and Dine~\cite{Affleck:1984fy,Dine:1995kz,Dine:1995uk}, see 
also~\cite{Allahverdi:1999je,Postma:2003gc}. 

It was argued that there could be possible non-perturbative effects~\cite{Olive:2006uw,Gumrukcuoglu:2008fk,Gumrukcuoglu:2009fj}, 
stemming from the $D$-terms 
of the potential for a rotating flat direction condensate. However it was shown in~\cite{Allahverdi:2008pf,Allahverdi:2006xh}
that such non-perturbative effects have no bearing for the decay of energy density in rotating flat direction(s).
 In the case of a rotating condensate all that can happen is a mere redistribution of the 
energy among the fields on the $D$-flat subspace.

There is a no-go theorem for a rotating condensate~\cite{Allahverdi:2008pf,Allahverdi:2006xh}, which states;
for MSSM flat direction(s), which are represented by gauge-invariant combinations of fields $\Phi_i=e^{i\alpha_i}\Phi_{i}$,
possible non-perturbative particle production from time-variation in the mass eigenstates caused by the $D$-terms:\\
{\vskip-0.5cm}

(1) cannot change the net particle number density in $\Phi_i$, denoted by $n_i=i\dot\Phi_{i}^{\ast}\Phi_{i}+h.c.$, and hence 
the total baryon/lepton number density stored in the condensate.\\
{\vskip-0.5cm}

(2) cannot decrease the total {\it comoving} particle number density in $\phi_i$, denoted by ${\tilde n}_i$, thus the total number density 
of quanta ${\tilde n}= \sum_{i}{{\tilde n}_i}$ in the condensate.  As a direct consequence of the conservation of energy density, non-perturbative effects will not increase the average energy of quanta $E_{\rm ave}$.\\
{\vskip-0.5cm}

The theorem also applies to elliptical trajectories, where the condensate will mainly contain particles (or anti-particles), but it will also 
contain a small mixture of anti-particles (or particles). The theorem is applicable for the subsequent evolution of the plasma formed 
after the phase of particle production. This implies that possible non-perturbative effects do not lead to the decay of a rotating 
condensate. They merely redistribute the energy which is initially stored in the condensate among the fields on the $D$-flat subspace~\cite{Allahverdi:2008pf}.

The marked difference between rotating and an oscillating flat direction case can be
understood from the trajectory of motion (i.e. circular for rotation versus linear for oscillation). An 
oscillating condensate $\phi$, whose trajectory of motion is a line, can be written as
\begin{equation} \label{real}
\phi = {\varphi \over 2} ~ {\rm exp} (i \theta) + {\varphi \over 2} ~ {\rm
exp} (-i \theta),
\end{equation}
and the conserved charge associated with the global $U(1)$
(corresponding to phase $\theta$) is given by
\begin{equation} \label{realn}
n = i {\dot \phi^*} \phi + {\rm h.c.} = 0 .
\end{equation}
This is not surprising since an oscillation is the superposition of two
rotations in opposite directions, which carry exactly the same number of
particles and anti-particles respectively. Therefore the net particle number
density stored in an oscillating condensate is zero.

Now consider non-perturbative particle production from an oscillating condensate.
One can think of this process as a series of annihilations among $N$ particles and $N$
anti-particles in the condensate, $N > 1$, into an energetic particle-anti-particle pair.
This is totally compatible with conservation of charge, see Eq.~(\ref{realn}); $n=0$ after
preheating as well as in the condensate.

On the other hand, a (maximally) rotating condensate
consists of particles or anti-particles {\it only}~\cite{Allahverdi:2008pf,Allahverdi:2006xh}. Conservation of
the net particle number density then implies that $N \rightarrow 2$
annihilations ($N > 2$) are forbidden: annihilation of particle (or
anti-particle) quanta cannot happen without violating the net particle number
density. Therefore the total number density of quanta will not
decrease, and the average energy will not increase~\footnote{Note that
an increase in the total particle number density, through creation of an
equal number of particles and anti-particles will be in agreement with
the conservation of the net particle number density. In this case the
resulting plasma will be even denser than the condensate.}.

Any possible non-perturbative particle production in the rotating condensate case 
will result in a plasma which is at least as dense as the initial condensate. All that can happen is 
a redistribution of the energy density among the fields on the $D$-flat subspace. These fields have masses comparable to the flat direction mass $m_0$, as they all arise from SUSY breaking. Since the average 
energy is $E_{\rm ave} \leq m_0$, the resulting plasma essentially consists of non-relativistic quanta. Its energy 
density $\rho = {\tilde n} E_{\rm ave}$ is therefore redshifted $\propto a^{-3}$.
The decay of a rotating condensate happens quite late~\cite{Allahverdi:2008pf}
\begin{eqnarray} 
H_{\rm dec}  \sim  m_0 \left({m_0 \over y \varphi_0}\right)~~~~({\rm m.~d.}) \,,~~~~
H_{\rm dec}  \sim  m_0 \left({m_0 \over y \varphi_0}\right)^{4/3}~~~~({\rm r.~d.}) \,, 
\end{eqnarray}
where ${\rm m.d.}$ corresponds to matter domination and ${\rm r.d.}$ corresponds to radiation domination, and $y$ is the
Yukawa coupling of the flat direction to MSSM matter. The decay happens essentially perturbatively for $y\varphi_0>m_0$, as envisaged by~\cite{Affleck:1984fy,Dine:1995kz,Dine:1995uk}


\subsubsection{SUSY thermalization}\label{FDAT}

Flat directions VEV can dramatically affect thermal history of the
universe. The reason is
that the MSSM flat direction VEV spontaneously breaks the SM gauge
group. The gauge fields of the broken symmetries then acquire a
SUSY conserving mass, $m_{\rm g} \sim g \vert \varphi \vert$,
from their coupling to the flat direction, where $g$ is a gauge
coupling constant. In which case a flat direction can
crucially alter thermal history of the universe by suppressing
thermalization rate of the reheat plasma. Note that, $m_{\rm g}$,
provides a physical infrared cut-off for scattering diagrams with
gauge boson exchange in the $t-$channel shown in Fig.~(A).  The
thermalization rate will then be given by (up to a Logarithmic
``bremsstrahlung'' factor):
\begin{equation} \label{chemical1}
\Gamma_{\rm thr} \sim \alpha^2 {n \over {\vert \varphi \vert}^2},
\end{equation}
where we have used $m^2_{\rm g} \simeq g^2 {\vert \varphi
\vert}^2$ where $\alpha\sim g^2/4\pi$. After the flat direction starts its oscillations at $H
\simeq m_0\sim {\cal O}(100)$~GeV, the Hubble expansion redshifts, ${\vert \varphi \vert}^2
\propto a^{-3}$, where $a$ is the scale factor of the FRW
universe. The interesting point is that, $n \propto a^{-3}(t)$, as well,
and hence $\Gamma_{\rm thr}$ remains constant while $H$ decreases for
$H(t) < m_0\sim {\cal O}(100)$~GeV. This implies that $\Gamma_{\rm thr}$ eventually catches up
with the expansion rate, even if it is initially much smaller, and
shortly after that the full thermal equilibrium will be achieved.
Depending on whether $m_0 > \Gamma_{\rm d}$ or $m_0 < \Gamma_{\rm d}$, different
situations will arise which we discuss separately~\footnote{ If two or more flat
directions with non-zero VEVs induce mass to the gauge bosons, then
$\vert \varphi \vert$ denotes the largest VEV.}.

\begin{itemize}

\item{ $\underline{m_0 > \Gamma_{\rm d}}$:~~~In this case the inflaton decays
after the flat direction oscillations start.  The inflaton
oscillations, which give rise to the equations of state close to
non-relativistic matter, dominate the energy density of the universe
for $H (t)> \Gamma_{\rm d}$. This implies that $a(t) \propto H^{-2/3}(t)$, and
$\vert \varphi \vert$ is redshifted $\propto H(t)$ in this period. We
therefore have, $\varphi_{\rm d} \sim \left({\Gamma_{\rm d} / m_0}\right) \varphi_0$,
where $\varphi_{\rm d}$ denotes the amplitude of the flat direction
oscillations at the time of the inflaton decay $H(t) \simeq \Gamma_{\rm d}$. 
By using the total energy density of the plasma; $\rho \approx 3 \left(\Gamma_{\rm d} 
M_{\rm P}\right)^2$, and  note that $\vert \varphi \vert$ and $n$ are both
redshifted $\propto a^{-3}(t)$ for $H < \Gamma_{\rm d}$,  after
using Eq.~(\ref{chemical1}), we find that complete thermalization
occurs when the Hubble expansion rate is~\cite{Allahverdi:2005mz}
\begin{equation} \label{thermal1}
H_{\rm thr} \sim 10 \alpha^2 
\left({M_{\rm P} \over \varphi_0}\right)^2 {m^2_0 \over m_{\phi}}\,.
\end{equation}
}

\item{ $\underline{m_0 < \Gamma_{\rm d}}$:~~~In this case the flat direction
starts oscillating after the completion of inflaton decay.  The
universe is dominated by the relativistic inflaton decay products for
$H(t) < \Gamma_{\rm d}$, implying that $a(t) \propto H(t)^{-1/2}$.
The number density of particles in the plasma is redshifted $\propto
H(t)^{3/2}$ until $H (t)= m_0\sim {\cal O}(100)$~GeV. Note that $n,~{\vert \varphi \vert}^2 \propto a(t)^{-3}$, and hence
$\Gamma_{\rm thr}$ remains constant, for $H(t) < m_0$. The reheat plasma
then thermalizes when the Hubble expansion rate is~\cite{Allahverdi:2005mz}
\begin{equation} \label{thermal2}
H_{\rm thr} \sim 10 \alpha^2 
\left({\Gamma_{\rm d} \over m_0}\right)^{1/2} \left({M_{\rm P} 
\over \varphi_0}\right)^2 {m^2_0 \over m_{\phi}}\,.
\end{equation}     
}
\end{itemize}


Since the kinetic equilibrium is built through $2 \rightarrow 2$
scattering diagrams as in Fig.~(A), which have one interaction vertex
less than those in $2\rightarrow 3$. Therefore the rate for
establishment of kinetic equilibrium will be $\Gamma_{\rm kin} \sim
\alpha^{-1} \Gamma_{\rm thr}$.  In SUSY case, the relevant time scales have 
an hierarchy~\cite{Allahverdi:2005mz}:
\begin{equation}
\Gamma_{\rm d} \gg \Gamma_{\rm kin} > \Gamma_{\rm thr}\,.
\end{equation}
The relative chemical equilibrium among different
degrees of freedom is built through $2 \rightarrow 2$ annihilations in
the $s-$channel with a rate $\sim \alpha^2 n/E^2 \ll \Gamma_{\rm
thr}$. Hence composition of the reheat plasma will not change until
full thermal equilibrium is achieved.
This implies that the universe enters a long period of quasi-adiabatic
evolution after the inflaton decay has completed. During this phase,
the comoving number density and (average) energy of particles remain
constant~\footnote{The decay of flat directions and their interactions with
the reheat plasma are negligible before the
universe fully thermalizes.}.



\subsubsection{Reheat temperature of the universe}
\label{RTOTU}

The temperature of the universe after it reaches full thermal
equilibrium is referred to as the reheat temperature $T_{\rm R}$. In
the case of MSSM, we therefore have~\cite{Allahverdi:2005mz,Allahverdi:2007zz}:
\begin{equation} \label{rehtemp}
T_{\rm R} \simeq \left(H_{\rm thr} M_{\rm P}\right)^{1/2}\,,
\end{equation}
where, depending on the details, $H_{\rm thr}$ is given by
Eqs.~(\ref{thermal1}) and~(\ref{thermal2}).  Since 
$H_{\rm thr} \ll \Gamma_{\rm d}$,  the reheat temperature 
is generically much smaller in MSSM ( or in a generic theory with gauge invariant flat directions ) 
than the standard expression $T_{\rm R} \simeq \left(\Gamma_{\rm d} M_{\rm P}\right)^{1/2}$, 
which is often used in the literature with an  assumption that  
immediate thermalization occurs after the inflaton decay.
Note that the reheat temperature depends very
weakly on the inflaton decay rate, for instance Eq.~(\ref{thermal2})
implies that $T_{\rm R} \propto \Gamma^{1/4}_{\rm d}$, while $T_{\rm
R}$ is totally independent of $\Gamma_{\rm d}$ in
Eq.~(\ref{thermal1}). Regardless of how fast the inflaton decays, the universe will not thermalize
until the $2 \rightarrow 3$ scatterings become efficient.  A larger $\varphi_0$ results in slower thermalization and
a lower reheat temperature, see the following table for some sample
examples~\footnote{If $H_{\rm thr} \geq \Gamma_{\rm d}$, the reheat temperature
in such cases follows the standard expression: $T_{\rm R} \simeq
\left(\Gamma_{\rm d} M_{\rm P}\right)^{1/2}$. This will be the case if
the flat direction VEV is sufficiently small at the time of the
inflaton decay, and/or if the reheat plasma is not very dilute. }.

\vspace*{7mm}
\begin{center}
\begin{tabular}{|r|r|r|}
\hline
VEV (in GeV) & $T_{\rm R}(\Gamma_{\rm d} = 10~{\rm GeV}<m_0$) & 
$T_{\rm R}(\Gamma_{\rm d} = 10^4~{\rm GeV}>m_0$) \\ \hline \hline
$\varphi_0 \sim M_{\rm P}$ & $3 \times 10^3$ & $7 \times 10^{4}$ \\
\hline
$\varphi_0 = 10^{-2} M_{\rm P}$ & $3 \times 10^5$ & $7 \times 10^6$ \\
\hline
$\varphi_0 = 10^{-4} M_{\rm P}$ & $3 \times 10^7$ & $7 \times 10^8$ \\
\hline
$\varphi_0 \leq 10^{-6} M_{\rm P}$ & $3 \times 10^9$ & $7 \times 10^{10}$ \\
\hline
\end{tabular} 
\vspace*{2mm}
\end{center}
\noindent
{ Table 1:}~ The reheat temperature of the universe for the
inflaton mass, $m_{\phi} = 10^{13}$ GeV, and two values of the
inflaton decay rate, $\Gamma_{\rm d} = 10,~10^4$ GeV (if the inflaton 
decays gravitationally,  we have $\Gamma_{\rm d} \sim 10$ GeV). The flat
direction mass is $m_0 \sim 1$~TeV. The rows show the values of
$T_{\rm R}$ for flat direction VEVs. Note that when the VEV of a flat direction
is $<10^{12}$~GeV, the flat direction can no longer delay the thermalization, 
and the reheat temperature, $T_{\rm R}\simeq (\Gamma_dM_{\rm P})^{1/2}$, remains a good 
approximation~\footnote{Majority of MSSM flat directions can take 
a VEV $\geq 10^{14}$~GeV, which are lifted by more than dimensional $6$ operators, for instance $udd$, $LLe$, etc. }. 
\vspace*{6mm}


\subsection{Quasi-adiabatic thermal evolution of the universe}
\label{deviation QATEOTU}

Before complete thermalization is reached the universe remains
out of equilibrium~\footnote{Right after the inflaton decay has completed the energy density of the
universe is given by; $\rho \approx 3 \left(\Gamma_{\rm d} M_{\rm
P}\right)^2$, and the average energy of particles is $\langle E
\rangle \simeq m_{\phi}$. For example, in a two-body decay of
the inflaton, we have exactly $E = m_{\phi}/2$. }.The deviation from full
equilibrium can be quantified by the parameter $''{\cal A}''$~\cite{Allahverdi:2005fq,Allahverdi:2005mz}, where
\begin{equation} \label{a} {\cal A} \equiv {3 \rho \over T^4} \sim 10^4
\left({\Gamma_{\rm d} M_{\rm P} \over m^2_{\phi}}\right)^2.  \end{equation}
Here we define $T \approx \langle E \rangle/3$, in accordance with
full equilibrium. Note that in full equilibrium, see Eq.~(\ref{full}),
we have ${\cal A} \approx g_{\ast}$ ($=228.75$ in the MSSM). 

One can also associate parameter ${\cal A}_i \equiv 3
\rho_i/T^4$ to the $i-$th degree of freedom with the energy density
$\rho_i$ (all particles have the same energy $E$, and hence $T$, as
they are produced in one-particle decay of the inflaton). Note that
${\cal A} = \sum_{i}{{\cal A}_i}$, and in full equilibrium we have
${\cal A}_i \simeq 1$. Note that ${\cal A}$ depends on the total
decay rate of the inflaton $\Gamma_{\rm d}$ and its mass $m_{\phi}$
through Eq.~(\ref{a}).  While, ${\cal A}_i$ are determined by the
branching ratio for the inflaton decay to the $i-$th degree of
freedom.  The composition of the reheat plasma is
therefore model-dependent before its complete thermalization. However,
some general statements can be made based on symmetry arguments.

During the quasi-adiabatic evolution of the reheated plasma, i.e., for
$H_{\rm thr} < H < \Gamma_{\rm d}$, we have~\footnote{One can express
${\cal A}_i$ in terms of a {\it negative} chemical potential $\mu_i$,
where ${\cal A}_i = {\rm exp}\left(\mu_i/T \right)$. Note that for a
large negative chemical potential, i.e., in a dilute plasma, the Bose
and Fermi distributions are reduced to the Maxwell-Boltzmann distribution and 
give essentially the same result. The
assignment of a chemical potential merely reflects the fact that the
number of particles remains constant until the number-violating
reactions become efficient. It does not appear as a result of a
conserved quantity (such as baryon number) which is due to some
symmetry. Indeed, assuming that inflaton decay does not break such
symmetries, the same chemical potential to particles
and anti-particles are the same.}~\cite{Allahverdi:2005mz}
\begin{equation} \label{kin}
\rho_i = {\cal A}_i {3 \over \pi^2} T^4\,, ~~~~ n_i = 
{\cal A}_i {1 \over \pi^2} T^4,~~~~~H \simeq {\cal A}^{1/2} \left({T^2 \over 3 M_{\rm P}}\right).
\end{equation}
In this epoch $T$ varies in a range $T_{\rm min} \leq T \leq T_{\rm
max}$, where $T_{\rm max} \approx m_{\phi}/3$ is reached right after
the inflaton decay.  Because of complete thermalization, $T$ sharply
drops from $T_{\rm min}$ to $T_{\rm R}$ at $H_{\rm thr}$, where the
conservation of energy implies that
\begin{equation} \label{FR}
T_{\rm R} = \left({{\cal A} \over 228.75}\right)^{1/4} 
T_{\rm min},~~\Longrightarrow ~~T_{\rm R} \leq  T_{\rm min}\,.
\end{equation}
%


\subsubsection{Particle creation in a quasi-thermal phase}
\label{QTPP}
In this section, we wish to create weakly interacting $\chi$ particles
from the scatterings of the MSSM particles during the quasi-thermal
phase of the universe. Recall the Boltzmann equation governing the number 
density of $\chi$ particles, which is given by~\cite{Jungman:1995df,Kolb:1988aj}:
\begin{equation} \label{chipro}
{\dot{n}}_{\chi} + 3 H n_{\chi} = \sum_{i \leq j} \langle v_{\rm rel} ~
\sigma_{ij \rightarrow \chi}\rangle n_i n_j. 
\end{equation}
Here $n_i$ and $n_j$ are the number densities of the $i-$th and
$j-$th particles, $\sigma_{ij \rightarrow \chi}$ is the cross-section
for producing $\chi$ from scatterings of $i$ and $j$, and the sum is
taken over all fields which participate in $\chi$ production. Also
$\langle ... \rangle$ denotes averaging over the distribution. 
Since the production of $\chi$ quanta will be Boltzmann suppressed if $T < m_{\chi}/3$.
Therefore, in order to obtain the total number of $\chi$ quanta produced from
scatterings, it will be sufficient to integrate the RH side
of Eq.~(\ref{chipro}) from the highest temperature down to $T_{\rm min}$.
The relic abundance of
$\chi$, normalized by the entropy density, $s$, is given by~\cite{Allahverdi:2005mz}
\begin{equation} \label{chidens}
{n_{\chi} \over s} \sim 10^{-5} \left({228.75 \over 
{\cal A}}\right)^{5/4} 
\sum_{i,j} 
\left[\int_{T_{\rm min}}^{T_{\rm max}}{{\cal A}_i {\cal A}_j 
{\langle v_{\rm rel}~\sigma_{ij \rightarrow 
{\chi}}\rangle}~ M_{\rm P}~ dT}\right]\,, 
\end{equation}
where we have used Eq.~(\ref{FR}).

\subsubsection{Gravitino production}
\label{GPQT}

For flat direction(s) VEV $\geq 10^{12}$~GeV,  slow thermalization 
results in a low reheat temperature, i.e $T_{\rm R}\leq 10^{9}$~GeV, 
which is compatible with the BBN bounds on
thermal gravitino production. However gravitinos are also produced
during the quasi-thermal phase prior to a complete thermalization of
the reheat plasma.  Generically gravitinos are produced from the 
scatterings of gauge, gaugino, fermion and sfermion
quanta with a cross-section $\propto 1/M^2_{\rm P}$.

During the quasi-thermal phase, the gauge and gaugino quanta 
have large masses $\sim \alpha^{1/2} \varphi_{\rm d}$ (induced by 
the flat direction VEV) at a time most
relevant for the gravitino production, i.e., when $H \simeq
\Gamma_{\rm d}$, therefore, they decay to lighter fermions and
sfermions at a rate $\sim {\alpha}^{3/2} \varphi^2_{\rm d}/m_{\phi}$,  
where $\alpha^{3/2} \varphi_{\rm d}$ is the decay width
at the rest frame of gauge/gaugino quanta, and $\varphi_{\rm
d}/m_{\phi}$ is the time-dilation factor. The decay rate is $\gg
\Gamma_{\rm d}$, thus gauge and gaugino quanta decay almost instantly
upon production, and they will not participate in the gravitino
production. As a consequence, production of the helicity $\pm 1/2$ states
will not be enhanced in a quasi-thermal phase as scatterings with a
gauge-gaugino-gravitino vertex will be absent~\footnote{ Otherwise 
gauge and/or gaugino quanta in the initial state (particularly
scattering of two gluons) have the largest production
cross-section~\cite{Ellis:1984eq,Moroi:1993mb,Moroi:1995fs,Bolz:2000fu}.}.

The following channels contribute to the gravitino production~\cite{Ellis:1984eq}: 
(a)~{\it fermion} + {\it anti-sfermion $\rightarrow$ gravitino} + {\it gauge field},~~
(b)~{\it sfermion} + {\it anti-fermion $\rightarrow$ gravitino} + {\it gauge field},~~
(c)~{\it fermion} + {\it anti-fermion} $\rightarrow$ {\it gravitino}+ {\it gaugino},~~
(d)~{\it sfermion} + {\it anti-sfermion} $\rightarrow$ {\it gravitino} + {\it gaugino}.

The total cross-section involves
cross-sections for multiplets comprising the LH (s)quarks $Q$, RH
up-type (s)quarks $u$, RH down-type (s)quarks $d$, LH (s)leptons $L$,
RH (s)leptons $e$ and the two Higgs/Higgsino doublets $H_u,H_d$. 
Since particles and anti-particles associated to the bosonic and fermionic components of
the multiplets which belong to an irreducible representation of a
gauge group have the same parameter ${\cal A}_i$. This implies that~\cite{Allahverdi:2005mz}
\begin{eqnarray} \label{final} 
\Sigma_{\rm tot} \equiv \sum^{3}_{i,j=1} ~ \sum^{2}_{a,b=1} ~ 
\sum^{3}_{\alpha, \beta = 1} 
{{\cal A}_{i,a,\alpha} {\cal A}_{j, b, \beta} ~  
\langle \sigma_{\rm tot} v_{\rm rel} \rangle} = \, \nonumber \\
{1 \over 32 M^2_{\rm P}} \sum {[6 \alpha_3 (2 {\cal A}^2_{Q} + 
{\cal A}^2_{u} + {\cal A}^2_{d}) + 
{9 \over 4} \alpha_2 (3 {\cal A}^2_{Q} + {\cal A}^2_{L} + {\cal A}^2_{H})} 
\, \nonumber \\
{+ {1 \over 4} \alpha_1 ({\cal A}^2_{Q} + 8 {\cal A}^2_{u} + 2 
{\cal A}^2_{d} + 3 {\cal A}^2_{L} + 6 {\cal A}^2_{e} + 3 {\cal A}^2_{H})}] \, ,
\end{eqnarray}
where $1 \leq i,j \leq 3$, $a,b =1,2$ and $1 \leq \alpha, \beta \leq 3$ are the flavor, weak-isospin
and color indices of scattering degrees of freedom respectively. Also
$\alpha_3$, $\alpha_2$ and $\alpha_1$ are the gauge fine structure
constants related to the $SU(3)_C$, $SU(2)_W$ and $U(1)_Y$ groups
respectively. The sum is taken over the three flavors of $Q,u,d,L,e$ and the two
Higgs doublets.  After replacing $\Sigma_{\rm tot}$ in
Eq.~(\ref{chidens}), and recalling that $T_{\rm max} \approx
m_{\phi}/3$, we obtain~\cite{Allahverdi:2005mz}
\begin{equation} \label{gravquasi}
{n_{3/2} \over s} \simeq \left(10^{-1} M^2_{\rm P} ~ \Sigma_{\rm tot}\right) 
\left({228.75 \over {\cal A}}\right)^{5/4} 
\left({T_{\rm max} \over 10^{10}~{\rm GeV}}\right) ~ 10^{-12}\,.
\end{equation}
Note that in full thermal equilibrium, $\Sigma_{\rm tot} =
\left(4 \pi/M^2_{\rm P}\right) \times \left(16 \alpha_3 + 6 \alpha_2 +
2 \alpha_1 \right) \simeq \left(10^{-1}/M^2_{\rm P}\right)$ (up to
logarithmic corrections which are due to renormalization group
evolution of gauge couplings).

Let us consider a 
simplistic scenario when the inflaton primarily decays into one flavor of LH (s)quarks.  
In this case ${\cal A}_Q = 1/24$ for the relevant flavor~\footnote{The total number of
degrees of freedom in one flavor of LH (s)quarks is $2~({\rm
particle-antiparticle}) \times 2~({\rm fermion-boson}) \times 2~({\rm
weak-isospin}) \times 3~({\rm color})$.}, while ${\cal A} = 0$ for the rest of the 
degrees of freedom. This results in a gravitino 
abundance: ${n_{3/2} / s} \simeq  ({{\cal A} / 228.75})^{3/4} ~
({T_{\rm max} / 10^{10}~{\rm GeV}}) ~ 10^{-12}$,
where ${\cal A}$ is given by Eq.~(\ref{a}) and note that $T_{\rm max}\gg T_{\rm R}$.
The largest value for $T_{\rm max} \simeq 10^{12}$ GeV, therefore the the tightest bound 
for unstable gravitinos come from BBN $\left(n_{3/2}/s\right) \leq
10^{-16}$ (arising for $m_{3/2} \simeq 1$ TeV and a hadronic branching
ratio $\simeq 1$) is satisfied if ${\cal A} \leq 10^{-6}$. Much weaker
bounds on ${\cal A}$ are found for a radiative decay. For example,
${\cal A} \leq 10^{-3}$ ($1$) if $m_{3/2} \simeq 100$ GeV ($1$ TeV).





\section{Generating perturbations with the curvaton}\label{sec:curvatonmodels}
%
%

\subsection{What is the curvaton ?}

The curvaton scenario is an interesting possibility~\cite{Linde:1996gt,Enqvist:2001zp,Lyth:2001nq,Lyth:2002my,Moroi:2001ct,Moroi:2002rd,Linde:2005yw}, 
see also~\cite{Mollerach:1989hu}, where the perturbations of more than one light scalar fields play important role during inflation. In many realistic examples of particle physics having more than one light scalar fields is quite natural, especially in the case of 
MSSM~\cite{Enqvist:2003gh}. However, as we shall argue below that within MSSM there are only handful of such good 
candidates for a curvaton~\cite{Allahverdi:2006dr,Enqvist:2002rf,Enqvist:2003mr}~\footnote{Sneutrino as a curvaton have been considered in Refs.~\cite{Mazumdar:2003bs,Mazumdar:2004qv,McDonald:2004by}, With an extension of MSSM, where right handed sneutrinos are introduced as SM gauge singlets. However sneutrino alone is not a $D$-flat direction, one must consider the $D$-term contribution of the potential also. There are also models of curvaton where the field is a pNGB~\cite{Dimopoulos:2003az,Dimopoulos:2003ii,Dimopoulos:2005bx}, or the curvaton action follows that of a 
Dirac-Born-Infeld~\cite{Li:2008fma,Zhang:2009gw}. However, it is uncertain how such models of curvaton would excite dominantly the SM degrees of freedom. }.

In the original  curvaton paradigm it was assumed that the curvaton is responsible
for generating the entire curvature perturbations, and the perturbations arising from the inflaton 
component are subdominant~\cite{Lyth:2001nq,Lyth:2002my,Moroi:2001ct,Moroi:2002rd}. However, there are variants where both inflaton and curvaton can contribute to the curvature perturbations, see for instance~\cite{Enqvist:2009zf,Erickcek:2009at}.  Irrespective of their origins the curvaton must possess the following properties:


\begin{itemize}

\item{Lightness of the curvaton:\\
During inflation the Hubble expansion rate is: $H_{inf} \gg m_\varphi$, where $m_\varphi$ is the effective mass of
the curvaton.  Since it does not cost anything in energy, therefore the quantum fluctuations are free to accumulate along a curvaton direction and form a condensate with a large VEV, $\varphi_0$.  During inflation, $V(\phi) \gg V(\varphi)$, where 
$\phi$ is the inflaton here. After inflation, $H \propto t^{-1}$, and the curvaton stays at a relatively large VEV due to large Hubble friction term until  $H \simeq m_\varphi$, when the curvaton starts oscillating around
the origin with an initial amplitude $\sim \varphi_0$. From then on
$\vert \varphi \vert$ is redshifted by the Hubble expansion $\propto
H(t)$ for matter dominated and $\sim H^{3/4}(t)$ for radiation dominated
Universe. The energy of the oscillating flat direction may eventually
start to dominate over the inflaton decay products. }

\item{Longevity of the curvaton and its dominance:\\
Furthermore, the curvaton must not evaporate due to thermal interactions from the plasma already created by the inflaton decay products~\cite{Postma:2002et,Allahverdi:2006dr}. The curvaton and the inflaton decay products can be weakly coupled, i.e. via non-renormalizable interactions. In such a case the curvaton can survive long enough and its oscillations can possibly dominate the energy density while decaying. It has been shown that MSSM flat direction can remain long lived~\cite{Allahverdi:2006xh,Allahverdi:2008pf}. However, if the curvaton does not dominate while decaying, and the curvaton decay products do not thermalize with the inflaton decay products, then this will lead to potentially large isocurvature fluctuations
\cite{Lyth:2002my,Lemoine:2009is,Moroi:2008nn,Endo:2003fr,Lyth:2003ip}. Note that the non-Gaussianity parameter, $f_{NL}$, is also constrained by the allowed isocurvature perturbations~\cite{Sasaki:2006kq,Bartolo:2003jx,Beltran:2008ei}. The curvaton can easily dominate the energy density if the inflaton decay products are dumped outside our own Hubble patch, which may be realizable in certain brane-world models with warped extra dimensions~\cite{Enqvist:2003qc,Enqvist:2004kg,Pilo:2004mg}. }
 
\item{Thermalization:\\
When the curvaton energy density does not dominate while decaying, then the curvaton decay products must
thermalize with that of the inflaton decay products, otherwise, there will be large remnant isocurvature perturbations. It is desirable that
both inflaton and curvaton excites {\it solely} the SM and/or MSSM degrees of freedom, which requires non-trivial 
construction on both the sectors~\cite{Allahverdi:2006dr}. }

\end{itemize} 

Curvaton paradigm with more than 2 light fields were also 
constructed in~\cite{Gupta:2003jc,Assadullahi:2007uw}, but they bear more uncertainties, especially when 
the curvatons belong to the hidden sector. There are various challenges to the curvaton paradigm,  which
we discuss below.

\begin{itemize}

\item{An absolute gauge singlet or a hidden sector curvaton:\\
An absolute gauge singlet curvaton's origin might lie within string theory~\cite{Moroi:2001ct,Moroi:2002rd}, or from a hidden 
sector~\cite{Dimopoulos:2002kt,Dimopoulos:2003az,Dimopoulos:2003ss,Lyth:2006nx,Pilo:2004mg,Mazumdar:2003vg,Bartolo:2002vf,Liddle:2003zw,Dimopoulos:2005bx,Dimopoulos:2004yb}. The modulus will generically couple to all other sectors presumably with a {\it universal coupling}, i.e. gravitationally suppressed interaction. Note that 
dumping entropy into non-SM-like sectors from the curvaton decay would lead to large isocurvature perturbations, if the curvaton is subdominant, and these non-SM like degrees of freedom do not thermalize with that of the inflaton decay products before
BBN~\cite{Lemoine:2009is}. 

If the inflaton is also a modulus or a hidden sector field, then it is important to make sure that the inflaton and the curvaton are extremely weakly coupled, otherwise, inflaton decay into curvaton can destroy the curvaton 
condensate prematurely before its dominance.}

\item{A SM gauge invariant curvaton:\\
MSSM provides gauge invariant curvaton 
candidates~\cite{Enqvist:2002rf,Postma:2002et,Enqvist:2003mr,Enqvist:2003qc,Mazumdar:2003bs,Enqvist:2004kg,Mazumdar:2004qv,BasteroGil:2003tj,McDonald:2003jk,McDonald:2004by,Riotto:2008gs,Kasuya:2003va,Allahverdi:2006dr}, 
which can decay into MSSM degrees of freedom directly. In this case the curvaton must ensure its longevity, and its dominance at the time of decay.  If the inflaton decay products are of MSSM/SM like quanta, then they can interact with the MSSM curvaton and can possibly destroy its coherence and longevity~\cite{Postma:2002et,Allahverdi:2006dr}. The advantage of an MSSM curvaton is that the 
couplings are that of the SM, therefore, predictions are robust. }

\end{itemize}

\subsection{Cosmological constraints on a curvaton scenario}

The main observable constraints for a curvaton scenario are: (1) isocurvature perturbations, and (2) primordial non-Gaussianity. 
The isocurvature perturbations are created when the curvaton fail to dominate the energy density 
while decaying and the curvaton decay proucts fail to thermalize with that of the inflaton decay products. There are 
potentially well motivated isocurvature perturbations one can create, CDM~\cite{Lyth:2002my,Kawasaki:1995ta,Beltran:2006sq,Beltran:2005xd,Beltran:2005gr,Lyth:2003ip}, baryon isocurvature  perturbations~\cite{Lyth:2002my,Mollerach:1989hu,Kawasaki:2001in,Koyama:1998hk,Hu:1994tj,Kodama:1986fg,Turner:1988sq,Dolgov:1992pu}, neutrino isocurvature perturbations were also considered in~\cite{Lyth:2002my}~\footnote{There is a way to avoid large isocurvature component 
provided the inflaton perturbations are subdominant during inflation, i.e. $(\delta \rho/\rho)_{inf}\ll 10^{-5}$. This can be achieved very well in many inflationary models~\cite{Enqvist:2009zf}, but the curvaton perturbations can create the required level of perturbations even if they do not dominate the energy density while decaying. This can be possible for a curvaton potential discussed in Eq.~(\ref{curv-pot-without-A}) for specific choices of $(n,~m_0)$, see Ref.~\cite{Enqvist:2009zf}.}.

For a conserved number density $\delta n_{i}=0$, where $i$ corresponds to baryons or CDM, the entropy perturbation
is given by~\cite{Wands:2000dp,Lyth:2002my}
\begin{equation}
{\cal S}_{i}=3\left(\zeta_{i}-\zeta\right)\,,
\end{equation}
where $\zeta$ is the total curvature perturbations, and $\zeta_{i}$ is the curvature perturbations in the $i$ component.
Recall that in a curvaton scenario the total curvature perturbations is given by; $\zeta=r\zeta_{\varphi}$, where 
$r=\rho_{\varphi}/\rho_r$ at the time of curvaton decay, and $\rho_r$ is the radiation energy density due to the  inflaton
decay products.

Let us now imagine, when the curvaton decays either into baryons (need not be the SM baryons), or non-relativistic
CDM, then $\zeta_{i}=\zeta_{\varphi}=(1/r)\zeta$. Combining this fact, the isocurvature perturbations yield:
\begin{equation}
{\cal S}_{i}=3\left(\frac{1-r}{r}\right)\zeta \,.
\end{equation}
However, if the curvaton does not decay into the baryons or CDM at all, then $\zeta_{i}=0$ and ${\cal S}_{i}=-3\zeta$, the isocurvature mode becomes three times larger than the curvature perturbations, which is ruled out by the current CMB and 
LSS data~\cite{Beltran:2004uv,Beltran:2005xd,Beltran:2005gr}, for axion isocurvature perturbations, see~\cite{Beltran:2006sq}.

In terms of a constrainable parameter, $\sqrt{\alpha}\equiv S_{i}/\zeta$, the constrains on $r$ read as:
\begin{equation}
0.98<~~r\approx1-\frac{\sqrt{\alpha}}{3}~~< 1.0\,,
\end{equation}
for $\alpha <0.0037$ at $95\%$~c.l.~\cite{Komatsu:2009kd,Komatsu:2008hk}. The fraction of energy densities, 
$r=\rho_{\varphi}/\rho_r$, also governs the non-Gaussianity generated by the curvaton. Thus the constraint 
on non-Gaussianity parameter $f_{NL}$ yields
\begin{equation}
-1.21 > f_{NL}\equiv \frac{5}{4r}>  -1.25\,,
\end{equation}
which is $1\sigma$ away from the central value of the quoted value from WMAP~\cite{Komatsu:2008hk}, which 
is $-9<f_{NL}< 111$ (at $95\%$ c.l.) In view of these bounds, it seems that the curvaton decay products cannot generate large CDM,
or baryon isocurvature perturbations, or they must not decouple from the thermal plasma created before by that of the inflaton. This 
could be achieved simply if the curvaton and the inflaton decay products excite the relevant degrees of freedom required for BBN. Otherwise, the curvaton must dominate while decaying and create all the SM or MSSM degrees of freedom. 
There are mixed curvaton-inflaton scenarios, where the curvature perturbations are contributed by both inflaton and 
curvaton~\cite{Erickcek:2009at}. In such models larger $f_{NL}$ is expected.


\subsection{Curvaton candidates}

Only handful of the curvaton candidates exist which belong to the observable sector, i.e. charged under the 
SM gauge group~\cite{Enqvist:2002rf,Postma:2002et,Enqvist:2003mr,Enqvist:2003qc,Mazumdar:2003bs,Enqvist:2004kg,Mazumdar:2004qv,BasteroGil:2003tj,Kasuya:2003va,Allahverdi:2006dr}. The SM Higgs can act as a potential curvaton, since Higgs is light compared  to any high scale of inflation, it would induce fluctuations $\sim H_{inf}/2\pi \langle h\rangle\sim 10^{-5}$, where $\langle h\rangle $ is the Higgs VEV during inflation. Validity of the SM Higgs potential beyond the electroweak scale makes it actually less attractive
candidate, besides it cannot dominate the energy density of the universe. Within MSSM, with two Higgses, it is possible to realize  a
curvaton scenario, where the inflaton energy density is dumped out of our own observable world, as a consequence the Higgses 
can dominate the energy density and create all the matter fields~\cite{Enqvist:2003qc,Enqvist:2004kg}. However this scenario will
work  {\it only} if the inflaton does not couple to the MSSM sector at all, which is very unlikely.


\subsubsection{Supersymmetric curvaton }

An important constraint arises from CMB temperature anisotropy involving the ratio of the perturbation 
and the background VEV of the curvaton, since this ratio is related to the curvature
perturbations~\cite{Enqvist:2001zp,Wands:2000dp,Lyth:2002my,Moroi:2001ct,Moroi:2002rd}. Provided the
perturbations do not damp during its evolution, strictly speaking for a quadratic potential, the final curvature 
perturbation is given by~\footnote{A detailed analysis of a curvaton scenario for a non-quadratic potential can be found 
in Refs.~\cite{Enqvist:2009eq,Enqvist:2009zf}.}:
\begin{equation}
\label{cobe}
\delta =\frac{\delta\varphi}{\varphi}= \frac{H_{inf}}{2\pi \varphi_{inf}}\sim 10^{-5}\,.
\end{equation}
For an MSSM flat direction curvaton, it is important to keep in mind that the they carry SM gauge couplings, therefore,
if the inflaton decay products create a plasma which has MSSM degrees of freedom then they would interact 
with the curvaton rendering thermal corrections to the curvaton potential inevitable. There are three issues which
have to be taken into account.


\begin{itemize}

\item{Curvaton must not have a renormalizable coupling to the inflaton, otherwise
curvaton cannot obtain large VEV during inflation. For instance, neither $H_uH_d$ nor 
$LH_u$ are good curvaton candidates, because a gauge singlet inflaton can couple to 
these flat directions through renormalizable interactions~\cite{Allahverdi:2007zz,Allahverdi:2005mz,Allahverdi:2006dr}.
The inflaton couplings to $LH_u$ or $H_uH_d$ ought to be very weak in order for them to be a curvaton candidate. }

\item{Curvaton must not induce a mass $\geq m_{\phi}/2$ to the inflaton decay 
products, otherwise, the two-body inflaton decay into MSSM quanta will be kinematically blocked.  
The inflaton decay will be delayed until the relevant flat direction has started its oscillation
and its VEV has been redshifted to sufficiently small values~\cite{Allahverdi:2007zz,Allahverdi:2005mz,Allahverdi:2006dr}. }

\item{The flat direction VEV of a curvaton must not break all of the SM gauge symmetry. 
This will affect the inflaton decay products to thermalize quickly, both kinetic
and chemical equilibrium of MSSM degrees of freedom will be delayed, and
furthermore, the curvaton oscillations will not be able to dominate the 
energy density~\cite{Allahverdi:2007zz,Allahverdi:2005mz,Allahverdi:2006dr}.}

\end{itemize}


Let us denote the flat direction superfield by $\varphi$. In the case of $u_id_jd_k$ (for $i,j,k=1,2,3,~j\neq k$) 
as a flat direction, $\varphi$ would represent the VEV. The MSSM superpotential can be rewritten as:
\begin{equation} \label{flatcoupling}
W_{\rm MSSM} \supset \lambda_1 { H}_u \varphi { \varphi}_1 + \lambda_2
{ H}_d \varphi { \varphi}_2 + \lambda_3 { L} \varphi { \varphi}_3,
\end{equation}
where ${ \varphi}_{1,2,3}$ are MSSM superfields, see Eq.~(\ref{mssm}), $L$ is denoted by $L_1, L_2$ or $L_3$, 
and $\lambda_1,~\lambda_2,~\lambda_3$ are the Yukawas. In general $m_{inf}\leq H_{inf}$, which is 
true for high scale models of inflation, and $\varphi_{inf}\sim 10^{5}H_{inf}$. 
Note that the VEV of the flat direction induces VEV dependent SUSY preserving masses to the MSSM particles,
$\sim \lambda_{1,2,3}\langle \varphi_{inf}\rangle$. Therefore, for the inflaton decay into MSSM quanta 
to be kinematically allowed, we require:
\begin{equation} \label{yukawabound1}
\lambda_{1},\lambda_2 \leq 10^{-5},
\end{equation}
if the $\underline{{ \Phi} { H}_u { H}_d}$ coupling is allowed by R-parity, and
\begin{equation} \label{yukawabound2}
\lambda_1,\lambda_3 \leq 10^{-5},
\end{equation}
if the $\underline{{ \Phi} { H}_u { L}}$ coupling is allowed, where $\Phi$  is the inflaton superfield. 

The above conditions 
considerably restrict the curvaton candidates within the MSSM, as only the first generations of (s)quarks and
(s)leptons have Yukawa coupling $\ls 10^{-5}$, while  first two generations have Yukawas $\geq 10^{-3}$.

Given these constraints an  acceptable flat directions should include {\it only} one ${ Q}$ or 
one ${u}$. The reason is that $D$- and $F$-flatness conditions for directions which
involve two or more ${ Q}$ and/or ${ u}$ require them to be of
different flavors (for details, see \cite{Gherghetta:1995dv}).  The
only flat directions with $\lambda_1 \leq 10^{-5}$ are as follows:
\vskip5pt
\noindent


\begin{itemize}

\item{$\underline{{ u} { d} { d}}$: This monomial represents
a subspace of complex dimension $6$~\cite{Gherghetta:1995dv}. $D$-flatness requires
that the two ${ d}$ are from different generations (hence at least
one of them will be from the second or third generation). This implies
that $\lambda_2 \geq 10^{-3}$, see Eq.~(\ref{flatcoupling}).  As a
consequence the two-body inflaton decay via the superpotential term; ${ \Phi} { H}_u
{ H}_d$ term will be kinematically forbidden, but $\Phi H_u L$ term will led to
the inflaton decay into MSSM degrees of freedom.
The VEV of $udd$ direction keeps $SU(2)_{\rm W}$ unbroken, so that the decay products of the inflaton with
$SU(2)_{\rm W}$ degrees of freedom can completely thermalize before the curvaton, $udd$, has a 
chance to dominate and decay.}
\vskip5pt
\item{$\underline{{ Q} { L} { d}}$: This monomial
represents a subspace of complex dimension $19$~\cite{Gherghetta:1995dv}. $F$-flatness
requires that ${ Q}$ and ${ d}$ belong to different
generations. Then, since ${ Q}$ and ${ d}$ are both coupled to
${H}_d$, Eq.~(\ref{flatcoupling}) implies that $\lambda_2 \geq
10^{-3}$.   The two-body inflaton decay via superpotential ${ \Phi} { H}_u
{ H}_d$ term will be kinematically forbidden, but the other superpotential
term is kinematically allowed, i.e. $\Phi H_uL$.  The VEV of $QLd$ directions completely 
break the $SU(2)_{\rm W} \times U(1)_{\rm Y}$, but leave a $SU(2)$ subgroup of the $SU(3)_{\rm C}$
unbroken. Therefore the associated color degrees of freedom can thermalize quickly to SM degrees of freedom, before
$QLd$ could dominate the energy density.}
\vskip5pt
\item{$\underline{{ L} { L} { e}}$~ This monomial represents
a subspace of complex dimension three~\cite{Gherghetta:1995dv}. $D$-flatness
requires that the two ${ L}$s are from different generations, while
$F$-flatness requires that ${ e}$ belongs to the third generations
(therefore all the three lepton generations will be involved).
A feasible curvaton candidate will be  ${ L}_2 { L}_3 { e}_1$ direction.
For this flat direction we have $\lambda_1 = 0$ and $\lambda_3
\sim 10^{-5}$ (see Eq.~(\ref{flatcoupling})). This implies that
two-body inflaton decay will proceed via the ${ \Phi} { H}_u
{ L}_1$ term without trouble. The flat direction VEV
breaks the electroweak symmetry $SU(2)_{\rm W} \times
U(1)_{\rm Y}$, while not affecting $SU(3)_{\rm C}$.}

\vskip5pt
\item{$\underline{{ L} { L} { d} { d} { d}}$: This
monomial represents a subspace of complex dimension
three~\cite{Gherghetta:1995dv}. $D$-flatness requires that the two ${ L}$s and
the three ${d}$s are all from different generations. This implies
that that $\lambda_2 \simeq 10^{-2}$ and $\lambda_1 = \lambda_3 = 0$
(see Eq.~(\ref{flatcoupling})). This direction breaks all of the SM gauge group. 
This results in late thermalization of the Universe~\cite{Allahverdi:2005mz,Allahverdi:2005fq} 
and the absence of thermal effects which, as a consequence, does not yield early oscillations of the flat
direction. }
\vskip5pt
\end{itemize}

Thus within MSSM the potential curvaton candidates could be
the ${u}_1 {d_2} {d_3}$ and ${Q}_1 L_3{d_2}$ (and possibly 
${L}_2 {L}_3 {e}_1$) flat directions. Because in all these two cases the 
inflaton can decay and produce a thermal bath, which will enable the curvaton
oscillations to dominate early during its oscillations, and the curvaton can receive finite temperature
thermal corrections. Without the latter the curvaton can never lead to dominate the energy density 
of the universe while decaying.


\subsubsection{The $A$-term curvaton and a false vacuum}
\label{CPA}
Let us consider an MSSM flat direction potential~\footnote{Similar potential also arises in Refs.~\cite{Chun:2004gx,Dimopoulos:2005bp}, where
the curvaton is a QCD axion in SUSY, and in~\cite{Dimopoulos:2002kt} where the curvaton is treated as a SM gauge singlet.}:
\begin{equation} \label{nonren}
V = m^2_0 {\vert \varphi \vert}^2 +
\lambda^2_{n} \frac{\vert \varphi \vert^{2(n-1)}} {M_P^{2(n-3)}}+
\left(A\lambda_{n}\frac{\varphi^{n}}{M_P^{n-3}}+h.c.\right)\,,
\end{equation}
where $\lambda_n\sim {\cal O}(1)$, $n \geq 4$, and $A \sim
m_0\sim {\cal O}(100~{\rm GeV}-1~{\rm TeV})$, and depends on a
phase. Note that we have not added the Hubble-induced corrections
to the mass and the A-term. They can be avoided for a simple choice of 
no-scale type K\"ahler potential, either motivated from R-symmetry~\cite{Dine:1995kz,Dine:1995uk}, shift symmetry or 
Heisenberg symmetry~\cite{Gaillard:1995az,Campbell:1998yi}. 

The radial and angular direction of the potential remains flat during
inflation, therefore they obtain random fluctuations. There will be equally
populated domains of Hubble patches, where the phase of the
$A$-term is either positive or negative. In either case, during inflation,
the flat direction VEV is given by:
\begin{equation}
\label{FVEV}
\varphi_{inf} \sim \left(m_0 M_P^{n-3}\right)^{1/n-2}\,.
\end{equation}
However there is a distinction between a positive and a negative phase
of the $A$-terms. The difference in dynamics arises after the end of
inflation. In the case of a positive $A$-term the flat direction starts
rolling immediately, but in the case of a negative $A$-term, the flat
direction remains in a {\bf false vacuum} at a  VEV which is given by
Eq.~(\ref{FVEV}). 

The mass of the flat direction around this false
minimum is very small compared to the Hubble expansion rate during
inflation, i.e. $(3n^2-9n+8)m_{0}^2\ll H_{inf}^2$ where $n>3$.  During
inflation the flat direction obtains {\it quantum fluctuations} whose
amplitude is given by Eq.~({\ref{cobe}). The flat direction can exit 
such a metastable minimum only if {\it thermal corrections} are taken into account.


\subsubsection{Thermal corrections to the curvaton}

\label{TCF}

The flat direction VEV naturally induces a mass $\sim\lambda \varphi_{inf}$ to the
field which are coupled to it, where $\lambda$ is a gauge or Yukawa
coupling. Note that $\varphi_{inf}$ is the initial VEV of the curvaton, after 
the end of inflation slides down gradually.  If there is a thermal bath with 
a temperature $T$ then there is a thermal corrections to the flat direction 
depending on whether $\lambda\varphi_{inf} \leq T$ or  $\lambda \varphi_{inf} > T$, 
different situations arise.
\vskip5pt
\noindent
\begin{itemize}
\item{$\underline{\lambda \varphi_{inf} \leq T}$:\\ 
Fields in the plasma which have a
 mass smaller than temperature are kinematically accessible to the
thermal bath. They will reach full equilibrium and result in a thermal
correction $V_{\rm th}$ to the flat direction potential
\begin{equation} \label{effmass1}
V_{\rm th} \sim + \lambda^2 T^2 {\vert \varphi \vert}^2 .
\end{equation}
The flat direction then starts oscillating, provided that $\lambda T > H$
\cite{Allahverdi:2006dr,Allahverdi:2007zz,Allahverdi:2005mz,Allahverdi:2005fq}. }

\vskip 5pt
\noindent
\item{$\underline{\lambda \varphi_{inf} > T}$:\\ 
Fields which have a mass larger
than temperature will not be in equilibrium with the thermal bath. For
this reason they are also decoupled from the running of gauge
couplings (at finite temperature). This shows up as a correction to
the free energy of gauge fields, which is equivalent to a logarithmic
correction to the flat direction potential~\cite{Allahverdi:2006dr}:
\begin{equation} \label{effmass2}
V_{\rm th} \sim  \pm \alpha ~ T^4 \ln \left({\vert \varphi \vert}^2\right)~,
\end{equation}
where $\alpha$ is a gauge fine structure constant. Decoupling of gauge
fields (and gauginos) results in a positive correction, while
decoupling of matter fields (and their superpartners) results in a
negative sign. The overall sign then depends on the relative
contribution of decoupled fields. As an example, let us consider the
${ H}_u { H}_d$ flat direction. This direction induces large
masses for the top (s)quarks which decouples them from the thermal
bath, while not affecting gluons and gluinos.  Therefore this leads to
a positive contribution from the free energy of the gluons.

Obviously only corrections with a positive sign can lead to flat
direction oscillations around the origin. Oscillations
begin when the second derivative of the potential exceeds 
the Hubble rate-squared which, from Eq.~(\ref{effmass2}), reads 
$\left(\alpha T^2/\varphi_{inf}\right) > H$.}

\end{itemize}


Note that thermal effects of the first type require that fields which
have Yukawa couplings to the flat direction are in full equilibrium,
while those of the second type require the gauge fields (and gauginos)
be in full equilibrium.  There are two main possibilities:

\begin{itemize}
\item
$\underline {{ u}_1 { d_2} { d_3}}$: $SU(2)_{\rm W}$ remains
unbroken in this case. This implies that the cor\-res\-pon\-ding gauge
fields and gauginos, $H_u$ and $H_d$ (plus the Higgsinos) and the LH
(s)leptons reach full thermal equilibrium. The back-reaction of $H_u$
results in a thermal correction $\lambda^2_1 T^2 {\vert \varphi
\vert}^2 $, see Eq.~(\ref{flatcoupling}), with $\lambda_1 \sim
10^{-5}$. The free energy of the $SU(2)_{\rm W}$ gauge fields result
in a thermal correction $\sim + \alpha_{\rm W} T^4 {\rm ln}
\left({\vert \varphi \vert}^2 \right)$. Note that the sign is positive
since the flat direction induces a mass which is $> T$ (through the
$d$) for the LH (s)quarks but not the $SU(2)_{\rm W}$ gauge fields and
gauginos.

\item
$\underline {{ Q}_1 { L} { d}}$: An $SU(2)$ subgroup
of $SU(3)_{\rm C}$ is unbroken in this case. Therefore only the
corresponding gauge fields and gauginos plus some of the (s)quark
fields reach full thermal equilibrium. Then the back-reaction of ${
u}_1$ and ${ d}_1$ results in a thermal correction $\left(
\lambda^2_1 + \lambda^2_2\right) T^2 {\vert \varphi \vert}^2$
according to Eq.(\ref{flatcoupling}), where $\lambda_1 \sim \lambda_2
\sim 10^{-5}$.  Note that  decoupling of a number of gluons (and gluinos)
from the running of strong gauge coupling results in a negative
contribution to the free energy of the unbroken part of $SU(3)_{\rm C}$.
\end{itemize}

Therefore, within MSSM and from the point of view of thermal effects,
the ${ u}_1 { d_2} { d_3}$ flat direction is the most suitable
curvaton candidate. The  ${ L}_2 {L}_3 { e}_1$ flat direction can obtain a large 
VEV . The $SU(3)_{\rm C}$ part of the SM gauge symmetry remains 
unbroken for this flat direction, and hence gluons, gluinos and (s)quarks 
will reach full equilibrium. We note that neither of ${ L}$ and ${ e}$ are coupled to the color
degrees of freedom. This implies that there will be no $T^2 \varphi^2$
or logarithmic correction to the flat direction potential in this
case. This excludes the ${ L}_2 { L}_3 { e}_1$ flat direction
from being a successful curvaton candidate.


\subsubsection{${u_1d_2d_3}$ as the MSSM curvaton}
\label{udd}

From the discussion in the previous section, it follows that thermal effects
will lead to a following potential:
\begin{equation}
\label{thermeff}
V_{\rm th} \sim \lambda^2_1 T^2 {\vert \varphi \vert}^2 +
\alpha_{\rm W} T^4 {\rm ln} \left({\vert \varphi \vert}^2\right),
\end{equation}
where $\lambda_1 \sim 10^{-5}$ and $\alpha_{\rm W} \sim 10^{-2}$.
Typically at  initial (larger) temperatures the first term dominates and at 
later (lower) temperatures the second term dominates. Note that the 
curvaton is in a metastable vacuum, only the temperature corrections can lift the
curvaton , therefore  we need $V_{\rm th} > m^2_0 \varphi^2_{inf}$, so that
the thermal effects will overcome the potential barrier. This yields:
\begin{equation} \label{cond1}
\alpha_{\rm W} T^4 > m^2_0 \varphi^2_{inf}.
\end{equation}
The thermal mass should be sufficient to trigger curvaton oscillations, otherwise,
the curvaton will remain in a metastable vacuum.
In order for the flat direction oscillations to start we must have
$d^2V_{\rm th}/d\vert \varphi \vert^2 > H^2$. This leads to:
\begin{equation} \label{cond2}
\alpha^{1/2}_{\rm W} \frac{T^2} {\varphi_{inf}} >H(T)\,,
\end{equation}
which always holds in a radiation-dominated phase where $H \simeq
T^2/M_{\rm P}$ (note that $\varphi_{inf} \ll M_{\rm P}$). This implies
that the ${ u}_1 { d_2} { d_3}$ direction starts oscillating once
the condition given in Eq.~(\ref{cond1}) is satisfied. This happens,
 when temperature of the Universe is given by
\begin{equation}
\label{tosc}
T_{osc} \sim \left(\frac{\varphi_{inf}} {10^{14}~{\rm GeV}}\right)^{1/2} \times
10^9~{\rm GeV}.
\end{equation}
Although, its a very  high temperature, the flat direction curvaton does not evaporate
due to the presence of a  thermal bath. Thermal scatterings are governed by 
$\lambda_1^2\varphi^2 \chi^2$, with $\varphi$ being the flat direction and, $\chi$
collectively denotes the fields in thermal equilibrium. The rate for
evaporation is proportional to $\lambda_1^4\sim (10^{-5})^4$, much feeble to
destroy the flat direction by evaporation.

The flat direction curvaton, $u_1d_2d_3$, is lifted by $n=6$
superpotential level, therefore $\varphi_{inf} \sim 3 \times 10^{14}$
GeV, see Eq.~(\ref{FVEV}).  In order to be a curvaton, the perturbations
should match the observed limit, therefore $H_{inf}\sim 2\pi\varphi_{inf}\times 10^{-5}\approx 10^{10}$~GeV.

The final reheat temperature is determined by the radial oscillations of the curvaton.
The curvaton oscillations dominate the energy density as the ambient temperature of the plasma 
redshifts below the $T_{osc}$ very rapidly due to the expansion of the universe. This is the {\it only}
example where the finite temperature effects can render the curvaton oscillations dominating while 
decaying~\cite{Allahverdi:2006dr}. The decay of $u_1d_2d_3$ curvaton will take place  via instant 
preheating discussed in section~\ref{RTCSB}.
The decay happens via SM gauge and Yukawa couplings and the decay products are the MSSM degrees of freedom.
There is no residual isocurvature fluctuations and the LSP dark matter can be created from the thermal bath.

Since $u_1d_2d_3$ as a curvaton dominates the energy density while decaying, there will be no 
significant non-Gaussianity. However as we shall show in the next section, there are MSSM flat directions
which do not dominate while decaying, they have a possibility to generate large non-Gaussianity, 
see for example~\cite{Riotto:2008gs,McDonald:2003jk}, where the
imaginary part of the MSSM flat direction curvaton is responsible for generating the perturbations. In these models 
the treatment for the inflaton decay products is not quite complete, it is not clear in which sector the inflaton decays.
Thermal corrections are important for the MSSM flat direction potential as we have seen in the above discussion.


\subsubsection{Models without A-term}

There are many viable candidates of curvaton within MSSM which do not use $A$-term.
These are flat directions which are lifted by hybrid operators, such as $W\sim \Phi^{n-1}\Psi/M^{n-3}$,
for such operators the $A$-term becomes dynamically negligible. A generic curvaton potential 
is given by~\cite{Enqvist:2002rf,Enqvist:2003mr,Enqvist:2003qc,Enqvist:2004kg}:
\begin{equation}\label{curv-pot-without-A}
V(\varphi)=m_{0}^2\varphi^2+\lambda^2_{n} \frac{\vert \varphi \vert^{2(n-1)}} {M_{\rm P}^{2(n-3)}}\,.
\end{equation}
The flat direction candidate, i.e $QuQuQue$ (lifted by $QuQuQuH_dee)$ by the superpotential 
term $n=9$ is the only viable term which can dominate the energy density of the universe at the time
of decay~\cite{Enqvist:2002rf,Enqvist:2003mr}. Note that the superpotential term which lifts the direction is a 
hybrid type and therefore the $A$-term vanishes, see the discussion after Eq.~(\ref{doesnotcontri}). In 
these models, it was assumed that prior to the curvaton 
domination the inflaton decays primarily into the hidden sector radiation, which is not a satisfactory assumption.

For $n=7$ the flat direction is $LLddd$ (lifted by $H_uLLLddd$), and for $n=6$ the flat direction $QdL$ 
(lifted by $QdLudd$) are allowed to be a curvaton candidate, provided the background equation of 
state is that of a stiff fluid at the time of curvaton oscillations. The flat directions which are lifted by $n=4$ or 
by $n=3$ can never dominate the energy density of the universe. This is due to the fact that their initial amplitude
of the oscillations is very low.

There is of course a way to cure flat directions which are lifted by $n=4$ by the renormalizable superpotential,
such as $H_uH_d$ (lifted by $(H_uH_d)^2$), and for $n=3$, the flat direction $NH_uL$, in the extension of the SM gauge group by 
$U(1)_{B-L}$, where $N$ is the right handed neutrino. They can dominate the energy density of the 
universe during oscillations, provided  the inflaton energy never gets dumped into the observable world~\cite{Enqvist:2004kg}. 

To summarize, when the curvaton does not dominate while decaying, then they can generate significant non-Gaussian
perturbations and isocurvature fluctuations, however, the results are model dependent, as it depends 
on the nature of the inflaton decay products and the choice of the flat direction.


\subsubsection{Curvaton and non-Gaussianity}
\label{See: Sec.CurvatonandNG}

One of the prime observable signature of a curvaton mechanism is to generate non-Gaussian perturbations of order 
$f_{NL}\sim {\cal O}(10-100)$.
However as we have seen earlier, in most of the models of inflaton and curvaton, the non-Gaussian perturbations 
are constrained by the isocurvature perturbations. Models where the curvaton dominates while decaying will
generate no large non-Gaussianity. For a realistic $A$-term curvaton model, $u_1d_2d_3$ will never generate 
large non-Gaussianity. The flat direction decays in couple of oscillations and its energy density dominate while decaying.

On the other hand, models without $A$-term curvaton  can potentially generate large non-Gaussianity, as many 
of the directions, $n=3, 4, 6, 7$, do not seem to dominate the energy density 
while decaying. There are however non-trivial constraints on such models in order to avoid large isocurvature perturbations.

\begin{itemize}

\item{Inflaton decay products must create the MSSM degrees of freedom. 
There should not be any trace of hidden degrees of freedom which
do not thermalize with that of the MSSM.  Otherwise the inflaton perturbations must be sub-dominant, i.e. 
$(\delta\rho/\rho)_{inf}\ll 10^{-5}$ during inflation~\cite{Enqvist:2009zf,Enqvist:2009eq}.}

\item{Provided the curvaton perturbations and the inflaton perturbations are of the same order, the curvaton must 
not create {\it large}  baryon asymmetry and/or dark matter through its decay.  In principle an 
{\it absolute gauge singlet} moduli curvaton can decay into gravitino or axino LSP. In such a case a stringent constraint on reheat temperature applies~\cite{Kawasaki:2004yh,Dine:2006ii,Endo:2006tf,Nakamura:2006uc,Kohri:2009ka}.}

\end{itemize}

In presence of non-renormalizable potential~\cite{Sasaki:2006kq,Enqvist:2008gk}, the $f_{NL}$ and 
$g_{NL}$ parameters  are given by:
\begin{eqnarray}
\label{modeqfNL}
f_{NL}&=&\frac{5}{4r}\left(1+\frac{\varphi_{0}\varphi_0^{\prime\prime}}{\varphi^{\prime 2}_0}\right)-
\frac{5}{3}-\frac{5r}{6}\,. \\
g_{NL} &=& \frac{25}{54}
\left[\frac{9}{4r^2}  \left(\frac{\varphi_{0}^2 \varphi_{0}^{\prime\prime\prime}}
{\varphi_{0}^{\prime 3}}+
\frac{3\varphi_{0} \varphi_{0}^{\prime\prime}}{\varphi_{0}^{\prime 2}}\right)
-\frac{9}{r}\left(1+\frac{\varphi_{0} \varphi_{0}^{\prime\prime}}{\varphi_{0}^{\prime 2}}
\right)+\frac{1}{2} \left(1-
\frac{9\varphi_{0} \varphi_{0}^{\prime\prime}}{\varphi_{0}^{\prime 2}}
\right)+10r + 3r^2\right]\,, \nonumber\\
\end{eqnarray}
where $\varphi_0$ is the initial amplitude of the curvaton field. For $r<< 1$, which is the ratio of curvaton and ambient 
energy densities, the above equations simplify to:
\begin{equation}
f_{NL} \simeq \frac{5}{4r} \left(1 +
\frac{\varphi_{0} \varphi_{0}^{\prime\prime}}{\varphi_{0}^{\prime 2}}\right)\,,~~~~~
g_{NL} \simeq \frac{25}{54}\left[
\frac{9}{4r^2}  \left(\frac{\varphi_{0}^2 \varphi_{0}^{\prime\prime\prime}}
{\varphi_{0}^{\prime 3}}+
3\frac{\varphi_{0} \varphi_{0}^{\prime\prime}}{\varphi_{0}^{\prime 2}}
\right)\right].
\end{equation}
For a quadratic potential, $f_{NL}\sim (5/4r)$ and $g_{NL}\sim -(10/3r)$, holds true and 
$g_{NL}\simeq -(10/3)f_{NL}$. In Ref.~\cite{Enqvist:2008gk,Enqvist:2009eq}, the authors have studied 
the departure from quadratic potential for a curvaton and obtained, $g_{NL}\sim {\cal O}(-10^4)-{\cal O}(-10^5)$
for $r\sim 0.01$. These values also depend on the non-renormalizable operator, $n$.  The constraints on isocurvature 
perturbations are extremely model dependent, as it depends on details of thermal history of the universe and the 
assumptions behind inflaton and curvaton sectors.


\subsection{Inhomogeneous reheating scenarios}

Inhomogeneous reheating scenarios were considered in Refs.~\cite{Kofman:2003nx,Dvali:2003em,Dvali:2003ar,Enqvist:2003uk,Mazumdar:2003va}. This idea is similar to the curvaton paradigm, the main difference here is that the isocurvature perturbations are now converted into curvature perturbations during the inflaton decay. Here we present a simple example within MSSM. For a SM gauge singlet 
inflaton, $\Phi$, the only renormalizable coupling to the MSSM is either 
$W\sim g \Phi H_uH_d$, or $g\Phi LH_u$ where $g\leq {\cal O}(1)$. There are other non-renormalizable couplings which appear in
the combinations with three superfields are~\cite{Allahverdi:2007zz}:
\begin{equation}
W\supset \frac{\Phi}{M_*} H_uQu+\frac{\Phi}{M_{*}}H_dQd +\frac{\Phi}{M_*}H_dLe + \frac{\Phi}{M_*}QLd +
\frac{\Phi}{M_*}udd+\frac{\Phi}{M_*}LLe\,.
\end{equation}
Where $M_*$ is a cut-off scale. Since $\Phi$ has a large VEV during inflation, therefore its direct 
couplings to $H_uH_d$ or $LH_u$ for a reasonable
range of coupling strength $g$ will render them heavy compared to the Hubble expansion rate, $g\langle \phi\rangle \geq H_{inf}$.
These fields will not obtain large quantum fluctuations, and will dynamically stay close to their respective minimum, 
$\langle H_u \rangle , \langle H_d\rangle, \langle L\rangle \approx 0$. Note here that it is assumed that inflation 
is driven by the large VEV. Therefore, fields coupled to either $H_u,~H_d$ or $L$ will remain massless by virtue of their zero VEV, and subject to large quantum fluctuations of order $H_{inf}/2\pi$ during inflation. It is possible to convert these fluctuations when the inflaton decays to the MSSM quanta 
at the time of reheating.

The effective coupling for the inflaton to decay either via $H_uH_d$ or $LH_u$ is given by:
\begin{equation}
g=g\left(1+\frac{\langle S\rangle}{M_{*}}+... \right)\,,
\end{equation}
where $S$ is the VEV of the field which couples to either $H_u,~H_d,~L$, 
then the inflaton decay will generate a quasi-thermal bath. The initial
decay width of the inflaton will generate a plasma with a temperature $T \propto \Gamma_{d}^{1/2}\propto g$.
Therefore, fluctuations in temperature induced by $\langle S\rangle \sim H_{inf}/2\pi$, would lead to
\begin{equation}
\frac{\delta T}{T}\sim \left. \frac{H_{inf}}{2\pi S}\right |_{decay}\,,~~~~~~~~{\rm or}~~~~~~~~\frac{\delta T}{T}\sim 
\left.\frac{H_{inf}}{2\pi M_*}\right |_{decay}\,,
\end{equation}
depending on whether the inflaton decays via non-renormalizable or renormalizable operators. In order to match the 
seed perturbations for CMB, $\delta T/T\sim 10^{-5}$, either the VEV of $S$ or the scale of new physics should 
be around $10^{5}H_{inf}$.

\section{String theory models of inflation}\label{sec:stringstheorymodels}
%
%

One of the best motivated framework of quantum gravity is the string theory. Therefore it is
natural to seek whether string theory can shed some light on inflation. There are many reviews dedicated to stringy 
inflation~\cite{Linde:2005dd,Kallosh:2007ig,HenryTye:2006uv,Burgess:2007pz,Cline:2006hu,Kachru:2003sx,McAllister:2007bg,Quevedo:2002xw,Baumann:2009ds,Baumann:2009ni}.  Since, there are many
models of inflation with large VEVs close to the Planck scale, which are particularly sensitive to the 
UV properties of the field theory, it is possible that string theory can provide some insight into the shape and stability of the potential. String theory also involves many degrees of freedom with equally large number of {\it physical} solutions, which makes it
vulnerable when it comes to making concrete predictions, such as selecting the right vacuum at low energies. It is hoped that cosmological 
observations along with particle phenomenology beyond the SM would help us constructing inflationary models~\footnote{With the discovery of branes~\cite{Polchinski:1995mt,Polchinski:1996na}, string theory has influenced string phenomenology and cosmology in a radical way, see~\cite{Quevedo:2002xw,Burgess:2007pz}. As a consequence, not all interactions see the same number of space-time dimensions. For instance, gravitons are allowed to propagate in the entire space-time dimensions, while the gauge fields including photons are localized on the brane position where the open strings must end~\cite{Polchinski:1995mt,Polchinski:1996na}. This has lead to a number of paradigms such as; {\bf a low string scale}: the string scale could be as low as TeV, the large extra dimensions could be as large as the micron scale~\cite{ArkaniHamed:1998rs,Antoniadis:1998ig,ArkaniHamed:1998nn,Antoniadis:1993jp}, or at intermediate scales~\cite{Burgess:1998px}, and  {\bf cosmological constant problem}: a possible solution emerges where it is possible to decouple $4d$ vacuum energy  from the $4d$ curvature~\cite{ArkaniHamed:2002fu,Carroll:2003db,Aghababaie:2003wz}.}. 

There are important avenues where string theory can actually help us in our understanding the inflation dynamics and various cosmological consequences.

\begin{itemize}

\item{UV completion:\\
The UV completion of an inflationary potential is related to the fact that whether the potential can be 
kept flat enough during the interval of slow-roll inflation or not at VEVs close to the cut-off of the theory. 
For instance, the slow-roll parameters, i.e. $\epsilon\leq 1, \eta\leq 1$, ought to be maintained during 
sufficiently large e-foldings of slow-roll inflation.  One particular example is the realization of chaotic inflation~\cite{Linde:1983gd} 
in string theory, but below the Planckian VEV, with the help of {\it assisted inflation}~\cite{Liddle:1998jc}. In string theory 
it is possible to realize $N$ number of string axions arising from partners of K\"ahler moduli, which can collectively drive 
inflation below the $4d$ Planck scale~\cite{Kanti:1999ie}. These axions have shift symmetry from $10d$ gauge invariance, 
which is broken non-perturbatively, and avoids the SUGRA $\eta $ problem, thus keeping the potential
sufficiently flat enough for the success of inflation~\cite{Dimopoulos:2005ac,Easther:2005zr}.   }

\item{Initial conditions:\\
String theory provides multiple vacua with all possible values of the cosmological 
constant~\cite{Bousso:2000xa,Susskind:2003kw,Denef:2004ze,Polchinski:2006gy}, for 
reviews, see~\cite{Douglas:2006es,Douglas:2004zg}.  There has been a new revelation in string theory after the advent of a  KKLT scenario~\cite{Kachru:2003aw}, where the moduli are stabilized in an 
anti de Sitter (AdS) space with a negative cosmological constant. The initial configuration is stabilized 
with the help of various flux vacua~\cite{Giddings:2001yu}. Various non-perturbative effects such as gaugino condensation 
and/or warped brane instantons lift the vacuum from AdS to de Sitter (dS) with a meta stable minimum with a life time greater than the present universe. These numerous vacua eliminate the problem of initial conditions for inflation~\cite{Susskind:2003kw,Clifton:2007en,Linde:2006nw}. Our patch of the universe could emerge from one 
such initial vacuum with a large cosmological constant. It is also quite  plausible that string theory would provide some hints on 
the nature of trans-Planckian problem~\cite{Martin:2000xs,Brandenberger:2000wr,Martin:2003kp,Kaloper:2002cs,Danielsson:2002kx}.
}

\item{Low primordial tensor to scalar ratio:\\
One of the bold predictions of string theory models of inflation is that the tensor to scalar ratio will be generically
small, i.e. $r< 0.001$. In order to obtain large tensor to scalar ratio, i.e. $r\sim 0.1$, one requires field values to 
be in the range, where $\Delta \phi\gg M_{\rm P}$~\cite{Lyth:1996im}.  In a stringy setup the scale of 
inflation is {\it always} below the $4d$ Planck scale, $M_{\rm P}$. Note that the {\it chaotic type inflation} is now driven with the help of {\it assisted inflation}~\cite{Dimopoulos:2005ac,Easther:2005zr}, where the largest tensor to scalar ratio could be detectable by the future experiments if $r<  0.13$. Similar arguments hold for brane inflation models, where the brane-anti-brane 
separation acts as an inflaton, but the tensor to scalar ratio remains quite small~\cite{Baumann:2006cd,Kallosh:2007cc}.}

\item{Cosmic (super)strings, localized gravity waves:\\
In brane-anti-brane case, inflation ends via annihilation of the branes. This process is likely to
generate cosmic strings~\cite{Sarangi:2002yt,Jones:2003da,Copeland:2003bj,Firouzjahi:2005dh}, see Sec.~\ref{See: Formation of cosmic (super)strings after brane inflation}. 
However their longevity is a model dependent issue~\cite{Copeland:2003bj}. Brane-anti-brane annihilation
would also lead to exciting gravity waves on sub-Hubble scales with a peak frequency governed by the string 
scale and a distinct sharp spectrum~\cite{Mazumdar:2008up}. If the string scale is close to TeV, then there is a possibility of detecting them in future gravitational wave observatories. }

\end{itemize}

Stringy models of inflation also bring various challenges, whose roots are tied to the origin of particle phenomenology.
One of the issues is pertaining to exciting the SM degrees of freedom, baryons and cold dark matter. Reheating and 
thermalization of the SM degrees of freedom in stringy models of inflation are poorly 
understood~\cite{Barnaby:2004gg,Frey:2005jk,Kofman:2005yz,Chen:2006ni,Langfelder:2006vd}. 


\begin{itemize}

\item{Embedding MSSM in a stringy setup:\\
There are tremendous progress in embedding MSSM in a stringy setup, with the help of intersecting 
branes in Type IIA/B theories. Typically these theories have quantized fluxes in presence of sources which
lead to a warped geometry. Embedding MSSM in a realistic warped geometry is not well understood yet.  Most of the 
constructions are done in a simple background geometry~\cite{Aldazabal:2000sa,Aldazabal:2000cn,Aldazabal:2000dg,Cvetic:2001nr,Cvetic:2001tj,Camara:2004jj, Blumenhagen:2006ci,Blumenhagen:2009gk}. Furthermore, besides the SM gauge group there are extra $U(1)$'s which appear in the spectrum, which can be advantageous for cosmology, such as a natural embedding for neutrino masses, leptogenesis, for a review see~\cite{Quevedo:2002xw}, or can bring numerous uncertainties in thermal history of the universe depending on what scales they are broken.}

\item{Inflationary sector:\\
It has been proven hard to embed inflationary sector within an observational sector, in most of the examples 
inflation and (MS)SM sectors are two distinct sectors. For all practical purposes inflaton remains in a hidden sector, 
whose couplings to the SM world remains unknown apriori.  In a warped geometry, with multi-throats it is not clear 
why only the SM or MSSM throat will be dominantly excited after the end of inflation, as required for the success of BBN. 
Furthermore, it has been argued that during inflation the effect of back reaction would alter the compactified geometry and all the throats below the inflationary scale~\cite{Frey:2005jk}.  }

\item{Graceful exit from a string landscape:\\
In a string landscape scenario, if the universe were to tunnel out of the false vacuum, then the universe
would be devoid of any entropy as the nucleated bubble would keep expanding forever with a negative spatial
curvature~\cite{Allahverdi:2007wh,Bousso:2000xa,Polchinski:2006gy}. Such a universe would have no place in a real world. Therefore, it is important that a bubble with an MSSM like vacuum must undergo the
last phase of slow-roll inflation, in order to gracefully exit the string landscape. Inflationary epoch should explain
the observed temperature anisotropy and also the right degrees of freedom upon reheating~\cite{Allahverdi:2007wh}.

}

\end{itemize}

\subsection{Moduli driven inflation}
\label{MDI}

Moduli are abundantly in large numbers in $4d$ effective theory, which can be used to construct inflation 
models. The potential
along the moduli remains massless in a SUSY limit. However, SUSY breaking, 
non-renormalizable superpotential terms, along with non-perturbative effects lift their flatness. These corrections are 
important to compute in order to keep the inflationary potential sufficiently flat. Some of these corrections
are understood recently in the context of a KKLT scenario~\cite{Kachru:2003aw}, which helps understanding the stabilization of all the complex structure moduli with the help of flux compactification~\cite{Giddings:2001yu}, leaving a single volume (K\"ahler) modulus which can be stabilized via non-perturbative effects.


\subsubsection{Basic setup}
\label{BS}

The basic setup for any string theory is the gravitational action in $10d$, which can be reduced to an effective $4d$
with a metric ansatz:
\begin{equation}\label{ansatz}
ds^2=e^{2A(x^{m})}\eta_{mu\nu}dx^{\mu}dx^{\nu}+g_{mn}dx^{m}dx^{n}\,,
\end{equation}
where $\mu,\nu$ runs for the non-compact space-time, $4d$, while $m,n$ runs for the compact $6d$.
Demanding that in $4d$ we arrive at  $N=1$ SUSY, $g_{mn}$ is required to be 
{\it Calabi Yau} manifold~\cite{Polchinski:1998rq,Green:1987sp,Green:1987mn}, a 
Ricci-flat metric with $SU(3)$ holonomy. The compactification scale of $6d$ is given by; $M_c\ll M_s$,
where $M_c$ is the compactification scale and $M_s$ is the string scale. The gravitational part of the action is given by:
\begin{eqnarray}
S&=& \frac{1}{2\kappa_{10}^2} \int d^{10}x \sqrt{-g_{10}}e^{-2\phi}R_{10}
= \frac{1}{2\kappa_{10}^2} \int d^{10}x \sqrt{-g_4}
e^{2A}e^{6u} e^{-2\phi}R_{4}\ \ \label{stringframe}
\end{eqnarray}
where $e^{6u}$ is the six dimensional volume of the compact metric and where 
$2\kappa_{10}^2 = (2\pi)^7\alpha^{\prime 4}$~\cite{Polchinski:1998rq}).  
Eq.~(\ref{stringframe}) does not give the usual Einstein-Hilbert term
for $g_{\mu\nu}$; therefore, a following transformation;
$g_{\mu\nu}= e^{6u} e^{-2\phi}g_{\mu\nu}$, leads to decouple the dilaton and the 
overall volume of the compactification. The string coupling is determined by $g_s=e^{\phi}$.
Then the effective action is the usual
\begin{equation}
\label{planck2}
S=\frac{1}{2\kappa_4^2}\int d^4x\sqrt{-g_E} R_E\ , \qquad \qquad
\frac{1}{\kappa_4^2} = \frac{V_6}{\kappa_{10}^2}\ ,\qquad  \qquad 
V_6 = \int d^6x \sqrt{\det g_{mn}}\, e^{2A}e^{-6u}\ .
\end{equation}
Note that now there can be a large hierarchy between 10d and 4d Planck 
constants, due to a large warp factor, $\propto e^A$, see~\cite{Randall:1999ee,Randall:1999vf, DeWolfe:2002nn,Frey:2003tf}.

For the rest of the discussion, an important ingredient will be  the $10d$ type IIB SUGRA, 
which arises as a low energy limit of type IIB string theory. The fields of IIB SUGRA are the metric, a complex
scalar, two 3-form field strengths, and a self-dual 5-form field strength.
The dilaton-axion scalar, $\tau=C_0+ie^{-\phi}$, combines the RR scalar with
the string coupling. It is  convenient to combine the R-R ($F_3=dC_2$) and NS-NS 
($H_3=dB_2$) 3-forms into a single complex field, $F_3=F_3-CH_3$. The 5-form, 
$\tilde F_5 = dC_4 -C_2 H_3$,  is neutral under self-duality, $\star \tilde F_5=\tilde F_5$, which 
is imposed as an equation of motion~\cite{Polchinski:1998rq}, for a review see~\cite{Grana:2005jc,Frey:2003tf}.

There are other non-perturbative stringy objects such as $D$ branes, a $Dp$ brane is an extended 
hypersurface over $p$ flat spatial dimensions in $p+1$ dimensional space-time, where open strings 
are free to move~\cite{Polchinski:1995mt,Polchinski:1996na}. Open strings do not propagate in the bulk, while the closed strings do.
In type IIB, the allowed number of p-branes are  $p=1,3,5,7,9$. The two $D$-branes interact by 
exchanging gravitons, dilatons and antisymmetric tensors, provided their separation is large compared to 
the string scale.

In a compact manifold the $D$-branes can modify the dynamics. They do so via gravitational 
field they create, which gives rise to the warping of the metric as shown in Eq.~(\ref{ansatz}). The presence 
of anti-symmetric tensor fields, for which the branes act as sources also lead to quantization of
three-form fluxes similar to the argument for monopoles~\cite{Giddings:2001yu}, i.e. 
\begin{equation}
\frac{1}{2\pi\alpha^{\prime}}\int_c H_3 \propto n_1,\,~\qquad \qquad \qquad 
\frac{1}{2\pi\alpha^{\prime}}\int_c F_3 \propto n_2\,,
\end{equation}
where $C$ is a 3-cycle, $n_1,~n_2$ are integers. The presence of fluxes induces scalar 
potential to the moduli, which can be used for fixing them. In this respect fluxes can break
some or all the residual SUSY in $4d$. Finally the branes can also wrap around 
a non-contractable surface, in particularly $D7$ branes, which fill 7 spatial dimensions also
extends into $4$ of the compact $6$ dimensions can wrap around 4-cycle in the internal dimensions.
The brane tension provides potential for the moduli, depending on how many times it wraps. 
There are also negative tension branes, known as orientifold planes, i.e. $O7$, required
to cancel to total charge in the internal dimensions created by the warping of $D7$ branes.
All these contributions lead to fix all the moduli, which are known as complex structure 
moduli, within type IIB compactifications~\cite{Giddings:2001yu}.


\subsubsection{KKLT scenario}
\label{KS}

In KKLT,  flux compactification lead to fixing all the moduli with a constant superpotential term, $W_0$, while
the volume modulus, $T$, is assumed to be fixed by the non-perturbative superpotential term in $4d$~\cite{Kachru:2003aw}:
\begin{equation}
W(T)=W_0+Ae^{-aT}\,,~~~~~K=-3\ln(T+\overline T)\,.
\end{equation}
where $a=2\pi/N,~A$ are constants, $K$ is the K\"ahler potential, which is assumed to be given at the tree level
arising from the compactification volume, $K=-2\ln(M_s^6V_6)$, where $M_s^6V_6\propto (T+\overline T)^{3/2}$.
The origin of non-perturbative superpotential term arises due to instanton contribution or the presence of $D7$ 
brane wrapping certain cycles of the internal dimensions carrying asymptotically free, $SU(N)$, gauge group.
Then the gaugino condensation along such a gauge group will induce a non-perturbative potential~\cite{Derendinger:1985kk,Dine:1985rz}.
Note that usually the K\"ahler potential also obtains non-perturbative corrections~\cite{Berg:2004ek,Berg:2005ja,Berg:2005yu}, 
however, for a choice of $W_0$ it is possible to neglect them~\footnote{There are two kinds of string corrections, 
string loop corrections in powers of $g_s\sim e^{\phi}$ and $\alpha^{\prime}\sim 1/M_s^2$ type corrections. It is known
that non-renormalization theorem forbids holomorphic superpotential to obtain corrections of 
either type~\cite{Grisaru:1979wc}. However, the K\"ahler potential is not similarly protected.}.
With the help of the above superpotential and the K\"ahler potential, the resulting minimum is found to 
be the SUSY one (for a pedagogical discussion, see~\cite{Burgess:2007pz}):
\begin{eqnarray}
D_TW(T_{m})&=&-aAe^{aT_m}-\frac{3(W_0+Ae^{-aT_m})}{T+\overline T_m}=0\,,\\
V(T_m,\overline T_m)&=&-\frac{3(W_0+Ae^{T_m})^2}{(T_m+\overline T_m)^3}=-\frac{(aAe^{-aT_m})^2}{3(T_m+
\overline T_m)}<0
\end{eqnarray}
where $T_m$ denotes the minimum. Note that the potential is adS in the minimum, and requires uplifting to get a dS
universe. As suggested by KKLT, this can be achieved by adding SUSY breaking contributions, 
such as anti-D3 brane ($\overline{D3}$), although breaks explicitly, its effect can be made small by the choice of brane tension, or in 
other words placing the $\overline{D3}$ brane in a warped geometry~\cite{Kachru:2003aw}, whose contribution turns out to be:
\begin{equation}\label{Uplifting}
V_{\bar D3}=\frac{C}{(T+\overline T)^2} \,,
\end{equation}
where $C>0$, given in terms of the brane tension and the six dimensional volume. The addition of $\overline{D3}$ 
lifts the potential keeping the minimum almost intact, $T\simeq T_m$, for which either the total potential vanishes or become dS. For  $|T|\rightarrow \infty$, potential $V\rightarrow 0$ leads to decompactification to $10d$. Thus, the potential is separated by the barrier which can keep the metastable minimum stable enough for the life time of the universe~\cite{Kachru:2003aw,Giddings:2004vr}.


\subsubsection{N-flation}
\label{NINF}

Amongst various realizations of assisted inflation, N-flation is perhaps the most interesting one~\footnote{
There are also realizations of assisted inflation via branes~\cite{Mazumdar:2001mm,Cline:2005ty} in a type IIB setting, with a large number of  Kaluza Klein modes~\cite{Kanti:1999ie}, exponential potentials arising from a generic string compactification~\cite{Copeland:1999cs}, 
and in a M-theory setting with the help of large number of NS-5 branes in \cite{Becker:2005sg}, etc. }. 
The idea is to have $N\sim 300(M_{\rm P}/f)\sim 10^{4}$ number of axions, where $f$ is the axion decay constant, of order 
$f\sim 0.1 M_{\rm P}^{-1}$. These axions can drive inflation simultaneously with a leading order 
potential~\cite{Dimopoulos:2005ac}:
\begin{equation}
V=V_0+\sum_{i}\Lambda_{i}^4\cos(\phi_{i}/f_{i})+...
\end{equation}
where $\phi_i$ correspond to the partners of K\"ahler moduli. The ellipses contain higher order 
contributions. The advantage of this potential is that there are {\it no} $H^2\phi_{i}^2$ type contributions at 
the lowest order provided there is a shift symmetry, therefore the SUGRA $\eta$ problem can be evaded. 
The axions have shift symmetry, which  are only broken at non-perturbative level. 

In string theory case, the assumption behind the potential is following. In a compactified framework, it is assumed 
that all the moduli are heavy and thus stabilized by prior dynamics, including that of the volume modulus. Only the 
axions of $T_i=\phi_i/f_i+iM_s^2R_i^2$ are light~\cite{Dimopoulos:2005ac}.  The assumption of 
decoupling the dynamics of K\"ahler modulus from the axions is still a debatable issue, see the discussion in~\cite{Kallosh:2007ig}. 
These axions can then obtain an exponential superpotential, $W\sim \sum_i W_i e^{2\pi T_i}$, correction 
similar to the KKLT setup, in addition to a constant superpotential piece, $W_0$. The shift symmetry is now 
protected by the choice of K\"ahler potential~\cite{Kallosh:2007ig}
\begin{equation}
K=-\ln\left[i\frac{C_{ijk}}{3!}(T_i-\overline T_i)(T_j-\overline T_j)(T_k-\overline T_k)\right]\,,
\end{equation}
where $C_{ijk}$ is the Calab-Yau intersection number. After rearranging the potential for the axions, 
$V\approx \sum V_i\approx \sum_{i}|\partial_{i}W|^2$, expanding them around
their minima and for a canonical choice of the kinetic terms, the Lagrangian simplifies to the lowest order in expansion:
\begin{equation}
{\cal L}=\frac{1}{2}\partial_{\mu}\phi_{i}\partial^{\mu}\phi_{j}-\frac{1}{2}m_{i}^2\phi_{i}^2+\cdots \,.
\end{equation}
The exact calculation of $m_i$ is hard, assuming all of the mass terms to be the same $m_{i}\sim 10^{13}$~GeV, and
$N> (M_{\rm P}/f)^2$, it is possible to match the current observations with a tilt in the spectrum, $n\sim 0.97$,
and {\it large} tensor to scalar ratio: $r\sim 8/{ N}_Q\sim 0.13$ for ${N}_Q\sim 60$. 


\subsubsection{Inflation due to K\"ahler modulus}
\label{IDKM}

Various realizations of modular inflation have been studied in the context of volume modulus, out of which 
the racetrack models~\cite{Pillado:2004ns,BlancoPillado:2006he,Lalak:2005hr},  
and the K\"ahler moduli inflation are the most popular ones~\cite{Conlon:2005jm,Bond:2006nc}. 
In the simplest version of racetrack inflation, the K\"ahler and the superpotential are given by:
\begin{equation}
K=-3\ln(T+\overline T)\,,~~~W=W_0+Ae^{-aT}+Be^{-bT}\,,
\end{equation}
where $A, B, a, b$ are calculable constant, and $T$ is a complex K\"ahler modulus. Similar to 
the KKLT scenario, it is also assumed that there is an uplifting of the minimum of the K\"ahler 
potential by $\overline{D3}$ brane, see Eq.~(\ref{Uplifting}). The potential is spanned in two real 
scalar directions with a complicated profile with many dS minima. For a certain choice of 
fine tuned parameters, it is possible to obtain sufficient slow-roll inflation near the saddle point
spanned in two real directions. The model can produce the spectral tilt $n_s\approx 0.95$.

A better racetrack model has been constructed with the help of two K\"ahler moduli~\cite{Denef:2004dm}, the
K\"ahler and  the superpotential are given by~\cite{BlancoPillado:2006he}:
\begin{eqnarray}
K &=&-2\ln\left[\frac{1}{36}\left((T_2+\overline T_2)^{3/2}-(T_1+\overline T_1)^{3/2}\right)\right]\,,\\
W &=&W_0+Ae^{-aT_1}+Be^{-bT_2}\,,
\end{eqnarray}
where there are 4 real scalars involved, i.e. $T_{1,2}=X_{1,2}+iY_{1,2}$. The uplifting of the potential
from adS to dS is given by: $\delta V\sim D/(X_2^{3/2}-X_1^{3/2})^2$.

In all these models the choice of $W_0$ is such that the corrections of the K\"ahler potential can be neglected.
This is due to the fact that $W_0$ is chosen small in order to explain the current cosmological constant.
However $W_0$ need not be small and in which case $\alpha^{\prime}$ corrections to k\"ahler potential
cannot be neglected~\cite{Conlon:2005jm,Bond:2006nc}. One such toy model with three moduli were constructed
where the K\"ahler and superpotential are given by:
\begin{eqnarray}
K &=&-2\ln\left[(T_1+\bar T_1)^{3/2}-k_2(T_2+\bar T_2)^{3/2}-k_3(T_3+\bar T_3)^{3/2}+\frac{\xi}{2}\right]\,,\\
W &=& W_0+\sum _{i=1}^{3}A_ie^{a_i T_i}\,,
\end{eqnarray}
where $A_i, a_i$ are constants. Of course, now there are more parameters and the full potential is hard to 
analyze, however, it is possible to imagine that only one of the modulus is driving inflation, say $T_3$, and rest 
of them are decoupled from the dynamics. The potential along the lightest modulus is then given by:
\begin{equation}
V\approx V_0- C(T_3+\bar T_3)^{4/3}\exp\left[-c(T_3+\bar T_3)^{4/3}\right]\,,
\end{equation}
for some positive constants $c$ and $C$. In all these class of models the spectral tilt comes out 
to be close to the observed limit, $n_s\sim 0.96$, with no significant gravity waves and no cosmic string strings are produced aftermath of inflation~\cite{Conlon:2005jm,Bond:2006nc}. In the above models, if the string loop corrections are included,
for instance the  Kaluza-Klein loop corrections to the potential arising from the wrapping of the $D7$ brane around a cycle, $T_3$, 
then the overall potential can be modified to admit a saddle point and the point of inflection~\cite{Cicoli:2008gp}. When the loop 
corrections dominate the potential, the potential with canonical kinetic term takes the form:
\begin{equation}
V_{inf}\approx V_0 +\frac{W_0^2}{\nu^2}\left(Ae^{-2\kappa \varphi}-\frac{B}{\nu}e^{-\kappa\varphi/2}+
\frac{C}{\nu^2}e^{\kappa\varphi}\right)
\end{equation}
where $A,~B,~C$ are tunable constants and $B$ could be chosen to be negative, $T_1=e^{\kappa \varphi}$, and $\kappa=2/\sqrt{3}$, and $\nu$ determines the large internal volume which determines the string scale, $M_s$. The field range for $\varphi$
is such that it can take large VEVs, i.e. $\varphi\sim {\cal O}(1-10)M_{\rm P}$, where the tensor to scalar ratio can be made appreciable, i.e. $r\sim 0.005$~\cite{Cicoli:2008gp}.


\subsection{Brane inflation}
\label{BI}

The position of various branes along with their motion can lead to inflation. Let us imagine that there is a gas of $Dp$ 
branes in $p+1$ space-time dimensions.  Their stress energy tensor leads to an equation of 
state~\cite{Brandenberger:2003ge}
\begin{equation}
{p}_p=\left[\frac{p+1}{d}v^2-\frac{p}{d}\right]\rho_d\,,
\end{equation}
where $p_p$ is the pressure and $\rho_p$ is the energy density of the gas of $p$ branes. In a relativistic limit,
$v^2\rightarrow 1$, a gas of branes behaves as a relativistic fluid with an equation of state, 
$w \equiv p_p/\rho_p=1/d$, while in the non-relativistic case, $v^2\rightarrow 0$, we obtain 
$w=-p/d$, a negative pressure which could lead to a 
power law inflation with a scale factor $a\propto t^{2/(d-p)}$ for $d=p+1$. This would inflate the 
entire $p+1$ dimensional bulk. A stringy realization would require that the dilaton be fixed at all times. A bound state of 
fractionally charged branes in $10d$ universe can also lead to a high entropy state, with an initial correlation 
length larger than the string scale, as discussed in~\cite{Chowdhury:2006pk}. These scenarios are helpful 
in setting up the initial conditions for the universe. First of all they provide a homogeneous Hubble patch with a 
large causal horizon (bigger than the string scale), where subsequent phases of inflation can take place.

Motion of much fewer branes can also lead to inflation, first realized in \cite{Dvali:1998pa,Burgess:2001fx,Burgess:2001vr,GarciaBellido:2001ky, Mazumdar:2001mm,Dvali:2001fw,Jones:2002cv,Herdeiro:2001zb,Dasgupta:2002ew,Hsu:2003cy,Hsu:2004hi,Firouzjahi:2003zy,Shandera:2006ax,Kachru:2003sx,Bean:2007hc}, for a 
review see~\cite{HenryTye:2006uv}. 
Consider a system of  $Dp-\overline{Dp}$ branes, where they interact via closed string exchanges between the branes,
i.e. the attractive gravitational (NS-NS), and the massive (R-R) interactions, see~\cite{Polchinski:1996na}, yields,
$V(y)\approx-\kappa_{10}^2T_p^2\Gamma((7-p)/2)(1/\pi^{(9-p)/2}y^{7-p})$, where $T_p=(2\pi\alpha^{\prime})^{(p+1)/2}$ 
is the $Dp$ brane tension, and $y$ is the inter-brane separation. For $p< 7$, the potential vanishes for large $y$.
At very short distances close to the string scale, there develops a tachyon in the spectrum, 
$\alpha^{\prime}m^2_{tachyon}=(y^2/4\pi^2\alpha^{\prime})-1/2$, which leads to 
annihilation of the branes, and a graceful end of branes driven inflation.

In a more realistic scenario, the branes have to be placed in a warped geometry. As a consequence of flux 
compactification, any mass scale, M, in the bulk  becomes $h_AM$, where $h_A\ll 1$, near the bottom of the warped 
throat. Thus the warping affects the overall normalization of the potential. It is assumed that a $D3$ brane is slowly
falling into the attractive potential of an $\overline{D3}$ brane placed at the bottom of the throat.
The sum total potential for a $D3-\overline{D3}$ brane potential is given by~\cite{HenryTye:2006uv}:
\begin{equation}
V(\phi)=\frac{1}{2}\beta H^2\phi^2+2T_3h_A^4\left(1-\frac{1}{N_A}\frac{\phi_A^4}{\phi^4}\right)+...\,,
\end{equation}
where $\phi=\sqrt{T_3}y$, the value of warping depends on the throat geometry, typically $h_A\sim 10^{-2}$, $N_A\gg 1$ is
the $D3$ charge on the throat, and $\beta\sim {\cal O}(1)$ arises due to the k\"ahler potential, which obtains contributions
from the brane positions.  The first term is reminiscent to the SUGRA $\eta$ problem, which plagues the brane inflation 
paradigm in general. A successful inflation would require $\beta \ll 1$, the inflationary predictions are very similar to that of 
the hybrid model of inflation.

There are some drawbacks of this scenario, the flatness of the potential is hard to obtain naturally, one
can try to modify the situation with dual formulation where instead of brane separation, one uses branes at 
angles~\cite{GarciaBellido:2001ky}, assisted inflation~\cite{Mazumdar:2001mm,Cline:2005ty}, or $D3$ brane falling towards $D7$ branes~\cite{Herdeiro:2001zb,Dasgupta:2002ew,Hsu:2003cy,Hsu:2004hi}. The issue of initial
condition is crucial for the brane inflation scenario to work, the position of a $D3$ brane has to be away from the bottom of the
throat, but there exists no stringy mechanism to do so. In a recent study \cite{Baumann:2006th,Baumann:2007np,Krause:2007jk}
an argument has been provided where it is possible to realize a slow-roll motion for a $D3$ brane where $D7$ brane is also extended 
in the bottom of the throat. In all these examples inflation happens near the {\it point of inflection}, which was first studied in the context of MSSM inflation~\cite{Allahverdi:2006iq,Allahverdi:2006we,Allahverdi:2008bt}.

Another variant of brane inflation has been discussed in the context of DBI (Dirac Born-Infeld) action~\cite{Silverstein:2003hf,Alishahiha:2004eh}, where a $D3$ brane rolls fast with almost a relativistic velocity, $v=16/27$, for a particular case of 
KS-throat~\cite{Klebanov:2000hb}. Inside a warped throat, the $4d$ metric is given by:
\begin{equation}
ds^2=h^2(y)(-dt^2+a(t)^2dx^2)+h^{-2}(y)g_{mn}(y)dy^mdy^n\,,
\end{equation}
The DBI action for a $D3$ brane is given by the brane position, $\phi(t)$, from the bottom of the throat:
\begin{equation}
S=-\int d^4x a^3(t)\left[T(\phi)\sqrt{1-\frac{\dot\phi^2}{T(\phi)}}+V(\phi)-T(\phi) \right]\,,
\end{equation}
where $T(\phi)=T_3h^4(\phi)$ and $h(\phi)$ is the warp factor depending on the brane position. 
For a slow-roll inflation, the action yields the standard kinetic term $\approx \dot\phi^2/2$. The potential
is given by a phenomenological mass term, and the coulomb potential between $D3$ and $\overline{D3}$ brane:
\begin{equation}
V(\phi)\approx \frac{m^2}{2}\phi^2+V_0\left(1-\frac{V_0}{4\pi^2v}\frac{1}{\phi^4}\right)\,,
\end{equation}
where $V_0=2T_3h^4_A$ is the brane tension at the bottom of the throat, where $h_A\sim 0.2-10^{-3}$.
The DBI inflation is quite similar to the K-inflation picture~\cite{ArmendarizPicon:1999rj}, where inflation is driven by a non-canonical 
kinetic term, a simple analogy can also be made with a fluid dynamical picture, where an equation of state
can be determined via:
\begin{equation}
p(\phi,{\cal K})=-T_3h^4(\phi)\sqrt{1-2{\cal K}/h(\phi)^4}+T_3h^4(\phi)-V(\phi)\,, \qquad 
\rho(\phi, {\cal K})=2{\cal K}p_{,{\cal K}}-p\,,
\end{equation}
where ${\cal K}=(\dot\phi^2/2T_3)$. The speed of sound is given by:
\begin{equation}
c_s^2=\frac{p_{,{\cal K}}}{\rho_{,{\cal K}}}=\frac{p_{,\cal K}}{p_{,\cal K}+2{\cal K}p_{,{\cal K K}}}
=1-2{\cal K}/h^4=\frac{1}{\Gamma^2}\,,
\end{equation}
where $\Gamma$ is a relativistic factor. Besides matching the amplitude of the perturbations and 
the scalar tilt, $n_s\sim 0.98$, there is a possibility of generating large non-Gaussianity towards the end of 
inflation. The value of $f_{NL}$ is determined by the relativistic motion of the brane, when $\Gamma \gg 1$,
$|f_{NL}|\approx 0.32 \Gamma^2$. For $|f_{NL}|<300$, gives $\Gamma \leq 32$~\cite{Alishahiha:2004eh,Creminelli:2005hu}. 
The tensor to scalar ratio depends on the choice of initial conditions and it drops significantly as 
$\Gamma\gg 1$~\cite{Lidsey:2006ia}.


\subsection{Reheating and thermalization}
\label{RAT}

In stringy models of inflation, reheating and thermalization of the SM or MSSM degrees of freedom are poorly
understood. This is mainly due to the fact that inflation happens in a hidden sector as far the MSSM sector 
is concerned. One problem for all these models is that they are bound to excite {\it possibly light non-SM} like degrees of 
freedom, which can pose several challenges for a successful BBN~\footnote{One possibility would be to dilute all of them via late inflation, and create the MSSM degrees of freedom afresh within an observable sector~\cite{Allahverdi:2007wh}.}.

There are also some observational virtues of stringy reheating. For instance, inflation driven by
$D3-\overline{D3}$ case has interesting consequences. Their annihilation leads to the production of 
$D1$ branes and fundamental $F1$ strings, which can be understood via tachyon 
condensation~\cite{Sen:1998ii,Sen:1998tt,Sen:2002nu,Witten:1998cd}.  The tachyon 
couples to $U(1)\times U(1)$ gauge theory 
associated with each brane, as it develops a VEV it breaks the gauge group spontaneously, which 
results in formation of $D1$ strings via Kibble mechanism and a confining flux tubes which become
the fundamental closed strings~\cite{Bergman:2000xf}. In Type IIB setup domain walls and monopoles are 
not excited, which correspond to $D0$ and $D2$ branes. The cosmic string tension, $\mu$, can be within
$10^{-13}< G\mu <10^{-6}$~\cite{Jones:2003da,Shandera:2006ax}. Brane-anti-brane annihilation also
excited gravity waves, whose peak frequency is determined by the string scale~\cite{Mazumdar:2008up}.
A string scale greater than TeV makes it impossible for such gravity waves to be detected in future.

Reheating in a warped geometry is a complicated process~\cite{Kofman:2005yz,Chialva:2005zy,Frey:2005jk}. 
Especially, in a multiple throat picture, where inflation happens in one throat and the SM is in another, there are 
many possibilities to reheat not only the SM throat, but also the other throats. 

\begin{itemize}

\item{Reheating various non-(MS)SM throats:\\
First of all, by no means it is guaranteed that the (MS)SM throat will be the only recipient of the inflaton 
energy density, there are other throats with similar cosmological constant, which are also 
reheated simultaneously  depending on the relative
warping between the SM throat and the other throats~\footnote{When the warping effects are small, as in the 
case of~\cite{Barnaby:2009wr}, reheating and preheating into the SM degrees of freedom have been studied systematically. 
The inflaton being the k\"ahler modulus couples to the SM sector and the moduli sector. The moduli eventually decay into the
SM sector. However the inflaton being a SM gauge singlet here, also couples to various hidden sectors and their couplings to the inflaton are largely unknown. It is therefore important to study the effects of reheating and preheating while taking into account of the hidden sectors along with the SM sector.}.
 }

\item{Stable KK modes and gravitons:\\
$D3$-$\overline{D3}$ annihilation also excites massive KK modes and gravitons. The heavy KK modes can decay into
the lightest KK mode and gravitons. The lightest KK mode can be absolutely stable at the bottom of the throat due to
conserved angular momentum~\cite{Kofman:2005yz}. The self annihilation of these KK modes is gravitationally suppressed, $\sigma \sim (L/R)^6(1/M_{\rm P}h^2)$, where $L$ is linear size of the $6d$ compactification volume and $R$ is the size of the
throat, $h$ is the relative warping~\cite{Kofman:2005yz,Chen:2006ni}, therefore once produced copiously during brane annihilation, these KK modes can overclose the universe. }

\item{Breakdown of an effective description of a (MS)SM throat:\\
Due to the hierarchy between inflation and (MS)SM  throat,  during inflation the (MS)SM throat will also be 
sensitive to inflation-induced vacuum fluctuations. The curvature scalar of the (MS)SM throat will obtain 
corrections of order, $R\sim 12 H^2_{inf}e^{-2A_{SM}}$, with $e^{-2A_{SM}}\sim (M_{\rm P}/M_{SM})^2\gg 1$ during 
inflation~\cite{Frey:2005jk}, which will render perturbative description of the (MS)SM  throat ineffective.  The (MS)SM 
throat will now be uplifted by warping given by $e^{A_{SM}}\sim H_{inf}/M_{\rm P}$~\cite{Frey:2005jk}.
Once inflation ends, the geometry of (MS)SM throat will start relaxing gradually before settling down to its original value.
The process of relaxation could give rise to a violent particle production and excitations of open and closed 
strings~\cite{Frey:2005jk}.}

\item{Reheating via tunneling:\\
The massive KK modes can decay into the (MS)SM throat via quantum 
tunneling~\cite{Dimopoulos:2001ui,Dimopoulos:2001qd}. There is a large uncertainty in the
tunneling rate, for a range of parameters given by the RR flux, $n_R\sim 10-100$, $6d$ 
compactified volume, $e^{4u}/g_s\sim 1-10^{3}$ and $g_s\sim 1/10$, the rate is roughly given by:
$\Gamma_{t}\sim {\cal O}(10-10^{10})(H_{inf}/M_{\rm P})^{3/2}H_{inf}$~\cite{Frey:2005jk}.
With the above tunneling rate the reheat temperature of the (MS)SM throat, 
$T_{\rm R}\sim \sqrt{\Gamma_t M_{\rm P}}$, will exceed the fundamental scale $M_{SM}$.
This would lead to excitations of the KK modes and gravitons in the (MS)SM throat. Again the
universe would be dominated a gas of non-relativistic KK particles.}

\end{itemize}

In all the cases, the reheating temperature is very close to the string scale, $T_{\rm R}\sim M_{SM}$, in the (MS)SM throat. 
This gives rise to thermal excitations of long open, and closed strings,  a phase reminiscence to the Hagedorn 
phase~\cite{Hagedorn:1965st}. Note, a very similar picture would unfold in any neighboring throats, 
(MS)SM throat is not a unique one.


\subsection{String theory landscape and a graceful exit}
\label{STLAGE}

The vacuum energy in a string theory landscape can be written as:
\begin{equation}
\label{bp}
V=M_{\rm P}^2\Lambda = M_{\rm P}^2\Lambda_0 +\alpha^{-2}\sum_i c_i n_i^2\ , 
\end{equation}
where $c_i\lesssim 1$ are constants and $n_i$ are flux quantum
numbers (note that $\Lambda$ has dimension ${\rm mass}^2$ 
in our notation as it enters the Einstein equation as $\Lambda g_{\mu\nu}$).  

All in all, string theory (from the
landscape point of view) could have from $10^{500}$ to even
$10^{1000}$ vacua \cite{Bousso:2000xa,Douglas:2004zg,Susskind:2003kw,Douglas:2006es},
with the vast majority of those having large cosmological constants.
In addition, our knowledge of the distribution of gauge groups over
the landscape suggests that one out of $10^{10}$ vacua have the SM
spectrum, at least in simple models
\cite{Blumenhagen:2004xx,Gmeiner:2005vz,Douglas:2006xy}.
Even if this fraction is much smaller on the whole landscape, there
are so many vacua that it seems likely very many will have a SM-like
spectrum.  From statistical arguments, most vacua should have 
badly broken SUSY, with large $F$-terms in the SUSY 
breaking sector.

 In addition, there exist vacua with large cosmological constant that are
``almost SUSY'' in the sense of \cite{Douglas:2004zg,Kallosh:2004yh}.
These vacua have vanishing (or nearly vanishing) $F$-terms, and their
cosmological constant and SUSY breaking are provided by a
$D$-term, such as that created by an anti-brane.  Upon decay of the
$D$-term, the remaining vacuum has small $\Lambda$ and low energy
SUSY.  The original model of \cite{Kachru:2003aw} is almost
SUSY in this sense because anti-D3-branes provide both the
cosmological constant and SUSY breaking.  There are also examples of 
toy ``friendly landscape'' models \cite{ArkaniHamed:2005yv}, in which the dynamics of $N$ scalars
create a landscape of vacua.  The landscape can also harbor 
negative $\Lambda$~\cite{Banks:2002nm}.  Also, \cite{Clifton:2007en}
has discussed the importance of negative $\Lambda$ vacua in possibly
separating parts of the landscape from each other.

It is expected that small jumps in $\Lambda$ with small bubble tension
$\tau$ to be the most common decays, which can in fact be quite rapid~\cite{Allahverdi:2007wh}.
In Ref.~\cite{HenryTye:2006tg} the authors have  argued that resonance can also play
an important role in tunneling across a landscape of many metastable
vacua. Also, the fact that there are many
possible decays, as emphasized by \cite{Freivogel:2004rd}, the whole
landscape, including the MSSM-like vacua, will be populated eventually
almost independently of initial conditions. 

One interesting way to exit the string landscape is through MSSM inflation.
Note that, when MSSM inflation starts, the ``bare'' cosmological
constant (that not associated with the MSSM inflaton) might still be
considerably larger than the present value.  This means that further
instanton decays should take place to reduce the bare cosmological
constant, and these decays should occur during MSSM inflation in order
to percolate efficiently.  Fortunately, MSSM inflation naturally
includes a self-reproduction (eternal inflation) regime prior to
slow-roll \cite{Allahverdi:2006iq,Allahverdi:2006cx,Allahverdi:2006we}, see Sec.~\ref{See: Inflating the bubble}.


\subsection{Other stringy paradigms}
\label{OSP}

There are other interesting paradigms, such as string gas cosmology and bouncing cosmology, 
for a review see~\cite{Brandenberger:2008nx}. The basic setup relies on stringy ingredients, such as 
statistical properties of a gas of strings and branes, and some aspects of string field theory. Their primary aim is to explain the seed perturbations for CMB without invoking cosmic inflation, see also~\cite{Gasperini:2002bn,Gasperini:1992em,Lidsey:1999mc}. However, all these scenarios suffer from the same symptoms, they require the universe to be exponentially large from the very beginning~\cite{Kaloper:1998eg}.


\subsubsection{String gas cosmology}
\label{STC}

In \cite{Brandenberger:1988aj}, Brandenberger and Vafa (BV) proposed a 
seemingly very natural initial condition for cosmology in string theory.
In BV cosmology, all nine spatial dimensions are compact (and toroidal
in the simplest case) and initially at the string radius.  The matter
content of the universe is provided by a Hagedorn temperature gas of
strings.  In addition to proposing a very interesting initial condition
and analyzing the thermodynamics of string at that point, however, BV 
argued that string theory in such a background provides a natural mechanism
for decompactifying up to three spatial dimensions (that is, allowing three
spatial dimensions to become macroscopic).  The BV mechanism works because
winding strings provide a negative pressure, which causes contraction
of the scale factor, as was shown explicitly 
in \cite{Tseytlin:1991xk,Tseytlin:1991ss}.  BV then gives a classical 
argument that long winding strings can only cross each other 
in three or fewer large spatial dimensions.  Therefore, since winding
strings freeze out quickly in four or more large spatial dimensions, 
the winding strings would cause re-collapse of those large dimensions.

The initial paper, \cite{Brandenberger:1988aj}, has inspired a broad literature.
One important generalization has been including branes in the 
gas of strings \cite{Maggiore:1998cz,Alexander:2000xv,Boehm:2002bm,
Alexander:2002gj}, and other space time topologies have also been
considered \cite{Easson:2001fy,Easther:2002mi}.  In particular, 
\cite{Brandenberger:2001kj} showed that interesting cosmological
dynamics happen when the expanding dimensions are still near the string
scale.

Importantly, several tests have been made of the BV mechanism for determining
the number of macroscopic dimensionality of space, see \cite{Sakellariadou:1995vk,Cleaver:1994bw,Easther:2003dd,Easther:2004sd,Danos:2004jz}. These tests are all
based on the fact that the winding strings will be unable to annihilate 
efficiently if their interaction rate, $\Gamma$, drops below the Hubble
parameter for the expanding dimensions. Based on simple arguments
from the low energy equations of motion and string thermodynamics it was 
demonstrated that the interaction rates of strings are negligible, so the
common assumption of thermal equilibrium cannot be applicable~\cite{Danos:2004jz,Skliros:2007cm}.


\subsubsection{Seed perturbations from a string gas}
\label{SPFSG}

Recently, a new structure formation scenario has been put forward in 
\cite{Nayeri:2005ck,Brandenberger:2006vv} . It was
 shown that string thermodynamic fluctuations in a quasi-static
primordial Hagedorn phase in $4d$, during which the temperature hovers
near its limiting value, namely the Hagedorn temperature, $T_H$ \cite{Hagedorn:1965st},
can lead to a scale-invariant spectrum of metric fluctuations. The crucial point 
of the mechanism is to note that provided three large spatial dimensions are compact,
the heat capacity $C_V$ of a gas of strings in thermodynamical equilibrium scales are $r^2$
with the radius of the box~\cite{Deo:1989bv,Deo:1991mp,Mitchell:1987th,Bowick:1985az,Bowick:1989us,Tye:1985jv}, 
then the heat capacity determines the root mean square mass fluctuations,
i.e. $\langle (\delta M)^2\rangle=T^2C_V$. With the help of Poisson equation, $\nabla ^2\Phi=4\pi G\delta \rho$,
and the definition of power spectrum, ${\cal P}_{\Phi}(k)\equiv k^3|\Phi(k)|^2$, one 
obtains~\cite{Nayeri:2005ck,Brandenberger:2006vv}:
\begin{eqnarray}
{\cal P}_{\Phi}(k) &=& 16\pi^2G^2k^{-1}|\delta \rho(k)|^2=16\pi^2G^2k^2(\delta M)^2(r(k)) \,, \nonumber \\
& \approx & 1920\pi^2c^{-1}G^2T_H^4(kr)^2\frac{T}{T_H(1-T/T_H)}\,,
\end{eqnarray}
where  $c$ is the velocity of light, $G$ is the Newton's constant and $T_H$ is the Hagedorn temperature.
The mean squared mass fluctuation $|\delta M|^2(r)$ in a sphere of radius $r(k)=k^{-1}$ is given by 
$|\delta \rho(k)|^2=k^3|\delta M|^2(r(k))$. The tilt in the spectrum is scale invariant and the fine tuning
in temperature has to be $\Delta T/T_H\sim 10^{-30}$ for $M_S\sim 10^{-10}M_{\rm P}$. The tensor
mode also requires similar level of fine tuning but with slight tendency towards a blue tilt, which could 
be a distinguishing feature of this setup~\cite{Nayeri:2005ck,Brandenberger:2006vv}.

To obtain this result, several criteria
for the background cosmology need to be satisfied.
First of all, the background equations must indeed admit a quasi-static
(loitering) solution.
Next, our three large spatial dimensions must be compact. It is
under this condition that \cite{Deo:1989bv,Deo:1991mp} the heat capacity $C_V$ as a 
function of radius $r$ scales as $r^2$. Thirdly, thermal equilibrium must be 
present over a scale larger than $1$mm 
during the stage of the early universe
when the fluctuations are generated. Since the scale of thermal
equilibrium is bounded from above by the Hubble radius, it follows that
in order to have thermal equilibrium on the required scale, 
the background cosmology should have a quasi-static phase. Finally,
the dilaton velocity needs to be negligible during the time interval
when fluctuations are generated~\cite{Kaloper:2006xw,Biswas:2006bs}~\footnote{It is 
not easy to satisfy all of the conditions required for the
mechanism proposed in \cite{Nayeri:2005ck,Brandenberger:2006vv} to work. In the context of a dilaton gravity
background, the dynamics of the dilaton is important. If the 
dilaton has not obtained a large mass and a fixed
VEV at a high scale, then it will be rolling
towards weak coupling at early times. This will lead to \cite{Tseytlin:1991xk,Veneziano:1991ek} a
phase in which the string frame metric is static, and thus the
string frame Hubble radius will tend to infinity, i.e. $H\approx 0$.}.


\subsubsection{Example of a non-singular bouncing cosmology}
\label{BONC}

In string theory, higher-derivative corrections to the Einstein-Hilbert
action appear already classically (i.e at the tree level), but we do
not preclude theories where such corrections (or strings themselves)
appear at the loop level or even non perturbatively.  From
string field theory~\cite{Mandelstam:1973jk,Kaku:1974zz,Green:1984fu,Witten:1985cc,Gross:1986ia,Gross:1986fk,Zwiebach:1985uq} (either light-cone or covariant) the
form of the higher-derivative modifications can be seen to be Gaussian,
i.e. there are $e^\Box$ factors appearing in all vertices (i.e. 
$(e^\Box\phi)^3$). These modifications can be moved to kinetic terms by field
redefinitions ($\phi\to e^{-\Box}\phi$). The non perturbative gravity
actions that we consider here will be inspired by such stringy kinetic terms~\cite{Biswas:2005qr}
\footnote{There are various discussions on singular bouncing cosmology in $4d$ in the context of 
`ek-pyrotic' and cyclic universe~\cite{Khoury:2001wf,Khoury:2001bz,Steinhardt:2001st,Steinhardt:2002ih}.
These models are interesting in their own right. However, constructing and stabilizing a background with a 
singular bouncing cosmology is a non-trivial task, see for some related 
discussions~\cite{Horowitz:2002mw,Craps:2002ii,Elitzur:2002rt,Horowitz:1991ap,Berkooz:2002je,Cornalba:2003kd}.}.

It was realized that if we wish to have both a ghost free and an asymptotically free theory 
of gravity~\footnote{While perturbative unitarity requires the theory to be ghost free, in order to 
be able to address the  singularity problem in General Relativity, it may be desirable to make 
gravity weak at short distances, perhaps even asymptotically free \cite{Weinberg:1976xy}. }, one has little choice
but to look into gravity actions that are non-polynomial in
derivatives, such as the ones suggested by string theory~\cite{Biswas:2005qr}:
\begin{equation} 
S=\int d^4x\ \sqrt{-g}F(R)\,, 
\label{action}
\end{equation} 
where
\begin{equation}
F(R)=R+\sum_{n=0}^{\infty}{c_{n}\over M_{\ast}^{2n}}R\Box^{n}R\,,
\label{non-pert}
\end{equation}
and $M_{\ast}$ is the scale at which non-perturbative physics
becomes important. $c_n$'s are typically assumed to be $\sim {\cal O}(1)$
coefficients.  It is convenient to define a function,
\begin{equation}
\Gamma(\lambda^2) \equiv \left(1-6\sum_0^{\infty}c_i\left[{\lambda\over M_{\ast}}
\right]^{2(i+1)}\right)\,.
\label{final-eqn}
\end{equation}
One can roughly think of $p^2\Gamma(-p^2)$ as  the modified inverse 
propagator for gravity (see \cite{Biswas:2005qr} for details). It was shown 
that if $\Gamma(\lambda^2)$ does not have any zeroes, then the action is ghost-free,
thus free from any classical instabilities, i.e. Ostrogradski instabilities (see \cite{Woodard:2006nt} for a review).

For homogeneous and isotropic cosmologies, where $a(t)$ is the scale factor, it is sufficient to look at the
analogue of the Hubble equation for the modified action (\ref{action},\ref{non-pert}).
Just as in ordinary Einstein gravity, here also the Bianchi identities (conservation equation) ensure 
that for FRW metrics, the field equation satisfies \cite{Schmidt:1990dh,Biswas:2005qr}
\begin{equation}
\tilde{G}_{00} \, = \, F_0R_{00}+{F\over 2}-F_{0;\ 00}-\Box F_0-
2\sum_{n=1}^{\infty}F_n\Box^n R-{3\over 2}
\sum_{n=1}^{\infty}\dot{F}_n\dot{(\Box^{n-1}R)} \, = \, T_{00}\,,
\label{tildeG00}
\end{equation}
where we have defined
\begin{equation}
F_m \, \equiv \, \sum_{n=m}^{\infty}\Box^{n-m}{ F\over \Box^n R}\,.
\end{equation}
It was shown in \cite{Biswas:2005qr} that Eq.~(\ref{tildeG00}) admits exact bouncing
solutions of the form
\begin{equation}
a(t) \, = \, \cosh\left[{\Lambda t\over \sqrt{2}}\right] \,.
\label{cosh}
\end{equation} 
in the presence of radiative matter sources and a non-zero cosmological constant.

A non-singular bouncing cosmology can also lead to a {\it cyclic inflation}~\cite{Biswas:2008kj,Biswas:2009fv}, where 
a {\it negative cosmological constant} plays a dominant role. The evolution gives rise to every cycle 
undergoing inflation on average~\cite{Biswas:2009fv}:
\begin{equation}
\langle H\rangle \equiv \frac{\int H dt}{\int dt}=\frac{1}{\tau_n}\ln\left(\frac{a_{n+1}}{a_n}\right)
\equiv \frac{{\cal N}_n}{\tau_n}\,,
\end{equation}
where $\tau_n$ is the time period of the $nth$ cycle. Imagining $\tau_n\approx \tau$, every cycle leads to moderate inflation
with the scale factor  increasing with every cycle; $a_{n+1}/a_{n}\approx \exp{\cal N}$. On average the Hubble expansion rate remains constant over many cycles. The exit of inflation happens when the universe leaves the negative cosmological constant to the positive cosmological constant via a dynamical scalar potential~\cite{Biswas:2009fv}.

\begin{acknowledgments}
It is a pleasure to thank Rouzbeh Allahverdi, Borut Bajc, Anders Basboll, Tirthabir Biswas, Robert Brandenberger, Cliff Burgess, Michele Cicoli, Hael Collins, Pratika Dayal, Jean-Francois Dufaux, Kostas Dimopoulos, Bhaskar Dutta,  Kari Enqvist, Andrew Frey, Juan Garcia-Bellido, Kazunori Kohri, Alex Kusenko, David Lyth, Tuomas Multamaki, Jerome Martin, Narendra Sahu, Philip Stephens, and Horace Stoica for fruitful discussions and comments.  A.M. is also thankful to SISSA, Niels Bohr Academy, and ULB Brussels for their kind hospitality, where various parts of the review have been written. A.M. is partly supported by the Marie Curie Research and Training Network ÒUniverseNetÓ (MRTN-CT-2006-035863). J.R. is funded in part by IISN and Belgian Science Policy IAP VI/11.
\end{acknowledgments}

\bibliography{infl-rept}


\end{document}